\PassOptionsToPackage{table}{xcolor}
\documentclass[acmsmall,screen, nonacm]{acmart}
\setcopyright{none}
\renewcommand\footnotetextcopyrightpermission[1]{}
\AtBeginDocument{%
  }

\usepackage{naive-ebnf}
\usepackage{amsmath}

\usepackage{amssymb}
\usepackage{bbm}

\usepackage{tikz}
\usetikzlibrary{external}
\tikzexternaldisable

\usepackage{graphicx}
\usepackage{tabularx}
\usepackage{microtype}
\usepackage{textcomp}
\usepackage{url}
\usepackage{wrapfig}
\usepackage{xspace}

\usepackage{hyperref}
\usepackage{mathtools}
\usepackage{relsize}
\usepackage{acro}
\usepackage{circledsteps}
\usepackage{enumitem}
\usepackage[most]{tcolorbox}

\newcommand{\acro}[1]{{\textsmaller[0.5]{#1}}}
\ifdefined\shortcite\relax\else\newcommand{\shortcite}[1]{\cite{#1}}\fi
\ifdefined\nd\relax\else\newcommand{\nd}[1]{#1{\acro{D}}}\fi
\newcommand{\graphblas}{Graph\acro{BLAS}\xspace}
\newcommand{\taco}{\acro{TACO}\xspace}

\DeclareMathOperator*{\argmin}{argmin}
\newcommand{\edge}{\acro{EDGE}\xspace}

\newcommand{\EDGEcaps}{EDGE\xspace}
\newcommand{\bfs}{\acro{BFS}\xspace}

\newcommand{\CPU}{\acro{CPU}\xspace}
\newcommand{\CSR}{\acro{CSR}\xspace}
\newcommand{\CSC}{\acro{CSC}\xspace}
\newcommand{\COO}{\acro{COO}\xspace}
\newcommand{\CSF}{\acro{CSF}\xspace}
\newcommand{\FPGA}{\acro{FPGA}\xspace}
\newcommand{\GPU}{\acro{GPU}\xspace}
\newcommand{\ODE}{\acro{ODE}\xspace}
\newcommand{\SSSP}{\acro{SSSP}\xspace}
\newcommand{\True}{\textsf{True}\xspace}
\newcommand{\False}{\textsf{False}\xspace}
\newcommand{\modifier}{\texttt{\edge{}-modifier\xspace}}
\newcommand{\designgoal}[1]{\textbf{design goal:~\ref{#1}}}

  {\color{red}}%
  {}

\usepackage{booktabs}

\usepackage{subcaption}

\definecolor{commentgreen}{RGB}{2,112,10}
\definecolor{eminence}{RGB}{108,48,130}
\definecolor{weborange}{RGB}{255,165,0}
\definecolor{frenchplum}{RGB}{129,20,83}
\definecolor{pastelpink}{RGB}{250,190,233}
\definecolor{pastelyellow}{RGB}{250,246,186}
\definecolor{pastelpurple}{RGB}{221,199,255}
\definecolor{pastelblue}{RGB}{174,198,207}
\definecolor{pastelgreen}{RGB}{119,221,119}
\definecolor{pastelteal}{RGB}{119,221,204}

\usepackage{alphalph}
\usepackage{etoolbox}

\newfloat{cascade}{hbtp}{lop}
\floatname{cascade}{Cascade}
\usepackage{mdframed}
\mdfsetup{
  linewidth=0.5pt,
  innerleftmargin=6pt,
  innerrightmargin=6pt,
  innertopmargin=2pt,
  innerbottommargin=2pt,
  skipabove=2pt,
  skipbelow=2pt
}

\newcommand{\cascadecomment}[1]{& \text{\normalfont\small\itshape\hfill #1} \notag \\}
\newcommand{\eqcomment}[1]{%
    \text{\scriptsize%
        \begin{tabular}[t]{@{}r@{}}[#1]\end{tabular}%
    }\notag\\%
}
\newenvironment{compactalign}
  {\begingroup
   \setlength{\abovedisplayskip}{0pt}%
   \setlength{\belowdisplayskip}{0pt}%
   \setlength{\abovedisplayshortskip}{0pt}%
   \setlength{\belowdisplayshortskip}{0pt}%
   \setlength{\jot}{1pt}%
   \align}
  {\endalign\endgroup}

\AtBeginDocument{%
  \AtBeginEnvironment{subequations}{%
  }
}

\tcbset{textmarkerold/.style={%
        enhanced,
        parbox=false,boxrule=0mm,boxsep=0mm,arc=0mm,
        outer arc=0mm,left=6mm,right=3mm,top=7pt,bottom=7pt,
        toptitle=1mm,bottomtitle=1mm,oversize}}

\tcbset{textmarker/.style={%
        enhanced,
        parbox=false,boxrule=0mm,boxsep=0mm,arc=0mm,
        outer arc=0mm,left=4mm,right=3mm,top=7pt,bottom=7pt,
        toptitle=1mm,bottomtitle=1mm,oversize}}

\newtcolorbox{hint_box}{textmarker,
     borderline west={6pt}{0pt}{pastelpurple},
     colback=pastelpurple!10!white}

\newtcolorbox{status_box}{textmarker,
     borderline west={6pt}{0pt}{pastelyellow},
     colback=pastelyellow!10!white}

\newtcolorbox{question_box}{textmarker,
     borderline west={6pt}{0pt}{pastelpink},
     colback=pastelpink!10!white}

\newtcolorbox{rq_box}{textmarker,
     borderline west={6pt}{0pt}{pastelblue},
     colback=pastelblue!10!white}

\newtcolorbox[auto counter]{sidebar_box}[2][]{textmarker,
    breakable,
    borderline west={6pt}{0pt}{pastelteal},
    colback=pastelteal!10!white,
    halign=justify,
    title=\textcolor{black}{\textbf{Inset~\thetcbcounter}~#2},
    title code={
      \path[fill=pastelteal!10!white] (title.south west) rectangle (title.north east);
      \path[draw=pastelteal,solid,line width=0.75mm]
      ([xshift=0mm]title.south west) -- ([xshift=0mm]title.south east);
      },
    nameref={#2},
    #1
}

\newtcolorbox{sidebar_box*}[2][]{textmarker,
    breakable,
    borderline west={6pt}{0pt}{pastelteal},
    colback=pastelteal!10!white,
    halign=justify,
    title=\textcolor{black}{\textbf{Inset}~#2},
    title code={
      \path[fill=pastelteal!10!white] (title.south west) rectangle (title.north east);
      \path[draw=pastelteal,solid,line width=0.75mm]
      ([xshift=0mm]title.south west) -- ([xshift=0mm]title.south east);
      },
    nameref={#2},
    #1
}

\newtcolorbox[auto counter]{bfs_box}[2][]{textmarker,
    enhanced,
    breakable,
    borderline west={6pt}{0pt}{pastelpurple},
    colback=pastelpurple!10!white,
    halign=justify,
    title=\textcolor{black}{\textbf{BFS Example Part~\thetcbcounter}:~#2},
    title code={
      \path[fill=pastelpurple!10!white] (title.south west) rectangle (title.north east);
      \path[draw=pastelpurple,solid,line width=0.75mm]
      ([xshift=0mm]title.south west) -- ([xshift=0mm]title.south east);
      },
    nameref={#2},
    #1
}

\newtcolorbox{bfs_box*}[2][]{textmarker,
    enhanced,
    breakable,
    borderline west={6pt}{0pt}{pastelpurple},
    colback=pastelpurple!10!white,
    halign=justify,
    title=\textcolor{black}{\textbf{BFS Example Part}:~#2},
    title code={
      \path[fill=pastelpurple!10!white] (title.south west) rectangle (title.north east);
      \path[draw=pastelpurple,solid,line width=0.75mm]
      ([xshift=0mm]title.south west) -- ([xshift=0mm]title.south east);
      },
    nameref={#2},
    #1
}

\newtcolorbox[auto counter]{example_box}[2][]{textmarker,
breakable,
borderline west={6pt}{0pt}{pastelpink},
colback=pastelpink!10!white,
halign=justify,
title=\textcolor{black}{\textbf{Example~\thetcbcounter}:~#2},
title code={
  \path[fill=pastelpink!10!white] (title.south west) rectangle (title.north east);
  \path[draw=pastelpink,solid,line width=0.75mm]
  ([xshift=0mm]title.south west) -- ([xshift=0mm]title.south east);
  },
nameref={#2},
#1
}

\newtcolorbox{example_box*}[2][]{textmarker,
breakable,
borderline west={6pt}{0pt}{pastelpink},
colback=pastelpink!10!white,
halign=justify,
title=\textcolor{black}{\textbf{Example}:~#2},
title code={
  \path[fill=pastelpink!10!white] (title.south west) rectangle (title.north east);
  \path[draw=pastelpink,solid,line width=0.75mm]
  ([xshift=0mm]title.south west) -- ([xshift=0mm]title.south east);
  },
nameref={#2},
#1
}

\newtcolorbox[auto counter]{alternative_box}[2][]{textmarker,
breakable,
borderline west={6pt}{0pt}{pastelblue},
colback=pastelblue!10!white,
halign=justify,
title=\textcolor{black}{\textbf{Alternative Notes~\thetcbcounter}:~#2},
title code={
  \path[fill=pastelblue!10!white] (title.south west) rectangle (title.north east);
  \path[draw=pastelblue,solid,line width=0.75mm]
  ([xshift=0mm]title.south west) -- ([xshift=0mm]title.south east);
  },
nameref={#2},
#1
}

\newtcolorbox{alternative_box*}[2][]{textmarker,
breakable,
borderline west={6pt}{0pt}{pastelblue},
colback=pastelblue!10!white,
halign=justify,
title=\textcolor{black}{\textbf{Alternative Notes}:~#2},
title code={
  \path[fill=pastelblue!10!white] (title.south west) rectangle (title.north east);
  \path[draw=pastelblue,solid,line width=0.75mm]
  ([xshift=0mm]title.south west) -- ([xshift=0mm]title.south east);
  },
nameref={#2},
#1
}
\tcbset{
  edgecodestyle/.style={
    enhanced,
    breakable=true,
    colback=white,
    colframe=black,
    boxrule=0.6pt,
    arc=0mm,
    outer arc=0mm,
    left=4mm,
    right=3mm,
    top=6pt,
    bottom=6pt,
  }
}

\tcbuselibrary{skins,breakable,listings}

\NewTCBListing{edgecodebox}{ O{python} }{
  edgecodestyle,
  listing only,
  listing options={
    language=#1,
    basicstyle=\footnotesize\ttfamily,
    numbers=left,
    numberstyle=\tiny,
    numbersep=4pt,
    breaklines=true,
    tabsize=2,
    showstringspaces=false,
    columns=fullflexible,
    keepspaces=true,
  }
}

\usepackage{color}

\DeclareMathOperator{\map}{\bigwedge}
\DeclareMathOperator{\red}{\bigvee}

\newcommand{\maptxt}{\textsc{Map}\xspace}
\newcommand{\reduce}{\textsc{Reduce}\xspace}
\newcommand{\populate}{\textsc{Populate}\xspace}

\newcommand{\maptmp}{\textsc{Map Temporary}\xspace}
\newcommand{\redtmp}{\textsc{Reduce Temporary}\xspace}

\newcommand{\subsubsubsection}[1]{\paragraph{#1}}

\newcommand{\maptmps}{\textsc{Map Temporaries}\xspace}
\newcommand{\redtmps}{\textsc{Reduce Temporaries}\xspace}

\newcommand{\merge}{$\mathsf{Merge}$\xspace}
\newcommand{\coordinate}{$\mathsf{Coordinate}$\xspace}
\newcommand{\compute}{$\mathsf{Compute}$\xspace}

\newcommand{\Exists}{\mathsf{Exists}\xspace}
\newcommand{\iterspec}{\texttt{iteration-spec}\xspace}

\makeatletter
\newcommand*{\inlineequation}[2][]{%
  \begingroup
    \refstepcounter{equation}%
    \ifx\\#1\\%
    \else
      \label{#1}%
    \fi
    \relpenalty=10000 %
    \binoppenalty=10000 %
    \ensuremath{%
      #2%
    }%
    ~\@eqnnum
  \endgroup
}
\makeatother

\newcommand{\setc}[2]{\ensuremath{\left\{\,#1 \;\middle|\; #2\,\right\}}}

\newcommand{\set}[1]{\ensuremath{\left\{\,#1\,\right\}}}
\newcommand{\tensor}[1]{\ensuremath{\mathit{#1}}}

\acmSubmissionID{XXX}

\author{Toluwanimi O. Odemuyiwa}
\email{todemuyiwa@ucdavis.edu}
\author{Serban D. Porumbescu}
\email{sdporumbescu@ucdavis.edu}
\author{Nandeeka Nayak}
\email{nandeeka@berkeley.edu}
\author{Michael Pellauer}
\email{mpellauer@nvidia.com}
\author{Joel S. Emer}
\email{emer@csail.mit.edu}
\author{John D. Owens}
\email{jowens@ucdavis.edu}

\begin{document}

\title{The EDGE Language: Extended General Einsums for Graph Algorithms}

\begin{abstract}
  In this work, we propose a unified abstraction for graph algorithms: the Extended General Einsums language, or \edge{}\@.
The \edge{} \emph{language} expresses graph algorithms in the language of tensor algebra, providing a rigorous, succinct, and expressive mathematical framework.
\edge{} leverages two ideas: (1) the well-known foundations provided by the graph-matrix duality, where a graph is simply a 2D \emph{tensor}, and (2) the power and expressivity of Einsum notation in the tensor algebra world.
In this work, we describe our design goals for \edge{} and walk through the extensions we add to Einsums in order to support more complex operations common in graph algorithms.
Additionally, we provide a few examples of how to express graph algorithms in our proposed notation.
Our hope is that a single, mathematical notation for graph algorithms will (1) allow researchers to more easily compare different algorithms and different implementations of a graph algorithm; (2) enable developers to factor complexity by separating the concerns of \emph{what} to compute (described with the extended Einsum notation) from the lower-level details of \emph{how} to compute; and (3) enable the discovery of different algorithmic variants of a problem through algebraic manipulations and transformations on a given \edge{} expression.

\end{abstract}

\maketitle

\section{Introduction}

From fraud detection in banks~\cite{Srivastava:2023:FDD, Pourhabibi:2020:FDS}, to our familial, social and professional connections~\cite{Kwak:2010:WTS, Kempe:2003:MSI}, to analyzing how various proteins impact disease expression~\cite{Gulbahce:2011:NMN}, graphs---or \emph{networks}---are ubiquitous in our daily lives and the infrastructure of society.
A graph is an abstract data structure that captures data entities and the complex connections between them.
Graph algorithms have been fundamental to answering questions about this data, allowing analysis that focuses on the relationship between various entities.
Given a specific problem (e.g., find the connected components or communities within a graph), there may typically be numerous algorithmic alternatives \emph{and} even a variety of choices for realizing the chosen algorithm as a software implementation or hardware module.
For example, the well-studied weakly connected-components problem---with algorithmic solutions earlier than 1973~\cite{Hopcroft:1973:AEA}---has grown to have hundreds of algorithms and implementations, each with its own time and space complexity.
Researchers and developers continue to discover new algorithms and specific implementations for well-studied algorithms even today~\cite{Liu:2022:SCC, Du:2023:MMB}.
Even the problem of finding the shortest path to all other entities in a graph given a starting point (single-source shortest path)---often considered a well-solved problem---still yields new implementations~\cite{Dong:2021:ESA}.

To invent a new algorithm, researchers often ``think really hard'' to determine the high-level steps.
Algorithms are then lowered to the implementation level, where a software developer or hardware designer may iterate on the implementation multiple times to optimize for runtime or energy efficiency.
Both algorithmic development and software or hardware implementation are hard; \textbf{is there a way to systematically think about inventing algorithms and implementing solutions?}

Additionally, different graph frameworks describe their implementations in imprecise natural language or pseudocode, at times using different terminology and program structure even when describing the same algorithm.
Moreover, most graph frameworks focus on highly performant implementations for a single target backend, with abstractions that do not enable fully exploring the algorithmic and implementation space.
This makes it difficult to directly compare different implementations, iterate on an implementation, and port one implementation from one environment to another.
\textbf{Is there a unified way of thinking about and expressing an implementation?}

An ideal approach to graph algorithms decomposes the problem into smaller pieces, each of which tools or developers can individually explore.
Overall, \textbf{is there a way to further factor the complex graph problem space into several, smaller concerns?}
Such an approach will enable developers to focus on specific aspects of a graph solution, transforming the design space into a tunable space with several degrees of freedom, some of which can be explored independently.

In this work, we take inspiration from the tensor algebra community, where
Einsum notation has served as a compact, declarative specification of tensor
computations. We cast graph problems into tensor algebra by extending Einsum
notation into \edge{}, the Extended General Einsums language: a language for
expressing graph algorithms as precise, declarative tensor computations.
\edge{} leverages two ideas: (1) graph-matrix duality, where a graph can be
represented as a 2D tensor, and (2) the separation of \emph{what} is computed
from \emph{how} it is computed, with mapping, data format, and backend choices
specified outside the Einsum~\cite{Kjolstad:2017:TTA, Nayak:2023:TDF,
Parashar:2019:TSA}.

\paragraph{The Tensor Algebra World}
We use the adjective ``extended'' to refer to the notation used by the programming language, compiler, and accelerator modeling community
(e.g., \taco{}~\cite{Kjolstad:2017:TTA}, Timeloop~\cite{Parashar:2019:TSA}, and TeAAL~\cite{Nayak:2023:TDF}),
as distinguished from the traditional mathematical expressions used in multilinear algebra\footnote{This includes the fields of statistical mechanics~\cite{Guo:2022:LRT}, quantum chemistry~\cite{Solomonik:2013:CTF}, Ricci calculus~\cite{Zund:1994:FDG, Laue:2020:SET}, and tensor contraction theory~\cite{Solomonik:2013:CTF}, among others.}.
For brevity, in this paper we will generally drop the ``extended'' word in ``extended Einsums'', and refer to this programming language/compiler/modeling community as the ``tensor algebra world''.
Einsums are the language that enables the compiler and modeling tools to precisely describe the computations present in the mathematical field of multilinear algebra.

\paragraph{Extending Einsums.} We begin by introducing the design of \edge{} and illustrating how traditional Einsums are extended to express graph algorithms, with examples throughout.
As much as possible, our extensions remain within the tensor machinery.
In contrast, prior linear-algebra-based approaches to graph computation (e.g., \graphblas) express these computations within the larger context of an imperative program (see \S~\ref{ssec:design-goals}).
Our extensions support expressing the following language features:
(1) user-defined data values and types;
(2) initialization of tensors;
(3) generic user-defined operators that express computation between two tensor values (\emph{\maptxt}, \S~\ref{sssec:map}), reduction of values (\emph{\reduce}, \S~\ref{sssec:reduce}), and the (potentially filtered) assignment of the output tensor (\emph{\populate}, \S~\ref{sssec:populate});
(4) a separation of computations on graph entities/tensor coordinates (\merge, \S~\ref{ssec:merge}) from graph values/tensor data (\compute, \S~\ref{ssec:merge});
(5) iterative algorithms (\S~\ref{ssec:iter}); and
(6) constraints that represent when particular computations occur based on certain conditions (rank variable expressions, \S~\ref{ssec:conditionals}).
Overall, this paper introduces \edge{} as a formal algebraic language for graph computation and establishes its semantic foundations.
While the framework naturally enables systematic exploration of algorithmic variants through algebraic transformation, a comprehensive study of this space---including performance characterization and automated exploration---is beyond the scope of this work and is addressed in complementary work.
Our focus here is on the language design, its formal semantics, and its ability to express and relate graph computations within a unified algebraic framework.

\paragraph{Contributions}
This paper makes the following contributions:
\begin{enumerate}
  \item \textbf{The \edge{} language.} We introduce \edge{}, an extension of Einstein summation notation that enables the expression of graph algorithms within a unified tensor algebra framework, including support for user-defined operators, iteration, and constraints.

  \item \textbf{Formal semantic foundation.} We define a set-theoretic and functional semantic model for \edge{}, providing a precise interpretation of programs and serving as a foundation for reasoning about correctness and transformation.

  \item \textbf{Algebraic formulation of graph computation.} We demonstrate
  how common graph algorithms can be expressed in \edge{} and introduce the
  \emph{eingebraic space}: the set of \edge{} expressions reachable from one
  another through algebraic transformation. In graph algorithms, this space
  captures variants of the same problem, such as push and pull \bfs{}, as
  algebraically related expressions within a single framework.
\end{enumerate}

The notation and abstractions presented in this paper form the first stage
of a larger \edge{} framework. In that framework, an \edge{} expression
serves as the precise, declarative specification of what is computed.
Future stages can then explore how that computation is realized: algebraic
manipulations can derive equivalent expressions, mapping specifications can
choose loop orders and tilings, data-format specifications can choose tensor
layouts, and backend specifications can bind the computation to software or
hardware targets. This paper focuses on the language design, formal
semantics, and expressive power needed to make that broader framework
possible.

\subsection{Roadmap}\label{ssec:roadmap}

This paper serves several audiences: it is a tutorial, a language
definition, a formal semantics, and a collection of case studies. \edge{}
is already being applied in multiple contexts, including graph analytics,
accelerator design, robotics, digital circuit simulation, and machine
learning workloads~\cite{
odemuyiwa:2026:MEB,
Nayak:2024:FML_micro,
Odemuyiwa:2025:FTF,
gilbert:2024:LEF,
zhu:2026:rsu,
Zhang:2025:transfusion,
Golden:2025:QRD,
odemuyiwa:2024:TeAALTutorial,
odemuyiwa:2025:TeAALTutorial,
Nayak:2023:TDF}.
This breadth motivates a comprehensive presentation of the language and its foundations.
We do not expect any single reader to need the entire paper on a first pass.

We use several visual conventions to help readers navigate the paper:
general examples appear in pink boxes; \bfs{} examples, which serve as the
running example, appear in purple boxes; design alternatives appear in grey
boxes; and optional mathematical or explanatory notes appear as insets.
Diagrams throughout the paper complement the discussion by illustrating
tensor operations and execution flows.

\begin{example_box*}[label=sidebar:example_box]{General Examples}
\end{example_box*}

\begin{bfs_box*}[label=sidebar:bfs-example]{\bfs{} Examples}
\end{bfs_box*}

\begin{alternative_box*}[label=sidebar:alt-example]{Design Alternatives}
\end{alternative_box*}

\begin{sidebar_box*}[label=sidebar:example]{Additional Information}
\end{sidebar_box*}

We suggest the following reading paths:
\begin{itemize}

  \hypertarget{roadmap:writing-edge}{%
  \item \textbf{Readers wishing to express graph algorithms (and beyond!) in \edge{}.}}
  Start with \S~\ref{ssec:walk-bfs}, which builds the language one feature
  at a time using \bfs{} as the primary running example. Read the \bfs{}
  example boxes closely to see how graph primitives such as gathering
  neighbors, filtering vertices, updating frontiers, and iteration appear in
  \edge{}. Then read \S~\ref{ssec:examples} for more complex examples, including
  \SSSP{} variants and concurrent connected components.

  \hypertarget{roadmap:tensor-algebra}{%
  \item \textbf{Readers with a tensor algebra background.}}
  Read \S~\ref{sec:background} and \S~\ref{ssec:design-goals} to understand
  how \edge{} relates to existing tensor algebra systems. Then focus on the
  extensions that go beyond traditional Einsum notation: user-defined
  operators (\S~\ref{ssec:computationspecs}), the separation of \merge{} from
  \compute{} (\S~\ref{ssec:merge}), iteration (\S~\ref{ssec:iter}), and
  \populate{} (\S~\ref{ssec:spopulate}). To see how \edge{} expressions can
  be manipulated algebraically, read \S~\ref{ssec:beyond-graphs} and
  Appendix~\ref{appendix:bf-derivation}.

  \hypertarget{roadmap:backend-designers}{%
  \item \textbf{Accelerator and backend designers.}}
  If your interest is mapping \edge{} workloads onto a tensor algebra
  accelerator or software backend, see also the
  \hyperlink{roadmap:tensor-algebra}{tensor algebra route}; your reading
  overlaps with it but with different emphasis. Read \S~\ref{sec:background},
  especially the discussion of tensor algebra concerns summarized in
  Table~\ref{tab:related}. Then focus on the language features that most
  directly affect dataflow, scheduling, and storage: \merge{} and \compute{}
  (\S~\ref{ssec:merge}), iteration (\S~\ref{ssec:iter}), and
  \populate{} (\S~\ref{ssec:spopulate}). \S~\ref{sec:5-mappings} sketches how
  \edge{} fits into a broader exploration framework.
  Several recent works build on \edge{} to explore the implementation
  space, including TeAAL~\cite{Nayak:2023:TDF, odemuyiwa:2024:TeAALTutorial,
  odemuyiwa:2025:TeAALTutorial}, FuseMax~\cite{Nayak:2024:FML_micro,
  Odemuyiwa:2025:FTF}, and related efforts~\cite{zhu:2026:rsu,
  odemuyiwa:2026:MEB}.

  \hypertarget{roadmap:implementing-edge}{%
  \item \textbf{Readers implementing \edge{} or a variant of \edge{}.}}
  Read \S~\ref{ssec:walk-bfs} for intuition about what the language is for,
  and \S~\ref{ssec:design-goals} for the constraints any implementation must
  respect---in particular, that every extension preserves the operational
  definition of an Einsum (\textbf{design goal~\ref{goal:ode}}). Then read
  \S~\ref{ssec:syntax} and \S~\ref{sec:sets}; together, the grammar and the
  formal semantics define the language. The semantics are written in a
  functional and set-theoretic style and should be treated as the primary
  reference for determining the meaning of an \edge{} expression.

  \hypertarget{roadmap:reference}{%
  \item \textbf{Readers looking for reference material.}}
  If you are not reading the paper linearly and instead need to interpret an
  \edge{} expression, \S~\ref{ssec:syntax} provides the grammar,
  Appendix~\ref{appendix:syntax-notes} offers additional syntax notes, and
  Appendix~\ref{appendix:merge} catalogs all 16 merge operators.

\end{itemize}

If none of these descriptions fit, start with
\hyperlink{roadmap:writing-edge}{the first route} and stop after
\S~\ref{ssec:examples}; this should provide enough intuition to identify
which remaining sections are relevant to your goals.

\section{Preliminaries}\label{sec:prelims}
This section briefly describes relevant graph and tensor algebra terminology.

\begin{wrapfigure}{l}{0.52\linewidth}
  \centering
  \includegraphics[width=.48\textwidth]{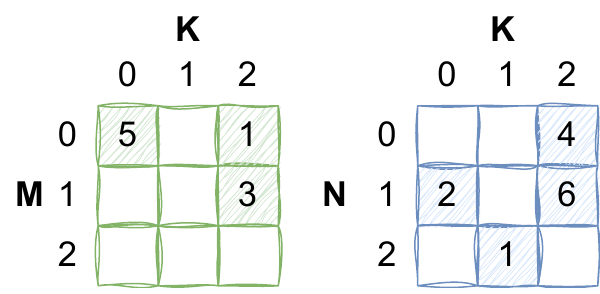}
  \caption{Uncompressed 2D tensors, $A$ and $B$, where locations containing zero are treated as empty in this example.
  This also shows the coordinate spaces of $A$ and $B$, whose points map to values in each tensor's data space.\label{fig:uncompressed}
  }
\end{wrapfigure}

\subsection{Graph Algorithms}\label{ssec:graph-algorithms}
A graph describes relationships between entities.
Suppose we have a graph, $G$.
It contains two entities: its \emph{vertex set} ($G.V$) and its \emph{edge set} ($G.E$)~\cite{CLRS:2022:ITA}.
$V$ is the set of all vertices in the graph, $|V|$ is the number of vertices in the graph, and $E$ is the set of all edges.
A vertex or node, $v$, in a graph is an entity possibly connected to other vertices by an edge, $e$.

We describe an undirected edge, $e \in E$, by an unordered set of vertices $\{s,d\}$, where $s,d \in V$, with the edge connecting vertices $s$ and $d$.
A directed edge is a tuple of ordered vertices, $(s,d)$, where $s,d \in V$, with the edge starting from $s$ and ending at $d$.
The starting point of an edge ($s$) is a \emph{source} or \emph{parent} vertex, while the ending point ($d$) is a \emph{destination} or \emph{child} vertex.

A given graph may be directed or undirected, where we can also describe an edge in an undirected graph as two directed edges: $(s,d)$ and $(d,s)$.
A graph may also be unweighted or weighted.
An \emph{unweighted} graph assumes all edges have a cost, or ``weight'' of one, while a \emph{weighted} graph stores a quantity on each edge.
A \emph{node-weighted} graph assigns weights to each of its vertices.
Finally, a graph may be dynamic; that is, its structure changes with time.
This may include a changing vertex set, edge set, edge weights, and/or node weights~\cite{Harary:1997:DGM}.

Several graph algorithms require a form of \emph{graph traversal}, where the neighbors of a vertex are successively explored~\cite{Merrill:2012:SGG, Graph500, Che:2009:RAB, Nai:2015:GUG}.
The \emph{active vertex/edge set}, or \emph{frontier}, refers to the subset of nodes and/or edges that are actively participating in the graph computation at a particular step in time~\cite{Shun:2013:LAL, Wang:2017:GGG}.
To explore the ideas between graph algorithms and tensor algebra, in this paper we will use breadth-first search (\bfs{}), a graph traversal algorithm, as a running example.
\bfs{} iteratively finds all vertices reachable from a starting node, recording the depth at which each vertex is found.
In \bfs{}, all vertices at a particular depth are discovered before any vertex with a larger depth.

\subsection{Tensors}\label{ssec:tensors}
Tensor algebra describes computations between \emph{tensors}, which are multidimensional views of data.
We employ the term \emph{rank} for the axes or indices of a given tensor~\cite{Nayak:2023:TDF}.
Thus, a \nd{3} tensor has three ranks, a \nd{2} tensor (or matrix) has two ranks, a \nd{1} tensor (or vector) has one rank, and a \nd{0} tensor is simply a scalar.
Let $A_{k, m}$ be a \nd{2} tensor named $A$.
It has two ranks, $M$ and $K$, each indexed by \emph{coordinates} $m$ and $k$.
The \emph{shape} of a rank is the maximum number of coordinates possible in that rank.
We assume the rank name is a variable that contains its shape, and by convention is specified in upper case.
The coordinate that indexes into a rank has the same variable name in lowercase.
Thus, for example, coordinate $m$ ranges from $0$ to $M-1$ unless otherwise specified.
The \emph{shape} of a tensor is the total maximum size of the tensor along each of its ranks.
Thus, $A$ has a shape of $M \times K$.
We refer to the total number of non-zeros (nnz) in the tensor as its \emph{occupancy}.
The example tensor $A$ in Figure~\ref{fig:uncompressed} has an occupancy of 3, while $B$ has an occupancy of 4.
We often describe tensors in terms of their densities: a \emph{dense} tensor has full or nearly full occupancy, while a \emph{sparse} tensor has low occupancy---a few nnz values relative to the shape of the tensor.

\begin{wrapfigure}{l}{0.56\linewidth}
  \centering

  \begin{subfigure}{\linewidth}
    \includegraphics[width=\linewidth]{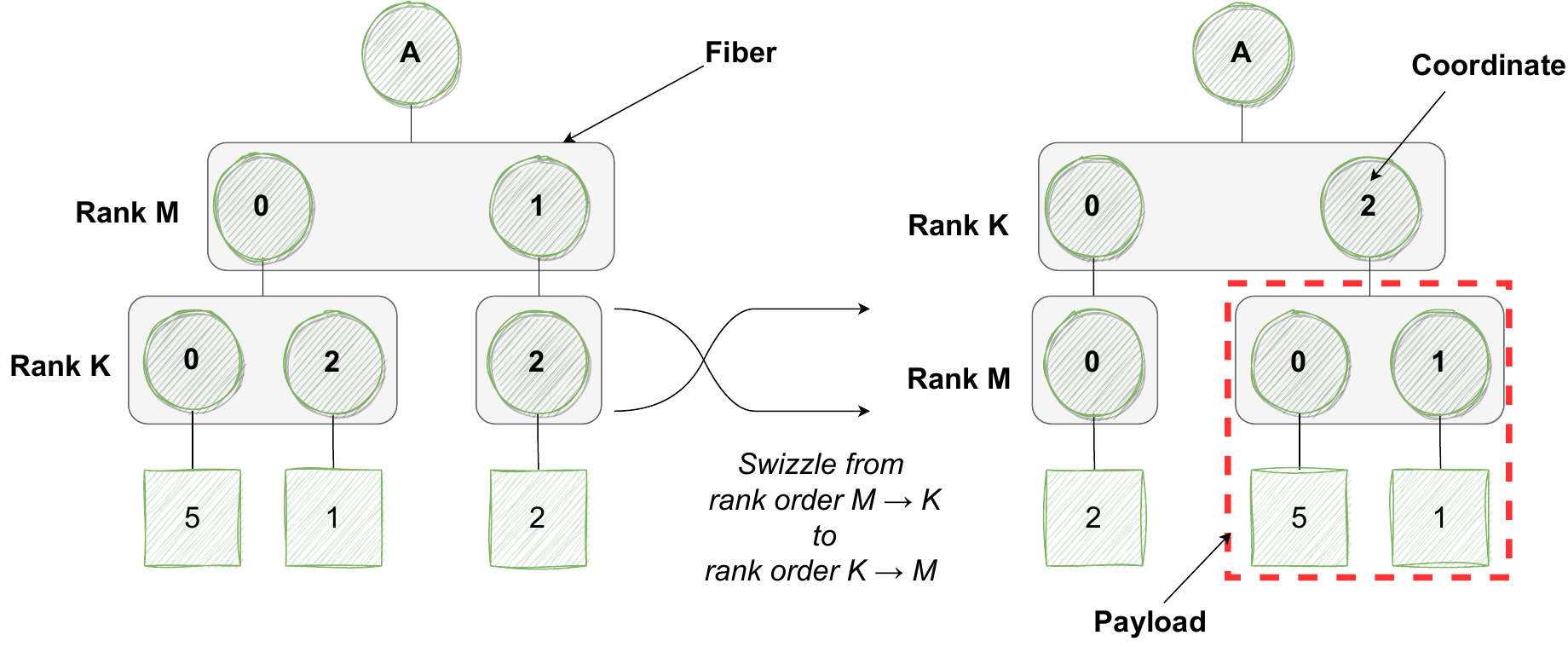}
    \caption{The two possible fibertree representations of $A$.}
    \label{fig:fibertree-a}
  \end{subfigure}

  \begin{subfigure}{\linewidth}
    \includegraphics[width=\linewidth]{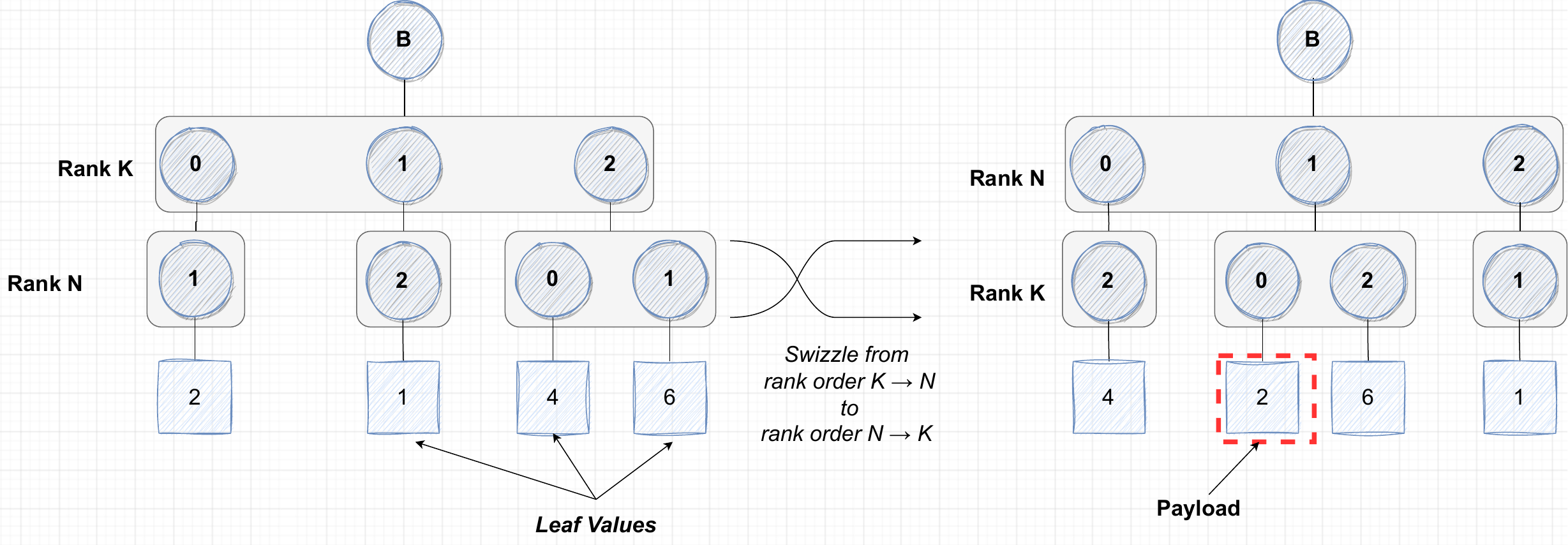}
    \caption{The two possible fibertree representations of $B$.}
    \label{fig:fibertree-b}
  \end{subfigure}

  \caption{Fibertree representations of the uncompressed tensors in Figure~\ref{fig:uncompressed}.}
  \label{fig:fibertree}
\end{wrapfigure}

The preceding terminology refers to abstract characteristics of a tensor; however, tensor algebra applications need to store the data in memory.
Traditionally, dense tensors are stored in an \emph{uncompressed} format: the memory location of each value matches its coordinate order.
Sparse tensors present an opportunity to reduce storage by using \emph{compressed} data formats such as Compressed Sparse Row/Column (\CSR/\CSC), Coordinate Format (\COO), and Compressed Sparse Fiber (\CSF)~\cite{Smith:2015:TMP}, amongst others~\cite{Chou:2018:FAF, muthu:2012:esc}.

To abstract away formatting details, we adopt the \emph{fibertree} terminology described by Sze et al.~\cite[Chapter 8]{sze:2020:epo}, with further details provided in the excellent fibertree description of Nayak et al.~\cite[Section 2.1]{Nayak:2023:TDF}.
A fibertree is a format-agnostic representation of a tensor.
A \emph{fiber} goes beyond the concepts of rows and columns in linear algebra to denote the set of entries for a particular rank for any $N$-dimensional tensor.
Each element in a fiber consists of a coordinate in the rank and its payload.
The payload consists of a reference to the next-level fiber or a leaf data value~\cite{Nayak:2023:TDF}.
The fibertree admits $N!$ possible representations, one for each possible \emph{rank order} of the tensor.
The rank order refers to ``the order of levels'' in the fibertree, going from top to bottom~\cite{Nayak:2023:TDF}.
To change the rank order of a tensor, one must apply \emph{rank swizzling} to the tensor.
Figures~\ref{fig:fibertree-a} and~\ref{fig:fibertree-b} show example fibertrees for tensors $A$ and $B$, and their corresponding representations when rank-swizzled.

\begin{figure}
  \centering
  \includegraphics[width=\textwidth]{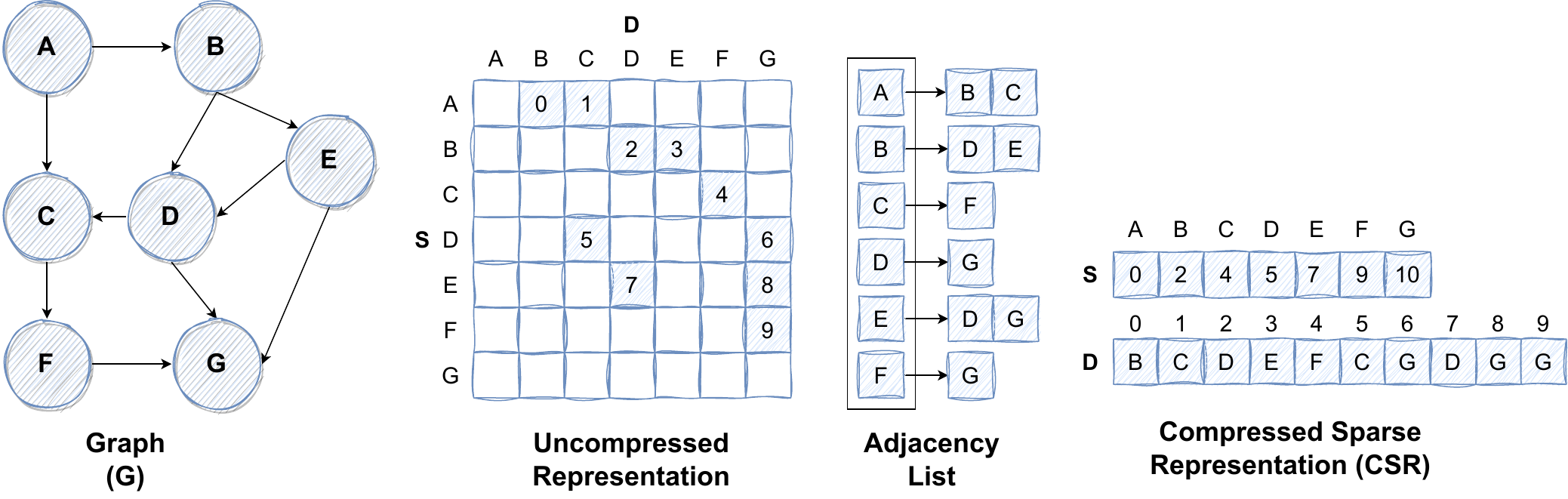}
  \caption{A directed graph, $G$, and some representations. The matrix values contain the edge ID, while the vertices are labeled alphabetically.\label{fig:exgraph}}
\end{figure}

\begin{wrapfigure}{l}{0.56\linewidth}
  \centering
  \begin{subfigure}{\linewidth}
    \includegraphics[width=\linewidth]{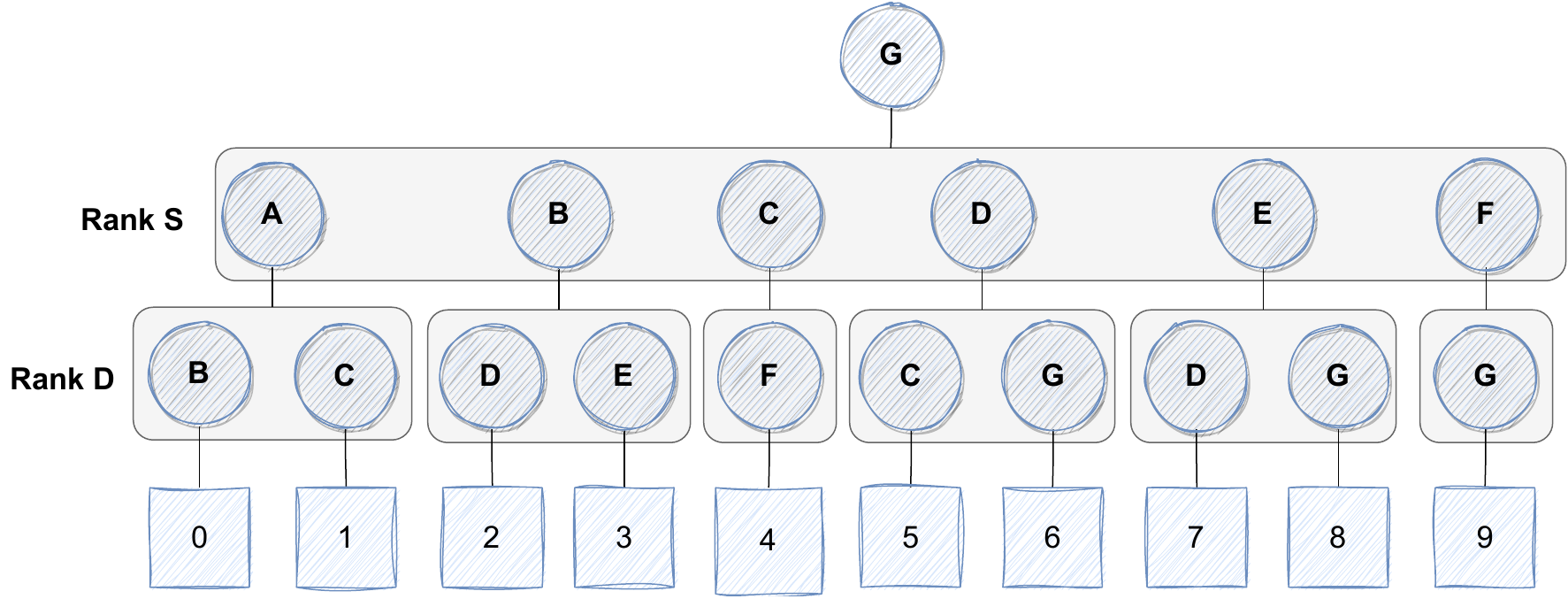}
    \caption{Here, we express $G$ as a tensor where the leaf values contain edge IDs.\label{fig:graph:fiber:edge}}
  \end{subfigure}
  \vfill
  \begin{subfigure}{\linewidth}
    \includegraphics[width=\linewidth]{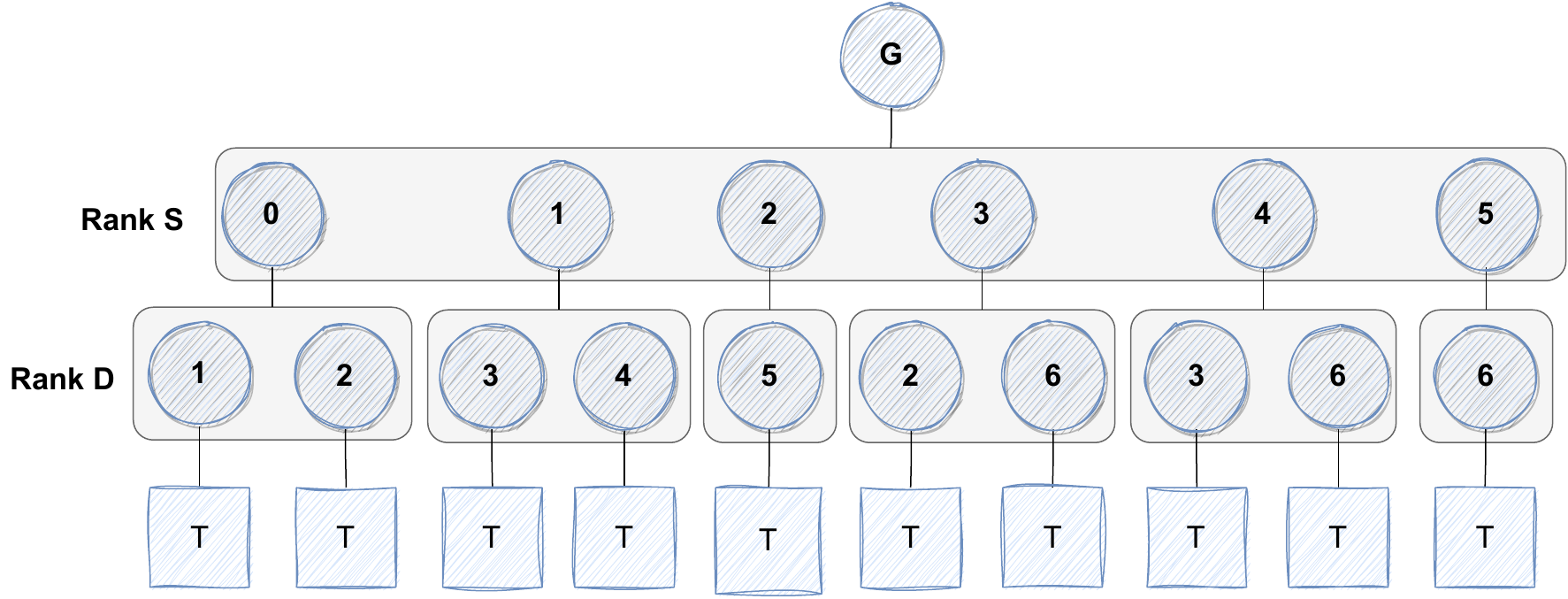}
    \caption{Here, we express $G$ as a tensor where vertex IDs are now numerical and the leaf values indicate whether there is an edge connecting vertex $s$ to vertex $d$, where $T$ is \True{} and $F$ (the empty value) is \False.\label{fig:graph:fiber:true}}
  \end{subfigure}
  \caption{The example graph in Figure~\ref{fig:exgraph} as two different fibertrees.\label{fig:graph:fiber}}
\end{wrapfigure}

Fibers with empty leaf values do not appear in the fibertree.
For example, in Figure~\ref{fig:fibertree-a} the fiber $m=2$ in $A$ is empty.
Likewise, the $M$ fiber at $k=1$ is also empty and thus does not appear in the fibertree representation (see swizzled representation in Figure~\ref{fig:fibertree-a}).
Additionally, the \emph{position} of a tensor value is its ordering in a zero-based numbering of the tensor's non-zero values, under a specific rank order.
For example, the value at coordinate $(1, 2)$ of tensor $A$, in Figure~\ref{fig:uncompressed}, has a position of 2, when following an $M$-major \emph{rank order} ($k$ varies faster than $m$).

A \emph{point} refers to a specific location in a tensor, identified by its coordinate tuple.
For example, point $(k=0,n=1)$ in $B$ (Figure~\ref{fig:uncompressed}) contains the data value of $2$.
Point $(k=0,n=0)$ in $B$ contains the data value of $0$.
A \emph{region} refers to a set of points in a tensor.
For example, the region $(k=0,n=:)$ refers to the entire $k=0$ fiber.
The region $(:)$ refers to the points of the entire tensor.
Thus, a point is a specific instance of a region.

A tensor has both a \emph{coordinate space} and a \emph{data space}.
Informally, the coordinate space is the set of possible points in the tensor, while the data space is the set of values the tensor may contain at those points.
We use these notions informally in this section.
In \edge{}, this abstraction is extended to support richer data types and user-defined computations over tensor values.
\S~\ref{sec:sets} defines these concepts precisely using set-theoretic semantics, including the treatment of tensor values, empties, and user-defined data types.

The $Z$ tensor has a coordinate space of $[0, M) \times [0, N)$, consisting of all possible $(m,n)$ tuple points, where $m$ and $n$ can each range over $[0,M)$ and $[0,N)$.
For brevity, we drop the range when describing coordinate spaces; thus, $Z$ has coordinate space $M \times N$.
Likewise, $A$ and $B$ have coordinate spaces of $M \times K$ and $K \times N$, respectively.
A \emph{projection} into a tensor maps a point or region in the coordinate space to the corresponding value in the tensor's data space.

We have found this abstraction useful in describing transformations on tensors, such as partitioning, without worrying about the format-specific implementations for those transformations~\cite{Odemuyiwa:2023:ASD, Nayak:2023:TDF, Hsu:2023:SAM}.
Table~\ref{tab:terms} summarizes some key tensor terminology.

\begin{table}[]
  \captionsetup{aboveskip=2pt}
  \caption{\label{tab:terms} Tensor Terminology, expanded from Odemuyiwa et al.~\cite[Table 1]{Odemuyiwa:2023:ASD}.}
  \centering
  \footnotesize
  \begin{tabular}{ll}
  \toprule
  Name        & Description \\
  \midrule
  tensor & a multidimensional view of data \\
  coordinate & the logical index of a value or fiber \\
  position & ignoring zero values, the index of a non-zero value or fiber under a specific rank ordering\\
  rank order & the ``major-order'' in which elements are laid out in memory \\
  point & the tuple of coordinates identifying a specific location\\
  region & a tuple of coordinates identifying a set of points\\
  coordinate space & the set of possible coordinate points in a tensor \\
  data space & the set of values a tensor may contain at its coordinate points \\
  occupancy & number of non-zero elements within a fiber or tensor \\
  uncompressed format & the position of each value matches its coordinate order \\
  \bottomrule
  \end{tabular}
\end{table}

\subsection{From Graphs to Tensors}
The matrix-graph duality~\cite{Kepner:2011:GAL} allows us to represent a graph as a \nd{2} tensor.
Figure~\ref{fig:exgraph} shows a graph and some of its various representations.
As with tensors, we can leverage the format-agnostic fibertree semantics to describe graphs---and other data structures---without considering \emph{how} they are stored in memory (Figure~\ref{fig:graph:fiber}).
\section{Background and Related Works}\label{sec:background}

Implementing a solution to a graph problem requires considering a large, complex problem space.
As an example, consider breadth-first search (\bfs{}), where implementations often decide between using a top-down (push), bottom-up (pull), or a hybrid, direction-optimized approach.
Push-\bfs{} first discovers neighbors of vertices in the frontier, then determines if those neighbors have been visited.
Pull-\bfs{} checks the set of all unvisited vertices, and determines if their parents are in the frontier.
A hybrid approach selects between the two approaches depending on the size of the frontier and the number of unvisited nodes at the current iteration.
In addition to deciding the overall algorithmic approach, developers must also decide which data structures to use for the graph, frontier, and any other data.
Each choice impacts performance: for example, a graph stored using a \CSR representation will have sequential access to the neighbors of a vertex, but many random accesses when finding the parents of a vertex.
Meanwhile, an uncompressed representation of the graph will take up more storage in memory, but provide efficient random accesses.
Other aspects to consider when implementing a graph solution include:
deciding computation order;
the ``unit of computation''~\cite{Coimbra:2021:AGP} including vertex-centric~\cite{Malewicz:2010:PAS, McCune:2015:TLA} or edge-centric~\cite{Zhou:2018:FFE, Roy:2013:XEC, Zhang:2018:ABG, Zhu:2020:WEC} computations;
parallelization and load-balancing strategies~\cite{Wang:2017:GGG, Brahmakshatriya:2021:CGA, Osama:2023:PMG, Osama:2022:EOP};
how to partition the graph~\cite{McCune:2015:TLA, Coimbra:2021:AGP, Zhang:2018:GAH};
and the target platform (\CPU, \GPU, \FPGA, etc.).

In the graph world, the dominant approach is to first implement a baseline solution, observe its performance, then update the baseline solution by changing different design aspects.
A developer iterates on this process until the implementation achieves the desired performance.
Most often, this approach does not separate graph algorithms from scheduling the computation of that algorithm, making full exploration of the combined design space a significant challenge.

Researchers have recognized this challenge with work that more efficiently searches the space of possible implementations.
For instance, GraphIt~\cite{Zhang:2018:GAH}, a graph framework for shared-memory systems, provides a menu of options for users (the schedule) after a user specifies the high-level description of the graph algorithm.
However, this is specific to shared-memory systems, and does not fully cover all the options in the complex problem space.
G2~\cite{Brahmakshatriya:2021:CGA} extends GraphIt to support a \GPU backend.
We view many of the proposed, \GPU-specific optimizations in G2 as platform-independent mapping choices in the Einsum framework.

When dealing with a complex problem space, a common design philosophy is \emph{separation of concerns}.
In this approach, designers factor the problem space into smaller, manageable subproblems that can be independently solved.

The tensor algebra space has successfully leveraged a separation-of-concerns design philosophy to factor its problem space into manageable parts.
It has also incorporated sparse tensor algebra into this philosophy, despite the additional complexity introduced by the irregular nature of sparse tensor workloads.
In order to leverage the separation-of-concerns approach already established in that field, we map graph algorithms to tensor algebra.
We now turn to the different concerns on which tensor algebra tools focus.

\subsection{Factoring Complexity in Sparse Tensor Algebra}\label{ssec:ta-soc}
In the sparse tensor algebra accelerator space, the design process involves using block diagrams, natural language, or pseudocode to describe the algorithm and model the architecture, followed by a custom simulator for evaluation, then an attempt at comparing to prior work~\cite{Nayak:2023:TDF}.
Features such as the sparse tensor format are ingrained into the architecture; for example, the MatRaptor~\cite{mat:2020:sri} accelerator relies on its custom $C^2SR$ data format (\CSR/\CSC-like) to mine parallelism, making it difficult to directly compare it with accelerators that are built around other data formats.
Even iterating over a design requires starting over at the algorithmic and modeling phase~\cite{Nayak:2023:TDF}.
Likewise, there is a similar pattern in the tensor algebra software space: given a tensor algebra problem, application developers often plunge directly into a highly performant, low-level implementation~\cite{Kjolstad:2020:STA, Kjolstad:2017:TTA}.
This mirrors the approach used by graph developers.

\begin{figure}
  \centering
  \includegraphics[width=\columnwidth]{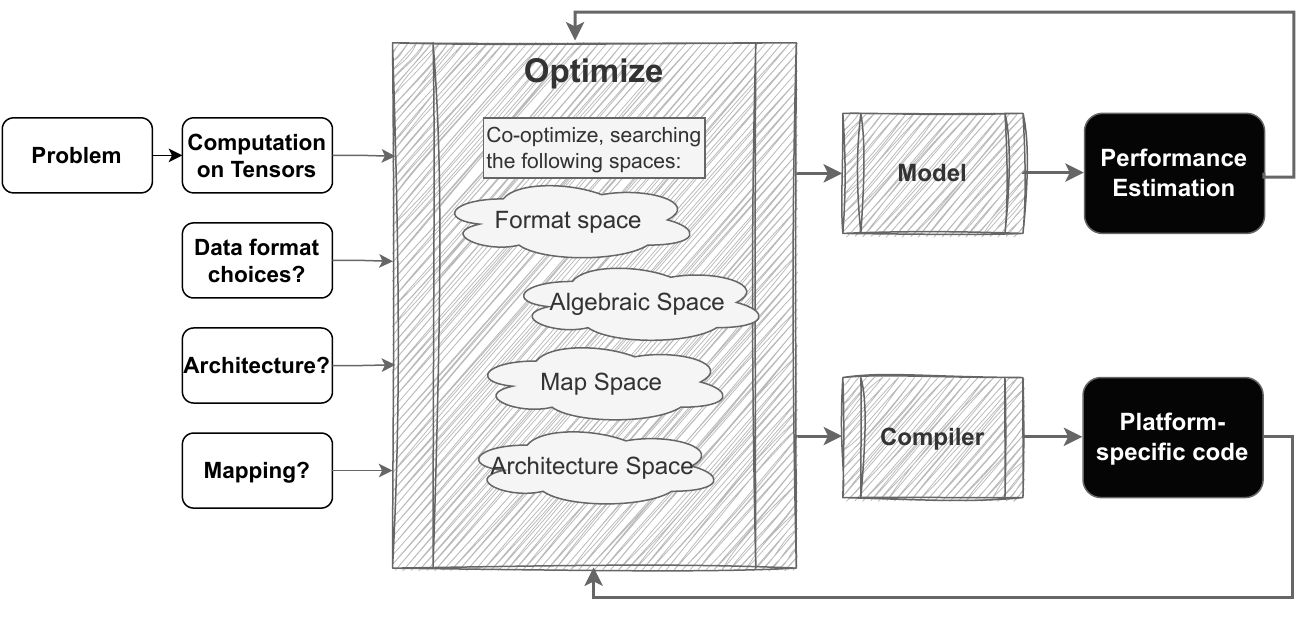}
  \caption{Our view of the complex design space in tensor algebra. Shaded boxes indicate \emph{processes} (searching the space), white/black boxes indicate inputs/outputs to the system.
  Given a problem, a tensor algebra implementation must try to optimize across the possible data formats, possible algebraic variants of the computation, and the mapping space, as well as selecting the best architecture.
  Tools tend to either model performance (for domain-specific accelerators), or generate platform-specific code. The results may be used to search for better points in the implementation space.\label{fig:ta-soc}
  }
\end{figure}

\begin{figure}
  \centering
  \includegraphics[width=\columnwidth]{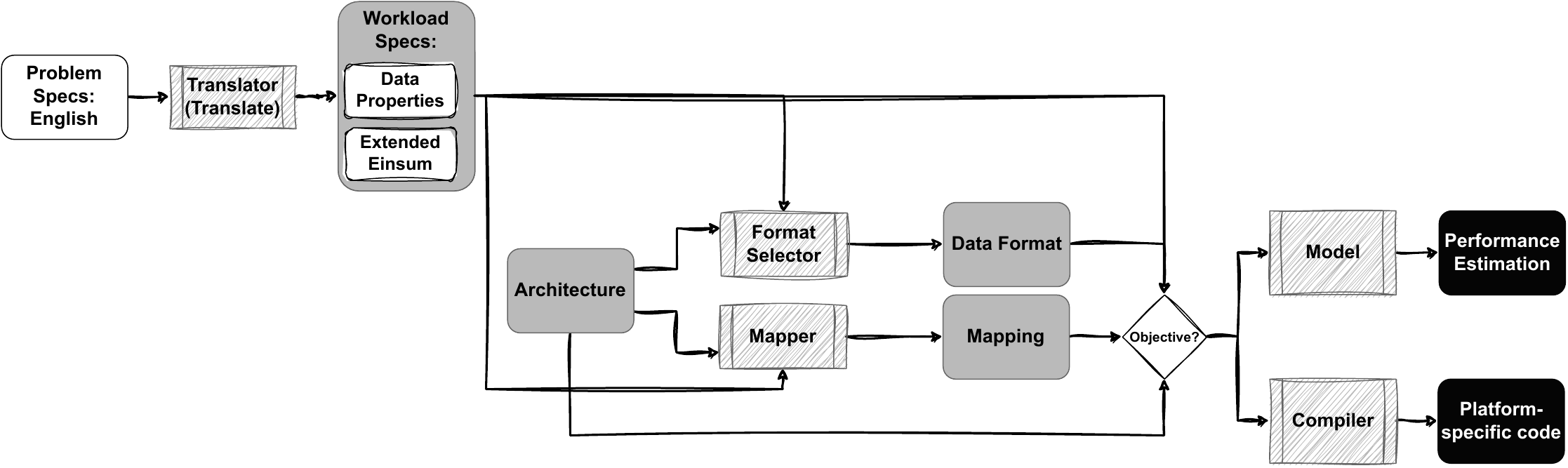}
  \caption{One common approach to factoring the design space in tensor algebra.
  Shaded boxes with side borders indicate \emph{processes}, plain shaded boxes indicate \emph{specifications} for a concern, while white/black boxes indicate inputs/outputs.
  Tools search subspaces to select a point within that space (e.g., a mapper selecting a mapping), then pass the selected point to the next subspace to inform its search (e.g., given a mapping, select the best data format).\label{fig:factored}%
  }
\end{figure}

\begin{figure}
    \centering
    \includegraphics[width=\columnwidth]{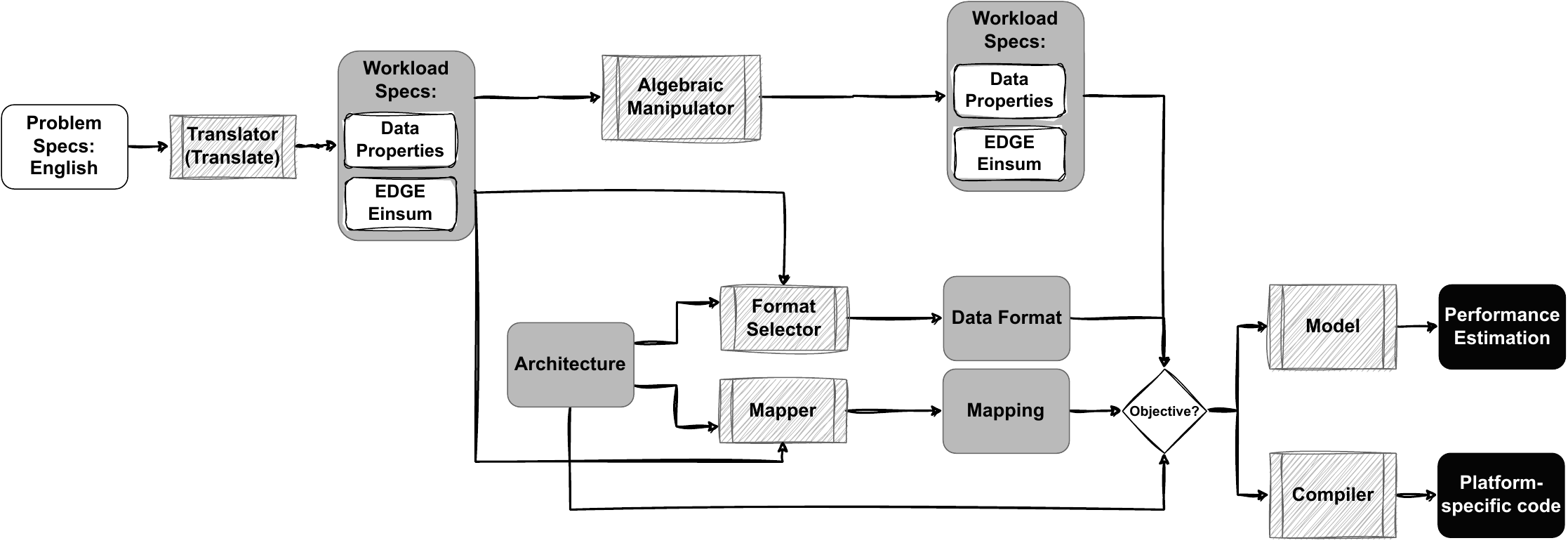}
    \caption{An instance of a factored design space with our extended general Einsums for graph algorithms.
    \edge{} introduces another concern to the factored space: algebraic manipulation of the Einsums.
    The algebraic manipulator produces a new Einsum that produces the same result as the original, input Einsum.
    By casting graph problems as extended Einsums, we can leverage the existing problem space found in tensor algebra.
    Shaded boxes with side borders indicate \emph{processes}, solid gray boxes indicate \emph{specifications} for a concern, while white/black boxes indicate inputs/outputs.\label{fig:factored:edge}
    }
  \end{figure}

\begin{table}
    \caption{Selected works in the tensor algebra space and the concerns on which they expand. Works are sorted by date of publication.
    \label{tab:related}}
    \resizebox{1.0\textwidth}{!}{
      \begin{tabular}{lcccccccccc}
      \toprule
      &
        \textbf{\begin{tabular}[c]{@{}c@{}}Representing \\ the \\ Workload\end{tabular}} &
        \textbf{\begin{tabular}[c]{@{}c@{}}Enabling \\ Algebraic \\ Manipulations\end{tabular}} &
        \multicolumn{1}{c}{\textbf{\begin{tabular}[c]{@{}c@{}}Representing \\ the \\ Mapping Space\end{tabular}}} &
        \multicolumn{1}{c}{\textbf{\begin{tabular}[c]{@{}c@{}}Exploration \\ of the \\ Mapping Space\end{tabular}}} &
        \multicolumn{1}{c}{\textbf{\begin{tabular}[c]{@{}c@{}}Representing \\ the \\ Data Format\end{tabular}}} &
        \multicolumn{1}{c}{\textbf{\begin{tabular}[c]{@{}c@{}}Exploring \\ the \\ Format Space\end{tabular}}} &
        \multicolumn{1}{c}{\textbf{\begin{tabular}[c]{@{}c@{}}Representing \\ the \\ Architecture\end{tabular}}} &
        \multicolumn{1}{c}{\textbf{\begin{tabular}[c]{@{}c@{}}Exploring \\ Architectures\end{tabular}}} &
        \multicolumn{1}{c}{\textbf{\begin{tabular}[c]{@{}c@{}}Performance \\ Modeling\end{tabular}}} &
        \textbf{\begin{tabular}[c]{@{}c@{}}Software \\ Implementation\end{tabular}} \\ \midrule
        \rowcolor{lightgray!25}\textbf{TACO~\cite{Kjolstad:2017:TTA}} &
        \begin{tabular}[c]{@{}c@{}}Dense/Sparse \\ Tensor Algebra\end{tabular} &
        &
        \multicolumn{1}{c}{X} &
        \multicolumn{1}{c}{} &
        \multicolumn{1}{c}{X} &
        \multicolumn{1}{c}{} &
        \multicolumn{1}{c}{} &
        \multicolumn{1}{c}{} &
        \multicolumn{1}{c}{} &
        X \\
      \textbf{Timeloop~\cite{Parashar:2019:TSA}} &
        \begin{tabular}[c]{@{}c@{}}Dense \\ Tensor Algebra\end{tabular} &
        &
        \multicolumn{1}{c}{X} &
        \multicolumn{1}{c}{X} &
        \multicolumn{1}{c}{} &
        \multicolumn{1}{c}{} &
        \multicolumn{1}{c}{X} &
        \multicolumn{1}{c}{} &
        \multicolumn{1}{c}{X} &
        \\
         \rowcolor{lightgray!25}\textbf{Maestro~\cite{Kwon:2019:URP}} &
        DNN &
        &
        \multicolumn{1}{c}{X} &
        \multicolumn{1}{c}{X} &
        \multicolumn{1}{c}{} &
        \multicolumn{1}{c}{} &
        \multicolumn{1}{c}{X} &
        \multicolumn{1}{c}{} &
        \multicolumn{1}{c}{X} &
        \\
      \textbf{GAMMA~\cite{Kao:2020:GAH}} &
        DNN &
        &
        \multicolumn{1}{c}{X} &
        \multicolumn{1}{c}{X} &
        \multicolumn{1}{c}{} &
        \multicolumn{1}{c}{} &
        \multicolumn{1}{c}{} &
        \multicolumn{1}{c}{} &
        \multicolumn{1}{c}{X} &
        \\
         \rowcolor{lightgray!25}\textbf{Sparseloop~\cite{Wu:2022:SAA}} &
        \begin{tabular}[c]{@{}c@{}}Sparse \\ Tensor Algebra\end{tabular} &
        &
        \multicolumn{1}{c}{} &
        \multicolumn{1}{c}{} &
        \multicolumn{1}{c}{X} &
        \multicolumn{1}{c}{} &
        \multicolumn{1}{c}{X} &
        \multicolumn{1}{c}{} &
        \multicolumn{1}{c}{X} &
        \\
      \textbf{CoSA~\cite{Huang:2021:CSC}} &
        DNN &
        &
        \multicolumn{1}{c}{} &
        \multicolumn{1}{c}{} &
        \multicolumn{1}{c}{} &
        \multicolumn{1}{c}{} &
        \multicolumn{1}{c}{} &
        \multicolumn{1}{c}{} &
        \multicolumn{1}{c}{X} &
        \\
         \rowcolor{lightgray!25}\textbf{Mind Mappings~\cite{Hegde:2021:MME}} &
        \begin{tabular}[c]{@{}c@{}}Dense \\ Tensor Algebra\end{tabular} &
        &
        \multicolumn{1}{c}{} &
        \multicolumn{1}{c}{X} &
        \multicolumn{1}{c}{} &
        \multicolumn{1}{c}{} &
        \multicolumn{1}{c}{} &
        \multicolumn{1}{c}{} &
        \multicolumn{1}{c}{X} &
        \\
        \begin{tabular}[c]{@{}l@{}}\textbf{CIN-P}\\ \textbf{Autoscheduler}\end{tabular} &
        \begin{tabular}[c]{@{}c@{}}Sparse \\ Tensor Algebra\end{tabular} &
        &
        \multicolumn{1}{c}{X} &
        \multicolumn{1}{c}{X} &
        \multicolumn{1}{c}{} &
        \multicolumn{1}{c}{} &
        \multicolumn{1}{c}{} &
        \multicolumn{1}{c}{} &
        \multicolumn{1}{c}{} &
        X \\
         \rowcolor{lightgray!25}\textbf{WACO~\cite{Won:2023:WLW}} &
        \begin{tabular}[c]{@{}c@{}}Sparse \\ Tensor Algebra\end{tabular} &
        &
        \multicolumn{1}{c}{} &
        \multicolumn{1}{c}{X} &
        \multicolumn{1}{c}{} &
        \multicolumn{1}{c}{X} &
        \multicolumn{1}{c}{} &
        \multicolumn{1}{c}{} &
        \multicolumn{1}{c}{} &
        X \\
      \textbf{LoopTree~\cite{Gilbert:2023:LEE}} &
        \begin{tabular}[c]{@{}c@{}}Dense\\ Tensor Algebra\end{tabular} &
        &
        \multicolumn{1}{c}{X} &
        \multicolumn{1}{c}{} &
        \multicolumn{1}{c}{} &
        \multicolumn{1}{c}{} &
        \multicolumn{1}{c}{X} &
        \multicolumn{1}{c}{} &
        \multicolumn{1}{c}{X} &
        \\
         \rowcolor{lightgray!25}\textbf{DOSA~\cite{Hong:2023:DDM}} &
        DNN &
        &
        \multicolumn{1}{c}{} &
        \multicolumn{1}{c}{X} &
        \multicolumn{1}{c}{} &
        \multicolumn{1}{c}{} &
        \multicolumn{1}{c}{} &
        \multicolumn{1}{c}{X} &
        \multicolumn{1}{c}{} &
        \\
      \textbf{TeAAL~\cite{Nayak:2023:TDF}} &
        \begin{tabular}[c]{@{}c@{}}Dense/Sparse\\ Tensor Algebra,\\ Cascades of Einsums\end{tabular} &
        &
        \multicolumn{1}{c}{X} &
        \multicolumn{1}{c}{} &
        \multicolumn{1}{c}{X} &
        \multicolumn{1}{c}{} &
        \multicolumn{1}{c}{X} &
        \multicolumn{1}{c}{} &
        \multicolumn{1}{c}{X} &
        \\
        \rowcolor{lightgray!25}\begin{tabular}[c]{@{}l@{}}\textbf{\edge{} Language}\\ \textbf{(This Paper)}\end{tabular} &
        \begin{tabular}[c]{@{}c@{}}Dense/Sparse\\ Tensor Algebra,\\ Graph Algorithms (and beyond)\end{tabular} &
        X &
        \multicolumn{8}{c}{$\xleftarrow{\hspace*{5.3cm}}$~\textbf{Enabled for graph algorithms through \edge{}}~$\xrightarrow{\hspace*{5.3cm}}$} \\ \bottomrule
      \end{tabular}
    } %
    \end{table}

Different works within the large space of work in this area focus on different aspects: (1) expanding the set of concerns, (2) enabling expressivity of a particular concern, or (3) exploring the design space enabled by the separation of concerns.
Figure~\ref{fig:ta-soc} shows our view of the complex tensor algebra space.
Starting with a problem specification, an implementation must somehow optimize over the space of possible tensor algebra representations (algebraic space), over different architectures, over different data formats, and over different space-time mappings.
To factor complexity, prior tensor algebra tools separate out the concerns, representing each concern with a specification or language (Figure~\ref{fig:factored}).
Rather than searching the complex space in Figure~\ref{fig:ta-soc}, various tools can now explore searching different points in the subspaces of each concern (see Figure~\ref{fig:factored}).
\edge{} expands on this concept by extending Einsums to support graph algorithms.

Now, in the domain of tensor algebra, most problems map to a single algorithm that solves that problem.
The graph world is different.
Even the most fundamental problems (e.g., finding the depth of each vertex from a single-source vertex) have various algorithmic variants (e.g., push vs.\ pull \bfs{}, or a variant that deletes edges from the graph~\cite{Green:2021:IDB, Zhang:2018:ABG})\@.
Thus, our work differs from tensor-algebra systems in adding one more concern: algebraic manipulation.
Our hypothesis (and experience) is that the different algorithmic variants that may solve a particular problem are algebraically equivalent, and that algebraic manipulation starting with one algorithmic variant can allow us to find both existing and novel variants.

We believe there are two aspects to the separation-of-concerns philosophy:
\begin{enumerate}
    \item What are the concerns? Then, given a concern, how does one \emph{specify} the concern?
    \item Since the problem has been factored into smaller subproblems, how does one \emph{search} the new, factored space for efficient implementations?
\end{enumerate}

With these in mind, we can now explore the different concerns of tensor algebra. These concerns will also apply to our characterization of graph computation as Einsums, and thus we will be able to leverage the insights from the tensor algebra community (and the corresponding separation of concerns) in graph computation.

\subsubsection{Concern: What is the computation?}\label{ssec:einsum}
To precisely specify computations, the tensor algebra community uses Einstein summation notation, or Einsums.
For example, suppose the problem description is ``do a linear projection of a set of vectors in one space to a set of vectors in another space''.
The traditional Einsum for this problem description (i.e., matrix-matrix multiplication) is:
\begin{equation}\label{eqn:einsum}
Z_{m,n} = \sum^{K}_{k=0} A_{m,k} \times B_{k, n}
\end{equation}

Current extended Einsum notation (or tensor index notation~\cite{Kjolstad:2017:TTA}) drops the $\sum$:
\begin{equation}\label{eqn:tin}
    Z_{m,n} = A_{m,k} \times B_{k, n}
\end{equation}
Although Einsums of this form have existed for several years (see NumPy~\cite{Harris:2020:APN} and PyTorch~\cite{Paszke:2019:PAI}), we recently formalized an \emph{operational definition of an Einsum} (\ODE) that formally describes how to evaluate a given expression~\cite[Section 2.2]{Nayak:2023:TDF}\label{term:ode}.
The discussion here gives the reader operational intuition.
\S~\ref{sec:sets} gives the precise set-theoretic semantics for \edge{}, including the definitions of iteration space, tensor coordinate spaces, data space sets, and empty values.

At an intuitive level, an Einsum expression involves an \emph{iteration space} and the coordinate spaces of the tensors appearing in the expression.
The iteration space is the set of coordinate tuples over which the computation ranges.
The tensor coordinate spaces are the spaces into which the expression projects to read input values and write output values.
\S~\ref{sec:sets} refines these notions formally.

In Equation~\eqref{eqn:tin}, the coordinate variables in the iteration space are given by the indices appearing in the subscripts of the Einsum expression.
Thus, Equation~\eqref{eqn:tin} specifies an \emph{iteration space} of $[0, M) \times [0, K) \times [0,N)$, where indices $m$, $k$, and $n$ range over $[0, M)$, $[0, K)$, and $[0, N)$, respectively.
For each $(m, k, n)$ \emph{point} in the iteration space, evaluating this Einsum projects into the coordinate spaces of $A$, $B$, and $Z$, retrieving values from $A$ and $B$ and accumulating the computed result into the corresponding output location in $Z$.
In this example, the computation multiplies values from $A$ and $B$ and reduces, through addition, all computed values that map to the same output $(m,n)$ point in $Z$.

The Einsum defines how each point in the iteration space maps to points in the coordinate spaces of the tensors.
Coordinate expressions in the subscripts of an Einsum indicate \emph{how} each iteration-space point maps to a tensor coordinate-space point (see \S~\ref{ssec:conditionals}).
For example, the following expression shifts the input values by one:
\begin{equation}\label{eqn:shift}
  Z_{m} = A_{m+1},
\end{equation}
In Equation~\eqref{eqn:shift}, the iteration space is $[0, M)$.
For each $m$ point in the iteration space, the expression projects to the $m$ point in $Z$ and the $m+1$ point in $A$.
Due to this coordinate expression, this computation will never access the $A_{0}$ point.
Additionally, if a projection maps to a tensor coordinate that does not exist---for example, when $m = M-1$, the projection $A_{(M-1)+1}$ (or simply, $A_M$) is out of bounds on $A$---we assume a returned empty value (e.g., $0$ in this example).

Overall, (1) the iteration space expresses the set of possible point tasks for a given Einsum, (2) the tensor coordinate spaces express where those tasks read and write tensor values, and (3) the Einsum expression maps points in the iteration space to points in those tensor coordinate spaces.

This operational view determines whether a particular expression or extension to Einsums is valid (see \S~\ref{ssec:design-goals}).
To execute a traditional Einsum, the implementation must (1) walk each point in the iteration space;
(2) for each point, retrieve the corresponding input values;
(3) perform the computation indicated by the right-hand side of the Einsum; and
(4) accumulate the result into the corresponding output location, reducing multiple contributions when necessary\label{def:ode}.
\edge{} extends this operational definition to support graph-oriented features such as user-defined values and operators, the separation of \merge{} and \compute{}, and iteration.
\S~\ref{sec:sets} defines this extended operational view precisely using set-theoretic semantics.

An Einsum simply specifies \emph{what} computation to perform, and not \emph{how} to perform that computation (e.g., the order in which to walk the iteration space).
Thus, the Einsum could also be written as
\begin{equation}\label{eqn:tin2}
    Z_{m,n} = A_{k, m} \times B_{k, n},
\end{equation}
because coordinate order does not matter.

Instead, iteration order is left to a different concern: \emph{mapping}~\cite{Parashar:2019:TSA}.
We leave other aspects of \emph{how} computation occurs in space and time to other concerns such as mapping, data format selection, and binding to a specific architecture (see Figure~\ref{fig:factored}).
In fact, Einsums enable a separation of concerns.
Tools such as Halide~\cite{Ragan-Kelley:2013:HAL} use Einsum-like expressions to declaratively represent the computation, then represent the aspects of how computation occurs into other concerns.
\edge{} builds on this operational definition, adding extensions that denote constraints that optimize computation (see \S~\ref{ssec:merge}).

  \paragraph{Prior Work: Extending Einsum Notation}\label{sec:cascade}
   To specify their workloads, both \taco~\cite{Kjolstad:2017:TTA} and Timeloop~\cite{Parashar:2019:TSA} use Einsums together with additional workload specifications such as tensor shapes.
   Timeloop extends this concept to allow \emph{affine} coordinate expressions, which broaden the scope of expressible computations to include (for example) convolution. %
   For instance, the following expression indicates a 1D convolution with a window size of $S$~\cite{Nayak:2023:TDF}:
   \begin{equation}\label{eqn:1d}
        Z_{q} = I_{q+s} \times F_{s},
   \end{equation}
   where the iteration space is $Q \times S$ and a reduction occurs over $s$.
   Sparseloop~\cite{Wu:2022:SAA}, an analytical model for sparse tensor algebra accelerators, in addition to Einsums, adds workload specifications for sparsity.
   Finally, our prior work TeAAL~\cite{Nayak:2023:TDF} introduces the concept of a \emph{cascade of Einsums}, which represents problems as a directed acyclic graph (DAG) of Einsum expressions.
   In \edge{}, these will appear as a sequence of sub-Einsum expressions.
   This is particularly relevant for graph problems that are composed of multiple computational steps.

\subsubsection{Concern: How do we walk the iteration space?}
The next step after specifying an Einsum is to specify a \emph{mapping}: how to walk the iteration space in both logical space and time.
A common approach is to lower the Einsum to an abstract loop nest program such as HiFiber~\cite{hifiber, Nayak:2023:TDF}, then explore various ways to transform that loop nest.
For example, both \taco{}~\cite{Kjolstad:2017:TTA} and Timeloop~\cite{Parashar:2019:TSA} represent the mapping space as a series of nested loop nests, either using the polyhedral model~\cite{Kjolstad:2017:TTA} or with a custom specification language~\cite{Parashar:2019:TSA}.
However, the loop nest approach does not adequately capture the space-time aspects of computation, such as uneven distributions of computation across space or time.
Works like Maestro~\cite{Kwon:2019:URP} adopt a \emph{data-centric} approach to representing the mapping space, which enables a description of mapping aspects such as multicasting data and pipelining computations~\cite{Kwon:2019:URP}.
Our current approach derives from TeAAL~\cite{hifiber, Nayak:2023:TDF}, which adopts a loop nest paradigm; however, future work will explore adopting a more data-centric mapping approach to better accommodate graph algorithms on architectures like \GPU{}s.

\paragraph{Prior Work: Defining the Mapping Space}
 One aspect of work in this area is defining the mapping space and the types of eligible transformations to go from one mapping to another.
 There are two aspects to mapping: transformations on the \emph{data} and transformations on the \emph{iteration space}.
 Transformations on data take the input tensor(s) and rearrange their layout in memory or at runtime in some way. For example, swapping the ranks of a tensor (rank-swizzling), reordering values, or partitioning the data into chunks all constitute data transformations.
Nayak et al.\ separate out transformations on data as \emph{content-preserving transformations}~\cite{Nayak:2023:TDF}.
Transformations on the iteration space describe how to walk the iteration space, such as processing particular points in parallel or sequentially, walking in a particular order, or dividing the iteration space into parallel chunks that are then each processed sequentially.

 The mapping space for \emph{dense} tensor algebra is well-supported due to compilation support for dense loop nests~\cite{Kjolstad:2017:TTA, Backus:1957:FAC, Zhou:2022:RRA}, including dependency analysis~\cite{Boulet:1998:SPD} and polyhedral analysis~\cite{Lamport:1974:PED, Boulet:1998:SPD, Ancourt:1991:SPD, Pouchet:2011:LTC}.
 On the other hand, \emph{sparse} tensor algebra presents new challenges, all stemming from the idea that an efficient implementation should not need to touch every point in the iteration space but instead only those points in the iteration space that map to non-empty tensor values~\cite{Kjolstad:2017:TTA}.
 Additionally, there is the added complexity of accessing various sparse storage formats, each of which has different access costs.

\paragraph{Prior Work: Searching the Mapping Space}

\begin{wrapfigure}{l}{0.60\linewidth}
\begin{edgecodebox}[Python]
for m1 in range(0, M1):
  for n1 in range(0, N1):
    for k1 in range(0, K1):
      for m0 in range(0, M0):
        for n0 in range(0, N0):
          for k0 in range(0, K0):
            Z[M0 * m1 + m0, N0 * n1 + n0] +=
              A[M0 * m1 + m0, K0 * k1 + k0] *
              B[K0 * k1 + k0, N0 * n1 + n0]
\end{edgecodebox}
\caption{An example mapping for tiled dense matrix-matrix multiplication.}
\label{fig:mm-tiled}
\end{wrapfigure}

Different mappings have different implications on temporal and spatial locality, reuse, and overall performance.
For example, suppose we have a three-deep sequential loop nest for dense matrix-matrix multiplication (\acro{GEMM}), without any tiling.
Equation~\ref{eqn:tin} represents the Einsum for this computation.
There are six different possible mappings.
The mappings with $m$ and $n$ in the outer loop nests are output-stationary, i.e., they enable perfect reuse on the output tensor.
However, they have poor reuse and temporal locality on the input tensors, where values in $A$ and $B$ are accessed $N$ times and $M$ times, respectively.
To improve reuse across the input tensors, an implementation may choose to tile the tensors~\cite{Hegde:2019:EAA, Odemuyiwa:2023:ASD}.
Figure~\ref{fig:mm-tiled} shows an example tiled \acro{GEMM} loop nest.
By adding a level of tiling, we increase the shape of the iteration space to $M1 \times N1 \times K1 \times M0 \times N0 \times K0$.
This also increases the number of possible mappings (based on loop order alone) to ninety!
We calculate this number by counting the number of possible tiled \acro{GEMM} loop nests, each with a different ordering. %
This does not account for the possible choices for tile sizes, or the possible choices on which loop levels should be parallel or sequential.

Thus, a huge space of tensor algebra work focuses on how to explore the enumerable mapping space based on a cost model.
Techniques include manually searching the space~\cite{Nayak:2023:TDF}, and automated searches such as random search~\cite{Parashar:2019:TSA}, integer linear programming~\cite{Huang:2021:CSC}, genetic algorithms~\cite{Kao:2020:GAH}, neural networks~\cite{Hegde:2021:MME}, and asymptotic cost analysis~\cite{Ahrens:2021:AST}.
Note that several of these require some concept of the underlying architecture, including the number of memory levels and compute levels in the memory and compute hierarchies.

\subsubsection{Concern: How is data stored?}
Performance is closely tied to the data format of tensors.
Since the operands in an Einsum are simply an algebraic abstraction of a tensor, they do not themselves indicate \emph{how} the tensors are stored in memory.
Choosing a data format requires considering the costs for iterative access, random access, and insertion or deletion of values.
Chou et al.\ provide a format abstraction that describes each rank of a tensor as a level-based format~\cite{Chou:2018:FAF}, and integrate this into the \taco{} compiler as a format specification language.
In this model, each rank level has encoding choices, including dense, compressed, singleton, range, and hashed~\cite{Chou:2018:FAF, Wu:2022:SAA}, among others.
For example, a format specification for the $A$ tensor may indicate that $m$ is uncompressed, and $k$ is compressed. This, in turn, maps to a \CSR data format~\cite{Chou:2018:FAF}.
These abstractions have been adapted in subsequent tensor algebra work~\cite{Parashar:2019:TSA, Nayak:2023:TDF, Odemuyiwa:2023:ASD, Hegde:2019:EAA, Wu:2022:SAA}.

\subsubsection{Concern: What is the architecture?}
To allow an implementation to execute on different architectures, the tensor algebra accelerator community defines separate \emph{architecture specifications}.
First introduced with Timeloop~\cite{Parashar:2019:TSA}, an architecture specification indicates the available accelerator compute units, on-chip network information, and memory hierarchy.

\subsubsection{Concern: How do we bind computation?}
Software frameworks take the first three concerns (Einsum, Mapping, and Format) and compile them into platform-specific code~\cite{Kjolstad:2020:STA}.
The accelerator community, however, has an additional concern, called \emph{binding}, where mapping and data choices are bound to physical hardware units like memory modules or compute units~\cite{Parashar:2019:TSA, Wu:2022:SAA, Nayak:2023:TDF, Gilbert:2023:LEE}.
A binding specification describes which data components reside at each storage level, how data is accessed (eagerly or lazily~\cite{Nayak:2023:TDF}), and how long data can reside there~\cite{Nayak:2023:TDF, Pellauer:2019:BEC}. For computation units, binding specifies the computation allowed on each compute component in the architecture specification.

\subsection{A Complex Optimization Space}
Figure~\ref{fig:ta-soc} shows the intertwined design space of tensor algebra, where, ideally, an optimizer can look at the combinatorial space of concerns and select the best design point.
As mentioned earlier, this is too complex, hence tensor algebra tools have instead focused on separating concerns and exploring the optimization space within those concerns (Figure~\ref{fig:factored}).
Given a problem, the standard process is to provide a specification for each of the concerns and interfacing logic that understands each of the concerns, then combine them to generate a (hopefully) performant implementation.
The accelerator community focuses on modeling performance in hardware.
A designer can then go on to synthesize a domain-specific accelerator, but now, iterating on the design involves simply changing one of the concerns.
Meanwhile, the software community focuses on compiling to platform-specific code, but now, iterating on an implementation involves changing the specifications of one of the concerns.
More recently, tools are trying to now search over multiple concerns to generate better design points.
The co-optimizer \acro{WACO}~\cite{Won:2023:WLW} searches the design space by taking into account both mapping \emph{and} data format, while DoSA~\cite{Hong:2023:DDM} co-optimizes by searching both the mapping space and the architecture space.
Table~\ref{tab:related} shows prior work and the concerns on which they focus.

\subsection{Graph Algorithms to Tensor Algebra}\label{sec:graph-algorithms-to-tensor-algebra}
Our work, \edge{}, seeks to extend Einsum notation to support graph algorithms.
The matrix-graph duality is a well-known concept in the graph algorithm community.
\emph{Graph theory} focuses on transforming graph problems into linear algebra problems by leveraging this duality, where linear algebra concepts like matrix decomposition, matrix inverses, and eigenvectors become relevant~\cite{Chung:1997:SGT}.
Our work does not focus on this aspect.

Our work is closer to \graphblas{}, which provides a framework for implementations built using high-performance building blocks based on linear algebra primitives~\cite{Kepner:2011:GAL, Kepner:2016:MFO}.
However, \graphblas{} does not separate concerns to the extent of tensor algebra tools: instead, in \graphblas{}, imperative variables are combined with linear algebra expressions and computations do not operate on tensors beyond \nd{2} matrices.
Additionally, unlike our \edge{} notation, \graphblas intertwines computation on vertex and edge entities with computations on their features (see \S~\ref{ssec:merge}). \acro{GSTACO}~\cite{dima:2023:gstaco} is similar to \edge{} in spirit, but still entangles concerns of mapping, format, and binding with the problem representation itself.

In our case, Einsum notation is strictly used as a \emph{language} or specification for graph algorithms, with certain rules (see \S~\ref{ssec:ta-soc}) on how the language can be manipulated.
By using Einsums, we can leverage the separation of concerns found in tensor algebra, and the corresponding space of work in that area.
If such a systematic approach can be used for graph algorithms, it will unlock a new space of principled design exploration and analysis of this problem domain.
To enable this, we \emph{extend} Einsums so that we can fully express various graph algorithms as tensor algebra computations.
We hope that future graph systems using our advances will leverage existing tensor algebra systems, whose tools will need to support searching a problem space that allows general user-defined data types and computations (\S~\ref{ssec:computationspecs}),
exploring algebraic variants of the Einsum (\S~\ref{sec:5-mappings}),
and incorporating optimizations based on computations occurring on coordinates (\S~\ref{ssec:merge}).

\section{Why Extend Einsums Further?}\label{ssec:design-goals}
While current Einsum notation is expressive, graph algorithms introduce new challenges that are not fully addressed by current notation.
Current Einsums do not fully denote, among other aspects, (1) user-defined computations beyond semirings, (2) user-defined data values, (3) partial updates of tensors, or (4) recursive and iterative computations.
We extend the language of Einsums with the following goals in mind:
\begin{enumerate}
    \item\label{goal:sepconcerns} The language should enable a separation of concerns. %
    \item\label{goal:ode} Every extension to Einsums must adhere to the operational definition of an Einsum (see \S~\ref{def:ode}).
    \item\label{goal:supports} As a direct consequence of design goal~\ref{goal:ode}, the extensions should still support traditional tensor algebra computations enabled by existing tools (e.g., \taco{}~\cite{Kjolstad:2017:TTA}, TeAAL~\cite{Nayak:2023:TDF} and Timeloop~\cite{Parashar:2019:TSA}).
    In this way, we hope to leverage prior work in the tensor algebra space.
    \item\label{goal:opt} Where possible, the language should expose key optimization opportunities to an underlying compiler or modeling tool. %
        Moreover, there should be as few restrictions as possible on a compiler processing the \edge{} language.
    \item\label{goal:simple} At the same time, the language should be as simple and elegant as possible, while precisely describing the computation. %
    \item\label{goal:powerful} The language should be expressive and powerful enough to capture \emph{as many as possible} graph algorithms (and other domains!) of interest. %
    \item\label{goal:manipulation} The language should enable transformations between equivalent expressions in clear, distinct steps.
\end{enumerate}

In addition to these main design goals, we also have the following lower-priority goals that we pursue only if they do not conflict with our primary goals: %

\begin{enumerate}[resume]
  \item\label{goal:math} The language should ``look'' like mathematics as much as possible.
  \item\label{goal:traditional} Where possible, one should be able to apply traditional mathematical rules when manipulating \edge{} expressions (see \textbf{design goal~\ref{goal:manipulation}}).
  \item\label{goal:applications} While aiming for expressivity, our goal is \emph{not} to be a general-purpose programming language.
  However, we hope to better express other traditionally irregular applications using \edge{}, for example (but not limited to) sparse LU factorization (a step commonly used to reorder graphs~\cite{Gaihre:2022:SSS}), various sorting algorithms, and common parallel primitives such as prefix-sum. %
\end{enumerate}

The \edge{} language is our answer to these design goals.
Graph algorithms contain computational patterns found in many other non-tensor applications; thus, while \edge{} initially targeted graph algorithms, we find that \edge{} also expresses a large range of algorithms beyond both tensor algebra and graph algorithms. These include: matrix decomposition, multi-head attention, Graph\acro{SAGE}, robotics motion planning, general neural network applications, and Cholesky, among others.
In fact, researchers have already begun using \edge{} to express their own workloads~\cite{huang:2024:MTG, gilbert:2024:LEF, golden:2025:RDM, won:2025:insum, Zhang:2025:transfusion, shwatal:2024:ipd, Nayak:2024:FML_micro, Odemuyiwa:2025:FTF}.

In the following sections, we describe the \edge{} language and when possible, indicate where we factor in specific design goals.

\tikzexternaldisable
\section{\EDGEcaps{} Features}\label{ssec:walk-bfs}
\bfs{} is one of the simplest graph algorithms to understand and analyze, and it serves as the basis for many more complex algorithms~\cite{CLRS:2022:ITA}.
We walk through different aspects of the \edge{} language, building the notation and extensions to Einsums step by step.
\bfs{} serves as our primary running example (see \S~\ref{sec:prelims}), demonstrating how \edge{} expresses full graph algorithms.
Additionally, when introducing individual language features, we sometimes use simpler tensor algebra expressions (such as \acro{GEMM}) to isolate and motivate each extension before applying it to \bfs{}.
Note that \edge{} expressions remain valid regardless of the density or sparsity of the involved tensors.
We present the final, complete \bfs{} \edge{} specification near the end of this section (\bfs{} Example~\ref{sidebar:full_edge_bfs}).
Overall, \edge{} consists of the following features:
\begin{itemize}
  \item Specifications for user-defined data values and types, as well as empty values (see \S~\ref{ssec:workloadspecs}, \textbf{design goal~\ref{goal:powerful}})
  \item Specifications that indicate how a user initializes tensor values (\S~\ref{sssec:initspecs}, \textbf{design goal~\ref{goal:powerful}})
  \item Specifications for generic user-defined operators (\S~\ref{ssec:computationspecs}, \textbf{design goal~\ref{goal:powerful}}), enabled through map and reduce actions, together with a default populate action (\S~\ref{ssec:mapreducepopulate}, \textbf{design goals~\ref{goal:ode},~\ref{goal:simple},~\ref{goal:powerful}})
  \item A separation of merge operations (operations on entities) from compute operations (operations on attributes) (\S~\ref{ssec:merge}, \textbf{design goals~\ref{goal:ode},~\ref{goal:simple},~\ref{goal:powerful}})
  \item Support for iteration and recursion (\S~\ref{ssec:iter}, \textbf{design goals~\ref{goal:sepconcerns},~\ref{goal:ode},~\ref{goal:simple},~\ref{goal:powerful}})
  \item An extended \emph{populate} action for more complex output updates, such as filtering or fiber-level writes (\S~\ref{ssec:spopulate}, \textbf{design goal~\ref{goal:powerful}})
  \item Rank variable expressions that enable conditionals and constraints on the iteration space (\S~\ref{ssec:conditionals}, \textbf{design goal~\ref{goal:powerful}})
\end{itemize}
\subsection{What problem are we solving?}\label{ssec:workloadspecs}
Before specifying the algorithm, we need to understand the problem:
what are the inputs and outputs?
What do they mean?
This concern corresponds to the ``data properties'' portion of the ``workload specifications'' in Figures~\ref{fig:factored} and~\ref{fig:factored:edge}.

\begin{bfs_box}[label=sidebar:bfs-problem]{\bfs{} Problem Description}
  The problem specification (in English) for \bfs{} is:
  \begin{itemize}
      \item Input: An unweighted graph with $|V|$ vertices and $|E|$ edges.
      \item Input: An initial set of starting vertices, which we call the input \emph{frontier}, along with their starting depths.
      \item Output: The set of vertices reachable from the input frontier, along with the corresponding depth for each vertex.
  \end{itemize}
\end{bfs_box}

\subsubsection{Tensor Declarations}\label{sssec:decspecs}
We begin by defining the tensors used in a computation.
The \edge{} language provides a separate specification for the properties of the involved tensors, indicating:
\begin{enumerate}
  \item The input and output tensor names.
  \item For each tensor, its corresponding \emph{rank names} and \emph{rank shapes}.
        To distinguish between the \emph{rank names} of a tensor and \emph{accessing} these ranks, we refer to the lower-case subscripts (rank accesses) as \emph{rank variables}.
        Rank names are indicated by superscripts.
        In any subsequent Einsum expression, the rank variables refer to accesses to these ranks, which must be listed in the same order as the rank names, regardless of the rank variable names.
        Since rank names and rank variables can be multi-letter identifiers, we use commas to delineate them in a list of rank names or variables; for example, $G_{sm, dm}$ contains two rank variables, $sm$ and $dm$.
  \item For each tensor, its \emph{data type} and \emph{empty value}.
        While tensor algebra data is nearly always numerical, graph algorithms use a wider range of data types, including tuple data types and key-value labels used in property graphs~\cite{Angles:2018:PGD, Purohit:2021:SPG}.
        Thus, in \edge{}, we allow any user-defined data type beyond the traditional integer, floating-point, and Boolean types used by tensor algebra~\cite{Kjolstad:2017:TTA, Parashar:2019:TSA}.
        \edge{} also allows any value (not just zero) to indicate ``empty''.
        Identifying the empty value exposes an optimization opportunity (\textbf{design goal~\ref{goal:opt}}), where a sparse tensor can compress away \emph{empty} values instead of only zero values (\S~\ref{ssec:tensors}).
        This extension is also supported by prior work~\cite{Henry:2021:CSA, Buluc:2017:TGC, Davis:2019:ASG}.
\end{enumerate}
We write tensor declarations as $\mathit{TensorName}^{\mathit{shape}} \rightarrow \mathit{datatype}, \text{empty}=\mathit{value}$, where the arrow ($\rightarrow$) separates the shape specification from the data type and empty-value specification.
\bfs{} Example~\ref{sidebar:bfs-datatype} provides an example for \bfs{}\@.

\begin{bfs_box}[label=sidebar:bfs-datatype]{\bfs{} Declaration Specification in \edge{}}
  \begin{subequations}\label{eqn:init}
    \begin{align}
      & \triangleright \text{Tensors} \notag\\
      G^{S\equiv|V|, D\equiv|V|} & \rightarrow \text{integer, empty}=0 \label{seqn:G}\\
      F^{S\equiv|V|} & \rightarrow \text{integer, empty}=\infty \label{seqn:F}\\
      P^{D\equiv|V|} & \rightarrow \text{Boolean, empty}=\False \label{seqn:P}
    \end{align}
  \end{subequations}

  Equation~\eqref{eqn:init} defines three tensors: $G$ (the graph), $F$ (the frontier), and $P$ (the set of visited nodes).
  Rank $S$ in $G$ has shape $|V|$, meaning it has a coordinate for every vertex in the graph.
  Likewise, rank $D$ has shape $|V|$.
  Each possible coordinate in $S$ refers to a source vertex, and each possible coordinate in $D$ refers to a destination vertex.
  The frontier $F$ and the set of visited nodes $P$ are both \nd{1} tensors, with shapes equal to the number of vertices in the graph.

  Although $G$ has \emph{rank names} $S$ and $D$ (indicated by superscripts), we may use different symbols to indicate \emph{rank accesses} (indicated by subscripts in expressions).
  In the expression $G_{q,r}$ (or equivalently $G_{sm, dm}$, using multi-letter names as described above), the rank variable $q$ accesses $G$'s $S$ rank, and the rank variable $r$ accesses its $D$ rank.

  Turning to data types and empty values: we assume that $G_{s,d}$ contains a value of $1$ whenever an edge exists between vertices $s$ and $d$, and $0$ otherwise.
  Here, the subscripts $s$ and $d$ indicate \emph{rank accesses}, that is, projections into the $G$ data space to retrieve the value at data point $(s,d)$.

  Tensor $F_{s}$ represents the active set of source vertices (the ``frontier'').
  The integer value of $F$ at each coordinate $s$ indicates the depth at which \bfs{} discovers the corresponding vertex.
  At a given coordinate $s$ in $F$, its \emph{value} is either empty if $s$ is not in the active set, or the corresponding integer depth otherwise.
  $F$ has an empty value of $\infty$ for vertices not yet discovered.

  Tensor $P_{d}$ is a Boolean tensor that indicates whether a vertex $d$ has been visited (\True) or not (\False), with an empty value of \False~\eqref{seqn:P}.
\end{bfs_box}

The declaration section provides guidance to an underlying compiler on how to cast data types during computation when operands have different data types (\textbf{design goal~\ref{goal:opt}}).
For example, suppose we declare the output tensor $Z$ in a \acro{GEMM} computation (Equation~\eqref{eqn:tin2}) to have a Boolean data type, while the input tensors $A$ and $B$ have integer data types.
An underlying compiler can either (1) cast the \acro{GEMM} computation to Boolean values or (2) optimize the multiplication and reduction steps to use Boolean checks instead.
Additionally, \edge{}'s explicit declaration section removes ambiguity from the main Einsum expressions with respect to tensor data types, shapes, and empty values (\textbf{design goal~\ref{goal:simple}}).

\subsubsection{Tensor Initializations}\label{sssec:initspecs}
The \edge{} language also provides a specification for \emph{initializations}, as shown in \bfs{} Example~\ref{sidebar:bfs-initspecs}.
By default, all tensors are initialized with their empty values.
The initialization section indicates which tensors are user-specified, as well as any algorithm-based initializations that must occur.
Algorithm-based initializations are themselves expressed as Einsum expressions (see \S~\ref{ssec:syntaxinit}).

\begin{bfs_box}[label=sidebar:bfs-initspecs]{\bfs{} Initialization Specification in \edge{}}
  In our \bfs{} example, the initialization specification is:
  \begin{subequations}\label{eqn:initialization}
    \begin{align}
      & \triangleright \text{Initialization} \notag\\
      G &= \langle\text{user-specified}\rangle \label{seqn:GI}\\
      F_{s:s\in \textit{id}} &= 0 \label{seqn:FI} \\
      P_{d:d\in \textit{id}} &= \True \label{seqn:PI},
    \end{align}
  \end{subequations}
  \noindent
  where we assume that $\textit{id}$ is a user-specified list of starting vertices.

  In this case, $G$ is user-specified.
  A program based on the \edge{} language may allow a user to load a file (such as a Matrix Market file~\cite{Boisvert:1996:MMW}) or construct a graph data structure programmatically.

  Expressions~\eqref{seqn:FI} and~\eqref{seqn:PI} are Einsums indicating that only coordinates $s$ in $F$ and $d$ in $P$ that are in the $\textit{id}$ list are initialized to non-empty values (initial depth $0$ and \True for the visited starting nodes, respectively).
  These Einsum expressions use our extended notation for \emph{conditionals}, which we describe in \S~\ref{ssec:conditionals}.
\end{bfs_box}

\subsection{Enabling General Computations}\label{ssec:computationspecs}
We extend Einsum expressions to support graph algorithms and to expose common graph optimizations to tools in the optimization space (\textbf{design goal~\ref{goal:opt}}).

Graph algorithms typically involve mathematical or logical operations beyond traditional multiplication and addition computations; thus, \edge{} allows user-defined computation (\textbf{design goal~\ref{goal:powerful}}), specifically:
\begin{enumerate}
  \item any user-defined \compute\ between two tensors,
  \item any user-defined \compute\ that gathers results into a single value, and
  \item any user-defined \compute\ that takes an input expression and places it into an output tensor.
\end{enumerate}
Additionally, tensors can contain any data type and can use any value to indicate emptiness.

\begin{alternative_box}[label=sidebar:alt-semiring]{Design Alternatives for Expressing User-Defined Computations}
\paragraph{\textbf{Enforcing Semiring Constraints:}}
One design alternative is to enforce specific algebraic properties on computations.
\graphblas maps computations to overloaded linear algebra operations described as semirings.
A semiring consists of a \emph{monoid} with an associated \emph{identity} value and a \emph{multiply operator}~\cite{Kepner:2016:MFO, Davis:2019:ASG}.

Scalar addition in matrix multiplication becomes a monoid: a binary operator with an identity value such that applying the operator to any value and the identity returns the value, and that is associative and commutative.
Any user-defined \compute\ overloading addition must satisfy these constraints.

The multiply operator is also binary, taking two input values to produce an output value.
The data type of the output value must match the data type of the monoid's identity.

With this approach, users can only express computations that can be cast as semirings.
Other computations---e.g., having one tensor be the exponent of another---are not expressible (\textbf{design goal~\ref{goal:applications}}).
In \edge{}, we allow general user-defined computations, while still preserving the ability to recognize and exploit semiring structure when it is present.
Our separation of \merge\ operations from \compute\ operations (\S~\ref{ssec:merge}) also admits expressions beyond semirings.
Semiring computations are therefore a \emph{subset} of the computations that our future algebra engine can exploit.
\end{alternative_box}

We now incrementally build the \edge{} notation.
To isolate each extension, we use simple tensor algebra expressions as minimal working examples, starting with \acro{GEMM}.
We return to \bfs{} once the notation is in place (see \bfs{} Example~\ref{sidebar:gather}).

\begin{enumerate}
  \item First, starting from Equation~\eqref{eqn:tin2}, which explicitly specifies multiplication, we introduce a general \emph{operation label} ``$\cdot$'' in the core expression:
  \begin{equation}\label{eqn:tin3}
    Z_{m,n} = A_{k, m} \cdot B_{k, n}.
  \end{equation}
  The operation label indicates that values accessed from the operands at a given iteration-space point participate in a user-defined \compute.

  \item In \edge{}, we keep the iteration-space structure and the \compute\ specification distinct.
  We therefore use a postfix \emph{modifier} (written after ``$::$'') to define the concrete \compute\ associated with the operation label:
  \begin{equation}\label{eqn:tin4}
    Z_{m,n} = A_{k, m} \cdot B_{k, n} :: \times.
  \end{equation}
  The expression to the left of ``$::$'' emphasizes the iteration space and coordinate accesses, while the modifier specifies the compute performed at each relevant point.

  \begin{alternative_box}[label=sidebar:alt-postfix]{Design Alternatives for Postfix Notation}
    Valid alternatives to postfix notation include infix and prefix notation.
    However, we find that both approaches clutter the core Einsum expression.
    Infix notation makes the \compute\ explicit between operands but becomes unwieldy as expressions grow more complex.

    Postfix notation separates the expression into two conceptual phases:
    \begin{enumerate}
      \item Left of ``$::$'': the iteration space and coordinate access variables.
      \item Right of ``$::$'': the \compute\ operations performed, as well as any constraints placed on the iteration space (see \S~\ref{ssec:merge}).
    \end{enumerate}
    Infix notation visually intertwines these phases, while prefix notation places the modifier before the iteration space has been defined.
  \end{alternative_box}

\item\label{item:mr}
  Many computations consist of both (i) \emph{pairwise} work at each iteration-space point and (ii) \emph{aggregation} of multiple contributions that target the same output coordinate.
  In \acro{GEMM}, for example, we perform $K$ pairwise multiplications for each $(m,n)$ and then aggregate those products by addition.
  \edge{} makes this structure explicit by introducing three actions:
  \begin{enumerate}
    \item a \maptxt\ action (denoted $\bigwedge$) that specifies the pairwise \compute,
    \item a \reduce\ action (denoted $\bigvee$) that specifies how multiple contributions are combined, and
    \item a \populate\ action that writes reduced results into the output tensor (\S~\ref{ssec:mapreducepopulate}).
  \end{enumerate}
  While in \acro{GEMM} the \populate\ action is simply assignment, naming it as a distinct action allows \edge{} to express computations where writing to the output is more involved, such as filtering results based on coordinates or updating an entire fiber rather than a single point (\S~\ref{ssec:spopulate}).
  Concretely, \edge{} uses the \maptxt\ and \reduce\ actions to produce \maptmps and aggregate them into \redtmps, while \populate\ materializes the final output.
  With this notation, we can express \acro{GEMM} in \edge{} as:
  \begin{equation}\label{eqn:tin5}
    Z_{m,n} = A_{k, m} \cdot B_{k, n} :: \bigwedge_k \times \bigvee_k +.
  \end{equation}

  The subscript on $\bigwedge_k$ indicates that the \maptxt\ action combines values sharing the same $k$ coordinate, while $\bigvee_k$ indicates that the \reduce\ action aggregates over the $k$ rank. We often drop these subscripts as they are redundant: at a given $(m, n, k)$ point in the iteration space, a pair of values from $A$ and $B$ will always be accessed, and will always share the same $k$ rank coordinate.

  In English, we read Equation~\eqref{eqn:tin5} as follows:
  ``For each $(m,n)$ coordinate\ in $Z$, the \maptxt\ action multiplies values from $A$ and $B$ with matching $k$ coordinates, the \reduce\ action aggregates the resulting products across all $k$, and the \populate\ action writes the reduced result into $Z$ (here, a simple assignment).''

  \item Certain computations benefit from skipping ineffectual work, that is, work that would involve empty values.
  In such cases, only a subset of the iteration space should be traversed, typically due to sparsity in one or more operands.
  We therefore separate \emph{merge}/\emph{coordinate} operations, which operate on operand coordinates, from \emph{compute} operations, which operate on operand values.
  At each iteration-space point, the \merge\ operation first inspects only the coordinates of the operands to decide whether the iteration-space point is effectual.
  If it is, the \compute\ operation processes the corresponding values (returning a computed value); otherwise the point is skipped entirely and the values are never fetched.
  Precisely specifying \merge\ operations therefore enables us to avoid ineffectual \compute\ operations.
  For \maptxt\ actions, \merge\ operations determine how operand coordinates are paired.
  For \reduce\ actions, \merge\ operations determine which coordinates contribute to an aggregation.
  In \acro{GEMM}, multiplication uses the \merge\ operation of \emph{intersection} (denoted by $\cap$)~\cite{Hegde:2019:EAA}, meaning that we only consider multiplications where both operands have non-empty values.
  Addition uses the \merge\ operation of \emph{union} (denoted by $\cup$)~\cite{Henry:2021:CSA}, meaning that we add when at least one operand is non-empty.
  \S~\ref{ssec:merge} provides further details on the \merge\ and \compute\ operators in \edge{}\@.
  Adding \merge\ operators yields the following \edge{} expression for \acro{GEMM}:
  \begin{equation}\label{eqn:tin6}
    Z_{m,n} = A_{k, m} \cdot B_{k, n} :: \bigwedge \times(\cap) \bigvee +(\cup).
  \end{equation}

  \item Many graph algorithms are iterative or recursive.
  Einsums naturally support iteration by introducing a \emph{generational} (also \emph{iterative}) rank for tensors that evolve over time.
  \S~\ref{ssec:iter} provides further details on generational ranks.
  We can express a computation in which the output of one iteration is used as an input to the next, for some number of iterations, as:
  \begin{equation}\label{eqn:tin7}
    \begin{split}
      Z_{i+1, m,n} &= A_{k, m} \cdot Z_{i, k, n} \\
      &:: \bigwedge \times(\cap) \bigvee +(\cup) \\
      \diamond: i \geq 5
    \end{split}
  \end{equation}
  where $\diamond$ denotes a stopping condition that halts iteration.
  \bfs{} uses a more meaningful stopping condition (an empty frontier) rather than a fixed iteration count; we return to this in \S~\ref{ssec:iter}.

  \item Finally, to express operations such as concatenation and conditionals, \edge{} allows arbitrary rank-variable expressions that determine how tensors are accessed.
  For example, the expression
  \begin{equation}\label{eqn:cond-access}
    Z_{m, n} = A_{m, f(n)}
  \end{equation}
  applies the user-defined function $f$ to the rank variable $n$ before accessing $A$, enabling patterns such as index remapping and conditional access (see \S~\ref{ssec:conditionals}).
\end{enumerate}

The remaining subsections formalize these extensions, beginning with the execution model for map, reduce, and populate in \S~\ref{ssec:mapreducepopulate}.

\subsection{Extension: Map, Reduce, and Populate Actions}\label{ssec:mapreducepopulate}
Every Einsum expression consists of a combination of one or more patterns:
\begin{enumerate}\label{list:patterns}
  \item\label{item:shared} shared indices that appear in multiple input tensors (e.g., $k$ in Equation~\eqref{eqn:tin6}),
  \item\label{item:reduce} indices that appear in one or more input tensors but not the output tensor (e.g., $k$ in Equation~\eqref{eqn:tin6}),
  \item\label{item:populate} indices that appear both in the input and the output (e.g., $m$ and $n$ in Equation~\eqref{eqn:tin6}), and
  \item\label{item:broadcast} indices that appear in the output tensor but in none of the input tensors (e.g., $p$ in the expression \inlineequation[eq:broadcast]{Z_{m, p} = A_{m}}).
\end{enumerate}

In conventional Einstein notation (as commonly used in tensor algebra), these patterns are typically paired with fixed operators: pattern~\ref{item:shared} indicates multiplication on the corresponding values (or an overloaded multiplication), pattern~\ref{item:reduce} indicates reduction (via overloaded addition) over the corresponding values, pattern~\ref{item:populate} indicates assignment to the output tensor, and pattern~\ref{item:broadcast} indicates repetition (where each value in $A_{m}$ is repeated $n$ times for a given $m$ for Equation~\eqref{eq:broadcast}).

Traditional Einsums conflate the patterns of computation with the actual computation operators.
Separating computations into \maptxt, \reduce, and \populate\ actions allows \edge{} to
factor a wide range of user-defined behavior into a small number of recurring
\emph{patterns}. Each action operates over different kinds of objects: \maptxt\
combines pairs of operand values, \reduce\ combines a set (or stream) of
intermediate values into one value per output coordinate, and \populate\
materializes results by writing, deleting, or leaving unchanged elements within
output fibers.
This separation exposes optimization opportunities (\textbf{design goal~\ref{goal:opt}}),
since an implementation can optimize the common \emph{pattern} independently of
the user-defined operators.
For example, a hardware designer may implement each of these actions as hardware primitives---where reduce is mapped to a reduction tree, map to a merger network, etc.---or even combine all three actions into a single network to reduce area overhead~\cite{Martinez:2023:FMD}.

\subsubsection{The \maptxt\ Action}\label{sssec:map}
\maptxt\ operates over operand values accessed at iteration-space points.
We denote \emph{map} actions with $\bigwedge$.
Each \maptxt\ action is associated with a specific operation label ($\cdot$), and the corresponding \compute\ operator appears in the modifier following that action.
When multiple operation labels appear in an expression, \edge{} distinguishes them using superscripts.

Operationally, each application of a \maptxt\ action at a given
iteration-space point produces a \emph{map temporary} (\maptmp).
A \maptmp\ represents the result of applying the map compute operator
to the operand values accessed at that iteration-space point.
These \maptmps are consumed by subsequent \reduce\ actions (and, ultimately, by
\populate), but we defer the formal definition to \S~\ref{sec:sets}.

For example, we describe element-wise multiplication across three vectors as:
\begin{align}\label{eqn:tin8}
  Z_{m} &= \bigl( A_{m} \cdot^1 B_{m} \bigr)_{m} \,\cdot^2\, C_{m} :: \bigwedge^1 \times \bigwedge^2 \times,
\end{align}
where $\bigwedge^1$ refers to a \maptxt\ action between $A$ and $B$ for operation label $1$ (that is, $\cdot^1$),
and $\bigwedge^2$ refers to a \maptxt\ action between the result of the first expression and $C$ for operation label $2$.

The parenthesized subexpression produces an \emph{anonymous tensor} with rank shape $M$ and rank variable $m$, as indicated by the subscript on the parentheses.
Intuitively, this anonymous tensor corresponds to the collection of \maptmps
produced by $\bigwedge^1$ at each $m$ point, which then serve as the left operand
to $\bigwedge^2$.

We require two \maptxt\ actions in this example because, for now, \edge{} enforces that each \maptxt\ action operates on exactly two operands (see Inset~\ref{sidebar:alt-patterns}).
Likewise, \edge{} expressions will always contain two input tensors and one output tensor, hence the need for \emph{anonymous} tensors.

\paragraph{Programming Model for Map:}
As an implementation walks the iteration space, the map action indicates which
regions of the data spaces of two operands to combine.
For each point in the iteration space, the Einsum indicates the data space accesses.
The implementation applies the map compute operator to the accessed values and
emits the resulting \maptmp.
For example, given a map action of $\bigwedge^1 \times$, we read the expression as:
``for a given point in the iteration space, multiply the corresponding data values
for the two operands of operation label $1$, producing a \maptmp.''

Likewise, for a map action of $\bigwedge^4 +$, we read the expression as:
``for a given point in the iteration space, add the corresponding data values
for the two operands of operation label $4$, producing a \maptmp.''

\begin{example_box}[label=sidebar:example-map2]{Element-Wise Addition}
  \begin{align}\label{eqn:ewadd}
    Z_{m, n} &= A_{m, n} \cdot B_{m, n} :: \bigwedge +,
  \end{align}

  Following the operational definition of an Einsum, an implementation will walk the $M \times N$ iteration space.
  For each $(m, n)$ point, the implementation projects into the data space and accesses the corresponding data values in $A$ and $B$.
  The data values correspond to the same $(m, n)$ point in both $A$ and $B$, thus, since the \emph{map} action is also on each $(m, n)$ point, the implementation combines the two accessed values through addition ($+$).
\end{example_box}

\begin{example_box}[label=sidebar:example-map3]{Outer Product}
The following Einsum---with \edge{} extensions---expresses the outer product of two vectors:
\begin{subequations}\label{eqn:cross}
  \begin{align}
      Z_{m, n} &= A_m \cdot B_n :: \bigwedge \times\\
  \end{align}
\end{subequations}
Read the expression as: ``For each $(m, n)$ point in the iteration space, multiply the corresponding data values and place them in the $(m, n)$ location in $Z$.''
\end{example_box}

\begin{wrapfigure}{l}{0.52\linewidth}
\vspace{-1.0\baselineskip}
\begin{edgecodebox}[Python]
for k in range(0, K):
    for m in range(0, M):
        Z[m] = min(Z[m], A[m, k])

# Another mapping option
for m in range(0, M):
    for k in range(0, K):
        Z[m] = min(Z[m], A[m, k])
\end{edgecodebox}
\vspace{-0.5\baselineskip}
\caption{Two example loop nests for Equation~\eqref{eqn:reduceex}.}
\label{fig:reduceex}
\end{wrapfigure}

\subsubsection{The \reduce\ Action}\label{sssec:reduce}
In the operational definition of a traditional Einsum, when an output point already contains a value, the Einsum will \emph{reduce} the value there with the computed value on the right-hand side (RHS) of the Einsum.
Thus, \reduce\ combines values that correspond to the same output point into a single result.

We denote \reduce\ actions with $\bigvee$.
A \compute\ operator must follow a reduce action---that is, the computation applies to the \reduce\ action.
Additionally, each \reduce\ action pertains to a tensor on the \acro{RHS} of the Einsum: either a single input tensor or the implicit, partial tensor created by an operation label ($\cdot$) expression (the \maptmp).
The compute operation for \reduce\ must be binary.

\paragraph{Programming Model for \reduce:}
Conceptually, the \reduce\ action consumes the stream of \maptmps produced by \maptxt\ and incrementally combines them into a single value
per output coordinate.
For each \maptmp, \reduce\ identifies the corresponding output
coordinate and merges the incoming value with the current reduction state using
the specified \merge\ and \compute\ operators.

Operationally, this can be viewed as a fold over the set of \maptmps associated
with a given output coordinate.
Each \maptmp\ contributes at most one update to the reduction state, and the
final reduced value is obtained once all relevant \maptmps have been processed. The final reduction state is a \redtmp. There are as many reduction states as there are output locations.

Because \reduce\ operates over \maptmps rather than directly over
tensor operands, all \maptxt\ actions that produce \maptmps for a given
iteration-space point must logically occur before the corresponding \reduce\
update for that point.
However, there are no ordering constraints across different iteration-space
points (see the two mapping choices in
Example~\ref{sidebar:example-red1}).
As a result, an implementation may freely interleave \maptxt\ and \reduce\ actions
across iteration-space points, enabling pipelined, streaming, or parallel
execution.

\begin{example_box}[label=sidebar:example-red1]{Single Operand Reduction}
Suppose we want to reduce the $k$ rank of a tensor by selecting the minimum value across $k$:
\begin{subequations}\label{eqn:reduceex}
\begin{align}
  Z_{m} &= A_{m, k} :: \bigvee \min .
\end{align}
\end{subequations}

As execution walks the iteration space ($M \times K$), it produces a stream of \emph{map temporaries} (here, the RHS value $A_{m,k}$ at each $(m,k)$ point) and incrementally combines them into a \emph{reduction temporary} (\redtmp) per output coordinate $m$.

Concretely, for each $(m,k)$ point, \reduce\ updates the current reduction state for $m$ by applying $\min$ to the incoming value $A_{m,k}$ and the current reduction state for $m$.
After all relevant $k$ values for a fixed $m$ have been processed, the resulting \redtmp\ holds the minimum value for that $m$ coordinate.
Finally, \populate\ (\S~\ref{sssec:populate}) materializes $Z$ by writing the \redtmp\ value into the corresponding output coordinate $Z_m$.

An \emph{implementation} may choose different loop orders (mapping choices) for iterating $M \times K$.
For example, iterating $m \rightarrow k$ fixes $m$ and streams all $k$ values for that $m$ through the reduction state before moving to the next $m$.
Iterating $k \rightarrow m$ instead fixes $k$ and streams all $m$ coordinates, updating each $m$ reduction state as it goes.
Both produce the same final \redtmp\ values (and thus the same materialized $Z$) as long as the implementation processes the same set of iteration-space points and applies $\min$ consistently.
Figure~\ref{fig:reduceex} shows two such loop nests.

  Note that these are just two possible implementations for the reduction; if, for instance, a particular compute operator for reduction is associative, then an implementation may choose to use a reduction tree or other parallel methods to execute the reduce action.
\end{example_box}

\begin{example_box}[label=sidebar:example-red2]{Reduction on a Compute Label}
  Suppose we have an expression that performs \acro{GEMM} on two tensors $A$ and $B$, then does an element-wise addition with a third tensor, $C$.
  \begin{align}
    Z_{m, n} &= \bigl(A_{k, m} \cdot^1 B_{k, n}\bigr)_{m, n} \,\cdot^2\,C_{m, n} :: \bigwedge^1 \times \bigvee^1 + \bigwedge^2 + \label{eqn:tin10}
  \end{align}
  Here, the \reduce\ action applies to the implicit, temporary \emph{anonymous tensor}  created by the first operation label ($\cdot^1$).
  Thus, \emph{for a given point in the iteration space}, an implementation must apply the \maptxt\ action first (multiplication between two data values), before applying the \reduce\ action.
  Note that this is the only constraint.
  The implementation may choose to create an entire $M \times N \times K$ partial tensor resulting from the \maptxt\ action on operation label $1$, then reduce the $k$ rank before applying the second operation label (element-wise addition with $C$, $\cdot^2$).
  Alternatively, an implementation may choose to process the entire Einsum point-by-point, multiplying two data values in $A$ and $B$ and adding the result to the current value in $C$'s $(m, n)$ location.
  Note that since the second operation label is addition, the addition with $C$ can only occur once.
\end{example_box}

In general, for a given tensor (explicit or anonymous), \reduce\ actions are applied in order of their appearance in the \edge{} expression.
Note that algebraic manipulations to the expression (see Figure~\ref{fig:ta-soc} and \S~\ref{sec:5-mappings}) may change the order depending on whether the \compute\ operators for the \reduce\ actions are commutative.

\begin{example_box}[label=sidebar:example-red3]{Order of Operations with Multiple Reduce Actions}

A given operation label ($\cdot$) in an \edge{} expression \emph{can only have ONE \reduce action}.
For example, to reduce over two different ranks with different compute operators, the expression must be split into two Einsums (a \emph{cascade} of Einsums~\cite{Nayak:2023:TDF}, which represents a DAG of individual Einsums):
  \begin{subequations}\label{eqn:tin11c}
    \begin{align}
        T_{p, k} &= A_{p, j, k}:: \bigvee \times \label{eqn:tin11ca}\\
        Z_{p}    &= T_{p, k} :: \bigvee + \label{eqn:tin11cb}
    \end{align}
  \end{subequations}
At a given $(p, j, k)$ point in the iteration space, the Einsum reduces into $T_{p, k}$ using the \compute operator of multiplication (Equation~\eqref{eqn:tin11ca}).
The Einsum can then reduce into $Z_p$ through addition ($Z_p$, Equation~\eqref{eqn:tin11cb}).
Note that an implementation may also choose to fully create the first temporary tensor, $T$, before $Z$, instead of processing them point-by-point.
\end{example_box}

\begin{example_box}[label=sidebar:example-red4]{Reading a Reduce Action}
  Suppose we want to reduce a tensor over two ranks, using addition.

  We specify the computation by indicating a single reduce action over both $j$ and $k$:
  \begin{subequations}\label{eqn:tin12b}
    \begin{align}
        Z_{p} = A_{p, j, k}:: \bigvee +
    \end{align}
  \end{subequations}
  We read the reduce action as follows: For a given $p$ region of the iteration space, gather all data values in the data space corresponding to all the possible $j$ and $k$ coordinates.
  Add the data values together and place in their corresponding $p$ location in the $Z$ tensor.
  More precisely, at a given point in the iteration space, add the \acro{RHS} value ($A_{p, j, k}$) to the current value in the \reduce\ state.
\end{example_box}

\subsubsection{The Default \populate\ Action}\label{sssec:populate}

\populate\ is the act of materializing (or \emph{populating}) the output tensor with values.
For a given point in the iteration space, the default populate action takes the currently computed value---the \redtmp, determined by the \acro{RHS} of the Einsum---and places it at the corresponding data point of the output element.
Note that the Einsum notation separates the \acro{LHS} and the \acro{RHS} of the Einsum by an assignment symbol ($=$).
Thus, the default \populate\ operation is simply assignment ($=$).
\emph{Assignment} assumes the initial state of the output tensor is entirely filled with empty values.
After fully computing the Einsum expression, \emph{assignment} (or the default \populate\ action) replaces the output tensor with the computed values.
This maps to pattern~\ref{item:populate} in \S~\ref{ssec:computationspecs}.

\begin{example_box}[label=sidebar:example-pop1]{Assignment as the Default Populate Action}
    As an example, suppose we have the following Einsum:
      \begin{align}
        Z_{m} = A_{m} \cdot B_{m}\label{eqn:pop1}
      \end{align}
  The assignment to $Z$ indicates that the computation on the \acro{RHS} completely fills (``populates'') the corresponding $m$ rank of the output.
  This assignment exists in every expression: every Einsum consists of an output tensor that the Einsum populates with the computation indicated on the \acro{RHS}\@.
\end{example_box}

Pattern~\ref{item:broadcast} in \S~\ref{ssec:computationspecs} highlights the situation where an output tensor may have a rank that does not appear in any of the input tensors.
Suppose the shape of that rank is $N$.
In such a scenario, the Einsum \emph{broadcasts} or copies the computed value on the \acro{RHS} to the corresponding output rank $N$ times.
Again, a simple assignment symbol ($=$) is enough to represent this pattern.

\begin{example_box}[label=sidebar:example-pop2]{Broadcast and the Default Populate Action}
  Equation~\eqref{eqn:pop2} shows an example:
      \begin{align}
        Z_{m, p} = A_{m} \cdot B_{m}\label{eqn:pop2}
      \end{align}

  Following the operational definition of an Einsum, each point $m, p$ in the iteration space computes a value on the \acro{RHS} and places it at the corresponding $m, p$ location in $Z$.
  But note that no input on the \acro{RHS} has a $p$ rank.
  Recall that the \reduce\ action contains as many states (or \redtmps) as the shape of the output tensor.
  This creates $M \times P$ \redtmps.
  An \edge{} interpreter may choose to repeat the \maptxt\ computation $P$ times, or simply \emph{broadcast} a given \maptxt\ to the $P$ reduction states.
  \populate\ then copies each $(m, p)$ \redtmp\ to the corresponding $Z$ location.
\end{example_box}

Every Einsum has a default \populate\ action, implied by the assignment symbol ($=$), where non-empty coordinates appearing on the right-hand side are copied to the matching rank on the left-hand side.
If coordinates on one side do not appear on the other side, then a reduction or broadcast \emph{must} occur (see \S~\ref{ssec:mapreducepopulate}, points~\ref{item:reduce} and~\ref{item:broadcast}).
Several problems need more complex populate operators, where the Einsum fills (populates) only the elements in the output tensor that satisfy a user-defined function, leaving the other elements unchanged.
We discuss this further in \S~\ref{ssec:spopulate}.

\subsubsection{Combining Actions}\label{sssec:combining_actions}
Overall, \edge{} factors user-defined functions into three classes of actions: \maptxt, \reduce, and \populate.
For \emph{a given point in the iteration space}, the order of actions is as follows: for a given operation label, process (1) map actions, then (2) reduce actions, then (3) populate actions. This order must be maintained for a point in the iteration space.

\begin{alternative_box}[label=sidebar:alt-patterns-gb]{Design Alternatives for Expressing Common Computational Patterns: \graphblas}
  \paragraph{\graphblas~\cite{Davis:2019:ASG, Kepner:2016:MFO}:} \graphblas{} provides two categories of primitives. First, a set of \emph{operators} used within expressions:
\begin{enumerate}
  \item overloaded multiplication, denoted $\bigotimes$,
  \item overloaded addition, denoted $\bigoplus$,
  \item the accumulation operator, denoted $\bigodot$ (overloaded element-wise ``update'' operator),
  \item the mask operator, denoted $<\!M\!>$, which the \graphblas machinery applies to the output as: $Z<M>$, and
  \item any user-defined unary operator.
\end{enumerate}
Second, a set of \emph{operations} applied to tensors:
\begin{enumerate}
  \item \emph{extract}, which assigns an output vector/matrix to a subset of the input vector/matrix,
  \item \emph{assign}, which assigns a subset of the output vector/matrix to the input vector/matrix, and
  \item \emph{transpose}, which transposes a vector/matrix.
\end{enumerate}
\graphblas{} further groups some of these operators into a \emph{semiring}, which consists of:
  \begin{enumerate}
    \item the monoid: an identity value and the overloaded addition ($\bigoplus$) operator for reduction, and
    \item the multiply operator: the overloaded multiplication operator ($\bigotimes$)
  \end{enumerate}

In \edge{}, both overloaded multiplication and addition are simply user-defined functions that a specification applies to map or reduce actions.
Likewise, the \graphblas{} mask operator, which applies a mask on the \acro{RHS} computation, is simply a specific instantiation of map, ($\bigwedge$) in \edge{}\@.

For example, to mask the \acro{GEMM} output using a mask tensor $M$, we can write the following \edge{} specification:
\begin{align}\label{eqn:mask1}
  Z_{m, n} &= \bigl(A_{k, m} \cdot^1  B_{k, n}\bigr)_{m, n} \cdot^2 M_{m, n} :: \bigwedge^1 \times \bigvee^1 + \bigwedge^2 \times,
\end{align}
where the map operation on operation label $2$ essentially masks the \acro{GEMM} computation from operation label $1$.
\edge{} does not need a separate class of computations to specify this!

Likewise, the \graphblas{} semiring is a specific instantiation of a combination of map and reduce actions: specifically, the \graphblas semiring ($\bigoplus, \bigotimes, I$) will always correspond to the following \edge{} specification pattern: %
\begin{align}\label{eqn:semiring}
  Z_{Z_{r}} &= A_{A_{r}} \cdot  B_{B_{r}} :: \bigwedge \otimes(\cap) \bigvee \oplus(\cup),
\end{align}
where $Z_{r}, A_{r}, B_{r}$ correspond to the rank variables of $Z$, $A$, and $B$, respectively.

The \graphblas{} accumulation operator, which updates an output tensor instead of entirely replacing it, maps to a combination of applying an iterative rank to the Einsum expression (see \S~\ref{ssec:iter}) and a compute operation between the old version of the output tensor and the computation that will update the tensor (see \S~\ref{ssec:iter}).

\edge{} naturally expresses other \graphblas{} operators---including \emph{extract}, \emph{assign}, and \emph{transpose}---as rank expressions that may constrain the iteration space (see \S~\ref{ssec:conditionals}).
\end{alternative_box}

\begin{alternative_box}[label=sidebar:alt-patterns]{Design Alternatives for Expressing Common Computational Patterns: Allow Any Pattern}
  \paragraph{Sparse Array Programming Model:} The sparse array programming model, proposed by Henry et al.~\cite{Henry:2021:CSA}, allows for generalized functions that must explicitly define the iteration space.
  There are no ``classes'' of operations; instead, in the most general case, a specification will contain user-defined functions that take as input any number of \emph{scalar} inputs and produce a scalar output.
  The compiler then applies these scalar functions to the tensors indicated by the corresponding Einsum expression.
  The model implicitly maps compute functions to ``element-wise operations, reductions, and broadcasts''~\cite{Henry:2021:CSA} through the index notation (see \S~\ref{list:patterns}).
  \edge{} can represent all of these patterns (see \S~\ref{ssec:mapreducepopulate}).

  Additionally, the array programming model goes beyond unary and binary operators to allow for any $n$-ary operator.
  For example, an $n$-ary operator may take in $n$ inputs, and always select the fifth input.
  In \edge{}, every compute operator is a binary operator.
  We maintain this invariant to make it easier for algebraic manipulation---such as commuting variables, distributing computations, etc.---in a system that uses \edge{} to find other algorithmic variants of a problem (see Figure~\ref{fig:ta-soc} and \S~\ref{sec:5-mappings}, and goal~\textbf{\ref{goal:simple}}).
\end{alternative_box}

\begin{bfs_box}[label=sidebar:gather]{Gather Neighbors of a Frontier}
  A key step in \bfs{}, and many other graph algorithms, is gathering the neighbors of a set of query vertices.
  Assuming the default multiplication/addition compute operators for now, in addition to the tensor declarations in Equations~\eqref{eqn:init} and~\eqref{eqn:initialization}, we can describe this phase in \edge{} as:
  \begin{align}\label{eqn:bfs_step}
        F1_{d} = G_{s, d} \cdot F_{s} :: \bigwedge \times \bigvee +
  \end{align}

  This Einsum specifies an $S \times D$ iteration space.
  For each $(s, d)$ point in the iteration space, the execution will project into the data space and retrieve the edge weight of edge $(s, d)$ from $G$, as well as the current depth of vertex $s$ from $F$ (data value of $F$).
  We will call the depth of vertex $s$ the ``vertex weight'' in future examples.
  If the current depth of $s$ is non-empty (i.e., it is not $\infty$---see \bfs{} Example~\ref{sidebar:bfs-datatype}), then each $(s, d)$ edge in $G$ for this given $s$ connects that $s$ vertex to a $d$ vertex.
  That is, the set of all $d$ coordinates for all $(s, d)$ points in $G$---where $s$ is fixed---is the set of all neighbors of the $s$ vertex.
  Execution will multiply the two values from $G$ and $F$, and place the value in the corresponding $d$ coordinate in the output frontier tensor, $F1$.
  The result is an integer value indicating the depth at which computation found the $d$ vertex.

  If multiple $s$ coordinates in the frontier connect to the same $d$ coordinate in $G$, then the Einsum in Equation~\eqref{eqn:bfs_step} will reduce the values through addition.
  In this particular example, the reduced values have no meaning, other than indicating the total number of parents/paths to each $d$ vertex.
  \bfs{} Example~\ref{sidebar:bfs-merge} replaces this operation with \texttt{min}, which simply indicates there is a path from $s$ to $d$.
  In graph terms, if multiple $s$ vertices in the frontier share the same destination vertex ($d$ coordinate in $G$), execution will reduce the computed values and store the reduced value at the $d$ vertex location for the output frontier $(F1)$.
\end{bfs_box}

\subsection{Extension: Separating Out Merge From Compute}\label{ssec:merge}
Most algorithms for graphs view edges and vertices as graph \emph{entities} that have associated \emph{attributes}~\cite[Chapter 22.1]{CLRS:2022:ITA}.
A graph entity refers to the components that make up a graph: either an edge or a vertex.
Each graph entity has an \acro{ID} (see \S~\ref{ssec:graph-algorithms}).
For example, the graph in Figure~\ref{fig:exgraph} contains vertices $A$ to $G$ (or when numerically labeled, vertices 0--6), and ten edges with IDs 0--9.
An \emph{attribute} describes a particular property of an entity.
For example, each vertex or edge may have an associated weight vector (common in road networks or graph neural networks).

\begin{bfs_box}[label=sidebar:bfs-entity]{Entities and Attributes in \bfs{}}
  In \bfs{}, each vertex has an attribute that changes throughout the course of the algorithm: its discovery depth.
  Aside from the query node, each vertex begins with a starting depth attribute of $\infty$, and ends with a final value of its minimum distance from the query node(s).

  The $\times$ computation for the map action in Equation~\eqref{eqn:bfs_step} does not make sense in this scenario; for \bfs{}, we need to set the depth of a destination vertex to the depth of its parent plus one (the default edge weight that connects that parent to the destination vertex).
  We can reformulate the \edge{} expression for the gather step as follows:

  \begin{align}\label{eqn:bfs_step2}
        F1_{d} = G_{s, d} \cdot F_{s} :: \bigwedge + \bigvee \min,
  \end{align}
  where, for a given $(s, d)$ point in the iteration space, the corresponding data values in $G$ and $F$ are \emph{added} together to generate a new depth value for that $(s, d)$ point.
  When multiple parent vertices ($s$ coordinates) share the same destination ($d$ coordinates), we can reduce over $s$ by selecting the minimum calculated depth for $d$ across all $s$ coordinates (\texttt{min} operation).
  In graph terms, given a destination vertex $d$, and new depth counts between that destination vertex and each of its parent $s$ vertices, we select and store the minimum depth.

  Overall, in the \edge{} formulation of \bfs{} found in Equation~\eqref{eqn:bfs_step2}, the $s$ and $d$ coordinate values identify a vertex \emph{entity}.
  Edge entities in this formulation consist of ($s, d$) tuples in $G$.
  If we number each element in the list of all non-empty $(s, d)$ tuples in $G$, then we can obtain the original edge IDs of $0\text{--}9$ found in Figure~\ref{fig:exgraph}.

  We revisit this expression with merge operators in \bfs{} Example~\ref{sidebar:bfs-merge}.
\end{bfs_box}

The promise of any sparse computation is that it need not touch every point in the iteration space.
For example, in element-wise multiplication between two sparse vectors, an efficient implementation will only access data values that exist in \emph{both} vectors.
(If we know a value does not exist in either vector, then we know the result of the multiplication will be zero, and thus we need not perform the multiplication.)
By \emph{intersecting} the coordinates to determine the pairs that contain non-empty values in both tensors, an implementation can determine the subset of the possible data values that will result in a meaningful product~\cite{Hegde:2019:EAA}, while avoiding ineffectual values.
When tensors are tiled or multidimensional, implementations can employ \emph{hierarchical} intersection, which eliminates unnecessary accesses to larger regions of the data space~\cite{Hegde:2019:EAA}.

Additionally, certain operations require only touching the entities and not the data values.
For example, simply gathering the neighbors of the set of vertices in $F$ (see \bfs{} Example~\ref{sidebar:bfs-entity}) does not actually require processing their data values.
It simply requires access to the shared non-zero $s$ coordinates in both $G$ and $F$, to determine which $d$ coordinates (neighbors) exist in $G$ for a given $s$ coordinate (parent) in $F$. In general, we can make our computations more efficient by (1) only accessing entities when necessary, and when accessing entities \emph{is} necessary, by (2) not accessing data values unless that is also necessary.

Let us take a closer look at the execution of Equation~\eqref{eqn:bfs_step2}.
It requires two steps: (1) gather neighbors of parent vertices ($s$ coordinates in $F$) by walking the iteration space and projecting into the data space, then (2) compute on the retrieved data values.
In \edge{}, we generalize these two steps into two types of operators: (1) the \emph{merge} operator and (2) the \emph{compute} operator.
We explain each below.

Earlier, we stated that every map and reduce action must be followed by a compute operator (see Sections~\ref{sssec:map} and~\ref{sssec:reduce}).
We now expand this: every map and reduce action consists of a \emph{merge operator} (\merge) and a \emph{compute operator} (\compute), denoted by $[\map|\red] <\textit{\compute}>(<\textit{\merge}>)$.

\subsubsection{What is compute?}\label{sssec:compute}
\textbf{Compute operators} process computations on \emph{data values} after an implementation projects into the data space.
For \maptxt\ actions, the \compute\ operator operates on a pair of data values at a given point in the iteration space.
For \reduce\ actions, at a given point in the iteration space, compute takes as input the current \reduce\ state, and the current \emph{computed} value on the RHS of the Einsum (the \maptmp) to generate a single value.
The user must specify an \emph{identity} value for each \compute\ operator. %
\edge{} allows any user-defined compute operator that meets these constraints.

\subsubsection{What is \merge?}\label{sssec:merge}
\textbf{Merge operators} process rank variables, culling which points an implementation needs to touch in the iteration space.
Without the merge operator, the \edge{} expression becomes a traditional tensor expression, where execution traverses the entire iteration space.
However, by separating out merge from compute, \edge{} enforces ``lazy'' computation, where an implementation accesses data values only if the corresponding iteration-space point indicates that the data value is non-empty.
This information (empty vs.\ non-empty points) is usually present in the metadata of a tensor~\cite{Hegde:2019:EAA, Odemuyiwa:2023:ASD}.

\edge{} applies \merge\ to a specified operation label.
At a given point in the iteration space, merge takes as input whether the corresponding tensor coordinates exist in the left-hand operand and whether the corresponding coordinates exist in the right-hand operand.
Given these inputs, \merge\ determines if the output coordinate will exist, and what the coordinate will be.
For example, given a map action over rank $k$ ($\bigwedge$) at a given point in the iteration space, the merge operator for this map action will check if the corresponding $k$ coordinate is empty or non-empty in both operand tensors, then using that merge operator, decide if the output will contain that $k$ coordinate.
If the output contains the $k$ coordinate, then the \edge{} machinery will apply \compute\ to the corresponding data values of the input operands.

An example merge operator is intersection ($\cap$), which need only touch points in the iteration space where a non-empty value exists for \emph{both} operands.
Another merge operator, union ($\cup$), need only touch points where \emph{at least one} of the operands is non-empty.
Overall, there are sixteen possible merge operators, which correspond to the sixteen truth tables possible given a binary operator that checks whether a given point contains an empty value or not in both operands.
Appendix~\ref{appendix:merge} lists the sixteen possible merge operators.

\begin{bfs_box}[label=sidebar:bfs-merge]{Intersection in the Gather Step}
  Note that if $G$ or $F$ is sparse in Equation~\eqref{eqn:bfs_step2}, there is no need to walk the entire iteration space.
  For example, if the frontier ($F_s$) contains a single vertex in its rank $s$, our operational definition of an Einsum states that evaluating an operation that involves $F_s$ will walk every $s$ point in the iteration space.
  However, an ideal, optimized implementation would only touch the points in the $s$ rank that are present in both $F$ and $G$\@.

  Let us expand Equation~\eqref{eqn:bfs_step2} to include merge operators:
  \begin{align}\label{eqn:bfs_step3}
    F1_{d} = G_{s, d} \cdot F_{s} :: \bigwedge_s +(\cap) \bigvee_s \min(\cup),
  \end{align}

  Now, the map action performs an intersection $(\cap)$ on the $s$ indices of $G$ and $F$, then applies the $+$ compute operator to the values that survive intersection ($\bigwedge_{s} +(\cap)$).
  In graph terms, we only compute the sum of vertex weight plus edge weight if the source vertex is in the frontier \emph{and} it is connected to the destination vertex by an edge. %
  If $F$ contains a single vertex, then an intersection merge operator would allow computation with the single-vertex frontier described above to only visit one point in the $s$ rank (one vertex in the graph $G$) rather than all points (all vertices).

  The reduce action looks at all the non-empty $s$ points ($\cup$), and applies the $\min$ operator to their values. %
\end{bfs_box}

\begin{alternative_box}[label=sidebar:alt-merge]{Design Alternatives for Merge}
  In \edge{}, we explicitly choose to input two Booleans for each point in the iteration space, where the Boolean corresponds to whether a coordinate exists or not (see truth tables in Appendix~\ref{appendix:merge}).
  These Booleans come from attributes rather than data values.

  A design alternative is to consider more complex inputs to determine how to cull the iteration space.
  For example, the sparse array programming model~\cite{Henry:2021:CSA} allows the equivalent of multi-operand merge operators by allowing the program specifier to indicate specific points at which to cull the iteration space.
  However, in keeping with \textbf{design goal~\ref{goal:simple}}, and noting our earlier constraint that all operation labels are binary (see \S~\ref{ssec:computationspecs}), \edge{} maintains that merge operators (at a given data point) are also binary.
  Additionally, enabling non-binary merge operators would allow \edge{} to support non-binary reduce operators (see Inset~\ref{sidebar:example-red1} and~\ref{sidebar:alt-patterns}); however, we find that we have not lost expressivity by constraining our operators to be binary (\textbf{design goal~\ref{goal:powerful}}).
  Thus, our merge operators are both simple and effective.
\end{alternative_box}

\subsubsection{Understanding Merge and Compute Operators}\label{sssec:mergemodel}
For graph algorithms, merge operators tend to correspond to operations on graph entities, while compute operators correspond to operations on attributes.
\edge{} \emph{parameterizes} the compute operator by a merge operator.

One can view the merge operator in three ways:
\begin{enumerate}
  \item\label{item:takeright}  The merge operator is necessary to fully \emph{define} the desired compute operation.
  For example, consider the unconventional ``take-left'' compute operator ($\leftarrow$), which, given two non-empty operands, selects the first, left-hand operand.
  However, suppose we want this computation to return an empty value if either operand is non-empty.
  To fully define this operation, a specifier needs to include the intersection ($\cap$) merge operator: $\map \leftarrow(\cap)$.

  \item The merge operator gives additional information about the compute operator.
  Merge tells one whether there is actual work to be done given particular values that are being given to the compute operator, i.e., is the operation \emph{effectual} (resulting in non-empty values) or \emph{ineffectual}?
  This information is often clear and implied by the normal use of the compute operation; for example, multiplication has an intersection ($\cap$) merge operator because multiplication by an empty value of zero is ineffectual.

  \item There are cases where one cannot infer the merge operation from the nominal compute operation.
  For example, traditional addition uses the union ($\cup$) merge operator, which needs at least one operand to be non-zero.
  However, suppose a computation requires addition that adds two values \emph{only} when both operands are non-zero.
  In this scenario, the program specifier will write an addition operation with an intersection merge operator: $\map +(\cap)$ (see Equation~\eqref{eqn:bfs_step3}).
  Here, specifying only the addition operator does not provide enough information to allow a compiler to generate the most efficient computation.

\end{enumerate}

Now, the operational definition of an Einsum needs the precise definition of the merge operation, the compute operation, and their outputs to fully specify computations on the iteration space and the data space.

\subsection{Unary Operators}\label{ssec:unary}
\edge{} allows a programmer to specify any user-defined unary function.
A unary operation takes as input a data value and transforms it in some way.
When applied to a tensor, the unary function individually transforms each element in the tensor, without consideration of the other elements in the tensor.

Unary functions are \emph{syntactic sugar} for a degenerate \maptxt: every unary function on some tensor $A$ decomposes into an Einsum with input tensors $A$ and $B$, where $B = 1$.
The \maptxt\ action uses a ``take-left'' ($\leftarrow$) \merge\ operator to select the $A$ tensor at every point in the iteration space. The \compute\ operator is simply the unary operator.

\begin{example_box}[label=sidebar:example-unary]{Applying Unary Functions}
A unary function transforms each element of a tensor.
For example, if we assume $B$ is a Boolean tensor, the following expression performs an element-wise $\text{AND}$ between $A$ and the complement of $B$ ($\neg B$).
  \begin{align}
        Z_{m} = A_{m} \cdot \neg B_{m} :: \bigwedge \text{AND} \label{eqn:unary}
  \end{align}
Here, for each $m$ point in the iteration space, an implementation will project into the data space and access the corresponding data values in $A$ and $B$.
For $B$, the unary function $\neg$ will return the complement of the current value at that $m$ point.

The above \edge{} expression is syntactic sugar for the following:
  \begin{align}
        Z_{m} = A_{m} \cdot^1 (B_{m} \cdot^2 \mathbf{1})_m :: \bigwedge^1 \text{AND} \bigwedge^2 \neg(\leftarrow) \label{eqn:unary2}
  \end{align}

\end{example_box}
\subsection{Execution Model of an \edge{} Einsum}\label{ssec:exec-model}
We are now ready to specify how to execute an \edge{} Einsum that contains \maptxt/\reduce\ actions (which, in turn, contain \merge/\compute operators).
\S~\ref{sec:sets} will further define the semantics.

We begin by walking the iteration space specified by the rank variables in the expression.
\begin{enumerate}
  \item  At \textbf{each point in that space}, process each specified operation label in the expression:
      \begin{enumerate}
        \item \textbf{Map:} Apply all \maptxt\ actions first, where, for each \maptxt\ action:

          \begin{enumerate}
            \item If specified, process the \merge.
                  Only ``touch'' points in the iteration space that survive \merge.
                  If the current iteration-space point does not survive \merge, immediately exit and move to the next point in the iteration space.
                  Overall, a failed \merge\ will abort the entire iteration-space point.
            \item If the current iteration-space point survives \merge, then project into the data space and apply \compute\ to the corresponding data values. Note that a performant implementation may combine the coordinate check in \merge\ with the data access. The execution model here describes the logical order of operations rather than a prescribed implementation.
          \end{enumerate}
          Each \maptxt\ action for a given point in the iteration space produces a \maptmp, consisting of the iteration space tuple and the computed value.

        \item \textbf{Reduce:} Apply \reduce:
          \begin{enumerate}
            \item Create as many \reduce\ states as the shape of the output tensor. At a given point in the iteration space, each state is indexed by a coordinate from the LHS\@.
            \item If specified, process \merge\ (see the \maptxt\ action).
            \item If a point survives \merge, reduce the corresponding data value on the RHS (the \maptmp\ value) with the current, corresponding state.
            If the current state is \emph{empty}, use the \compute\ operator's identity value.
          \end{enumerate}
          The final, reduced state is a \redtmp, which consists of the output coordinate and the reduced value.

      \end{enumerate}
    \item \textbf{Populate:} Populate the corresponding point in the output tensor. By default, this is simple assignment; see \S~\ref{ssec:spopulate} for the extended form. %
\end{enumerate}

\begin{example_box}[label=sidebar:example-map]{Merge and Compute in \acro{GEMM}} %
  Let us return to the \acro{GEMM} example:
  \begin{align}\label{eqn:tin13}
    Z_{m, n} &= A_{k, m} \cdot B_{k, n} :: \bigwedge \times(\cap) \bigvee +(\cup),
  \end{align}

  We walk each $(m, n, k)$ point in the $M \times N \times K$ iteration space.
  Note that the \emph{order} in which we walk this space will be determined by the mapping (not discussed in this work, see TeAAL~\cite{Nayak:2023:TDF}).
  At each point, we will first apply \maptxt.
  The intersection merge operator ($\cap$) evaluates whether the $k$ coordinate is non-empty in both $A$ and $B$ for the given $(m, n, k)$ point.
  If the $k$ coordinate is non-empty, we can then apply the compute operator of multiplication ($\times$) between $A$ at data point $(m, k)$ and $B$ at data point $(n, k)$.
  This computation will produce a \maptmp\ at iteration point $(m, n, k)$.

  For all the generated \maptmps, we need to reduce them with the specified \reduce\ action.
  For a given $(m, n, k)$ point in the iteration space, we update the corresponding $(m, n)$ \reduce state, applying the \reduce action to the current \maptmp\ value and the current $(m, n)$ \reduce state.

  Note that the Einsum does not actually specify the overall order in which to walk the iteration space; it only specifies the order of operations when looking at a specific \emph{point} in the iteration space.
  Thus, an implementation may first apply the \maptxt\ action to \emph{all} points in the iteration space, producing a \maptmp\ tensor of shape $M \times N \times K$, before reducing over the $K$ rank of that tensor.
  An implementation may also choose to instead incrementally multiply and reduce data values at each point in the iteration space.

  Additionally, although we described \merge\ and \compute\ as point-by-point steps, an implementation may even choose to apply a \merge\ operation to all iteration-space points first, before applying the \compute\ operator.
  For example, an implementation of the \acro{GEMM} expression above may first intersect the \emph{entire} $k$ rank of $A$ and $B$ by gathering all non-empty coordinates of $A$ and $B$, performing a set intersection between them, and producing an output stream of $k$ coordinates that are shared by both $A$ and $B$.
  Now, execution need only access regions of the $M \times N \times K$ space that contain $k$ coordinates that survived intersection.
\end{example_box}

\begin{example_box}[label=sidebar:multi-rank]{Merge in Multi-Rank Map Actions}
  Let us add an intersection merge operator to the outer product example:
    \begin{align}\label{eqn:cross-map}
        Z_{m, n} &= A_m \cdot B_n :: \bigwedge_{m, n} \times(\cap)
    \end{align}
  At any given point in the iteration space, computation will proceed only if the corresponding $m$ and $n$ coordinates in $A$ and $B$, respectively, both contain non-empty values.
  The addition of this merge operator allows an implementation to decide whether to perform the computation or not based only on information from graph entities, potentially reducing references into their attributes. %

\end{example_box}

\begin{bfs_box}[label=bfs:fullstep]{Describing One Iteration of \bfs{} in \edge{}}
  We now have all the notational tools to fully express the \emph{first iteration} of \bfs{} in \edge{}:
  \begin{subequations}\label{eqn:bfs_step_full}
    \begin{align}
      & \triangleright \text{Extended Einsum} \notag\\
      T_{d}  &= G_{s, d} \cdot F_{s} :: \bigwedge_{s} +(\cap) \bigvee_{s} \min(\cup) \label{seqn:advance}\\
      F1_{d} &= T_{d}    \cdot \neg P_{d} :: \bigwedge_{d} \leftarrow(\cap) \label{seqn:filter} \\
      P1_{d} &= P_{d}    \cdot F1_{d} :: \bigwedge_{d} \text{OR}(\cup)  \label{seqn:update}
    \end{align}
  \end{subequations}

  We can describe each step from the ``graph world'' perspective and from the Einsum perspective:
  \begin{enumerate}
    \item
      \textbf{Graph View:} Equation~\eqref{seqn:advance} gathers the neighbors of vertices in the frontier ($F$), updates the depth of each discovered neighbor by one, and removes any duplicate depths for the same neighbor by selecting the minimum depth.

      \textbf{Einsum View:} Equation~\eqref{seqn:advance} creates an $S \times D$ iteration space.
      We intersect $F$ and $G$ on $s$.
      That is, we only project into the data space when $F$ and $G$ both contain non-empty $s$ coordinates (i.e., $F$ has one or more query vertices and those vertices exist as parent vertices in $G$).
      After projecting into the data space, we add the current depth in $F$ at coordinate $s$ to the edge weight in $G$ at point $(s, d)$.
      We place the resulting value at coordinate $d$ in the output tensor, reducing using the $\min$ operator when multiple depth values for $d$ exist (due to multiple non-empty $s$ coordinates in $F$ connecting to the same $d$ coordinate in $G$).
      $T_{d}$ contains the temporary results of this step, where each non-empty $d$ vertex/coordinate has a data value equal to one more than the data value of its parent in $F$.

    \item
      \textbf{Graph View:}   Equation~\eqref{seqn:filter} takes the gathered neighbors and filters out neighbors that have already been visited, i.e., neighbors that appear in $P$.
      This produces a new output frontier, $F1_{d}$.

      \textbf{Einsum View:} In Einsum terms, we apply the unary complement operator ($\neg$) to $P$, such that accessing the Boolean data value of $P$ at point $d$ returns the complement of that value.
      In fact, we can interpret the entire expression of $\neg P_{d}$ as a tensor whose values are the complement of $P$.
      For this Einsum, we only project into the data space when both $T$ and $\neg P$ contain non-empty values (intersection merge operator).
      Once in the data space, we apply the special, ``take-left'' compute operator, which selects the data value of the first operand as the output value.

    \item
      \textbf{Graph View:}
      Finally, Equation~\eqref{seqn:update} updates the set of visited nodes and places this update in a new tensor $P1_d$.

      \textbf{Einsum View:} At each $d$ point in the iteration space, if either $P$ or the new frontier, $F1$, is non-empty (union merge operator---$\cup$), project into the data space.
      The compute operator is the Boolean $\text{OR}$ function; since the output is Boolean, the Einsum machinery enforces that the integer value at $F1_{d}$ (see \bfs{} Example~\ref{sidebar:bfs-datatype}) will be cast to a Boolean value of \True when a non-empty value exists, and \False, otherwise.
      Thus, the final output tensor, $P1$, will contain non-empty values (i.e., no \False values) at $d$ locations where $P$ was non-empty as well as where $F1$ is non-empty.
      That is, $P1$ is an updated copy of $P$, with new values from the new frontier.
  \end{enumerate}
\end{bfs_box}

\begin{alternative_box}[label=sidebar:alt-linalg]{Design Alternatives for Merge vs.\ Compute}
Linear algebra tools that support graph algorithms tend to conflate computations on graph entities with computations on graph attributes.
For example, \graphblas represents the ``gather'' step of \bfs{}---which gathers the neighbors of vertices in a frontier---as a single, bulk step of matrix-vector multiplication between the adjacency matrix of the graph $G$ and the vector frontier, $F$.
\edge{} essentially factors this into two components: first gather the neighboring vertices of parents in the frontier (operating on \emph{entities}), then update the \emph{values} or attributes of those neighbors.

We now present three more cases where the design alternative of conflating merge with compute does not suffice.

\paragraph{\textbf{Case 1:}} Matrix-vector multiplication is simply a series of dot-products between the columns of $G$ and the vector frontier.
In \graphblas, the semiring for \bfs{} uses $\min$ for reduction, $+$ (addition) for multiplication, and a ``zero value''---which is \graphblas' equivalent of \edge{}'s empty value---of $\infty$~\cite{Kepner:2016:MFO}.
In \edge{}, merge operators simply take as input whether the coordinates of a rank are empty or not in either tensor operand.
This enables an \edge{} implementation to potentially optimize computation by removing ineffectual computations, looking only at the coordinates before processing data values.
However, in \graphblas, this optimization is not as obvious.
Optimizations are fully at the mercy of the underlying data format: an uncompressed format for $G$ and $F$ would result in a \graphblas implementation accessing each element of a given column in $G$ and each element of $F$, regardless of whether or not the computation will be effectual.
Meanwhile, a \CSR data format for $G$ that uses $s$ as the rows enables a \graphblas implementation to skip any $d$ vertices (column coordinates in \CSR) that are not present in the \CSR representation.
In \edge{}, we are explicitly telling the compiler ``find a way to track when coordinates exist, and leverage the existence of coordinates to cull the iteration space''.
A compiler can then choose to select the appropriate data format and optimization strategy based on the optimization opportunity the \edge{} expression exposes (\textbf{design goal~\ref{goal:opt}}).

\paragraph{\textbf{Case 2:}} Additionally, note that $G$'s empty value is 0, while $F$'s empty value is infinity ($\infty$) (see \bfs{} Example~\ref{sidebar:bfs-datatype}).
Consider the expression for the ``gather'' step, which intersects the $s$ coordinates of $G$ and $F$ for the map action, repeated here:
\begin{align}
  T_{d}  &= G_{s, d} \cdot F_{s} :: \bigwedge_{s} +(\cap) \bigvee_{s} \min(\cup) \label{seqn:excap1}
\end{align}

Suppose we replace the intersection with union:
\begin{align}
  T_{d}  &= G_{s, d} \cdot F_{s} :: \bigwedge_{s} +(\cup) \bigvee_{s} \min(\cup) \label{seqn:excup2}
\end{align}

In the first example, the intersection operator enforces that projection into the data space occurs only when an $s$ coordinate is present both in $F$ (with non-infinity values) and in $G$ (with non-zero values).
For the second example, the \emph{union} ($\cup$) operator now enforces that projection into the data space occurs when at least one of the operands is non-empty at a given $s$ coordinate.
In many instances of \bfs{}, the $s$ rank of $G$ will be dense---that is, nearly every vertex in the graph has neighbors (and thus is a parent).
Now, an implementation walking the iteration space will project into the data space for nearly every $s$ coordinate---regardless of the sparsity of $F$.
Thus, the \edge{} specification can clearly indicate the optimization opportunities available to an underlying compiler or implementation (\textbf{design goal~\ref{goal:opt}}).
Meanwhile, \graphblas cannot differentiate between the two scenarios.

\paragraph{\textbf{Case 3:}} Finally, suppose we want to perform \acro{GEMM}, but with addition replacing the multiplication step:
  \begin{align}
    Z_{m,n}  &= A_{k,m} \cdot B_{k,n} :: \bigwedge_{k} +(\cup) \bigvee_k +(\cup)\label{seqn:exadd3}
  \end{align}
Here, the map action uses a union operator ($\cup$), since that is the natural merge operator for addition.

Now, suppose we want the addition (for the map action) to only apply when both operands are non-empty.
If one of the operands contains an empty value, the output should be zero.
That is, if $A$ has a non-empty value at $(k, m)$ but $B$ is empty at $(k, n)$, the output at $(m, n)$ should be zero instead of the $A$ value alone.
This is simple to represent in \edge{} (\textbf{design goal~\ref{goal:simple}}):
\begin{align}
  Z_{m,n}  &= A_{k,m} \cdot B_{k,n} :: \bigwedge_{k} +(\cap) \bigvee_k +(\cup)\label{seqn:exadd4}
\end{align}
Here, we replace the ``natural'' union operator for addition  with an intersection operator.
Equations~\eqref{seqn:exadd3} and~\eqref{seqn:exadd4} will produce different results.
\end{alternative_box}

\subsection{Extension: Expressing Iteration}\label{ssec:iter}
Many graph algorithms, including \bfs{}, are iterative or recursive.
These algorithms also tend to update data structures on each iteration.
Additionally, this computation pattern is not typical for traditional tensor algebra.

In \bfs{} Example~\ref{bfs:fullstep} (Equation~\eqref{eqn:bfs_step_full}), note the use of $F1$ and $P1$, instead of directly updating the frontier tensors $F$ and $P$\@.
Compilers often benefit from static single-assignment form (\acro{SSA})~\cite{Cooper:2023:IR} (\textbf{design goals:~\ref{goal:opt},~\ref{goal:simple}}).
Thus, \edge{} requires that Einsums also be \acro{SSA}\@.

\begin{bfs_box}[label=bfs:cascade]{Specifying Multiple Iterations of \bfs{}}
  Equation~\eqref{eqn:bfs_step_full} is only one iteration of \bfs{}\@.
  A full \bfs{} computation repeats until the frontier is empty.
  However, as previously mentioned, Einsums must be in static single-assignment form.
  One way to express this is as a cascade of Einsums (\S~\ref{sec:cascade}), where each iteration is a sequence of gather, filter, and update steps:

  \begin{subequations}\label{eqn:bfs_cascade}
    \begin{align}
      & \triangleright \text{Extended Einsum} \notag\\
      T1_{d}  &= G_{s, d} \cdot F_{s} :: \bigwedge_{s} +(\cap) \bigvee_{s} \text{ANY}(\cup) \label{seqn:advance1}\\
      F1_{d} &= T1_{d}    \cdot \neg P_{d} :: \bigwedge_{d} \leftarrow(\cap) \label{seqn:filter1} \\
      P1_{d} &= P_{d}    \cdot F1_{d} :: \bigwedge_{d} \text{OR}(\cup)  \label{seqn:update1} \\
      T2_{d}  &= G_{s, d} \cdot F1_{s} :: \bigwedge_{s} +(\cap) \bigvee_{s} \text{ANY}(\cup) \label{seqn:advance2}\\
      F2_{d} &= T2_{d}    \cdot \neg P1_{d} :: \bigwedge_{d} \leftarrow(\cap) \label{seqn:filter2} \\
      P2_{d} &= P1_{d}    \cdot F2_{d} :: \bigwedge_{d} \text{OR}(\cup)  \label{seqn:update2} \\
      \ldots&\notag
    \end{align}
  \end{subequations}
  However, Equation~\eqref{eqn:bfs_cascade} is unwieldy.
  How do we indicate the number of times to repeat the cascade?
  How do we codify the pattern of gather, filter, and update in a way that is easy to read?
\end{bfs_box}

Traditional matrix algebra and \graphblas{} are limited to at most two dimensions.
Einsums impose no limits on the number of dimensions of a tensor.
\edge{} leverages this: %
to represent iteration, \edge{} adds a special \emph{generational} (or iterative) rank, $I$, to the iteration space.
Any tensor that changes from iteration to iteration will also include this dimension.
Thus, simply by adding a rank, we are able to naturally represent a key component of many graph algorithms while maintaining the operational definition of an Einsum (\textbf{design goals~\ref{goal:ode},~\ref{goal:simple}}).

When a generational rank is present in an \edge{} Einsum, we assume the expressions repeat until a stopping condition, indicated by $\diamond: <\text{Boolean Expression}>$, where the stopping condition can be any expression that returns \True or \False\label{par:stopping_condition}.
When walking the iteration space, the stopping condition simply indicates the shape of $I$, which may not be known until runtime.

\begin{bfs_box}[label=bfs:cascadeiter]{Specifying All the Iterations of \bfs{}}
  In the case of \bfs{}, both the $F$ tensor and the set of visited nodes ($P$) change on each iteration, and therefore include the $i$ rank.

    The new \edge{} expression, with the generational rank, is:
    \begin{subequations}\label{eqn:bfsi}
      \begin{align}
        & \triangleright \text{Extended Einsum} \notag\\
        T_{i, d}  &= G_{s, d} \cdot F_{i, s} :: \bigwedge_{s} +(\cap) \bigvee_{s} \min(\cup) \label{seqn:advancei}\\
        F_{i+1, d} &= T_{i, d} \cdot \neg P_{i, d} :: \bigwedge_{d} \leftarrow(\cap) \label{seqn:filteri} \\
        P_{i+1, d} &= P_{i, d} \cdot F_{i+1, d} :: \bigwedge_{d} \text{OR}(\cup)  \label{seqn:updatei} \\
        &\diamond: ||F_{i+1} || \equiv 0 \label{seqn:stop}
      \end{align}
    \end{subequations}

    In our \bfs{} example, several tensors have a generational rank (every use of $i$).
    Einsums~\eqref{seqn:advancei},~\eqref{seqn:filteri}, and~\eqref{seqn:updatei} repeat until a stopping condition (indicated by $\diamond$) is satisfied.
    Equation~\eqref{seqn:stop} indicates the \emph{occupancy} or number of non-zeros in the current active vertex set.
    Computation stops once that occupancy is zero.
    One can view adding a generational rank to $F$, for example, as a form of collapsing the $F1, F2, \ldots$ tensors in the unrolled cascade of Einsums (Equation~\eqref{eqn:bfs_cascade}) into a single, \nd{2} tensor, whose generational rank, $I$, is unbounded.
\end{bfs_box}

Note that a compiler can perform liveness analysis on generational tensors, such as those in Equation~\eqref{eqn:bfsi} (see \textbf{design goal~\ref{goal:opt}}).
In this case, such an analysis would reveal that only two copies of the frontier are needed at any given time, leading to software optimizations like using two copies of the frontier and alternating between them (``ping-ponging'')~\cite{Sha:2019:GBG, Wang:2017:GGG}.

For a given point in the iteration space (e.g., one containing $i+1$), determining the corresponding value for the output requires a value from a different point in the iteration space (e.g., one containing $i$).
This point-by-point dependency is the only constraint the \edge{} machinery places on computation.
An actual implementation may choose to take a bulk-synchronous approach, which processes all points for a given $i$ region in the iteration space, before incrementing $i$.
Other implementations may instead choose an asynchronous, or fine-grained synchronicity approach, such as skipping ahead and buffering computation at a particular point until dependencies are resolved, or having certain points of the iteration space computed before others, etc.

Overall, our choice to represent iteration and dependencies using generational ranks does not limit an underlying implementation to a particular approach (\textbf{design goal~\ref{goal:opt}}).
We desire this flexibility for \edge{}; it enables a separation of concerns between expressing \emph{what} computation to perform at each point in the iteration space and \emph{how} to actually walk the iteration space and map the points both in time and in space (to hardware/software units like threads or processing engines) (\textbf{design goals~\ref{goal:sepconcerns},~\ref{goal:opt}, and~\ref{goal:powerful}}).

Finally, we note that at Einsum granularity, a cascade of Einsums is \acro{SSA}.
However, during execution of a single Einsum, the process is not \acro{SSA} due to reduction or rank-granularity populate (see \S~\ref{ssec:spopulate}).
This introduces an implementation constraint on dependent operations since the dependent operation must wait for the final update (from reduction or populate) or operate speculatively.

\begin{bfs_box}[label=bfs:any]{Replacing Minimum with $\text{ANY}$}
 In Equation~\eqref{eqn:bfsi}, notice that we filter vertices that have already been visited (Equation~\eqref{seqn:filteri}).
 Thus, the output frontier for the current iteration, $F_{i+1, d}$, will always contain either the empty value ($\infty$) or depth values equal to $i+1$.
 Likewise, the input frontier for the current iteration, $F_{i, d}$, will always contain either the empty value or depth values equal to $i$.
 When performing the gather step (Equation~\eqref{seqn:advancei}), since the depth values will always increase by one ($G$'s values are all 1 or 0), we can replace the $\min$ compute operator for the reduce action with $\text{ANY}$:
  \begin{subequations}\label{eqn:bfsior}
      \begin{align}
        & \triangleright \text{Extended Einsum} \notag\\
        T_{i, d}  &= G_{s, d} \cdot F_{i, s} :: \bigwedge_{s} +(\cap) \bigvee_{s} \text{ANY}(\cup) \label{seqn:advanceiany}\\
        F_{i+1, d} &= T_{i, d} \cdot \neg P_{i, d} :: \bigwedge_{d} \leftarrow(\cap) \label{seqn:filteriany} \\
        P_{i+1, d} &= P_{i, d} \cdot F_{i+1, d} :: \bigwedge_{d} \text{OR}(\cup)  \label{seqn:updateiany} \\
        &\diamond: ||F_{i+1} || \equiv 0 \label{seqn:stopany}
      \end{align}
    \end{subequations}

    The compute operator $\text{ANY}$ selects any of the non-zero values in rank $s$ when reducing over $s$.
    This is a powerful manipulation of the original Einsum that used $\min$ (Equation~\eqref{eqn:bfsi}): %
    rather than walking every $s$ coordinate to ensure computation is selecting the minimum for a given $d$ coordinate,
    an implementation can optimize reduction by exiting early once \emph{any} of the $s$ coordinates return a non-empty value. This ``early-exit'' optimization is common in high-performance \bfs{} implementations~\cite[Section 5.3]{Yang:2018:IPE} (\textbf{design goals~\ref{goal:opt} and~\ref{goal:manipulation}}).
    By representing \bfs{} as an Einsum, it becomes easier for an Einsum writer to notice the possibility of replacing $\min$ with $\text{ANY}$.
    This manipulation is a \emph{manual} manipulation, as it goes beyond the capabilities of a compiler\label{manual}.
\end{bfs_box}

\begin{alternative_box}[label=sidebar:alt-iter]{Design Alternatives for Expressing Iteration}
An alternative approach to expressing iteration is to place the Einsum expression for one step of computation inside a \verb|for| loop.
In fact, \graphblas does exactly this: it places computation within a loop nest, and steps out of the matrix abstraction to encode depth (or ``level'' information) as a separate, scalar variable~\cite[Figure 2]{Davis:2019:ASG}.
Implicitly, this imposes a bulk-synchronous implementation and compute order, where all computations for all query nodes and neighbors must occur for this iteration first, before moving to the next iteration.
This approach entangles the concerns of \emph{what} the computation is with \emph{how} computation maps in space and time (\textbf{design goal~\ref{goal:sepconcerns}}).

\end{alternative_box}

  \begin{bfs_box}[label=sidebar:full_edge_bfs]{Full \edge{} Specification for \bfs{}}
    Finally, we can express the full \edge{} specification for \bfs{}\@:
    \begin{subequations}\label{eqn:edge_bfs_full}
      \begin{align}
        & \triangleright \text{Tensors} \notag \\
        G^{I, S\equiv|V|, D\equiv|V|} &\rightarrow  \text{integer, empty=}0 \label{seqn:Gf}\\
        F^{I, S\equiv|V|} &\rightarrow \text{integer, empty=}\infty \label{seqn:Ff}\\
        T^{I, D\equiv|V|} &\rightarrow \text{integer, empty=}\infty \label{seqn:Tf}\\
        P^{I, D\equiv|V|} &\rightarrow \text{Boolean, empty=\False} \label{seqn:Pf} \\
        & \triangleright \text{Initialization} \notag\\
        G &\rightarrow \langle\text{user-specified}\rangle \label{seqn:GIf}\\
        F_{0, s:s\in id} &= 0 \label{seqn:FIf} \\
        P_{0, d:d\in id} &= \True \label{seqn:PIf} \\
        & \triangleright \text{Extended Einsum} \notag\\
        T_{i, d}  &= G_{s, d} \cdot F_{i, s} :: \bigwedge_{s} +(\cap) \bigvee_{s} \text{ANY}(\cup) \label{seqn:advancef}\\
        F_{i+1, d} &= T_{i, d} \cdot \neg P_{i, d} :: \bigwedge_{d} \leftarrow(\cap) \label{seqn:filterf} \\
        P_{i+1, d} &= P_{i, d} \cdot F_{i+1, d} :: \bigwedge_{d} \text{OR}(\cup)  \label{seqn:updatef} \\
        &\diamond: ||F_{i+1} || \equiv 0 \label{seqn:stopf}
      \end{align}
    \end{subequations}

    Now, tensor $F_{i, s}$ represents the active set of source vertices for the $i$'th iteration (i.e., at depth $i$).
    $P_{i, d}$ indicates if a vertex $d$ has been visited by the start of iteration $i$.
    The output $F_{i+1, d}$ denotes the resulting set of discovered vertices at depth $i+1$ (and their corresponding depths).
    The output $P_{i+1, d}$ denotes the total set of visited vertices after $i$ iterations.
    The initialization section (Equations~\eqref{seqn:FIf}--\eqref{seqn:PIf}) now specifies the state of $F$ and $P$ on the first iteration ($i = 0$).

  Equation~\eqref{seqn:advancef} gathers the neighbor list of the frontier $(F)$, then adds the current depth of each source in the frontier to the edge weight of each neighbor (recall that the adjacency graph $G$ stores a 1 for each pair of connected vertices). %
  Equation~\eqref{seqn:filterf} removes any neighbors that have already been visited, thus generating the new frontier.
  Once we gather neighbors, we apply the compute $+$ operator to the current depth of active vertices (values in $F$) and the resulting $(s,d)$ points in $G$.
  Since neighbors ($d$ vertices) may belong to multiple sources, we need to select which source updates a neighbor.
  We do this using the reduce action, which merges the neighbor lists into a new active vertex set by reducing over the $s$ rank.
  In this particular expression, the $\text{ANY}$ operator selects any one of the duplicate sources.
  We intersect the resulting temporary tensor---$T_{i, d}$---with the complement of the $P$ tensor ($\neg P$).
  Since $P$ is a boolean tensor, and the output needs to contain the new active vertex set with their depths, the compute operator is a ``take-left'' operation ($\leftarrow$), which selects the value in the first operand.
  This corresponds to filtering out neighbors that have already been visited.
  Finally, Equation~\eqref{seqn:updatef} updates the visited tensor with the results of the new active set of vertices, in a similar manner as the single-step Equation~\eqref{seqn:update} in \bfs{} Example~\ref{bfs:fullstep}. %
\end{bfs_box}
\subsection{Extension: Expanding the Populate Action}\label{ssec:spopulate}
\begin{figure}
  \centering
  \includegraphics[width=\textwidth]{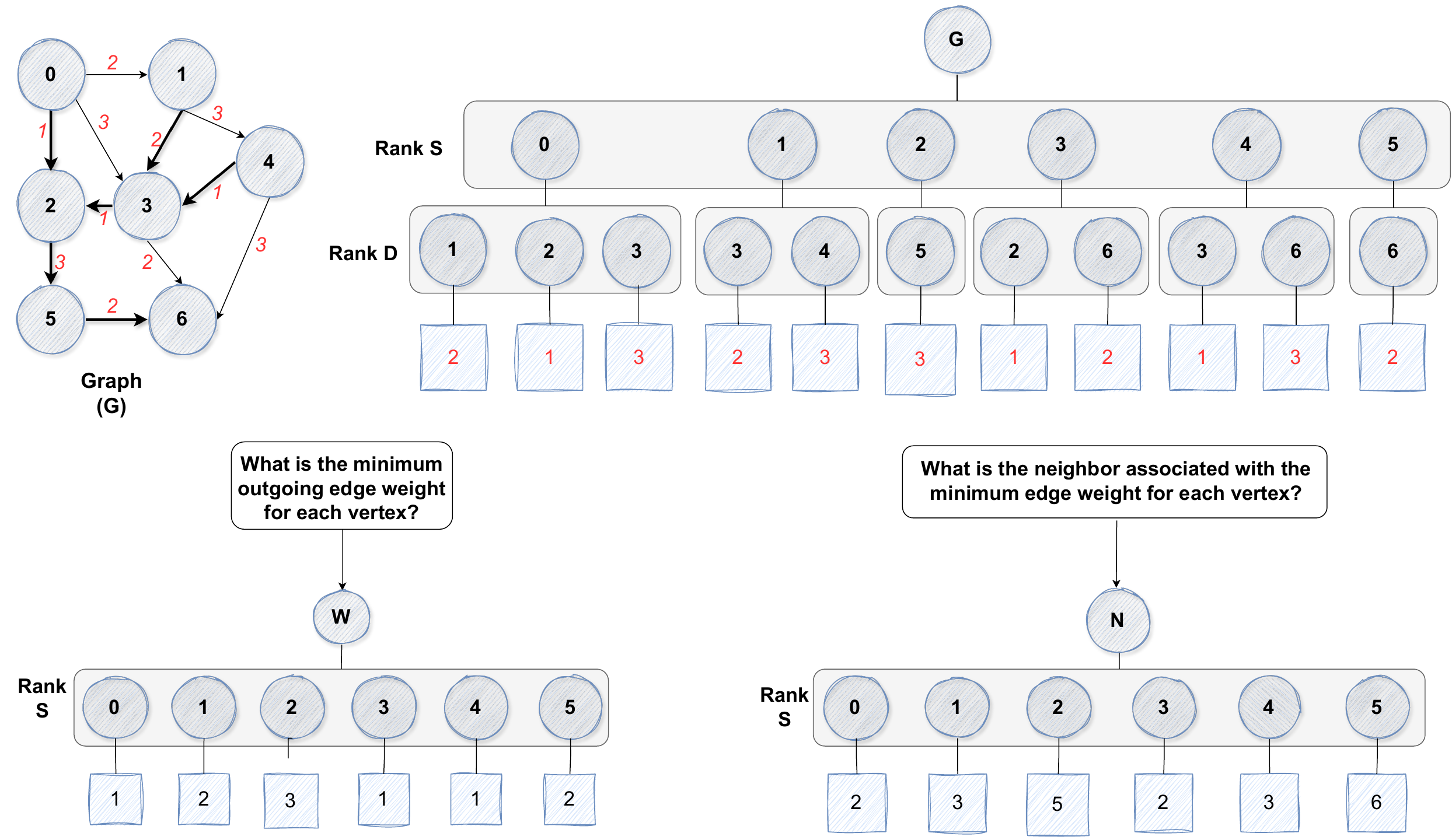}
  \caption{ (Top Left) A directed graph, $G$, with edge weights in \textcolor{red}{red}. (Top Right) A fibertree representation of the graph, $G_{s, d}$. The leaf values contain the edge weight. (Bottom Left) Resulting tensor, $W$, that records the minimum outgoing edge weight for each vertex in the graph (see Equation~\eqref{eqn:red3}). (Bottom Right) Resulting tensor, $N$, that records the \emph{destination vertex ID} of the outgoing edge with the minimum edge weight.\label{fig:weight-ex}}
\end{figure}

\begin{figure}
  \centering
  \includegraphics[width=\textwidth]{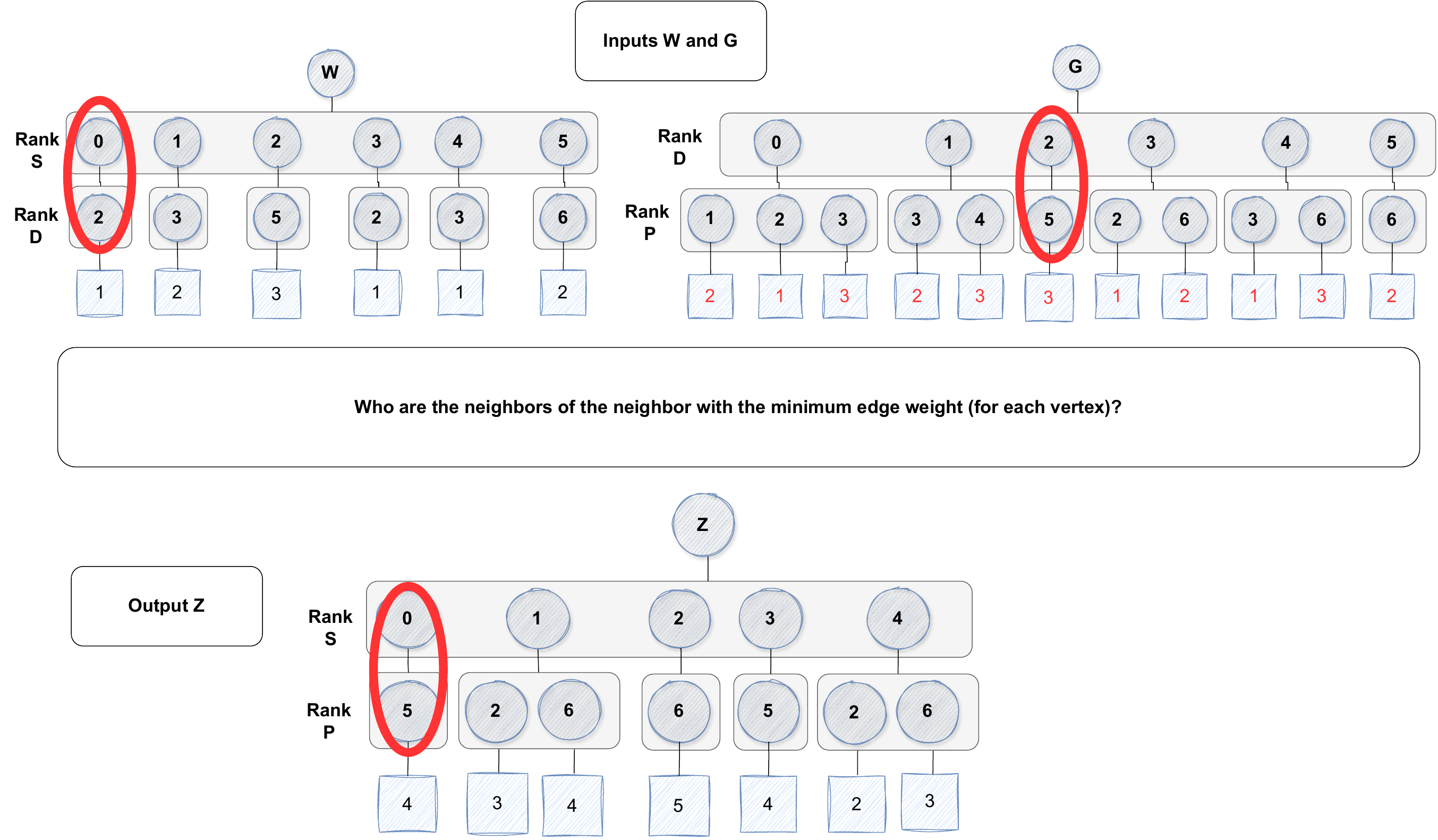}
  \caption{The two input graphs, $W$ and $G$, involved in the Einsum expression that finds the neighbors of the minimum-weighted vertex
  (see Equation~\eqref{eqn:min-neigh}).
  Red circles indicate the intersection of the $d$ rank, when $s = 0$, between $W$ and $G$. The result is in output tensor $Z$.
  $Z$ does not contain vertex $s=5$ as its minimum-weighted neighbor, vertex $6$, has no neighbors.
  The values in $Z$ indicate the cost of going from vertex $s$ to vertex $p$ through the minimum-weighted neighbor of $s$.
\label{fig:argmin-populate}}
\end{figure}

\begin{figure}
  \centering
  \includegraphics[width=.85\textwidth]{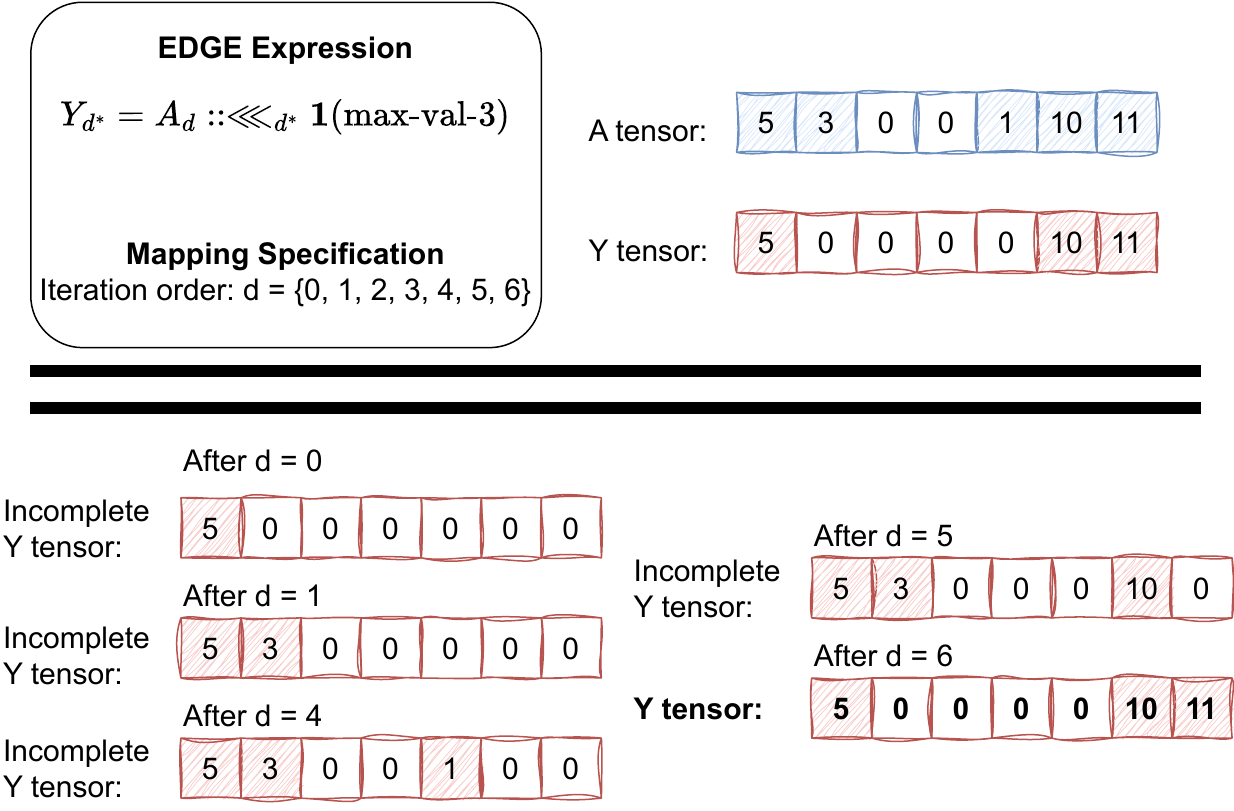}
  \caption{Top half: An \edge{} expression using populate, where $Y$ is a filtered version of $A$ that contains only $A$'s maximum three values (Equation~\eqref{eqn:max-val-ex}).
  We show example input tensors $A$ and $Y$, and a specific mapping specification on how execution should walk the $D$ iteration space. Bottom half: The state of $Y$ and its mutable $d$ rank after execution has processed certain points in the iteration space.\label{fig:AY-ex}}
\end{figure}

\begin{figure}
  \centering
  \includegraphics[width=\textwidth]{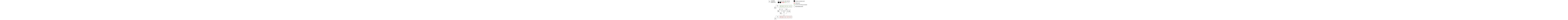}
  \caption{Execution steps after processing iteration-space point $d=1$ from the problem in Figure~\ref{fig:AY-ex}. Dots are points in the iteration space (dots). Black points indicate execution has already occurred, dots with red x's indicate the Einsum culled that point (due to sparsity), and green indicates the current point execution is processing (d=4). We first skip all culled points in the iteration space. This leads us to touch iteration-space point $d=4$ (Step 1.). At this point, we can project into the Einsum expression data space and perform compute on the \acro{RHS} (Step 2.). To process populate, we first retrieve the fiber corresponding to the mutable rank of the output tensor---in this case $d^*$ (3a.), then retrieve the current iteration-space point---in this case $d=4$ (3b.), then retrieve the computed \acro{RHS} value at the current point---in this case, the value at $A_4$ (3c.), and finally, we apply the coordinate operator of selecting the maximum three values of the current partial output fiber $Y_d$ and the current data point on the \acro{RHS} ($A_4$) (3d). This process updates the output tensor (4.)\label{fig:AY-steps}}
\end{figure}

Up to this point, we have assumed that when we compute something, we copy the result directly into the corresponding output location.
This ``default populate action'' (defined earlier in \S~\ref{sssec:populate}) is sufficient for many graph algorithms, including \bfs{}\@.
As it turns out, different algorithms may require more flexible ways to write into the output; for example, selectively updating an output location based on the current values at other output locations, or maintaining an output fiber that adheres to some property (e.g., keep only the lowest 3 values). We characterize these as variations on populate.
\edge{} extends Einsums to support these variations, while still maintaining the operational definition of an Einsum (\textbf{design goal~\ref{goal:ode}}).

Thus far, we have expanded on two \edge{} actions:
\begin{enumerate}
  \item \maptxt, which, at a given iteration-space point, retrieves data space points on the RHS, computes on them, and places them at an output data space point on the LHS\@; and,
  \item \reduce, which, at a given iteration-space point, retrieves the current \reduce state and the computed data space point on the RHS and  merges the two, updating the \reduce state with the new value.
\end{enumerate}
Both actions only retain information related to \emph{the current iteration-space point}.
Several algorithms use computations that require information on data related to other iteration-space points.
\populate\ addresses this limitation.
For example, what if we want the output rank to store only the result of the minimum three non-empty coordinates?
Or, what if we want the output rank to contain the maximum three computed values?
These questions cannot be computed through map or reduce actions alone, both of which, by definition, apply their merge and compute operators point-by-point in the iteration space and read/update the corresponding \emph{points} of the data space.

To enable this new class of computation, \edge{} introduces the specialized \emph{\populate} action (in contrast to the default populate, i.e., simple assignment, described in \S~\ref{sssec:populate}).
Populate goes beyond standard assignment.
In \edge{}, populate is denoted by $\lll$, and the specialized populate is written as follows: \[\lll_\textit{mutable\_rank\_list}\,<\!\textit{\compute}\!>(\textit{\coordinate}).\]
The subscript on $\lll$ indicates the mutable rank variable(s) on which it operates.
The rank variables will always be an output rank, with an asterisk superscript, ``$*$'', to indicate the Einsum will modify the rank or ranks indicated by the mutable rank list.

At a given iteration-space point, \populate\ retrieves the computed data space point on the RHS, and the entire, current output \emph{fiber} (not point) corresponding to the mutable rank on the LHS\@.

\populate\ uses two user-defined operators, each with different roles:
\begin{enumerate}
  \item \coordinate\ operator: this is a user-defined operator that takes as input both the rank variables and their corresponding data values ($\textit{coordinate\_operator}$). Since populate handles a \emph{fiber}, this operator goes beyond the merge operator used by \maptxt\ and \reduce\ to enable it to process the set of coordinates and values in both the fiber and the data space point on the RHS\@.
  This allows the populate action to update the output based on a user-defined operation on the \emph{coordinates} in addition to the values.
  \item The compute operator: As with \maptxt and \reduce, \populate\ allows any user-defined compute operator, which specifies how to process the \emph{value} being placed in the output.
\end{enumerate}

\subsubsection{An Example: Using \reduce\ versus \populate}
\begin{example_box}[label=sidebar:example-edgeweight]{Finding the Minimum Edge Weight For Each Vertex}
Suppose we want to express the computation where, for each source in a graph, we want to select the minimum outgoing edge weight (see Figure~\ref{fig:weight-ex}, Bottom Left).
Given the input graph $G$, we can return a weight vector $W$ that stores the minimum edge weight for each source vertex, $s$.
If the edge weights represent the capacity along that link (edge) in a network, such a computation is useful in graph problems like the minimum-cut problem or single-source shortest paths (SSSP) in weighted graphs.
The \edge{} computation for this expression is:
\begin{align}
  W_{s} = G_{s, d} :: \bigvee_{d} \min(\cup)\label{eqn:red3}
\end{align}
Equation~\eqref{eqn:red3} reduces over $d$ for each $s$ coordinate, selecting the $(s, d)$ value that is the smallest ($\min$ compute operator) across all the $d$ coordinates ($\cup$ merge operator).
The output, $W_{s}$, contains the minimum edge weight for each source in the graph.
Figure~\ref{fig:weight-ex} (bottom left) shows an example graph and the corresponding weight tensor.

This operation tells us the minimum outgoing edge weight at each vertex. However, it does not tell us \emph{to which} neighbor the edge weight corresponds.
\end{example_box}

Suppose we want to record \emph{to which} neighbor the minimum edge weight corresponds (see Example~\ref{sidebar:example-edgeweight}).
The classic operator for this type of query, where we want to know where the solution occurs as opposed to what the solution is, is the $\argmin$ compute operator:

\begin{example_box}[label=sidebar:example-noargmin]{Why not use $\argmin$?}
  Example~\ref{sidebar:example-edgeweight} does not tell us \emph{to which} neighbor the edge weight corresponds.
  The compute $\min$ operator returns the minimum value, while $\argmin$  returns the coordinate to which the minimum value belongs.
  Note that using $\argmin$ instead of $\min$ as the compute operator would return the coordinate as a \emph{value}:
  \begin{align}
    N_{s} = G_{s, d} :: \bigvee_{d} \argmin(\cup)\label{eqn:red7}
  \end{align}
  Here, for a given $s$ coordinate, the compute operator finds the minimum value across the $d$ rank and returns---\emph{as a data value}---the corresponding $d$ coordinate that gave that minimum value.
  That is, the corresponding $s$ location in $W$ now contains, as a data value, the $d$ vertex that had the minimum edge weight.
  Figure~\ref{fig:weight-ex} (bottom right) shows an example graph and the corresponding weight tensor.
\end{example_box}

Oftentimes, however, we need the $d$ vertex as a \emph{coordinate} in the output.

For example, we may want to find the neighbors of the minimum $d$ vertex in the next sub-Einsum expression that uses $W$ as an input, using this ``gather'' expression:
\begin{align}\label{eqn:min-neigh}
  Z_{s, p} = G_{d, p} \cdot W_{s, d} :: \bigwedge +(\cap) \bigvee \min(\cup).
\end{align}
Here, $W_{s, d}$ contains a single non-zero value for each $s$ fiber, where the corresponding non-zero $d$ coordinate is the neighbor with the minimum-weighted edge of $s$.
Intersecting the graph, $G$, with $W$ retrieves the neighbors ($p$ coordinates) of each $d$ vertex in $W$.
The final result, $Z$, contains the neighbors of the minimum-weighted neighbor of each vertex $s$ in the graph.
Figure~\ref{fig:argmin-populate} shows an example graph ($G$), the desired $W$ tensor, and the corresponding $Z$ output that records the path cost from an $s$ vertex to $p$ vertex through $s$'s minimum-weighted neighbor.

This computation requires $d$ to appear in $W$ as a coordinate and not a value.
The tensor output, $W$, must have a shape of $S \times D$, where each point in $W_{s, d}$ contains a data value of the edge weight if $d$ is the minimum-weighted neighbor of $s$.
Notice we cannot write the following, which uses the reduce action, as it breaks the operational definition of an Einsum (ODE) (\textbf{design goal~\ref{goal:ode}}):
  \begin{align}
    W_{s, d} = G_{s, d} :: \bigvee_{d} \min(\cup) \hspace{1cm}\textit{this violates the ODE}\label{eqn:red3b}
  \end{align}
The ODE mandates that in this Einsum, the reduction rank variable \emph{cannot} appear in the output.
Why is this?
Ignoring the \modifier{}, the expression indicates an $S \times D$ iteration space.
As we walk the iteration space and project into the data space, values in $G$ at each point in the space directly populate the corresponding point in $W$.
Now let us consider the reduce action after the \modifier{}: for each $(s, d)$ point, reduce the corresponding RHS value into the corresponding LHS location.
However, in this expression, every $(s, d)$ point on the RHS corresponds to exactly one $(s, d)$ point on the LHS (due to the operational definition of an Einsum).
There is no reduction!

Another way to view this expression is as follows: for each $s$, the reduce action says we must look at the $d$ rank and select the minimum \emph{value} across all non-empty values of $d$.
The $d$ rank will disappear.
Yet this expression indicates the $d$ rank is present in the output.
At which location in the output tensor should we place the computed minimum values? We will return to how we address this problem in a little while, but begin by discussing the mechanism by which we will solve it: the populate action.

For a given point in the iteration space, the default populate action (i.e., assignment) modifies the corresponding point in the data space of the output tensor.
Now we introduce a more general populate action: for a given point in the iteration space, the general populate action \emph{may modify other points} (within a specific fiber) in the output.
We must now look at an entire fiber of the output tensor to determine how to update it based on computation at a point on the \acro{RHS} of the expression.
In \edge{}, we call the rank of the fiber that may be modified the ``mutable rank'', demarcated by an asterisk ($*$) on the rank in the output tensor.
For example, if populate can alter the $d$ fiber of an output tensor, $Y$, the \acro{LHS} of the Einsum will look as follows:
\begin{equation}\label{eqn:max-val-ex}
  Y_{d^*} = A_{d} :: \lll_{d^*} \mathbbm{1}(\text{max-val-3})
\end{equation}
Here, populate ($\lll$) will modify the $d$ fiber, hence we subscript $Y$ by $d^*$.
We read Equation~\eqref{eqn:max-val-ex} as ``for each $d^*$ point in $Y$, check if it contains one of the maximum three values of all values seen thus far. Remove any values that are not in the maximum three values.''
This particular expression returns a vector, $Y$, that only contains values at $d$ locations that correspond to the maximum three values in $A$.
The compute operator, $\mathbbm{1}$, is a ``pass-through'' computation; it copies the values that survive the coordinate operation to the output.
Figure~\ref{fig:AY-ex} shows an example input tensor $A$, the corresponding desired final output $Y$, and the state of $Y$ after touching specific points in the iteration space.

\begin{example_box}[label=sidebar:example-populate]{Finding the maximum three values in a tensor.}
Using Equation~\eqref{eqn:max-val-ex} as an example (see Figure~\ref{fig:AY-steps}), at a specific $d$ \emph{point} in the iteration space (e.g., $d = 4$), populate takes as input the corresponding computed \emph{point} of the corresponding \acro{RHS} expression.
In this example, the computed point is simply a projection into the data space of $A$. %
Populate also takes as input the corresponding current fiber of the \emph{output} data space on the LHS---or the ``partial'' fiber ($Y_{d^*}$ in this example) at the current point in execution.
We call this fiber ``partial'' as it is not the final state of the output fiber.
The fiber is finalized when execution touches all points in the iteration space.

Populate outputs a modified partial fiber, consisting of a set of non-empty coordinate locations and their corresponding values (output $Y$ fiber after each $d$ point in Figure~\ref{fig:AY-ex}).
Overall, at a given point in the iteration space, populate results in an output fiber with updated values at locations both at the current data space point and optionally at other data space points.
In this example, from one iteration point to the next, populate will result in an output fiber that contains the maximum three values seen \emph{thus far}.
By the time execution has touched all points in the iteration space, the output fiber will contain the maximum three values of the input.

Note that certain computations may require both the Einsum and a mapping specification (such as iteration order) to fully specify the computation.
An example is when the coordinate operator outputs different results depending on the order of traversal.
\end{example_box}

\subsubsection{Programming and Execution Model for Populate}
\paragraph{Programming Model:}
From the programming model perspective, populate processes a region of the iteration space, and outputs an updated region.
The coordinate operator considers the coordinates and values of the LHS fiber, and the coordinate and value of the corresponding RHS point.
The coordinate operator then outputs a set of coordinates to delete and a set of coordinates to write to for the updated LHS fiber.
For each ``write'' coordinate, the \emph{compute} operator then accepts the RHS value (\redtmp) as input and computes a value to place at the output location.

\paragraph{Execution Model for Populate:}

Given an \edge{} expression, \emph{for a given point in the iteration space}, an implementation must first process all map and reduce actions (see \S~\ref{sssec:mergemodel}).
This generates an implicit, ``shadow tensor'' (set of \redtmp s) with data coordinates corresponding to each point in the output, and data values corresponding to the computation at the processed points in the iteration space.
This shadow tensor will contain empty values at data points that execution has not processed yet in the iteration space (e.g., after $d=0$ in bottom half of Figure~\ref{fig:AY-ex}, every other $d$ point in $Y$ contains empty values of $0$).

The default populate action will simply walk the iteration space created by this shadow tensor and copy the corresponding data values to the output locations.
If a more complex populate action is present, the following will occur:
\begin{enumerate}
  \item The coordinate operator of the populate action takes as input the current point in the iteration space, the corresponding data value for this point on the \acro{RHS} (i.e., the \redtmp), all the non-empty \emph{coordinates} of the mutable rank(s) (indicated by the mutable rank variable list), and the corresponding data values for those coordinates in the current partial output fiber.
  \item The coordinate operator will then use the input coordinates and/or the input data values to process which coordinates will now be present in the mutable rank populate is processing.
  That is, the coordinate operator will return a new list of valid coordinates for that fiber. It will also return a list of coordinates to \emph{delete} from the fiber.
  \item The compute operator, for each valid coordinate, computes on the \redtmp value and returns a value.
  \item The final output fiber contains non-empty locations corresponding to the coordinate list produced by the coordinate operator, with values corresponding to the computed values.
\end{enumerate}
Overall, for a given point in the iteration space, map and reduce only modify the given data space points corresponding to the iteration-space point.
On the other hand, populate may modify other data space points that are not related to the current iteration-space point.

For example, in Figure~\ref{fig:AY-ex} (bottom half), touching iteration-space point $d=5$ results in an update of output locations $d=4$ and $d=5$, even though $d=4$ has no direct relationship (through rank variable expressions in the subscripts) to the current iteration point.

Note that although we describe populate in a point-by-point manner, execution can parallelize populate, by processing iteration-space points in parallel.
An underlying implementation will need to resolve simultaneous updates to the output fiber in a way that is consistent with the computation specified by the Einsum.
\subsubsection{More Populate Examples}
\begin{example_box}[label=sidebar:example-argmin]{Finding the Neighbor Vertex With Minimum Edge Weight}
We can now properly express the intent of the (incorrect) Equation~\eqref{eqn:red3b} as:
  \begin{align}
    W_{s, d^*} = G_{s, d} :: \lll_{d^*} \mathbbm{1}(\text{min-val-1}) \label{eqn:pop3}
  \end{align}

Here, we assume the coordinate operator---``$\text{min-val-1}$''---is a user-defined function that returns the $d$ coordinate corresponding to the minimum value of the current state of $W$ (at a specific $s$) and $G_{s, d}$ for the current iteration-space point.
If multiple values have the same minimum value, we assume the function has a way of selecting one (such as randomly picking one of the values).

What is this Einsum saying?
At each $(s, d)$ point in the iteration space, retrieve the corresponding data space value of $G$, and the current $d$ fiber (both non-empty coordinates and values) in $W$ for the given $s$.
Then process populate, by first applying the coordinate operator, which returns a valid set of $d$ coordinates that corresponds to the minimum value seen across the data value for $G_{s, d}$ and the current $d$ fiber of $W$ (at $s$).

Now, execution applies the compute operator---``pass-through''---which simply keeps the data value corresponding to the valid coordinate outputted from the coordinate operator.
\end{example_box}

To actually read Equation~\eqref{eqn:pop3}, one can follow the following recipe:%
\begin{enumerate}
  \item Starting from \emph{the tensor on the left of the \modifier{}}, look at all the non-empty points in the \emph{mutable\_rank\_list} of the output.
  \item For each non-empty $s$ input region of the iteration space, consider each $d$ point. %
  \item For the corresponding \emph{values} of the partial output, and the corresponding value in the input $d$ point, select the minimum value.
  Place this minimum value in its matching $d$ location in the output.
  \item Set the output value of all other $d$ locations to empty values.
\end{enumerate}

\begin{example_box}[label=sidebar:example-noargmin2]{Path Through a Minimum-Weighted Neighbor}
  With the populate operator in Example~\ref{sidebar:example-argmin}, we can now fully express the cost of going from each vertex $s$, to every neighbor  2 hops away ($p$) if there exists a 2-hop path that passes through the minimum-weighted neighbor of $s$:
  \begin{subequations}
    \begin{align}
      & \triangleright \text{Tensors} \notag\\
      G^{S\equiv|V|, D\equiv|V|} & \rightarrow \text{integer, empty}=0 \label{seqn:noG}\\
      W^{S\equiv|V|, D\equiv|V|} & \rightarrow \text{integer, empty}=0 \label{seqn:noW}\\
      Z^{S\equiv|V|, D\equiv|V|} & \rightarrow \text{integer, empty}=0 \label{seqn:noz}\\
      & \triangleright \text{Extended Einsum} \notag\\
      W_{s, d^*} &= G_{s, d} :: \lll_{d^*} \mathbbm{1}(\text{min-val-1}) \label{eqn:noarg1}\\
      Z_{s, p} &= G_{n, p} \cdot W_{s, n} :: \bigwedge_{n} +(\cap) \bigvee_n \min(\cup) \label{eqn:noarg2}
    \end{align}
  \end{subequations}
  In Equation~\eqref{eqn:noarg1}, we find the neighbor with the minimum edge weight for each source vertex in $G$.
  Tensor $W$ represents the subgraph of $G$ that only contains sources and their corresponding neighbors with the minimum edge weight (of all their original neighbors).

  Equation~\eqref{eqn:noarg2} then finds the neighbors of those neighbors, where the $d$ rank in $W$ of the previous sub-Einsum (Equation~\eqref{eqn:noarg1}) is now renamed to the $n$ rank.
  The $p$ rank represents the neighbors of these neighbors.
  The output tensor, $Z$, now contains the original source vertices in the graph ($s$), and the neighbors of each $s$'s minimum-weighted neighbor.
  Each $Z_{s, p}$ value contains the distance from $s$ to $p$.
  Note that the $\min$ compute operator on the reduce action does not have an effect, since the $n$ rank has an occupancy of one from the populate action in the previous sub-Einsum.
  This expression precisely describes the example in Figure~\ref{fig:argmin-populate}.
\end{example_box}

\begin{example_box}[label=sidebar:example-argmin2]{Find the Top Three Edge Weights}
We can extend the computation in Equation~\eqref{eqn:max-val-ex} (Example~\ref{sidebar:example-populate}) to express selecting the top three edge weights and their corresponding neighbors:
  \begin{align}
    W_{s, d^*} = G_{s, d} :: \lll_{d^*} \mathbbm{1}(\text{max-val-3})\label{eqn:pop4}
  \end{align}
Now, for a given $s$ coordinate in $W$, the $d$ fiber will contain the maximum three values. %
\end{example_box}

\begin{example_box}[label=sidebar:example-neighid]{Finding the Maximum Neighbor ID}
Suppose we want to select the \emph{\acro{ID}} of the maximum neighbor for each source vertex in a graph. The corresponding \edge{} Einsum is as follows:
  \begin{align}
    W_{s, d^*} = G_{s, d} :: \lll_{d^*} \mathbbm{1}(\text{max-coord-1})\label{eqn:pop5}
  \end{align}
Equation~\eqref{eqn:pop4} takes advantage of the flexible \emph{coordinate operator}.
For each $(s, d)$ point in the iteration space, the populate action will check if the current $d$ coordinate for that point is greater than the $d*$ coordinate(s) present in the incomplete output tensor $W$ (where $s$ is fixed to the $s$ portion of the iteration point.)
If it is, populate fills the old $d*$ coordinate in $W$ with an empty value, and updates the corresponding $d*$ coordinate (that matches $d$) with the value at $G_{s, d}$. %
The compute operator $\mathbbm{1}$ simply means ``pass-through'', that is, it copies the value whose coordinate(s) survived the coordinate operation into the output.
\end{example_box}

\begin{example_box}[label=sidebar:example-default]{The Default Populate Action}
Finally, note that we can express the default populate operator---or assignment ($=$)---explicitly as:
  \begin{align}
    W_{s, d^*} = G_{s, d} :: \lll_{d^*} \mathbbm{1}(\mathbbm{1}),\label{eqn:pop6}
  \end{align}
where all $d$ coordinates and values are directly passed-through (the $\mathbbm{1}$ function) to the output.
\end{example_box}

\begin{example_box}[label=sidebar:example-populate-transform]{A Common Transformation Pattern for Populate}
Some populate expressions modify a single output point based on a set of input values.
We can recast such expressions in terms of a cascade of two Einsums that use map and reduce actions.
For example, we can transform Equation~\eqref{eqn:noarg1} (the populate form from Example~\ref{sidebar:example-noargmin2}), which uses populate to get the vertex with the minimum-weighted incoming edge, as follows:
\begin{subequations}
\begin{align}
  T_{s} &= G_{s, d} :: \bigvee_{d} \min(\cup) \label{eqn:popred1} \\
  W_{s, d} &= G_{s, d} \cdot T_{s} :: \bigwedge_{s} \text{return-if-equal}(\cap). \label{eqn:popred2}
\end{align}
\end{subequations}

Here, Equation~\eqref{eqn:popred1} finds the value of the minimum-weighted outgoing edge for each $s$ vertex, using a reduce action on $d$ ($\bigvee_{d} \min(\cup)$).
Equation~\eqref{eqn:popred2} then uses a map action between $G$ and $T$ to filter out edges (or $(s,d$) pairs) whose value is not equal to the minimum-weighted edge for the corresponding $s$ vertex.
The compute operation of ``$\text{return-if-equal}$'' checks if two values are equal and returns that value if they are, otherwise it returns the empty value.

Note, however, that this transformation only works if each $s$ vertex has only one neighbor ($d$) with the minimum-weighted edge.
If multiple neighbors have the same minimum-weighted edge, Equation~\eqref{eqn:popred2} will return all of those neighbors.
Populate, on the other hand, can specify in its coordinate operator how to resolve which neighbor to select.

This is a transformation pattern to go from specific instances of populate to a combination of reduce and map.
We can also transform an Einsum in the opposite direction if an expression contains the pattern in Equations~\eqref{eqn:popred1}--\eqref{eqn:popred2}.
Such patterns are useful in algebraic manipulations of Einsums, when generating one Einsum from another (\textbf{design goal~\ref{goal:manipulation}}).
\S~\ref{ssec:sssp} shows an example of such a manipulation, deriving a single-source shortest path algorithm from another algorithm.
\end{example_box}

The concept of populate is new and potentially powerful, and previous work in tensor algebra does not support this.
However, as shown in the previous examples, graph algorithms \emph{need} populate.
Another relevant application for populate is in graph sampling---a common step in graph neural networks and database applications---which selects vertices and edges from a given graph~\cite{Leskovec:2006:SLG}. %
In this case, the coordinate operator of populate is now a general user-defined sampling function (rather than just $\text{max-val}$, etc.,) that determines which vertices/edges to include in the final, sampled graph.%

\subsubsection{Map/Reduce vs.\ Populate}

Both map and reduce actions use a merge operator, which, for a given iteration point, checks if a given coordinate exists or not in the involved input tensor operands.
If the coordinate exists, it passes both that coordinate (for all involved tensors) and the corresponding data values to the compute operand.
Populate uses a coordinate operator, which, for a given iteration point, looks at the coordinate and value corresponding to the computed \acro{RHS} point and the list of coordinates and values corresponding to the partial output region.
The operator returns a list of valid coordinates, which it passes on to the compute operator for processing of the data values.
We consider merge to be a special type of coordinate operator: it checks if a data value is an empty value to determine if the coordinate exists or not.

\textbf{Overall, populate allows us to do two new things not possible with map/reduce: (1) operate on coordinates themselves, and (2) modify a partial output fiber.}
One should consider populate if a graph computation is performing complex operations on the vertex IDs of a graph or the \emph{coordinates} of a tensor.
Examples include sampling vertices or filtering out which vertices to use based on the vertex IDs themselves.
Additionally, any computation that scatters updates, where multiple inputs may update the same output, may benefit from populate.

Note that because populate takes a list of coordinates as one of its inputs (from the partial output fiber), simple binary merge operators no longer suffice for specifying populate's behavior.
Hence, the coordinate operator allows any user-defined function that operates on coordinates as though they are values.

\subsection{Extension: Enabling Rank Variable Expressions}\label{ssec:conditionals}
In \edge{}, tensor subscripts can contain expressions, termed \emph{rank variable expressions}.
These expressions enable features such as incorporating conditional computations into the Einsum (common in graph algorithms, see \S~\ref{ssec:examples}), selecting a subregion of a tensor, strided accesses, and enabling ``windowing'' operations---such as those found in convolution.
The previous examples in this section use only rank variables to access the tensors.
There are two types of rank variable expressions:
\begin{enumerate}
  \item Expressions on \emph{how} to project into the data space given a point in the iteration space.
  This involves any expression that maps the rank variable, which walks the iteration space, to a point in the data space.
  For example, $A_s$ maps the $s$ variable in the iteration space to the corresponding $s$ point in data space (accessing $A$).
  The expression $A_{s+5}$ maps the $s$ variable in the iteration space to the $s+5$ point in the data space.
  \edge{} places no restrictions on the types of expressions allowed, as long as the expression is a function that maps the rank variable from one point to another.
  Convolution (see Equation~\eqref{eqn:1d}) is an example of such an expression.

  \item\label{item:conds} %
  Expressions that filter which regions of the iteration space an Einsum touches (i.e.,\ \emph{conditional rank variable expressions}), essentially creating a subspace of the iteration space.
   Einsums contain this type of expression when the problem explicitly requires filtering of points in the iteration space.
  For example, suppose we want to create the upper triangle of a given \nd{2} tensor.
  The corresponding \edge{} Einsum is:
    \begin{align}
      L_{s: s < d, d} = G_{s, d},\label{eqn:constrain1}
    \end{align}
  where one can read the rank expression $s: s < d$ as ``$s$, such that $s$ is less than $d$''.
  The Einsum machinery enforces that \textbf{a constraint applied to the rank variable of one tensor} (e.g., $L$ in this case), \textbf{applies to all tensors containing that rank variable.}
  This is because the constraint is modifying the \emph{iteration space}, which impacts which regions of the data space to access.
  By convention, we will generally write constraints on the rank variables of the output tensor.

  Equation~\eqref{eqn:constrain1} places a constraint on which points are valid in the iteration space.
  Without the constraint, the valid region of the iteration space is $S \times D$, or more precisely:
  \begin{subequations}\label{eqn:space}
    \begin{align}
      s &\in [0, S)\\
      d &\in [0, D)
    \end{align}
  \end{subequations}

  With the constraint, the valid region of the iteration space is now:
  \begin{subequations}\label{eqn:space2}
    \begin{align}
      s &\in [0, d)\\
      d &\in [0, D)
    \end{align}
  \end{subequations}

  Note that manipulations are possible on the given iteration space expressions (\textbf{design goal~\ref{goal:manipulation}}).
  In Equation~\eqref{eqn:space2}, when $d$ is 4, $s$ can only take on coordinate values from $0$ to $3$ (inclusive).
  When $d$ is 5, $s$ can only take on coordinate values from $0$ to $4$.
  From the perspective of the operational definition of an Einsum, the constraints in Equation~\eqref{eqn:space2} state the following: for a given $(s, d)$ point, we consider this point valid if and only if $s < d$.
  Thus, the constraint can be rewritten as:
  \begin{align}\label{eqn:space3}
    0 \leq s < d < D.
  \end{align}

  Note the default constraints on any rank variable are $0$ to its shape (see Equation~\eqref{eqn:space}).
  Since $S \equiv D \equiv |V|$, we can update the constraint in Equation~\eqref{eqn:space3} to:
  \begin{align}\label{eqn:space4}
    0 \leq s < d < S.
  \end{align}

  Using the transitive property, we rewrite the above constraint to express the valid domain of $s$ and $d$ separately:
  \begin{subequations}\label{eqn:space5}
    \begin{align}
      s &\in [0, S)\\
      d &\in (s, D),
    \end{align}
  \end{subequations}
  and its corresponding Einsum:
    \begin{align}
      L_{s, d:d>s} = G_{s, d}.\label{eqn:constrain4}
    \end{align}
\end{enumerate}

\subsubsection{Selective Computations}
By definition, \textbf{if an Einsum does not touch a point in the iteration space due to rank variable expressions, the corresponding output data space point is by default empty.}
For example, the following expression results in empty values along $Z$'s diagonal:
\begin{align}\label{eqn:emptyeqn}
 Z_{m, n} = A_{m, n: n \neq m}
\end{align}

\edge{} supports the assignment of different expressions in different regions of the iteration space, using case statements:
\begin{equation}\label{eqn:cases}
  Z_{m, n} = \begin{cases}
                A_{m, n} & n \neq m \\
                B_{k, m} \cdot C_{k, n} :: \bigwedge_k \times(\cap) \bigvee_k +(\cup) & \text{otherwise}
            \end{cases}
 \end{equation}
Each branch of a case statement is itself a valid \edge{} expression, with its own modifier if needed.
At each point in the iteration space, Equation~\eqref{eqn:cases} selects the appropriate sub-Einsum expression.
Here, the diagonal of $Z$ now contains the dot product of fibers in $B$ and $C$ when $m = n$.
Refer to Equation~\eqref{eqn:sugar} for an alternate way to express this computation using map, reduce and rank variable expressions.

\subsection{Rank Variables as Tensors}\label{ssec:rank-as-tensor}
\edge{} allows rank variables to appear as tensors within an expression.
Suppose we have the following expression:
\begin{align}
  Z_{m, n} = A_{m, n} \cdot m :: \bigwedge \times(\cap). \label{eqn:scalingrank}
\end{align}
Here, each $m$ scales each $n$-fiber of $A$ (at fixed $m$).
For example, when $m = 2$, this expression multiplies all the values in the $n$-fiber of $A$ at $m=2$ by 2.
In the execution model, at each point in the iteration space, we project into the data space of $A$, and cast the $m$ coordinate as a value for use in the map action.
At a higher level, one can consider the $m$ term in the expression to simply be a scalar whose value changes at each point in the iteration space.

\subsection{Concluding Thoughts}\label{ssec:features-conclusion}
Although both the specialized populate extension (\S~\ref{ssec:spopulate}) and conditional expressions (\S~\ref{ssec:conditionals}) are entirely new to the Einsum world (to the best of our knowledge), \emph{they still adhere to the ODE} (\textbf{design goal~\ref{goal:ode}}). %
These two extensions are extremely powerful in expressing graph computations that filter certain vertex or edge IDs or perform certain operations only when certain conditions are met (\textbf{design goal~\ref{goal:powerful}}).
Graph computations, unlike tensor computations, require filter operations and conditional expressions.
To support these needs, we refine and expand the operational definition of an Einsum to separate computation into three phases: map, reduce, and populate.
The next section delves into the precise semantics.
\S~\ref{ssec:examples} then describes more complex algorithms that use these extensions.

\section{Semantics using Set Theory}\label{sec:sets}

In this section, we define the semantics of \edge{} using set theory.
For clarity, we present the definitions for a single Extended Einsum with
two input tensors, aliased to $A$ and $B$, and one output tensor, $Z$.
Cascades are interpreted as sequences of such Einsums.

\paragraph{Motivation.}
Prior work presented an operational definition of an Einsum~\cite{Nayak:2023:TDF, Nayak:2024:FML_micro},
and earlier sections of this paper have shown how to reason about \edge{} expressions.
However, these descriptions do not fully specify the mathematical meaning of
an \edge{} expression.\footnote{When in doubt about the meaning of an
\edge{} expression, the set-theoretic definitions in this section are
authoritative.}

We therefore give a set-theoretic semantics for Extended Einsums (\edge{}).
We use set theory because it is widely accessible and provides a common
mathematical language for the application domains where \edge{} is already
in use, including robotics, RTL simulation~\cite{zhu:2026:rsu}, graph
algorithms, machine learning~\cite{andrulis:2026:faf, Nayak:2024:FML_micro, Nayak:2023:TDF, Odemuyiwa:2025:FTF},
and iterative linear algebra~\cite{Golden:2025:QRD}.
Set theory provides the minimal common mathematical machinery needed to
describe \edge{}'s semantics across this range of users.
Our target audience is readers with a college-level foundation in discrete
mathematics, as well as programming-language experts.
\subsection{Sets}
Extended Einsums operate over several sets, and making these sets explicit is
the first step toward defining the semantics of \edge{}.
The following discussion introduces the sets used to describe the iteration
space, tensor coordinates, and corresponding data values.
Table~\ref{tab:key-set} provides a summary and
Figure~\ref{fig:is-to-cs-function} (in \S~\ref{ssec:tensor}) visualizes
how points in the iteration space, rank coordinate spaces, and data values
relate to each other.

\begin{table}[htbp]
\centering
\caption{Key Set Definitions for \edge{}.}
\label{tab:key-set}
\small
\begin{tabularx}{\linewidth}{@{} l l l X @{}}
\toprule
\textbf{Term} & \textbf{Notation} & \textbf{Category} & \textbf{Definition} \\
\midrule
Rank variable
  & $r_d \in R_d$
  & Iteration
  & A variable that determines the iteration space; by default
    $R_d = \mathbbm{W}$, but may be user-defined
    (e.g., $r \in \mathbbm{R}$). \\[4pt]
Rank variable set
  & $R_d$
  & Iteration
  & The set from which a rank variable is drawn. \\[4pt]
Iteration space
  & $\tensor{IS}$
  & Iteration
  & The Cartesian product $R_0 \times R_1 \times \cdots \times R_{D-1}$;
    defines the region over which an Einsum computes. \\[6pt]
Rank coordinate
  & ---
  & Coordinate
  & A variable that indexes into a specific rank of a tensor. \\[4pt]
Rank coordinate set
  & $\tensor{RCS}^T_R$
  & Coordinate
  & The set of all valid coordinates for rank $R$ of tensor $T$.
    May be dense, sparse, or user-defined. \\[4pt]
Rank shape
  & $RS$
  & Coordinate
  & The cardinality $|\tensor{RCS}^T_R|$ of a rank coordinate set. \\[4pt]
Dense rank coord.\ set
  & $\tensor{DenseCS}_R$
  & Coordinate
  & The default coordinate set $\setc{r}{r \in [0, RS)}$ for dense tensors. \\[6pt]
Tensor coordinate space
  & $\tensor{CS}^T$
  & Tensor
  & Cartesian product of all rank coordinate sets of $T$; the set of all
    valid index tuples into $T$. \\[4pt]
Point
  & ---
  & Tensor
  & An $N$-tuple element of $\tensor{CS}^T$; uniquely addresses one
    position in an $N$-dimensional tensor. \\[4pt]
Rank name
  & e.g., $P,\,Q,\,M$
  & Tensor
  & A unique identifier for a rank within a specific tensor. Distinct from
    rank variables even when the same letter is used. \\[4pt]
Einsum coord.\ space
  & $\tensor{ECS}$
  & Tensor
  & Set of ordered points for each tensor in an Einsum:
    $\bigl(\tensor{CS}^Z,\,\tensor{CS}^A,\,\tensor{CS}^B\bigr)$ \\[6pt]
Data space set
  & $\tensor{DS}^T$
  & Data
  & The set of data values tensor $T$ can contain; user-defined
    (e.g., integers, reals, strings). \\[4pt]
Empty value set
  & $E^T$
  & Data
  & A singleton subset of $\tensor{DS}^T$ identifying the ``empty'' value
    of $T$ (e.g., $\{0\}$ for integers); $E^T \subseteq \tensor{DS}^T$. \\
\bottomrule
\end{tabularx}
\end{table}

\paragraph{Notation convention.}
Set names in this section follow a consistent scheme.
The \textbf{base} identifies the kind of set ($\tensor{CS}$ for coordinate
space, $\tensor{RCS}$ for rank coordinate space, $\tensor{DS}$ for data
space, $\tensor{IS}$ for iteration space, $\tensor{ECS}$ for Einsum
coordinate space).
A \textbf{superscript} names the tensor the set belongs to
(e.g., $\tensor{CS}^Z$ is ``the coordinate space of tensor $Z$'').
A \textbf{subscript} names a restriction, either to a particular rank
(e.g., $\tensor{RCS}^T_R$ is ``the rank coordinate space of rank $R$ of
tensor $T$'') or to a refinement of the set itself
(e.g., $\tensor{DS}^T_{ne}$ is ``the non-empty data space of tensor $T$'').
Scalars (such as the rank shape $RS$) are not sets and are written without
decoration.
When defining sets, we use the notation $\{x \mid P(x)\}$, read as ``the set
of all $x$ such that $P(x)$ is true''.
In this section, a \emph{point} may refer either to a point in the iteration
space or to a point in a tensor coordinate space; we qualify the phrase when
the distinction matters.

\subsubsection{Iteration Space and the Rank Variable Set}\label{sssec:rvs}
\emph{Rank variables} determine the iteration space over which an Einsum
iterates, equivalently the region over which computations can occur.

Given an iteration space with $D$ ranks, we define the iteration space,
$\tensor{IS}$, as the Cartesian product of $D$ different rank variable sets:
\begin{align}
\tensor{IS} = R_0 \times R_1 \times R_2 \times \cdots \times R_{D-1}.
\end{align}

A rank variable, $r_d$, is an element in a rank variable set.
A rank variable set may be user-defined, but by default, its elements are
whole numbers:
\begin{align}
r_d \in R_d,
\qquad
R_d = \mathbbm{W}.
\end{align}
This does not mean an implementation must enumerate all of $\mathbbm{W}$.
Rather, the rank variable set defines the space a rank variable may be
drawn from, while rank variable expressions, tensor coordinates, and
\merge{} guards determine which points are actually evaluated.

For example, a user-defined rank variable set may allow
$r \in \mathbbm{R}$ or some other space.
Indeed, work in the extended Einsum area already allows continuous
indices~\cite{won:2024:CTA}.

Suppose we have the following Einsum:
\begin{equation}\label{eqn:GEMM}
Z^{P, Q}_{m, n} = A^{R,S}_{m, k} \times B^{T,U}_{k, n}.
\end{equation}
Readers should recognize this as a matrix-matrix multiplication
(\acro{GEMM}).

This expression has three rank variables: $m \in M$, $n \in N$, and
$k \in K$.
Thus, its iteration space, $\tensor{IS}_{\text{GEMM}}$, is
$M \times N \times K$, where $m \in M \subset \mathbb{W}$
(with similar definitions for $K$ and $N$).

\subsubsection{Rank Coordinate Sets}\label{sssec:rank_coordinate_set}
A rank coordinate is a variable that indexes into a specific rank of a tensor.
An $n$-dimensional tensor contains $n$ ranks, each with a unique name.
A coordinate uniquely identifies a location within a specific rank of a
specific tensor.
We call the set of possible coordinates for a given rank the
\textbf{rank coordinate set}.
The \textit{rank shape} is the cardinality of the coordinate set.

In Equation~\ref{eqn:GEMM}, each tensor has two ranks, whose unique names we
indicate in the superscript.
Here, $Z$ has two ranks, with the first rank named $P$ and the second rank
named $Q$.

\begin{example_box}[label=sidebar:dense_rcs]{Dense Rank Coordinate Sets}
Oftentimes, in dense cases, we implicitly define the coordinate set of a rank
through the shape.
Given a rank $R$ with shape $\tensor{RS}$, we can define the following dense
rank coordinate:
\[
\tensor{DenseCS}_R := \{r \mid r \in [0, RS)\}
\]

In the \emph{tensor declaration} region of a fully-specified \edge{}
expression, we often write the following:
  \begin{align}\label{eqn:dense_rcs}
    &\triangleright \text{Tensors} \notag \\
    A^{K=3, M=5} &\rightarrow  \text{Integer, empty=}{0}
  \end{align}
This indicates two rank names, $K$ and $M$, and refers to the following rank
coordinate sets:
    \begin{align}
    \tensor{RCS}^A_K &= \setc{k}{k \in [0, K)}  \\
    \tensor{RCS}^A_M &= \setc{m}{m \in [0, M)}
    \end{align}
\end{example_box}

\begin{example_box}[label=sidebar:sparse_rcs]{Sparse Rank Coordinate Sets}
In the sparse case, not every coordinate within the $[0, RS)$ range exists.
Suppose we have a rank, $R$, where only four coordinates exist
(i.e., will contain non-empty fibers):

\[
\tensor{RCS}^A_M := \{0, 1, 5, 7\}
\]

In \edge{} expressions, we often do not explicitly specify the coordinate
space for sparse tensors and only indicate that
$\tensor{RCS}^A_R \subseteq \tensor{DenseCS}_R$.
Syntactically, the notation is similar to the dense case
(Example~\ref{sidebar:dense_rcs}), but we can also explicitly specify it:
  \begin{align}\label{eqn:sparse_rcs}
    &\triangleright \text{Tensors} \notag \\
    A^{K=3, M=8} &\rightarrow  \text{Integer, empty=}{0} \\
    &\ \ \tensor{RCS}^A_M = \{0, 1, 5, 7\}
  \end{align}
\end{example_box}

Since rank coordinate sets can be user-defined, all sorts of interesting
custom coordinate sets can exist (see Example~\ref{sidebar:abc_rcs}).

\begin{example_box}[label=sidebar:abc_rcs]{User-Defined Rank Coordinate Sets}
In graph algorithms, vertices may be given non-integer ids, such as vertex
`a', `b', `c', etc.
To specify this coordinate space, one can explicitly state the set of the
corresponding rank in the tensor declaration:
\begin{align}
&\triangleright \text{Tensors}\\
&G^{S=|V|, D=|V|} \rightarrow \ldots \\
&\ \ \tensor{RCS}^G_S =
\{\mathtt{a},\; \mathtt{b},\; \mathtt{c},\; \mathtt{d},\; \mathtt{e}\},
\qquad
\tensor{RCS}^G_D = \tensor{RCS}^G_S
\end{align}
Here, the $S$ rank of $G$ is indexed by character coordinates
(as is the $D$ rank of $G$).
\end{example_box}

\subsubsection{Tensor Coordinate Space Sets}
A tensor coordinate space, always associated with a specific tensor, is the
set of all rank coordinate tuples for that tensor.
It is the Cartesian product of all rank coordinate sets associated with a
given tensor.
Given an $N$-tensor, an element in the coordinate space set
\emph{of that tensor} is an ordered tuple of $N$ coordinates, where each
coordinate corresponds to a specific rank in the tensor, ordered according
to the order in which the ranks are declared in the tensor.
For instance, in Equation~\ref{eqn:GEMM}, $Z$'s tensor coordinate space is
$P$ followed by $Q$\@.
Each coordinate in the tuple is drawn from the rank coordinate set for the
corresponding tensor rank.

The coordinate space set for a tensor $T^{M, N, P, Q, \ldots}$ with ranks
$M$, $N$, $P$, $Q, \ldots$, is the Cartesian product of the rank coordinate
sets for those ranks:
\begin{align}
\tensor{CS}^{T}
=
\tensor{RCS}^T_M
\times
\tensor{RCS}^T_N
\times
\tensor{RCS}^T_P
\times
\tensor{RCS}^T_Q
\times \ldots
\end{align}

Explicitly enumerating its elements, $\tensor{CS}^T$ is the set of all tuples
$(m, n, p, q, \ldots)$ where each coordinate is drawn from the rank coordinate
set of the corresponding rank:
\[
\tensor{CS}^T :=
\setc{(m, n, p, q, \ldots)}
{
m \in \tensor{RCS}^T_M
\;\wedge\;
n \in \tensor{RCS}^T_N
\;\wedge\;
p \in \tensor{RCS}^T_P
\;\wedge\;
q \in \tensor{RCS}^T_Q
\;\wedge\; \ldots
}.
\]
We call each such $N$-tuple in $\tensor{CS}^T$ a \emph{point};
when needed, we distinguish this from a point in the iteration space.

\subsubsection{Data Space Set}
A tensor potentially contains a data value of a specific user-defined data type
at every point in the coordinate space of the tensor.
The \emph{data space set} consists of all data values a tensor \emph{can}
contain (that is, the \emph{codomain} of a tensor, see
\S~\ref{ssec:tensor}).
Let $\tensor{DS}^T$ be the data space of tensor $T$.
Then,
\begin{align}
\tensor{DS}^T := \set{\text{user-defined set of data values}}
\end{align}

Additionally, we distinguish the \emph{empty value set}, $E^T$, which is a
singleton subset of $\tensor{DS}^T$ containing the user-specified value
considered ``empty'' or ``null'' for the tensor.
Let $P_T$ be a user-defined predicate over $\tensor{DS}^T$ with exactly one
satisfying value. Then:
\begin{align}
E^T := \setc{x \in \tensor{DS}^T}{P_T(x)},
\qquad
|E^T| = 1.
\end{align}
Moreover, $E^T \subseteq \tensor{DS}^T$.

Future work may expand on this by allowing multiple ``sentinel'' values in
place of the empty set.

\begin{example_box}[label=sidebar:data_spaces]{Data Types and Empty Values}
A common empty value in traditional tensor algebra is $0$
(for integer data types).
Then, we can write the following \edge{} tensor declaration:

\begin{align}
&\triangleright \text{Tensors} \notag \\
A^{K=3, M=8} &\rightarrow  \text{Integer, empty=}{0} \\
\end{align}

This corresponds to the following sets:

\begin{align}
\tensor{DS}^A &= \setc{v}{v \in \mathbb{W}}\\
E^A  &= \setc{x}{x \equiv 0}
\end{align}

We can read these as follows: ``the data space of $A$ is every $v$ in the
space of whole numbers'', and ``the empty value set of $A$ is a single value
$x$ equivalent to $0$''.
\end{example_box}
\subsection{Functions and Transformations}

A function associates elements in an input set (the \emph{domain}) with
elements in an output set (the \emph{codomain})~\cite{schmidt:1997:DSM}.
More precisely, each element in the domain of a function is associated with
exactly one output element; for partial functions, the domain is a subset of
the input set.
Throughout this section, partial functions appear often enough that we use the
arrow $\rightharpoonup$ as a visual reminder that the function is partial
(``defined for some, but not necessarily all, elements of its input set''),
reserving $\rightarrow$ for total functions (defined for all elements in the
input set).

The \emph{graph} of a function is a set of ordered pairs, where the first
element in the pair is the input element, and the second element is the
associated output element.

The following uses functions to describe the semantic behavior of different
components in an \edge{} expression.

\subsubsection{Rank Variable Expression}

A \emph{rank variable expression} (RVE) is an expression that, when evaluated
at a point in the iteration space ($\tensor{IS}$), produces a rank coordinate
value for a single, specific rank of a single, specific tensor.
It is many-to-one: given a point in the iteration space, it produces exactly
one rank coordinate value, but multiple points in the iteration space may
produce the same rank coordinate value.
Each tensor rank has its own RVE\@.

For a tensor $T$ with rank $R$, we write the corresponding RVE as
$\Gamma^T_R$.
For a point in the iteration space,
\begin{align}
(rv_0, rv_1, \ldots, rv_{D-1}) \in \tensor{IS},
\end{align}
we have
\begin{align}
\Gamma^T_R(rv_0, rv_1, \ldots, rv_{D-1}) = r.
\end{align}
If $r \in \tensor{RCS}^T_R$, then $r$ is a valid coordinate for rank $R$ of
tensor $T$.
If $r \notin \tensor{RCS}^T_R$, then the value produced by the RVE lies
outside the rank coordinate set for that tensor rank.

That is, a rank variable expression takes as input a point in the iteration
space and returns a rank coordinate value.
For brevity, we will write $is$ to denote a generic point in the iteration
space:
\[
is = (rv_0, rv_1, \ldots, rv_{D-1}) \in \tensor{IS}.
\]

\begin{example_box}[label=sidebar:gemm_rve]{RVEs in GEMM}
The GEMM expression in Equation~\ref{eqn:GEMM} has six ranks
(two for each tensor).
Each rank is associated with a rank variable expression (RVE) that sends a
point in the iteration space to a rank coordinate value.

The iteration space for GEMM is:
\[
\tensor{IS} = M \times N \times K.
\]

\begin{enumerate}
\item Tensor $Z$ has two ranks, $P$ and $Q$. The RVEs select the corresponding
rank variables:
\[
\Gamma^Z_P(m,n,k) = m,
\quad
\Gamma^Z_Q(m,n,k) = n.
\]

\item Tensor $A$ has ranks $R$ and $S$, accessed via:
\[
\Gamma^A_R(m,n,k) = m,
\qquad
\Gamma^A_S(m,n,k) = k.
\]

\item Tensor $B$ has ranks $T$ and $U$, accessed via:
\[
\Gamma^B_T(m,n,k) = k,
\qquad
\Gamma^B_U(m,n,k) = n.
\]
\end{enumerate}

Overall, in GEMM, every RVE is a trivial projection.
Each RVE simply picks one component out of the tuple $(m, n, k)$ in the
iteration space.
\end{example_box}

\begin{example_box}[label=sidebar:conv_rve]{RVEs in Convolution}
Suppose we want to express a 2D convolution of an $8 \times 6$ output from a
$10 \times 12$ input and a $4 \times 4$ filter, with no padding and unit
stride.
The corresponding \edge{} expression is:
\begin{align}
\triangleright \text{Tensors}\\
Z^{M=8, N=6}  & \rightarrow \text{Integer, empty}=0  \\
A^{U=10, V=12} & \rightarrow \text{Integer, empty}=0 \\
B^{H=4, J=4} & \rightarrow \text{Integer, empty}=0 \\
& \triangleright \text{Extended Einsum}\notag \\
Z_{p, q} &= A_{p+s, q+r} \cdot B_{s, r} :: \bigwedge \times
\end{align}

Semantically, this \edge{} Einsum defines the iteration space as:
\begin{align}
\tensor{IS} = S \times R \times P \times Q.
\end{align}

It has six RVEs, one for each rank in the Einsum:
\begin{enumerate}
\item For the $M$ rank of tensor $Z$:
\[
\Gamma^Z_M(s, r, p, q) = p.
\]

\item For the $N$ rank of tensor $Z$:
\[
\Gamma^Z_N(s, r, p, q) = q.
\]

\item For the $U$ rank of tensor $A$:
\[
\Gamma^A_U(s, r, p, q) = p+s.
\]

\item For the $V$ rank of tensor $A$:
\[
\Gamma^A_V(s, r, p, q) = q+r.
\]

\item For the $H$ rank of tensor $B$:
\[
\Gamma^B_H(s, r, p, q) = s.
\]

\item For the $J$ rank of tensor $B$:
\[
\Gamma^B_J(s, r, p, q) = r.
\]
\end{enumerate}

Unlike GEMM, two of the RVEs here ($\Gamma^A_U$ and $\Gamma^A_V$)
combine multiple variables in the iteration space through addition.
This is how \edge{} expresses the sliding-window access pattern of
convolution: each rank coordinate value for $A$ is a separate function of
the iteration space ($\Gamma^A_U$ produces $p+s$, $\Gamma^A_V$ produces
$q+r$), and not just a projection of a single variable in the iteration
space.
\end{example_box}

An RVE may produce a coordinate value that lies outside the rank coordinate
set of the tensor rank it indexes.
The consequence depends on whether the RVE belongs to an input tensor or the
output tensor.

For an input tensor, an out-of-range coordinate value means that the tensor
access is undefined at that coordinate.
The corresponding existence predicate evaluates to \texttt{False}, and the
\merge operator determines whether the action produces no value or proceeds
using the tensor's empty value.
This is useful for boundary, halo, and padding-like behavior.

For the output tensor, the output RVEs must produce a valid coordinate in
$\tensor{CS}^Z$.
If the output coordinate is outside $\tensor{CS}^Z$, then there is no output
location for that point in the iteration space, and that point produces no
\maptmp.

Thus, in Example~\ref{sidebar:conv_rve}, if $q+r \geq 12$ (since $V$ has a
shape of 12), the access to $A_{p+s,q+r}$ is outside the coordinate space of
$A$.
This does not by itself invalidate the output coordinate $(p,q)$.
Instead, $\Exists_A(p+s,q+r)$ evaluates to \texttt{False}; the \merge
operator then determines whether \maptxt produces no value or proceeds using
the empty value of $A$.

\subsubsection{Einsum Coordinate Set}

Given RVEs, we can define the \emph{Einsum Coordinate Space}
($\tensor{ECS}$) for a particular Einsum.
An element of $\tensor{ECS}$ is an ordered tuple of tensor coordinate tuples:
\[
(cs^Z, cs^A, cs^B).
\]
The tuple is ordered as follows: the output tensor coordinate, followed by
the first input tensor coordinate, and then the second input tensor
coordinate.

For convenience, we write $\Gamma^T(is)$ for the tuple of RVE outputs for
all ranks of tensor $T$ at a point $is$ in the iteration space.
For example, if tensor $T$ has ranks $R_0,\ldots,R_{N-1}$, then:
\[
\Gamma^T(is)
=
\bigl(
  \Gamma^T_{R_0}(is),
  \ldots,
  \Gamma^T_{R_{N-1}}(is)
\bigr).
\]

For an Einsum with output tensor $Z$ and input tensors $A$ and $B$,
$\tensor{ECS}$ is the set of coordinate tuples produced by the RVEs at
points in the iteration space whose output coordinate is valid:
\begin{align}
\tensor{ECS}
:=
\setc{
  (cs^Z, cs^A, cs^B)
}{
  \begin{array}{l}
  \exists is \in \tensor{IS}
  \text{ such that }
  cs^Z = \Gamma^Z(is)
  \;\wedge\;
  cs^Z \in \tensor{CS}^Z
  \\[2pt]
  {}\wedge\;
  cs^A = \Gamma^A(is)
  \;\wedge\;
  cs^B = \Gamma^B(is)
  \end{array}
}.
\end{align}

The output coordinate $cs^Z$ must be a valid point in $\tensor{CS}^Z$.
However, the input coordinates $cs^A$ and $cs^B$ may lie outside
$\tensor{CS}^A$ or $\tensor{CS}^B$.
When this happens, the corresponding tensor existence predicate evaluates to
\texttt{False}.

\begin{example_box}[label=sidebar:ecs]{ECS for GEMM}
For the GEMM expression in Equation~\eqref{eqn:GEMM}, the RVEs produce:
\[
cs^Z = (m,n),
\qquad
cs^A = (m,k),
\qquad
cs^B = (k,n).
\]
Thus, the Einsum Coordinate Space for this expression is:
\begin{align}
\tensor{ECS}_{\text{GEMM}}
=
\setc{
  \bigl((m,n),\; (m,k),\; (k,n)\bigr)
}{
  \begin{array}{l}
  (m,n,k) \in \tensor{IS}_{\text{GEMM}}
  \\[2pt]
  {}\wedge\;
  (m,n) \in \tensor{CS}^Z
  \end{array}
}.
\end{align}
\end{example_box}

\begin{example_box}[label=sidebar:ranks_vars]{Rank Names vs. Rank Variables}
In the examples thus far, we have taken care to provide distinct rank names
and rank variables within each Einsum.
Oftentimes, however, an Einsum expression may use the same variable to
represent both the rank \emph{name} and a rank \emph{variable}.
Regardless, \emph{rank names are unique to each tensor} and are
\emph{distinct from rank variables}.

For example, the GEMM expression in Equation~\ref{eqn:GEMM} may be rewritten
as:
\[
Z^{M, N}_{m, n} = A^{M, K}_{m, k} \times B^{K, N}_{k, n}.
\]
Here, rank variable $m$ is distinct from the $M$ rank of $Z$, which is also
distinct from the $M$ rank of $A$.
\end{example_box}

\subsubsection{Einsum Projection}\label{ssec:ep}

Given RVEs, the \emph{Einsum Projection} ($\tensor{EP}$) evaluates those
RVEs at a point in the iteration space and produces the corresponding output
and input tensor coordinate tuples.
The output coordinate tuple must lie in $\tensor{CS}^Z$ for $\tensor{EP}$ to
be defined at that point.
Input tensor coordinate tuples may lie outside their tensor coordinate spaces;
when they do, the corresponding tensor existence predicate evaluates to
\texttt{False}.

Thus, $\tensor{EP}$ is a partial function:
\begin{align}
\tensor{EP} : \tensor{IS} \rightharpoonup \tensor{ECS}.
\end{align}
For a point $is \in \tensor{IS}$, the projection is:
\begin{align}
\tensor{EP}(is)
=
\bigl(
  \Gamma^Z(is),
  \Gamma^A(is),
  \Gamma^B(is)
\bigr),
\end{align}
when $\Gamma^Z(is) \in \tensor{CS}^Z$.
If $\Gamma^Z(is) \notin \tensor{CS}^Z$, then $\tensor{EP}$ is undefined at
$is$.

To capture how a particular Einsum's RVEs associate points in the iteration
space with tensor coordinates, we use the graph of $\tensor{EP}$.
In set theory, this graph is the set of input/output pairs realized by
$\tensor{EP}$ for that Einsum.

\begin{example_box}[label=sidebar:ecs_conv]{Einsum Coordinate Space with RVEs}
For 2D convolution (Example~\ref{sidebar:conv_rve}), the Einsum projection
produces tuples of the form:
\[
\bigl((p,q),\; (p+s,q+r),\; (s,r)\bigr).
\]
The input coordinates $(p+s,q+r)$ and $(s,r)$ may lie outside their tensor
coordinate spaces; if they do, the corresponding existence predicate
evaluates to \texttt{False}.

Replacing each coordinate with its RVE and expanding, the Einsum projection
creates the following graph:
\begin{align}
\textit{graph}(\tensor{EP})
=
\setc{
  \Bigl(
    is,\;
    \bigl(
      (p,q),\;
      (p+s,q+r),\;
      (s,r)
    \bigr)
  \Bigr)
}{
  \begin{array}{l}
  is = (s,r,p,q) \in \tensor{IS}
  \\[2pt]
  {}\wedge\;
  p \in \tensor{RCS}^Z_M
  \;\wedge\;
  q \in \tensor{RCS}^Z_N
  \end{array}
}.
\end{align}
\end{example_box}

Each rank variable expression is local to the rank it accesses.
Einsum projections provide a global view of rank accesses, and thus enable
algebraic manipulations of rank variable expressions that impact other rank
variable expressions within the Einsum.

\begin{example_box}[label=sidebar:alg_rve]{Eingebraic Manipulation of RVEs}
Let us introduce two new variables, $c = p+s$ and $d = q+r$, drawn from new
rank variable sets $C$ and $D$, respectively.
Given these expressions, then $p = c-s$ and $q = d-r$.
Substituting these terms into the Einsum projection produces the following:
\begin{align}
\textit{graph}(\tensor{EP})
=
\setc{
  \Bigl(
    is,\;
    \bigl(
      (c-s,d-r),\;
      (c,d),\;
      (s,r)
    \bigr)
  \Bigr)
}{
  \begin{array}{l}
  is = (s,r,c,d) \in \tensor{IS}
  \\[2pt]
  {}\wedge\;
  c-s \in \tensor{RCS}^Z_M
  \;\wedge\;
  d-r \in \tensor{RCS}^Z_N
  \end{array}
},
\end{align}
where the iteration space is now $S \times R \times C \times D$.
The corresponding, transformed Einsum is:
\begin{align}
    Z_{c-s, d-r} &= A_{c, d} \cdot B_{s, r}.
\end{align}
\end{example_box}

\subsubsection{Specifying Iteration Space Constraints}\label{ssec:sis}

\edge{} allows us to specify constraints on the shape of the iteration space.
For example, triangular iterations (such as traversing the lower triangle of
a matrix) require the iteration space to exclude certain points that the
Cartesian product of rank variable sets would otherwise admit.

Output RVEs and explicit predicates narrow down the set of points in the
iteration space that participate in an Einsum.
If an output RVE produces an out-of-bounds coordinate at a given point in the
iteration space, that point produces no \maptmp.
If an input RVE produces an out-of-bounds coordinate, the corresponding tensor
access is undefined and its existence predicate evaluates to \texttt{False}.

We also allow a user to constrain the space explicitly: the user states a
condition over variables in the iteration space, and points failing the
condition are excluded.

\edge{} captures this by letting users attach predicates to rank variables,
using the notation
``$\tensor{rv} : \langle \text{predicate} \rangle$'',
where ``$:$'' denotes a predicate over the iteration space
(see \S~\ref{ssec:conditionals}).
While such constraints may appear locally within tensor subscripts, they
globally restrict the set of valid points in the iteration space.
Semantically, the predicate restricts the set of points in the iteration
space in the graph of the Einsum projection
(see Example~\ref{sidebar:rve_predicates}).

\begin{example_box}[label=sidebar:rve_predicates]{Predicates in Rank Variable Expressions}
The following Einsum only accesses the lower triangular region of the $B$
tensor:
\begin{align}
Z_{m, n} = A_{m, k} \cdot B_{n, k: k < n}
\end{align}
Here, the second rank of $B$ is accessed via rank variable $k$ such that
$k < n$.
The predicate $k < n$ restricts which points in the iteration space
participate in the Einsum.
Points $(m,n,k)$ with $k \geq n$ are excluded from the graph of
$\tensor{EP}$; they do not produce \maptmp elements and are not observed by
\reduce.
This does not make the output coordinate $(m,n)$ invalid.
Rather, for each output coordinate $(m,n)$, \reduce combines only the
contributions from points $(m,n,k)$ where $k < n$.

The corresponding Einsum projection is then:
\begin{align}
\textit{graph}(\tensor{EP})
=
\setc{
  \Bigl(
    is,\;
    \bigl(
      (m,n),\;
      (m,k),\;
      (n,k)
    \bigr)
  \Bigr)
}{
  \begin{array}{l}
  is = (m,n,k) \in \tensor{IS}
  \\[2pt]
  {}\wedge\; k < n
  \end{array}
}.
\end{align}

This indicates that only $(m,n,k)$ points in the iteration space where
$k < n$ participate in the Einsum.
\end{example_box}

\subsubsection{Tensor}\label{ssec:tensor}

\begin{figure}
  \centering
  \includegraphics[width=\linewidth]{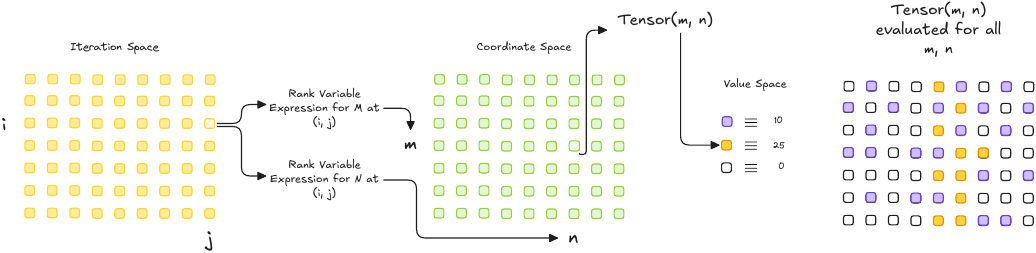}
  \caption{Two rank variable expressions transform points in the $I \times J$
  iteration space to points in the $M \times N$ coordinate space of tensor
  $\tensor{Tensor}$. The tensor itself takes a point in the coordinate space
  to a value in the data space.}
  \label{fig:is-to-cs-function}
\end{figure}

Let the non-empty data space be the data space without the empty value set:
\begin{align}
\tensor{DS}^T_{ne} &\subseteq \tensor{DS}^T \\
\tensor{DS}^T_{ne} &= \tensor{DS}^T \setminus E^T.
\end{align}

A tensor, $T$, is a \emph{partial} function that goes from its tensor
coordinate space to its \emph{non-empty} data space:
\begin{align}
T : \tensor{CS}^T \rightharpoonup \tensor{DS}^T_{ne}.
\end{align}

For a given point:
\begin{align}
T(p,q,r,\ldots) = dv,
\qquad dv \in \tensor{DS}^T_{ne}.
\end{align}

Note how this is a general enough abstraction to admit any underlying
implementation, whether fibertrees~\cite{Nayak:2023:TDF}, a hash table that
represents a tensor with many ranks but repeated values, sparse structures,
and so on.
Figure~\ref{fig:is-to-cs-function} visualizes an example tensor as a
function.

\begin{example_box}[label=sidebar:tensor_decl]{Semantics of the \edge{} Tensor Declaration Region}
We can now state the semantic meaning of \edge{}'s tensor declaration section.
Suppose we have the following declarations:
\begin{subequations}
\begin{align}
\triangleright \text{Tensors}\notag\\
Z^{M=5, N=6} & \rightarrow \text{Integer, empty}=0 \\
A^{K=10, M=5} & \rightarrow \text{Integer, empty}=0 \label{td:A}\\
B^{K=10, N=6} & \rightarrow \text{Integer, empty}=0.\label{td:B}
\end{align}
\end{subequations}

These declarations induce the following sets and functions:
\begin{enumerate}
\item There is a tensor $Z$, defined as a function
\[
Z : \tensor{CS}^Z \rightharpoonup \tensor{DS}^Z_{ne}.
\]

\item The coordinate space of $Z$ is the Cartesian product of its rank
coordinate spaces:
\[
\tensor{CS}^Z = \tensor{RCS}^Z_M \times \tensor{RCS}^Z_N.
\]

\item The declaration $M=5$ is syntactic sugar for a rank name paired with its
rank coordinate space. That is,
\[
\tensor{RCS}^Z_M = \setc{m}{m \in [0,5)},
\qquad
\tensor{RCS}^Z_N = \setc{n}{n \in [0,6)}.
\]

\item The data space of $Z$ is the set of integers:
\[
\tensor{DS}^Z = \setc{v}{v \in \mathbb{Z}}.
\]

\item The declared empty value set of $Z$ is the singleton containing $0$:
\[
E^Z = \{0\}.
\]

\item $Z$ is only defined where coordinates map to non-empty values.
\end{enumerate}

We can similarly derive the semantics for lines~\eqref{td:A}--\eqref{td:B}.
\end{example_box}

\subsubsection{Fibers and Subtensors}

At times, we may wish to refer to a subtensor
(or \emph{fiber}~\cite{Nayak:2023:TDF}) of a tensor.
Within our functional view of tensors, this notion is naturally expressed via
partial application (currying).

Given an $N$-dimensional tensor, specifying $L < N$ rank coordinates
partially evaluates the tensor function and yields a subtensor over the
remaining ranks.

For example, consider a five-dimensional tensor $Z^{P,Q,R,S,U}$:
\begin{enumerate}
\item The fully applied expression
\[
Z(p,q,r,s,u)
\]
evaluates to a single data value, $dv \in \tensor{DS}^Z_{ne}$.

\item We can partially apply the tensor by fixing specific rank coordinates:
\[
Z(p,q,r,s)
\]
For a fixed input $(p_i,q_i,r_i,s_i)$, it evaluates to a one-dimensional
subtensor indexed by the rank coordinate space $\tensor{RCS}^Z_U$:
\[
Z(p_i,q_i,r_i,s_i)
= \setc{(u,dv_u)}{u \in \tensor{RCS}^Z_U\;\wedge\; dv_u \in \tensor{DS}^Z_{ne}}.
\]
Equivalently, we may write this as a unary function:
\[
Z_{p_i,q_i,r_i,s_i}(u) = dv_u,
\qquad u \in \tensor{RCS}^Z_U.
\]
In tensor algebra terminology, this corresponds to fixing the $P,Q,R,S$
ranks and extracting a one-dimensional fiber along the $U$ rank
(a $U$-fiber).

\item The partially applied expression
\[
Z(p,q,s)
\]
fixes only the $P,Q,S$ ranks.
For a given $(p_i,q_i,s_i)$, it evaluates to a two-dimensional subtensor
indexed by the Cartesian product of rank coordinate spaces
$\tensor{RCS}^Z_R \times \tensor{RCS}^Z_U$:

\begin{align}
Z(p_i,q_i,s_i)
&=
\Bigl\{
  \bigl((r,u), dv_{r,u}\bigr)
  \mid
  (r,u) \in \tensor{RCS}^Z_R \times \tensor{RCS}^Z_U\;\wedge\;
  dv_{r,u} \in \tensor{DS}^Z_{ne}
\Bigr\}.
\end{align}

Equivalently, this subtensor may be viewed as a function
\[
Z_{p_i,q_i,s_i}(r,u) = dv_{r,u},
\qquad (r,u) \in \tensor{RCS}^Z_R \times \tensor{RCS}^Z_U.
\]

In tensor algebra terms, this operation fixes the $P,Q,S$ ranks and returns a
two-dimensional subtensor spanning the remaining $R \times U$ rank coordinate
spaces.
\end{enumerate}

\subsubsection{Existence Check}\label{sssec:existcheck}

Tensors are \emph{partial} functions: not all coordinates produce non-empty
values.
To determine whether a particular coordinate yields a non-empty value, we
define the \emph{existence predicate} for a given tensor, $\Exists$.

Semantically, $\Exists_T(c)$ evaluates to \texttt{True} iff $T$ is defined at
coordinate $c$, and to \texttt{False} otherwise:
\begin{align}
\Exists_T(c) := \bigl(c \in \mathrm{dom}(T)\bigr).
\end{align}
Since $\mathrm{dom}(T) \subseteq \tensor{CS}^T$, if $c$ is outside
$\tensor{CS}^T$, then $\Exists_T(c)$ evaluates to \texttt{False}.
Since the codomain of $T$ is $\tensor{DS}^T_{ne}$, if
$c \in \mathrm{dom}(T)$, then evaluating $T(c)$ yields a non-empty value.

The graph of $\Exists$ over a set $C$ of coordinate tuples queried for
tensor $T$ is given by:
\begin{align}
\textit{graph}_C(\Exists_T)
=
\setc{
  \bigl(c, b\bigr)
}{
  c \in C \;\wedge\; b \equiv \bigl(c \in \mathrm{dom}(T)\bigr)
}.
\end{align}

\begin{sidebar_box}[label=sidebar:total-vs-partial]{Total vs.\ Partial Tensor Semantics}
There are two mathematically equivalent ways to model tensors with empty
values:

\textbf{Total-function approach.}
A tensor may be viewed as a total function
\[
T : \tensor{CS}^T \rightarrow \tensor{DS}^T,
\]
where the data space includes a distinguished empty value $e^T$ with
$E^T = \{e^T\}$.
Under this model, every tensor coordinate is defined, and sparsity is
represented implicitly by mapping many coordinates to $e^T$.
This interpretation aligns naturally with dense array semantics.

\textbf{Partial-function approach.}
Alternatively, a tensor may be viewed as a partial function
\[
T : \tensor{CS}^T \rightharpoonup \tensor{DS}^T_{ne},
\]
whose domain consists only of coordinates associated with non-empty values.
Here, sparsity is represented explicitly by the absence of a mapping.

We choose the partial-function approach as it enables existence checks
(for \merge operators) without implying an implementation
(\designgoal{goal:opt}).
That is, an implementation may choose to realize the domain check
($c \in \mathrm{dom}(T)$) in any way it chooses.

Specifically, in the total-function approach, an implementation must touch the
coordinate space \emph{and} the data space of a given tensor to determine if a
particular location in a tensor is empty.
In the partial-function approach, an implementation can implement existence by
similarly checking the tensor values, or by encoding it in the metadata
(e.g., the COO tensor format simply does not include coordinates for empty
tensor locations).
\end{sidebar_box}
\subsection{\maptxt, \reduce, and \populate Actions}\label{action:overview}

In the following subsections, we fix an \edge{} Einsum with two input tensors,
$A$ and $B$, an output tensor, $Z$, and present the semantics of each action
as operating over this Einsum.

\begin{figure}
  \centering
  \includegraphics[
    height=0.6\textheight,
    keepaspectratio
  ]{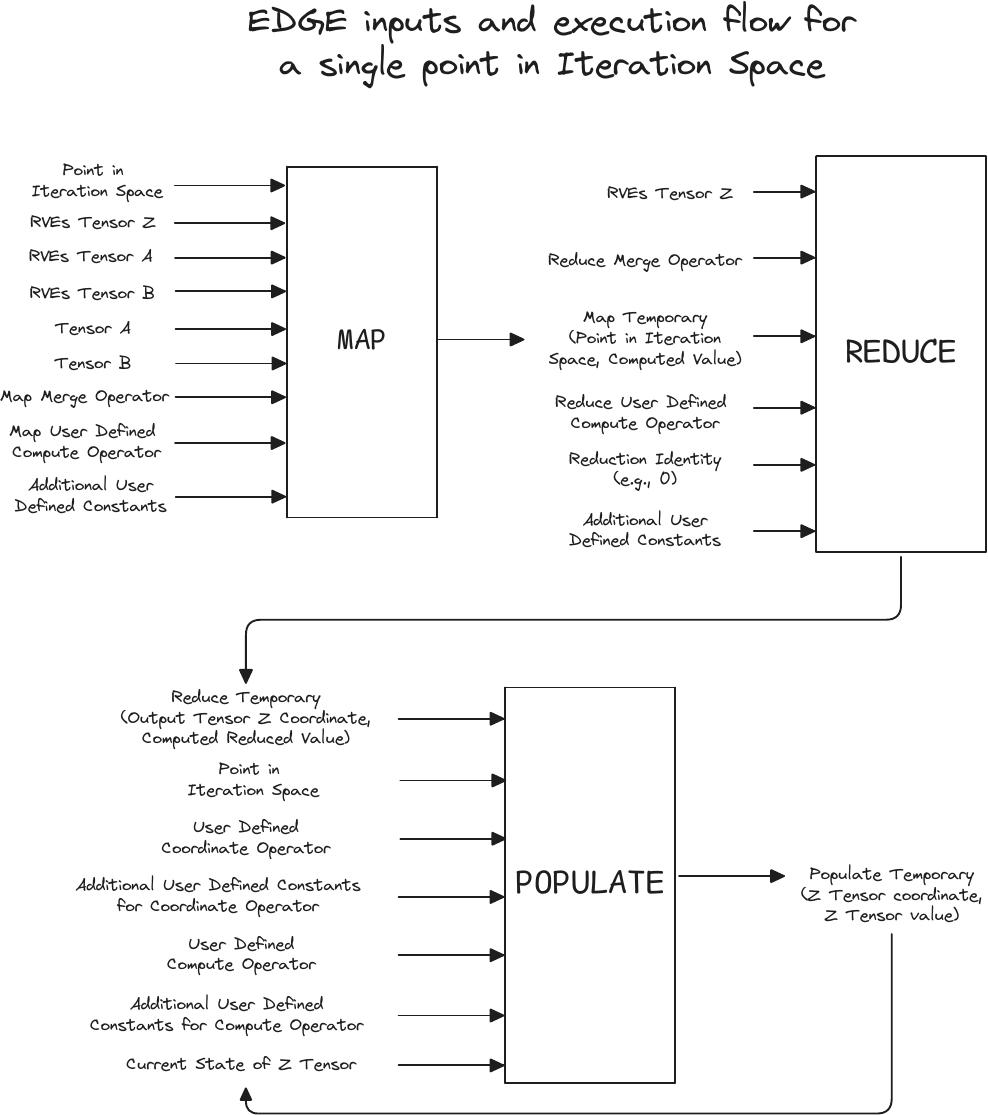}
  \caption{Execution flow of an \edge{} Einsum with input tensors $A$, $B$
  and output tensor $Z$, evaluated at a single point in the iteration space.
  The figure highlights how a single point in the iteration space passes
  through rank variable expressions to produce coordinates in the tensor
  coordinate space, and how those coordinates produce a value in the data
  space.}
  \Description{EDGE inputs and execution flow for a single point in the
  iteration space. At a point, Map executes, then Reduce, then Populate,
  with various other inputs to the three modules.}
  \label{fig:edge-single-point-exec}
\end{figure}

\begin{figure}
  \centering
  \includegraphics[
  width=\linewidth]{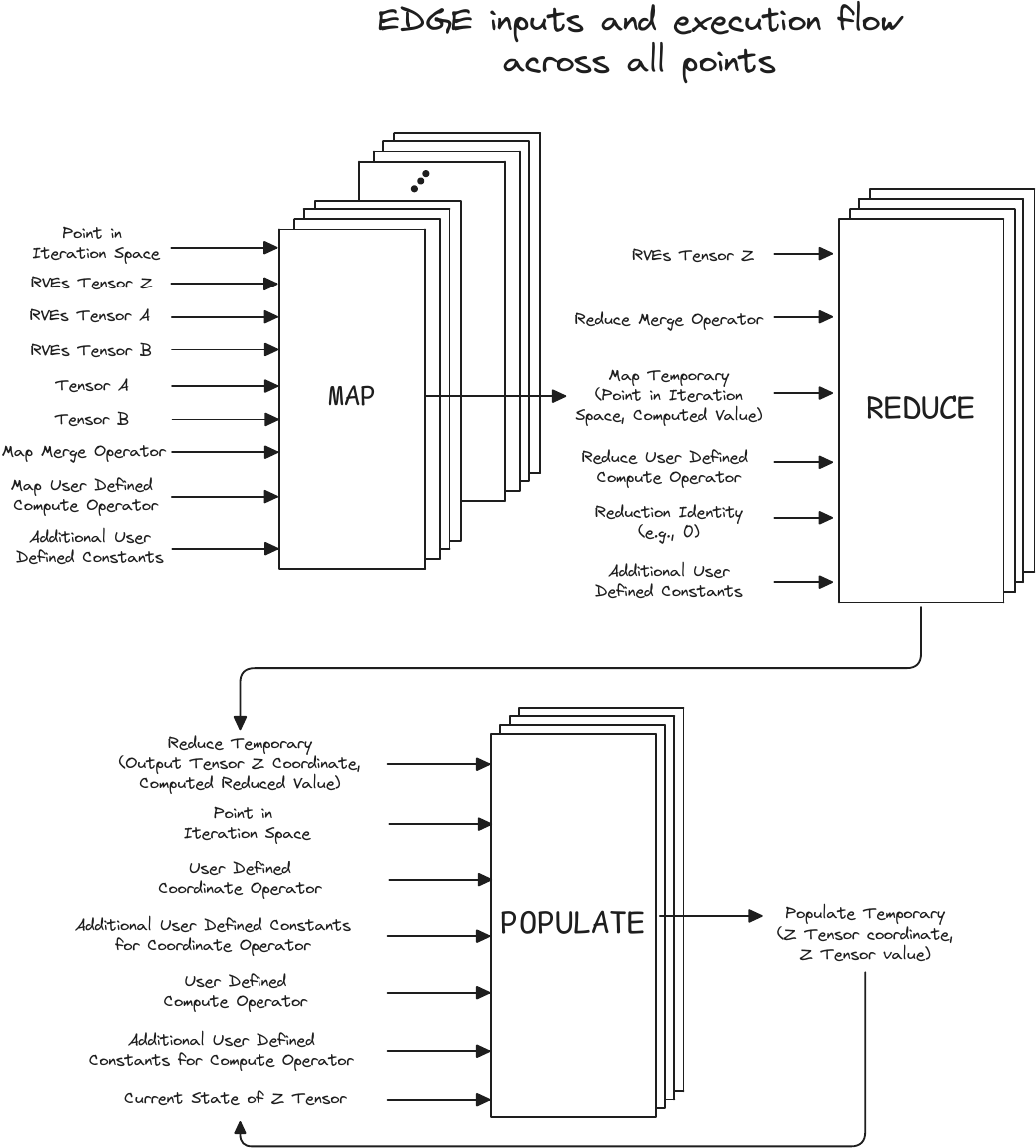}
  \caption{Execution flow of an \edge{} Einsum evaluated over the entire
  iteration space.
  The diagram illustrates how points in the iteration space are projected
  through rank variable expressions into tensor coordinate spaces, and how
  values are produced and propagated across all points.}
  \Description{EDGE inputs and execution flow across all points in the
  iteration space. Multiple invocations of Map can invoke the same Reduce
  action.}
  \label{fig:edge-all-points-exec}
\end{figure}

\begin{figure}[t]
  \centering
  \includegraphics[
    width=0.6\linewidth,
    keepaspectratio
  ]{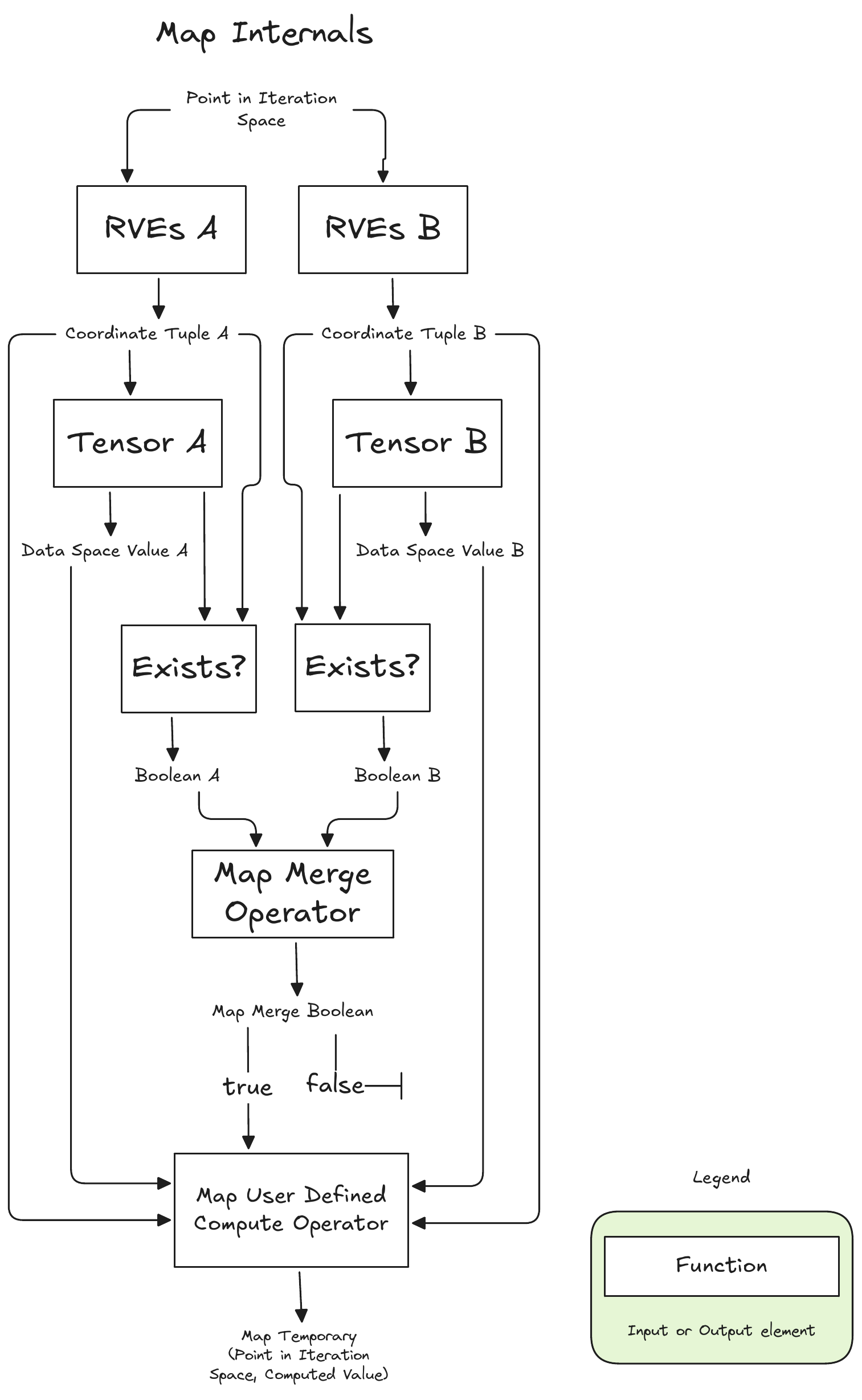}
  \caption{\maptxt: Operand values accessed at a point in the iteration
  space are combined to produce a \maptmp, which is then forwarded to the
  subsequent \reduce action.}
  \Description{Internal structure of the EDGE Map operator, showing how
  tensor operand values at a single point in the iteration space are
  accessed, combined, and emitted as a map temporary that is consumed by
  Reduce.}
  \label{fig:edge-map-internals}
\end{figure}

The following sections step through the semantics of \maptxt, \reduce, and
\populate.
Figures~\ref{fig:edge-single-point-exec}--\ref{fig:edge-all-points-exec}
provide a high-level view of how \maptxt feeds into \reduce and finally
\populate.
Although we present the steps sequentially, an implementation of \edge{} is
free to use any strategy as long as dependency constraints (represented by
the arrows) are maintained.
Thus, one implementation may choose to execute \maptxt for \emph{all} points
in the iteration space (potentially in parallel), before executing the
remaining actions; another may choose to generate a \maptmp for a single
point in the iteration space, then immediately have \reduce consume it; and
still another may choose an asynchronous and pipelined approach where all
actions are in flight for different regions of the iteration space.
The following semantics simply specify the \emph{meaning} and behavior of
these actions (\designgoal{goal:sepconcerns}).

\subsection{\maptxt}\label{action:map}

Given a single point in the iteration space, \maptxt\ determines whether a
computation occurs at that point and, if so, applies the \compute\ operator
to the corresponding tensor values of $A$ and $B$.
The result is an intermediate value consumed by subsequent actions
(e.g., \reduce).
The \emph{execution model} applies \maptxt\ independently to every point in
the iteration space.

Before diving in, we introduce two operators that appear throughout
\maptxt, \reduce, and \populate:

\begin{description}
    \item[\merge.]
    The \merge\ operator takes two Booleans (the existence of the left
    operand and the right operand at the current point in the iteration
    space) and returns a single Boolean.
    In \maptxt, this Boolean acts as a guard: if \merge\ returns
    \texttt{True}, \maptxt proceeds to \compute; if \merge\ returns
    \texttt{False}, \maptxt produces no value at this point.
    In later actions, such as \reduce, the same Boolean output is interpreted
    according to that action's semantics.
    \merge\ may be any of the 16 Boolean functions of two variables
    (see Appendix~\ref{appendix:merge}); common choices include intersection
    ($\cap$) and union ($\cup$).

    \item[\compute.]
    The \compute\ operator is a user-defined function that combines operand
    values into a single output value.
\end{description}

\paragraph{Inputs and Outputs.}
\maptxt\ is a function that takes as input
(1) a point in the iteration space,
(2) the rank variable expressions (RVEs) for each tensor in the Einsum,
(3) the input tensors $A$ and $B$,
(4) a \merge\ operator, and
(5) a user-defined \compute\ operator, with optional user-defined constants.

The output, when \maptxt is defined at this point in the iteration space,
is the computed value for that point.
When paired with the point in the iteration space as a \maptxt\ ``coordinate'',
the (point, value) tuple forms a \maptmp element.

We first describe the operational steps performed by \maptxt, then present
a precise semantic definition using set-theoretic notation.

\subsubsection{\maptxt\ Steps}
\begin{enumerate}
\item \maptxt\ first applies the Einsum projection to the point in the
iteration space, producing a tuple of tensor coordinates in the Einsum
coordinate space:
\begin{align}
\tensor{EP}(rv_0, rv_1, \ldots, rv_{D-1})
&= (cs^Z, cs^A, cs^B), \notag\\
\text{where } &cs^Z \in \tensor{CS}^Z,\;
cs^A \in \tensor{CS}^A,\;
cs^B \in \tensor{CS}^B.
\end{align}

\item Given the tensor coordinates, \maptxt\ applies the existence
predicate ($\Exists$) to the input tensors $A$ and $B$, producing two
Boolean values.

\item The \merge\ operator consumes these two Booleans and produces a single
Boolean value that determines whether this point in the iteration space
contributes a computation at all.
Isolating this decision in a two-input Boolean operator lets us express all
16 possible combinations uniformly (see Appendix~\ref{appendix:merge}).

\begin{align}
\text{\merge} : \mathbb{B} \times \mathbb{B} \rightarrow \mathbb{B}.
\end{align}

\item If \merge\ evaluates to \texttt{True}, \maptxt\ activates the
\compute\ operator.
If \merge\ evaluates to \texttt{False}, \maptxt\ produces no value at this
point in the iteration space; the \maptxt\ function is \emph{undefined}
there.
We define the following guard predicate:

\begin{align}
\mathsf{Guard}(is)
:=
\text{\merge}\bigl(
  \Exists_A(cs^A),
  \Exists_B(cs^B)
\bigr).
\end{align}

\item The \compute\ operator consumes
(1) the evaluated tensor values of $A$ and $B$ at the current point in the
iteration space,
(2) their corresponding coordinate tuples,
(3) the point in the iteration space, and
(4) any user-defined constants.
The \compute\ operator is user-defined and returns a value in the data space
of $Z$.
\begin{align}
\text{\compute} :
\tensor{DS}^A \times \tensor{DS}^B
\times \tensor{CS}^A \times \tensor{CS}^B
\times \tensor{IS}
\rightarrow \tensor{DS}^Z.
\end{align}
If the evaluated value of either $A$ or $B$ is undefined (that is, the
\merge\ operator evaluated to \texttt{True} even though one operand was
empty), \compute\ substitutes the corresponding empty value as the operand.
If the computed value lies in the empty space of the output ($dv \in E^Z$),
the \maptmp\ element is simply omitted at this point: the partial function
\maptxt\ is left undefined here.

\item Finally, the \maptmp\ element consists of an ordered pair: the point
in the iteration space and the computed value.
That is, the \maptmp\ element is a tuple of the point in the iteration space
passed as an input to \maptxt, and the corresponding computed output value
for that point.
\end{enumerate}

\subsubsection{Set Theory Notation for \maptxt}

\maptxt is a partial function over the iteration space that produces a value.
When parameterized on \compute, \merge, the set of RVEs, and tensors, its
domain/codomain mapping is:
\begin{align}
\maptxt_{A,B,\Gamma^A,\Gamma^B,\text{\merge},\text{\compute}}
: \tensor{IS} \rightharpoonup \tensor{DS}^Z_{ne}.
\end{align}

The \maptmp\ set is a set of ordered pairs, where each pair consists of a
point in the iteration space and the corresponding computed value.
Note that \maptxt\ is itself a tensor, whose graph we name the \maptmp.
Specifically, when indexed by a point in the iteration space, the \maptmp\
will return the corresponding computed value generated by \maptxt.

\begin{definition}[Map Semantics]
Fix an Einsum with input tensors $A$ and $B$, output tensor $Z$, rank variable
expressions $\Gamma^A$ and $\Gamma^B$, a merge operator (\merge), and a
compute operator (\compute).

The \maptxt\ action is defined as a function
\begin{align}
\maptxt : \tensor{IS} \rightharpoonup \tensor{DS}^Z_{ne}.
\end{align}

Let $e^A \in E^A$ and $e^B \in E^B$ denote the empty values of $A$ and $B$.
For each point $is \in \tensor{IS}$, let
\begin{align}
(cs^Z, cs^A, cs^B) = \tensor{EP}(is),
\end{align}
and define:
\begin{align}
dv_A(is)
&:=
\begin{cases}
A(cs^A), & \text{if } \Exists_A(cs^A), \\
e^A, & \text{otherwise},
\end{cases}
\\
dv_B(is)
&:=
\begin{cases}
B(cs^B), & \text{if } \Exists_B(cs^B), \\
e^B, & \text{otherwise}.
\end{cases}
\end{align}

The graph of \maptxt is then:
\begin{align}
\textit{graph}(\maptxt)
=
\{
  \bigl(cs^{\maptxt}, dv\bigr)
  \;\mid\;
  &is \in \tensor{IS},\;
  cs^{\maptxt} = is,\;\notag\\
  &(cs^Z, cs^A, cs^B) = \tensor{EP}(is),\notag\\
  &\text{\merge}\bigl(\Exists_A(cs^A), \Exists_B(cs^B)\bigr)
    = \texttt{True},\notag\\
  &dv = \text{\compute}\bigl(
    dv_A(is), dv_B(is), cs^A, cs^B, is
  \bigr),\notag\\
  &dv \notin E^Z
\},\\
\maptmp = \textit{graph}(\maptxt).
\end{align}
\end{definition}

\subsection{Reduce}\label{action:reduce}

The output tensor's rank variable expressions determine when multiple
right-hand side computed values (i.e., \maptmp values) correspond to the
same output location.
\reduce gathers results from \maptxt based on the \emph{output} location.
These values need to be \emph{combined} in some manner: \reduce specifies
how to combine results.

In \edge{}, \reduce\ does not directly place results in the output tensor
(see \populate).
Instead, \reduce\ iterates over the same iteration space as \maptxt,
queries whether \maptxt\ produced a non-empty value at each
point in the iteration space, and aggregates the resulting intermediate values
according to the coordinate space of the output tensor.

\paragraph{Inputs and Outputs.}
\reduce\ takes as input:
(1) a point in the iteration space;
(2) the \maptmp tensor, which may be queried at each point in the iteration space;
(3) a \merge\ operator;
(4) the rank variable expressions for $Z$ ($\Gamma^Z$); and
(5) a user-defined \compute\ operator, which additionally takes as input
an identity element for \reduce\ and any user-defined constants.

\reduce\ maintains internal state indexed by the coordinate space of the
output tensor ($Z$).
Conceptually, there is one reduction state per output tensor coordinate.
The output of \reduce\ is a reduction temporary (\redtmp) element, an
ordered pair consisting of an output tensor coordinate and the final
computed, reduced value.
The \redtmp is a ``shadow'' of the final output tensor, $Z$
(see Inset~\ref{sidebar-shadow}).

\begin{sidebar_box}[label=sidebar-shadow]{Shadow Tensors and the \edge{} Execution Model}
In our original \emph{operational definition of an Einsum}, computed
values are placed directly into the output tensor location, ``reducing''
if another value is already present.

In \edge{}, the ``output tensor'' is instead realized as a \emph{shadow
tensor}, where computed values (\maptmp values) are placed into a location
in the shadow (i.e., the reduction state $\tensor{RState}$, which
eventually becomes the \redtmp), reducing with whatever value is already
there.

The \populate action then either copies this shadow tensor (in the default
assignment case) or applies a final transformation to its contents before
materializing the output tensor.

This execution model remains consistent with the operational definition of
an Einsum, while cleanly separating the responsibilities of computation,
aggregation, and placement.
\end{sidebar_box}

\begin{figure}[t]
  \centering
  \includegraphics[
    width=\linewidth,
    keepaspectratio
  ]{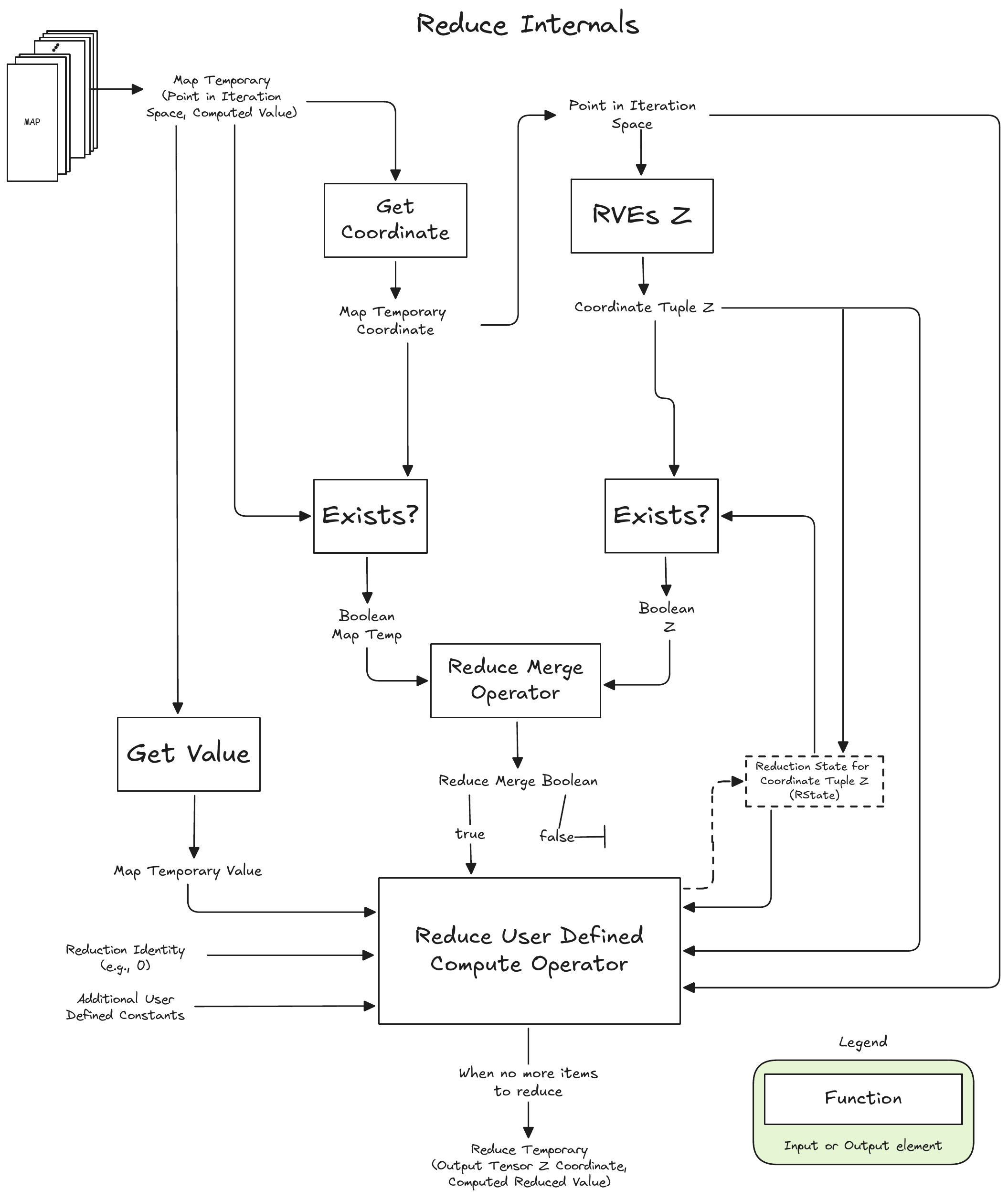}
  \caption{\reduce: for each output tensor coordinate $cs^Z$, \reduce\
  combines the current reduction state $\mathsf{RState}(cs^Z)$ with the
  incoming \maptmp\ value using the user-specified \merge\ and \compute\
  operators. The result updates $\mathsf{RState}(cs^Z)$. When no further
  \maptmp\ elements remain for that coordinate, $\mathsf{RState}(cs^Z)$
  contributes a final reduced value to \redtmp.}
  \label{fig:edge-reduce-internals}
\end{figure}

\subsubsection{\reduce\ Steps}\label{ssec:reduce_steps}

For each output tensor coordinate $cs^Z \in \tensor{CS}^Z$, \reduce\
maintains a corresponding reduction state, denoted
$\tensor{RState}_{cs^Z}$.

\begin{enumerate}
    \item For a given point in the iteration space, \reduce applies the RVEs for
    $Z$ to retrieve the corresponding coordinate
    ($cs^Z = \Gamma^Z(is)$).

    \item \reduce checks whether the current value of the reduction state is non-empty.  If this is the first time \reduce is processing
    this output coordinate, the current value of the reduction state is
    initialized to the identity element for the given \compute operator.

    \item Likewise, for that same point in the iteration space, \reduce applies an
    existence predicate to the \maptmp. That is, \reduce checks whether
    $\maptmp(is)$ returns a defined (non-empty) value. Note that \maptxt\
    \emph{must} have executed at this point in the iteration space for
    \reduce to process the \maptmp:
    \[
      \Exists_{\maptmp}(is) := \bigl(is \in \mathrm{dom}(\maptmp)\bigr).
    \]

    \item The resulting Booleans are passed as
    input to the \merge operator. As with \maptxt, \merge is
    user-specified and is one of 16 operators
    (see Appendix~\ref{appendix:merge}).
    The first/left-hand operand to \merge originates from the current value
    of $\tensor{RState}$, and the second/right-hand operand to \merge
    originates from the \maptmp.

    \item If \merge evaluates to \texttt{True}, \reduce activates the
    \compute operator and updates the reduction state.
    If \merge evaluates to \texttt{False}, the reduction state for this
    output coordinate is set to the empty value of $Z$.
    This is because, in \reduce, the \merge operator determines not
    only whether the incoming \maptmp value is used, but also whether the
    current reduction state remains valid after observing this
    point in the iteration space.
    For example, with an intersect \merge operator, a missing \maptmp value
    (i.e., empty) invalidates the accumulated reduction state for that
    output coordinate.

    As with \maptxt, we treat \merge as a guard:
    \begin{align}
    \mathsf{Guard}_{\reduce}(is)
    :=
    \text{\merge}\bigl(
      b_s,
      b_m
    \bigr),
    \end{align}
    where $b_s$ indicates whether the current value of the reduction state is non-empty, and $b_m$ indicates whether $\maptmp$ is defined at this point in the iteration space.

    \item The \compute operator consumes the current value of the reduction
    state and the evaluated \maptmp value at this point in the iteration space.

    If $\maptmp(is)$ is defined, then the evaluated \maptmp value is
    $\maptmp(is)$.
    If $\maptmp(is)$ is undefined but \merge evaluates to \texttt{True},
    then \compute receives the empty value of $Z$ for the \maptmp operand.
    This mirrors the behavior of \maptxt: when a \merge operator permits
    computation despite a missing operand, the missing operand is
    represented by the corresponding empty value.

    The \compute operator combines these values using a user-specified
    operation that takes as input
    (1) the current value of the reduction state,
    (2) the evaluated \maptmp value,
    (3) the coordinate tuple for $Z$,
    (4) the point in the iteration space,
    (5) the identity value for the reduction occurring
    (e.g., 0 for addition), and
    (6) any user-defined constants.

    \item The $\tensor{RState}_{cs^Z}$ is updated with the result of
    \compute.

    \item This process repeats incrementally for each \maptmp associated
    with the same output coordinate.
    When no further \maptmp elements remain for a given output coordinate,
    \reduce considers the corresponding reduction state as final.

    \item The resulting \redtmp is an ordered pair consisting of the output
    coordinate $cs^Z$ and the final value stored in
    $\tensor{RState}_{cs^Z}$.
    If that value is empty, \reduce does not return a value for the given
    $cs^Z$.
\end{enumerate}

When the reduction operator is associative and commutative, \reduce\ is
equivalent to a fold over the \maptmp\ elements corresponding to each output
coordinate.
In this case, the final reduced value is independent of the order in which
\maptmp\ elements are processed.

If the reduction operator is not associative and commutative, then the
result may depend on the order in which the corresponding \maptmp\ elements
are processed.
In that case, \edge{} guarantees only that the result is equivalent to
applying the reduction operator to some valid ordering of those elements.
If a particular ordering is required for correctness or reproducibility, the
user must specify that ordering explicitly at the mapping level of the
separation of concerns.

\subsubsection{Set Theory Notation for \reduce}\label{ssec:reduce_semantics}
At a high level, \reduce\ iterates over the iteration space, queries
\maptmp, and uses the rank variable expressions for $Z$ to identify the
corresponding output coordinate.
For each output coordinate, \reduce\ maintains a reduction state and updates
that state according to the user-specified \merge\ and \compute operators.
After all points in the iteration space have been processed, the non-empty final
states form the \redtmp.

\begin{definition}[Reduce Semantics]
Fix an Einsum with output tensor $Z$, a reduction merge operator \merge,
and a reduction compute operator \compute.

Let $\textit{graph}(\maptxt) \subseteq \tensor{IS} \times \tensor{DS}^Z$
denote the graph of \maptxt, and let
\begin{align}
\Gamma^Z : \tensor{IS} \rightarrow \tensor{CS}^Z
\end{align}
be the group of RVEs that extract the output tensor coordinate associated
with a point in the iteration space.

The \reduce\ action maintains a state function
\begin{align}
\mathsf{RState} : \tensor{CS}^Z \rightharpoonup \tensor{DS}^Z.
\end{align}
The state is created lazily for an output coordinate when that coordinate
is first processed.
When created, its initial value is the identity for the given \compute
operator.
After that point, $\mathsf{RState}$ may store either a non-empty reduced
value or the empty value $e^Z \in E^Z$.

Let $\mathsf{RState}^{old}$ denote the reduction state before processing
the current point in the iteration space, and let $\mathsf{RState}^{new}$ denote
the reduction state after processing it.

Let $e^Z \in E^Z$ denote the empty value of $Z$, and let
$\mathbbm{1}_{\reduce}$ denote the identity value for the reduction.

At a point in the iteration space, the output coordinate is:
\begin{align}
cs^Z &= \Gamma^Z(is).
\end{align}

The current value of the reduction state is:
\begin{align}
dv_s
:=
\begin{cases}
\mathsf{RState}^{old}(cs^Z),
&\text{if } cs^Z \in \mathrm{dom}(\mathsf{RState}^{old}), \\
\mathbbm{1}_{\reduce},
&\text{otherwise}.
\end{cases}
\end{align}

The Boolean inputs to the \merge operator are:
\begin{align}
b_s &:= \bigl(dv_s \notin E^Z\bigr), \\
b_m &:= \Exists_{\maptmp}(is).
\end{align}

The value consumed from \maptmp\ at this point in the iteration space is:
\begin{align}
dv
:=
\begin{cases}
dv', &\text{if } (is, dv') \in \textit{graph}(\maptxt), \\
e^Z, &\text{otherwise}.
\end{cases}
\end{align}

The reduction state is updated incrementally according to:
\begin{align}
\mathsf{RState}^{new}(cs^Z)
:=
\begin{cases}
\text{\compute}\bigl(dv_s, dv\bigr),
&\text{\merge}(b_s, b_m) = \texttt{True}, \\
e^Z,
&\text{otherwise}.
\end{cases}
\end{align}
All other reduction states are left unchanged during this update step.

After this point in the iteration space is processed,
$\mathsf{RState}^{new}$ becomes the current reduction state for subsequent points in the iteration space.

Let $\mathsf{RState}^{final}$ denote the reduction state after all points in the iteration space have been processed.
Then \reduce\ returns:
\begin{align}
\redtmp
=
\setc{
  \bigl(cs^Z, dv\bigr)
}{
  cs^Z \in \mathrm{dom}(\mathsf{RState}^{final})
  \;\wedge\;
  dv = \mathsf{RState}^{final}(cs^Z)
  \;\wedge\;
  dv \notin E^Z
}.
\end{align}
\end{definition}

\subsection{Populate}

The presence of ``$*$'' annotations in the left-hand side of an
expression determines which fibers are mutable with the tensor.
A ``$*$'' annotation on a given rank indicates that \populate\ may
update multiple coordinates within the corresponding fiber, rather than
writing to a single output coordinate.
This fiber-level update is necessary for computations where the fate of
one coordinate depends on the other coordinates in the same fiber, such
as selecting the top-$k$ values, keeping only the minimum-weighted neighbor,
sampling a subset of vertices, or deleting old entries while inserting
new ones.

Suppose we have the following three output tensors on the left-hand side
of three different Einsums:
\begin{subequations}
\begin{align}
Z_{m, n, q} = \ldots \label{mnq} \\
Z_{m, n, q^*} = \ldots \label{mnq*}\\
Z_{m, n^*, q^*} = \label{mn*q*}\ldots
\end{align}
\end{subequations}

We will use these three \edge{} expressions to walk through the semantics
of \populate.
The first scenario (Equation~\ref{mnq}, no ``$*$'') is assignment, or
``default'' \populate (see Inset~\ref{sidebar-default}).

\paragraph{Inputs and Outputs.}
\populate\ takes as input
(1) a point in the iteration space;
(2) the \redtmp tensor;
(3) the rank variable expressions for $Z$;
(4) a user-defined \coordinate\ operator, which additionally takes as
input the iteration-space point ($is$) and any user-defined constants;
(5) a user-defined \compute\ operator, which additionally takes as input
any user-defined constants and the iteration-space point ($is$); and
(6) the current output tensor $Z$.

$Z$ is a tensor that the \edge{} expression \emph{creates}.
We define $Z$ by iteratively updating its graph through successive
applications of \populate.
\populate\ outputs an updated $Z$ (as $Z'$) replacing either a single
tuple $(cs^Z, dv)$ in the ``default'' assignment mode, or \emph{fiber}
of such tuples.

\begin{figure}[t]
  \centering
  \includegraphics[
    width=\linewidth,
    keepaspectratio
  ]{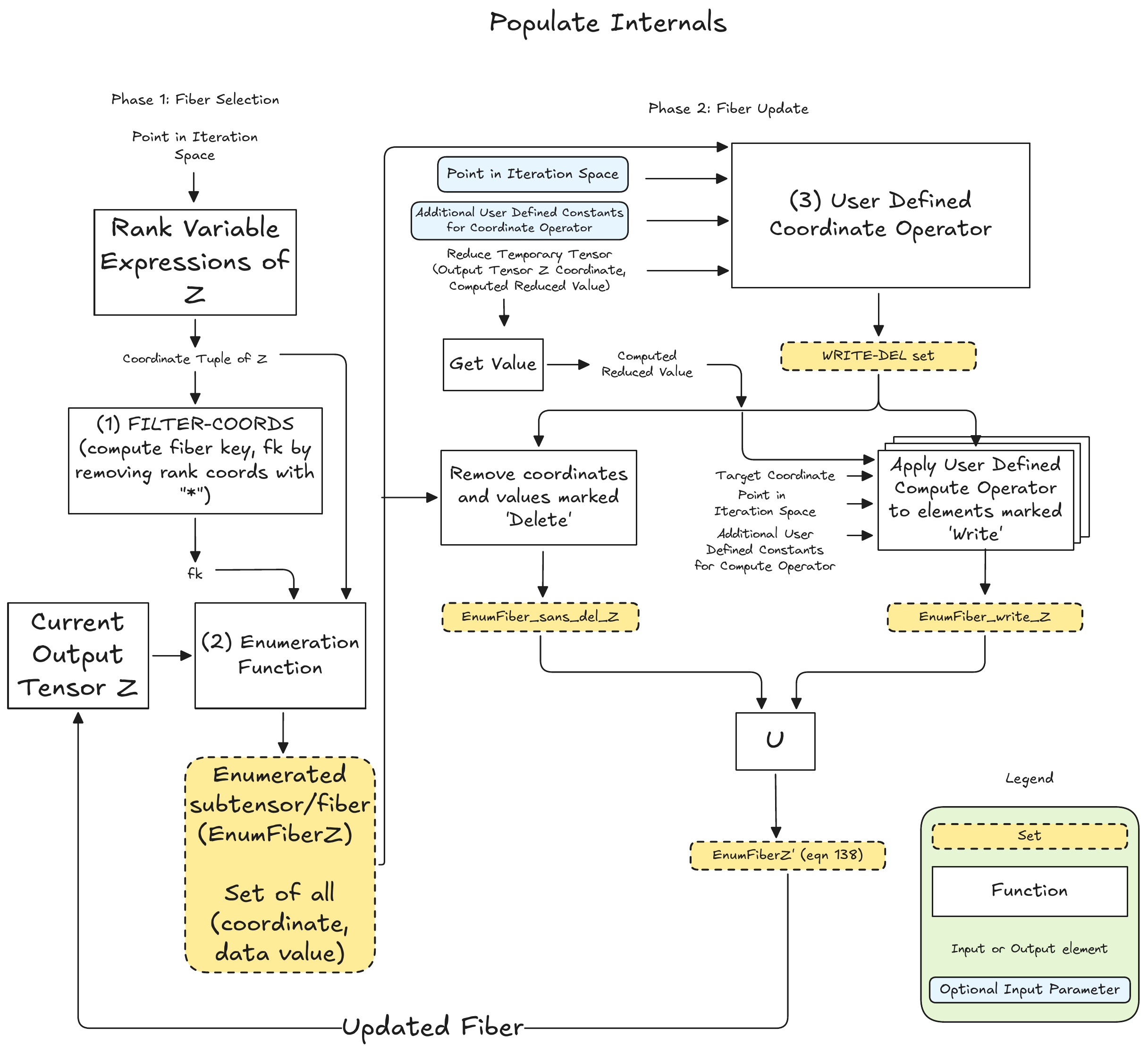}
  \caption{\populate: given the current state of the output tensor $Z$
  and an incoming \redtmp\ element, \populate\ selects the corresponding
  fiber (using the fiber key derived from ``$*$''-marked ranks), applies
  a user-defined \coordinate\ operator to produce a
  \textsc{Write}/\textsc{Delete}/\textsc{None} action label for each
  coordinate in the fiber, and then applies the \compute\ operator to
  produce the new values at \textsc{Write} locations. The updated fiber
  replaces the original in $Z$.}

  \label{fig:edge-populate-internals}
\end{figure}

\subsubsection{\populate\ Steps}\label{ssec:populate_semantics}

\paragraph{Fiber-key extraction.}

\begin{enumerate}
\item Given a coordinate tuple of $Z$, \populate\ computes a
\emph{fiber key} ($\mathsf{fk}$) by removing any rank coordinate
marked with ``$*$'' in the \edge{} expression.
Intuitively, the fiber key is the part of the output coordinate
that stays fixed while \populate\ updates the mutable fiber.
Equivalently, it names which fiber of the output tensor is being
updated in this round of \populate.
We denote this operation by the function \textsc{filter-coords}.

\begin{example_box}[label=example:fiberkey]{Fiber Key Examples}
For the default-assignment expression
$Z_{m,n,q}$ in Equation~\ref{mnq}, no ranks are marked with ``$*$''.
Thus, the tuple $(m,n,q)$ remains unchanged, yielding
$\mathsf{fk} = (m,n,q)$.

For the $Q$-fiber expression
$Z_{m,n,q^*}$ in Equation~\ref{mnq*}, the mutable rank is $q$.
Thus, the tuple $(m,n,q)$ yields $\mathsf{fk} = (m,n)$.

For the $NQ$-fiber expression
$Z_{m,n^*,q^*}$ in Equation~\ref{mn*q*}, the mutable ranks are
$n$ and $q$.
Thus, the tuple $(m,n,q)$ yields $\mathsf{fk} = (m)$.
\end{example_box}
\end{enumerate}

\paragraph{Fiber selection.}

\begin{enumerate}
\setcounter{enumi}{1}
\item Given the filtered coordinate tuple (fiber key), \populate\ indexes
the current output tensor $Z$ to retrieve the corresponding fiber
(subtensor).
However, instead of fibers, \populate\ uses the set of
$\bigl((m_i, n, q), dv\bigr)$ tuple pairs.
That is, \populate\ uses the \emph{subset} of the $Z$ tensor that matches
the subtensor.
We term this subset the \emph{enumerated subtensor/fiber}
($\mathsf{EnumFiber}_Z$).

\begin{example_box}[label=example:pop-fibers]{Fiber Keys to Fibers}
For Equation~\ref{mnq}, this yields a single value: $Z(m,n,q)$.
For Equation~\ref{mnq*}, this yields a $Q$-fiber:
$Z(m_i,n_i) = Z_{m_i,n_i}(q)$.
For Equation~\ref{mn*q*}, this yields an $N \times Q$ subtensor
(an $NQ$-fiber):
$Z(m_i) = Z_{m_i}(n,q)$.
In the case of the $NQ$-fiber, the subtensor is a set of
$\bigl((n, q), dv\bigr)$ pairs.
\end{example_box}
\end{enumerate}

\begin{sidebar_box}[label=sidebar:fibers]{The \textsc{EnumFiber} Function}
\textsc{EnumFiber} is useful because \populate\ must know not only
which values are in a mutable fiber, but also their full coordinates in
the output tensor.
The full coordinates are needed to decide which existing entries to
delete, which empty coordinates to write, and how to splice the updated
fiber back into the graph of $Z$.

Suppose we have the expression
\[
Z_{m, n, q^*} = A_{m, n, q} :: \lll \;\ldots
\]

Here, the fiber key is $\mathsf{fk} = (m, n)$.
Fixing a point $(m_i, n_i)$, the corresponding fiber consists of all
values of $Z$ whose full coordinate tuples match $(m_i, n_i)$ on the
non-$*$ ranks:
\[
Z(m_i, n_i) = \{(q, dv)\},
\]
where $m$ and $n$ are fixed and $q$ varies.

The formal enumerated fiber representation used by \populate\ preserves
the full coordinate tuple for each element:
\[
\mathsf{EnumFiber}_Z(m_i, n_i)
=
\setc{ \bigl((m_i, n_i, q), dv\bigr) }
      { q \in \tensor{RCS}^Z_Q }.
\]
\end{sidebar_box}

\paragraph{Coordinate operator.}

\begin{enumerate}
\setcounter{enumi}{2}

\item \populate\ applies a user-defined \coordinate\ operator to the
retrieved enumerated fiber of $Z$ and the incoming \redtmp.
The \coordinate\ operator takes as input
(1) the current enumerated fiber,
(2) the coordinate tuple and value contained in the \redtmp\ for this
iteration-space point, and optionally
(3) the iteration-space point ($is$) and
(4) any user-defined constants.

It returns a \textsc{write-del} set, whose elements are pairs of
coordinate tuples and actions drawn from
$\{\texttt{Write}, \texttt{Delete}, \texttt{None}\}$.
The cardinality of this set matches that of the input fiber.

We impose the following semantic constraint on the \coordinate\ operator:
\textbf{a coordinate tuple may be marked as \texttt{Write} only if, at
this iteration-space point ($is$), the coordinate maps to the empty value
of $Z$ ($E^Z$) in $Z$.}
In other words, at this point in the iteration space, one cannot put a
value at a coordinate location in $Z$ that already contains a value.
This prevents overwriting existing values within a single round of
\populate.
The user is responsible for defining \coordinate\ so that this constraint
holds; an implementation of \edge{} should raise an error (at compile or
runtime) if a violation is detected.
The invariants that justify this constraint are detailed in
Inset~\ref{sidebar:temporal_constraint}.

\begin{sidebar_box}[label=sidebar:temporal_constraint]{Constraints on Each Round of Populate}
We enforce the following constraint during a single round of fiber updates.

We want:
\begin{enumerate}
  \item Writes only target locations empty at the start of the round
  \item A location deleted in this round cannot also be written in this round
  \item Future rounds may write to locations deleted in this round
\end{enumerate}

Let:
\begin{itemize}
  \item $S$ be the current state
  \item Let $\mathrm{Loc}$ be the set of all locations (occupied and
  unoccupied) in the output fiber
  \item A round is the current point in the iteration space for which
  populate is occurring
  \item A ``pre-state'' is the state of the mutable fiber before a round
  of writes/deletions occur for this round of populate
  \item $\mathrm{dom}(S)
  \subseteq \mathrm{Loc}$ be the set of occupied locations
  \item $D \subseteq \mathrm{Loc}$ be the delete set
  \item $W \subseteq \mathrm{Loc}$ be the write set
\end{itemize}

The key idea is that writes are only checked against the pre-state. So
the invariant for (a) is:
\[
W \cap \mathrm{dom}(S) = \varnothing
\]

and the invariant for (b) is:
\[
D \cap W = \varnothing
\]

Together:
\[
W \cap \mathrm{dom}(S) = \varnothing \wedge D \cap W = \varnothing
\]

We can simplify further. If writes must be empty in the pre-state, then
deletions do not matter for that check. Therefore, the constraints are
logically independent:
\begin{itemize}
\item $W$ must avoid currently occupied locations
\item $W$ must avoid deletions in the same round
\end{itemize}
Because deleted locations are occupied in the pre-state, we can express
our temporal condition equivalently as:
\[
W \cap \big(\mathrm{dom}(S) \cup D\big) = \varnothing
\]
Writes must avoid modifying anything that either already exists or is
being deleted this round.
\end{sidebar_box}
\end{enumerate}

\paragraph{Fiber update.}

\begin{enumerate}
\setcounter{enumi}{3}

\item For each element in \textsc{write-del} marked \texttt{Delete},
\populate\ removes the corresponding coordinate and value from the
current enumerated $Z$ fiber.
The resulting enumerated $Z$ fiber, now with zero or more elements
removed, is the \emph{post-delete fiber},
denoted $\mathsf{EnumFiber}^{\mathsf{sans\_del}}_Z$.
Here, $\mathsf{sans\_del}$ denotes the state of the fiber after the
\texttt{Delete} actions have been applied.

\item Elements marked \texttt{None} leave the enumerated $Z$ subtensor
unchanged.

\item For each element in \textsc{write-del} marked \texttt{Write}:
\begin{enumerate}
    \item \populate\ invokes the user-defined \compute\ operator to
    produce a value.
    The \compute\ operator takes as input
    (1) the target coordinate,
    (2) the value from the \redtmp,
    (3) the iteration-space point ($is$), and
    (4) any user-defined constants,
    and returns a single data value.
    The target coordinate is passed explicitly because a single iteration
    space point may produce multiple \texttt{Write} actions (one per
    coordinate in the output fiber); \compute\ may produce a different
    value at each target coordinate.
    The \compute\ operator is invoked once for each coordinate marked as
    \texttt{Write} in the previous step.
    \item The value returned by \compute\ will be written at the
    corresponding target coordinate (that was marked \texttt{Write}).
\end{enumerate}

\item We refer to the set of values produced by \compute, together with
their associated target coordinates, as the \emph{write fiber},
denoted $\mathsf{EnumFiber}^{\mathsf{write}}_Z$.
\end{enumerate}

\paragraph{Updated output tensor.}

\begin{enumerate}
\setcounter{enumi}{7}

\item \populate\ produces an updated subtensor for $Z$ by applying the
delete and write fibers.
The updated tensor $Z'$ may subsequently be used as input to later
\populate\ invocations.
\end{enumerate}

Overall, the \populate\ action preserves the semantics of \reduce by
ensuring that every value written to the output tensor originates from a
corresponding \redtmp.
This follows from the structure of \populate: the \coordinate\ operator
may choose where to write, but it does not create data values.
The only data value available to the \compute\ operator is the value
carried by the incoming \redtmp\ element, together with the target
coordinate, the iteration-space point, and any user-defined constants.
Thus, for every value written into $Z$, there exists a corresponding
\redtmp\ element whose value was used to produce that write.

\subsubsection{Set Theory Notation for \populate}
\label{ssec:pop_def}

At a high level, \populate\ updates the graph of the output tensor ($Z$)
by selecting a fiber, determining how that fiber should be modified, and
producing an updated tensor that reflects those changes.
Each invocation of \populate\ processes a single reduction temporary
(\redtmp).

\paragraph{Preliminaries.}
Let the current output tensor be
\[
Z : \tensor{CS}^Z \rightharpoonup \tensor{DS}^Z_{ne}.
\]
Let $(cs^R, dv_r)$ denote the \redtmp element produced by \reduce\ and
accessed by \populate\ at iteration-space point $is$.

\paragraph{Fiber-key extraction.}
Some ranks of $Z$ may be marked with ``$*$'' in the \edge{} expression.
This indicates that \populate\ should operate over a fiber (i.e., a
subtensor over the remaining non-``$*$'' coordinates) rather than a
single coordinate.
To select the fiber key, we define the \emph{filter-coords} function:

Let $Z$ have ranks $(cs_0, cs_1, \ldots, cs_{k-1})$, and let
\[
\mathcal{F} \subseteq \{0, \ldots, k-1\}
\]
denote the indices of ranks \emph{not} marked with ``$*$''.
The fiber-key function is defined as:
\begin{align}
\textsc{filter-coords} : \tensor{CS}^Z \rightarrow \tensor{CS}^Z_{\mathcal{F}}, \\
\textsc{filter-coords}(cs^Z) := (cs_i)_{i \in \mathcal{F}}.
\end{align}

\paragraph{Fiber selection.}
Given a fiber key $\mathsf{fk} \in \tensor{CS}^Z_{\mathcal{F}}$,
we define a \emph{fiber-selection} function that returns the set of
\emph{enumerated} coordinate-value pairs associated with that fiber:

\begin{align}
\mathsf{EnumFiber}_Z(\mathsf{fk})
=
\setc{(cs^Z, dv)}{
  cs^Z \in \tensor{CS}^Z \;\wedge\;
  \textsc{filter-coords}(cs^Z) = \mathsf{fk} \;\wedge\;
  Z(cs^Z) = dv
}.
\end{align}
This representation retains full coordinate tuples for each element in
the fiber, rather than partial coordinate tuples that include only the
rank coordinate of the varying rank.

\paragraph{Coordinate operator.}
The \coordinate\ operator determines how the selected fiber should be
modified in response to the incoming \redtmp.
The operator has the following mapping:
\begin{align}
\text{\coordinate} :
\mathsf{EnumFiber}_Z(\mathsf{fk})
\times (\tensor{CS}^Z \times \tensor{DS}^Z)
\times \tensor{IS}
\rightarrow
\mathcal{P}\!\left(
  \tensor{CS}^Z \times \{\texttt{Write}, \texttt{Delete}, \texttt{None}\}
\right),
\end{align}
where $\mathcal{P}$ is the power set. That is, the output of
\coordinate\ is a set, the \textsc{write-del} set. Note that the
\textsc{write-del} set is a set of (coordinate, action) pairs, where an
action may be \texttt{Write}, \texttt{Delete}, or \texttt{None}; in other
words, it includes \texttt{None} entries in addition to Writes and Deletes.

\begin{align}
\mathsf{WriteDel}_{\mathsf{fk}}
=
\text{\coordinate}\bigl(
  \mathsf{EnumFiber}_Z(\mathsf{fk}), (cs^R, dv_r), is
\bigr)
\\
\mathsf{List_{Del}}
=
\setc{cs^Z}{
  (cs^Z, \texttt{Delete}) \in \mathsf{WriteDel}_{\mathsf{fk}}}
\\
\mathsf{List_{Write}}
=
\setc{cs^Z}{
  (cs^Z, \texttt{Write}) \in \mathsf{WriteDel}_{\mathsf{fk}}}
\end{align}

Each element of $\mathsf{WriteDel}_{\mathsf{fk}}$ specifies an action
(\texttt{Write}, \texttt{Delete}, \texttt{None}) for a coordinate in the
enumerated fiber.
$\mathsf{List_{Del}}$ and $\mathsf{List_{Write}}$ are sets containing the
corresponding coordinate tuples.

\paragraph{Fiber update.}
First, all coordinates marked for deletion are removed:
\begin{align}
\mathsf{EnumFiber}^{\mathsf{sans\_del}}_Z(\mathsf{fk})
=
\mathsf{EnumFiber}_Z(\mathsf{fk})
\setminus
\setc{(cs^Z, dv)}{
  cs^Z \in \mathsf{List_{Del}}
}.
\end{align}
The superscript $\mathsf{sans\_del}$ denotes the state of the fiber after
the \texttt{Delete} actions have been applied.

Next, coordinates marked \texttt{Write} produce new values using a
user-defined \compute\ operator:
\begin{align}
\text{\compute} :
\tensor{CS}^Z \times \tensor{DS}^Z \times \tensor{IS}
\rightarrow \tensor{DS}^Z.
\end{align}

The resulting enumerated write fiber is:
\begin{align}
\mathsf{EnumFiber}^{\mathsf{write}}_Z(\mathsf{fk})
=
\setc{
  (cs^Z, \text{\compute}(cs^Z, dv_r, is))
}{
  cs^Z \in \mathsf{List_{Write}}
}.
\end{align}
The updated fiber is then:
\begin{align}
\mathsf{EnumFiber}_{Z'}(\mathsf{fk})
=
\mathsf{EnumFiber}^{\mathsf{sans\_del}}_Z(\mathsf{fk})
\cup
\mathsf{EnumFiber}^{\mathsf{write}}_Z(\mathsf{fk}).
\end{align}

\paragraph{Updated output tensor.}
At a given iteration-space point $is$, \populate\ operates on exactly one
fiber.
Let
\[
cs^Z = \Gamma^Z(is),
\qquad
\mathsf{fk} = \textsc{filter-coords}(cs^Z)
\]
be the output coordinate and corresponding fiber key induced by $is$.

The updated output tensor $Z'$ is defined by replacing only this fiber and
leaving all other fibers unchanged:
\begin{align}
\textit{graph}(Z')
=
\bigl(
  \textit{graph}(Z) \setminus \mathsf{EnumFiber}_Z(\mathsf{fk})
\bigr)
\cup
\mathsf{EnumFiber}_{Z'}(\mathsf{fk}).
\end{align}

\begin{sidebar_box}[label=sidebar-default]{Assignment (``Default'' Populate)}
Assignment is a specific instance of populate.
Suppose we have the following expression:
\[
Z_{m, n} = \mathcal{R}_{m, n}
\]
where each element in $\mathcal{R}$, representing the \redtmp, is copied
to the corresponding $(m, n)$ location in $Z$.

For a given $(m_i, n_i)$ point in the iteration space (where $i$
represents this instance of $(m, n)$), \populate does the following:
\begin{enumerate}
\item Apply \textsc{filter-coords}, yielding a fiber key of $(m_i, n_i)$.
\item Access the current output tensor ($Z$) at $(m_i, n_i)$ and form the
\emph{enumerated fiber}. Since there is only a single point,
$\mathsf{EnumFiber}_Z(m_i,n_i)$ is a set with the following, single
element: $\big((m_i, n_i), Z(m_i, n_i) \big)$.
\item Apply the \coordinate operator to (1)
$\mathsf{EnumFiber}_Z(m_i, n_i)$ along with (2) the $(m_i, n_i)$ point
and corresponding data value ($R$ at $(m_i, n_i)$) for the RHS \redtmp.
The iteration-space point $(is)$ and user-defined constants are unused
for assignment operations.
The \coordinate operator returns a \textsc{write-del} set that contains
a single element: $\big((m_i, n_i), \textsc{write}\big)$. This indicates
that \populate will write to point $(m_i, n_i)$ in $Z$.
\item Apply the \compute operator. For assignment, this is simply the
``pass-through'' operator, which returns the value from the \redtmp
($R$ at $(m_i, n_i)$). This forms an
$\mathsf{EnumFiber}^\mathsf{write}_Z(m_i,n_i)$.
\item Update $Z$ with the $\mathsf{EnumFiber}^\mathsf{write}$.
\end{enumerate}
Overall, the steps above correspond to placing the computed value on the
RHS (\redtmp) into the corresponding point on the LHS ($Z$).
\end{sidebar_box}

\begin{example_box}[label=example:multi-reduce]{Using Populate for Data-Dependent Selection}
Suppose an input fiber $I_m$ contains several candidate values, and we
want to select one value from that fiber.
For example, the selected value could be the minimum, the maximum, or
the first value satisfying a user-defined predicate.
This selection is data-dependent: the coordinate operator must examine
the fiber before deciding which coordinate should remain live.
We can express this pattern of computation using \populate:
\begin{subequations}
\begin{align}
Z_{m^*} = I_m :: \lll_m \mathbbm{1}(\textsc{dependent\_select}) \\
Y = Z_m :: \bigvee \text{ANY}(\mathbbm{1})
\end{align}
\end{subequations}

The coordinate operator \verb|dependent_select| is one possible
implementation of this pattern.
It is not a fixed primitive of \edge{}; rather, it stands for a
user-defined coordinate operator whose internal selection policy may be
chosen by the programmer.
For example, the selection policy could keep the minimum value, keep the
maximum value, or keep all values satisfying a predicate.
The following pseudo-code shows the common structure of such an operator:
\begin{center}
\label{alg:dependent-select}
\begin{minipage}{0.85\linewidth}
\begin{edgecodebox}[python]
def dependent_select( lhs_enum_fiber = (z_m), rhs_tuple = (m, I_m.val)):
  # Form a new enumerated fiber that combines the
  # lhs_enum_fiber with the rhs_tuple
  combined_enum_fiber = combine(lhs_enum_fiber, rhs_tuple)

  # Look at all the elements in the combined fiber and select which
  # coordinates will be included in the new output fiber.
  # This select can be any function
  live_coords = select(combined_enum_fiber)

  # Retrieve coordinates that existed before this update
  z_coords = coords(z_m)

  # Which coordinates are we deleting?
  delete_list = list(set(z_coords) - set(live_coords))

  # Did we introduce any new coordinates compared to the previous coords?
  edit_list = list(set(live_coords) - set(z_coords))

  # What is unchanged?
  none_list = list(set(live_coords)) - edit_list

  return (none_list, delete_list, edit_list)
\end{edgecodebox}
\end{minipage}
\end{center}

\end{example_box}

\begin{example_box}[label=example:filter-even]{Filter Even Values}
Using populate, we can filter a fiber to include only even values:
\begin{subequations}
\begin{align}
Z_{m^*} = I_m :: \lll_m \mathbbm{1}(\textsc{filter\_even}) \\
\end{align}
\end{subequations}
\verb|filter_even| follows the same pattern as
\verb|dependent_select| (Example~\ref{example:multi-reduce}), with the
following \verb|select| function:\footnote{These are meant to be
\emph{example implementations}; optimizations are possible.
For example, instead of filtering even values from the combined fiber,
one can simply check the incoming value (\lstinline|I\_m.val|) and create
a combined fiber if that value is even, otherwise keep the old, $Z$ fiber
as the combined fiber.}
\begin{center}
\begin{minipage}{0.85\linewidth}
\begin{edgecodebox}[python]
def even_select(combined_enum_fiber):
  coords = []
  for coord_tuple, val in combined_enum_fiber:
    if is_even(val):
      coords.append(coord_tuple)
  return coords
\end{edgecodebox}
\end{minipage}
\end{center}
\end{example_box}

\section{\EDGEcaps Syntax}\label{ssec:syntax}
In this section, we summarize the syntax of the \edge{} language.
This includes the concrete grammar, required components, and default specifications introduced throughout \S~\ref{ssec:walk-bfs}.
Figure~\ref{fig:grammar} shows the grammar for the \edge{} language, including tensor declaration, initialization, and Extended Einsums.
Tables~\ref{tab:defaults} and~\ref{tab:action-defaults} summarize the default operators and actions when the corresponding symbols are omitted.

\begin{figure}
\centering
\begin{tcolorbox}[colback=white, colframe=black, boxrule=0.5pt, left=0pt]
    \scalebox{0.71}{%
        \begin{minipage}{1.41\textwidth}%
            \input{tex/notes/main.ebnf-cropped}
        \end{minipage}%
    }
\end{tcolorbox}
\caption{Grammar for the \edge{} language.
\label{fig:grammar}}
\Description{Grammar for the \edge{} language.}
\end{figure}

\begin{table}[h]
    \caption{Default specifications for each operator. An action cannot appear in an expression without an accompanying \compute{} operator (see Table~\ref{tab:requireds}), so there is no entry for \compute{} operators here.
    }
    \label{tab:defaults}
    \resizebox{1.0\textwidth}{!}{
        \begin{tabular}{llll}
            \toprule
            \textbf{When this is not present:} & \textbf{Then the default operator is:} & \textbf{Shorthand Example} & \textbf{Fully Specified Version} \\ \midrule
          \rowcolor{lightgray!25}\begin{tabular}[c]{@{}l@{}}\merge{} Operator \\ (in \maptxt{})\end{tabular} &
              \begin{tabular}[c]{@{}l@{}}pass-through ($\mathbbm{1}$)\end{tabular} &
              $Z_{m, n} = A_{m, n} \cdot B_{m, n} :: \map \times$ &
              $Z_{m, n} = A_{m, n} \cdot B_{m, n} :: \map \times(\mathbbm{1})$
               \\
          \begin{tabular}[c]{@{}l@{}}\merge{} Operator \\ (in \reduce{})\end{tabular} &
              \begin{tabular}[c]{@{}l@{}}pass-through ($\mathbbm{1}$)\end{tabular} &
              $Z_{m} = A_{m, k} :: \bigvee +$ &
              $Z_{m} = A_{m, k} :: \bigvee +(\mathbbm{1})$
               \\
          \rowcolor{lightgray!25}\begin{tabular}[c]{@{}l@{}}\coordinate{} Operator \\ (in \populate{})\end{tabular} &
              \begin{tabular}[c]{@{}l@{}}pass-through ($\mathbbm{1}$)\end{tabular} &
              $Z_{m^*} = A_m :: \lll_{m^*} \mathbbm{1}$ &
              $Z_{m^*} = A_m :: \lll_{m^*} \mathbbm{1}(\mathbbm{1})$
              \\
          \bottomrule
            \end{tabular}
    }
\end{table}

\subsection{Tensor Declaration}
The tensor declaration region of an \edge{} specification (\S~\ref{ssec:workloadspecs}) declares the rank names, shapes or the rank coordinate set (see \S~\ref{sssec:rank_coordinate_set}), data types, and empty values of all tensors used in the Extended Einsum region (see \bfs{} Example~\ref{sidebar:full_edge_bfs} and Equation~\eqref{seqn:Gf}--\eqref{seqn:Pf}).
If a tensor rank is unbounded, such as a generational rank, no shape is specified. These declarations remove ambiguity about the size of the iteration space, the meaning of tensor coordinates, and the behavior of operators on empty values.

\subsection{Initialization}\label{ssec:syntaxinit}
The initialization region specifies the values of input tensors. Any input tensor not initialized in this section is assumed user-specified.
All \emph{output} tensors are initially empty.
Any valid extended Einsum expression may appear in this section as the right-hand side of an input-tensor assignment (see Inset~\ref{sidebar:full_edge_bfs}, Equation~\eqref{seqn:FIf}).

\subsection{Extended Einsum}
A fully specified extended Einsum is a cascade of Einsum expressions.
\edge{} also supports nested cascades, extending the cascade model we introduced in Nayak et al.~\cite{Nayak:2023:TDF}.
This is useful for expressing modular computations that feed into a larger computation, such as in Louvain's method~\cite{Blondel:2008:FUC} and simple concurrent connected components~\cite{Liu:2022:SCC}.

\subsubsection{Cascade of Einsums}
A cascade consists of a sequence of Einsum expressions. If a cascade is iterative, we additionally specify (a) a generational rank variable, (b) its initial value, and (c) its rank variable set (\S~\ref{sssec:rvs}). If these are omitted, the default generational rank variable is $i$, the default initial value is $0$, and the default rank variable set is
\begin{equation}
RV_i \;=\; \{\, i \mid i \in [0, I)\ \wedge\ \neg(\diamond : \langle \textit{Boolean Expression} \rangle(i)) \,\}.
\end{equation}
The diamond predicate excludes generations for which the predicate evaluates to \texttt{true}.
The full \edge{} \bfs{} expression in Equation~\eqref{eqn:edge_bfs_full} uses these default specifications.

 \subsubsection{An Einsum Expression}
 Each expression contains the iteration space specifications (\iterspec{}), followed by the corresponding specifications for \maptxt{}, \reduce{}, and \populate{} actions (in the \modifier{}).
 In Equation~\eqref{seqn:advancef}, the \iterspec{} is the portion containing the input and output tensors, their corresponding rank variable expressions, and the operation label ($\cdot$): ($T_{i, d}  = G_{s, d} \cdot F_{i, s}$).

\paragraph{Anonymous Tensors}
An \edge{} Einsum normally has two input operands and one output tensor.
Expressions with more than two input operands are written using \emph{anonymous tensors}.
In such cases, precedence must be explicit.
Thus, every binary operation is enclosed in parentheses.
Each parenthesized subexpression denotes a partial tensor that may be used by an enclosing expression.

For example, using algebraic manipulation (see \S~\ref{sec:5-mappings}), we can express both neighbor gathering (Equation~\eqref{seqn:advancef}, and $\cdot^1$ in Equation~\eqref{eqn:combined}) and neighbor filtering (Equation~\eqref{seqn:filterf}, and $\cdot^2$ in Equation~\eqref{eqn:combined}) as a single Einsum step:
\begin{align}\label{eqn:combined}
    F_{i+1, d}  = ( G_{s, d} \cdot^1 F_{i, s})_{i, d} \cdot^2 \neg P_{i, d} :: \bigwedge^1 +(\cap) \bigvee^1 \text{ANY}(\cup) \bigwedge^2 \leftarrow(\cap),
\end{align}

\begin{table}[h]
\caption{Default specifications for each action. When an action is omitted, \edge{} applies the default action shown here. Required accompanying operators are summarized in Table~\ref{tab:requireds}.}
\label{tab:action-defaults}
\resizebox{1.0\textwidth}{!}{
  \begin{tabular}{cccc}
    \toprule
    \textbf{When this is not present:}
    & \textbf{Then the default action is:}
    & \textbf{Shorthand Example}
    & \textbf{Fully Specified Version} \\ \midrule
    \rowcolor{lightgray!25}\begin{tabular}[c]{@{}c@{}}\maptxt{} Action\\ (between 2 operands)\end{tabular}
    & $\bigwedge \times(\mathbbm{1})$
    & $Z_{m,n} = A_{m, n} \cdot B_{m, n}$
    & $Z_{m,n} = A_{m, n} \cdot B_{m, n} :: \bigwedge_{m, n} \times(\mathbbm{1})$ \\
    \reduce{} Action                                                             & $\bigvee +(\mathbbm{1})$
    & $Z_{m} = A_{m, k}$
    & $Z_{m} = A_{m, k} :: \bigvee_k +(\mathbbm{1})$ \\
    \rowcolor{lightgray!25}\populate{} Action
    & $\lll \mathbbm{1}(\mathbbm{1})$
    & $Z_{m} = A_m $
    & $Z_{m^*} = A_m :: \lll_{m^*} \mathbbm{1}(\mathbbm{1})$
    \\ \bottomrule
  \end{tabular}
}
\end{table}

\subsubsection{Actions in an Einsum Expression}

Every \maptxt{} and \reduce{} action \emph{must} correspond to an \emph{operation label} ($\cdot^{\langle{\textit{binary\_label}}\rangle}$) in the \texttt{iteration-specs}.
Recall that a single-operand Einsum (or ``unary'' Einsum) is syntactic sugar for a degenerate \maptxt{} (see \S~\ref{ssec:unary}).
Thus, in a single-operand Einsum containing no operation labels, \reduce{} by default refers to operation label 1.

 Meanwhile, a two-operand Einsum may contain a \reduce{} action on an operation label:
 \begin{equation}\label{eqn:degree2}
    \textit{MaskedDegree}_{s} = G_{s, d} \cdot \text{Mask}_{s}:: \bigwedge \text{AND}(\cap)  \bigvee_d +(\cup).
  \end{equation}
  This expression first filters out the vertices in the graph that are in a mask tensor ($\map$) before gathering the degrees of the filtered graph ($\red$).

  If we wanted to gather the degrees of all $s$ vertices first, and then mask the result, we would need to specify a cascade of Einsums:
  \begin{subequations}\label{eqn:degree3}
    \begin{align}
    \textit{Degree}_{s} &= G_{s, d} :: \bigvee_d +(\cup)\label{eqn:degree} \\
    \textit{MaskedDegree} &=  \textit{Degree}_{s} \cdot \text{Mask}_{s} :: \bigwedge \text{AND}(\cap) \label{eqn:masked}
    \end{align}
  \end{subequations}
 Equation~\eqref{eqn:degree} finds the degree of every source vertex in $G$ by reducing over the $d$ rank (where \reduce{}, by default, refers to operation label 1).
 The next Einsum (Equation~\eqref{eqn:masked}) selects from which sources ($s$ rank variable) to extract the degree.

  Note that in the first expression (Equation~\eqref{eqn:degree2}) we perform less than or equal to the number of additions in the second expression (Equation~\eqref{eqn:degree3}) since the first cascade only reduces over masked $s$ elements rather than over all $s$ elements.
  Future work will explore how to automate the algebraic manipulations necessary to go from one form to the other (\textbf{design goal~\ref{goal:manipulation}}, see \S~\ref{sec:5-mappings}).

 \subsubsection{Merge and Coordinate Operators}
 Table~\ref{tab:actions-summary} summarizes the inputs and outputs of \merge{}, \coordinate{}, and \compute{} operators.

 When an Einsum expression uses multiplication and addition operators in the traditional sense (as in \acro{GEMM}, Equation~\eqref{eqn:tin}), we can drop the \maptxt{}, \reduce{}, and \populate{} actions (\textbf{design goal~\ref{goal:supports}}).
 Note that the default \emph{\merge} operator for such an Einsum is the ``pass-through'' operator ($\mathbbm{1}$), which touches all coordinates regardless of whether they contain empty values or not (see Appendix~\ref{appendix:merge}, item~\ref{item:pass}).
 \merge{}/\coordinate{} operators are optional, while \compute{} operators must be specified (Table~\ref{tab:requireds}).
 If a particular operation label does not contain a \merge{} operator, the \edge{} machinery applies the \compute{} operator to every point in the iteration space (default pass-through \merge{} operator).

 \label{item:empty_test}For example, the gather step in \bfs{} (\bfs{} Example~\ref{sidebar:full_edge_bfs}, Equation~\eqref{seqn:advancef}) \emph{needs} the intersection ($\cap$) operator on the \maptxt{} action to execute efficiently.
 Intersection will ignore all $s$ points in the iteration space where $F_{s}$ contains an empty value of $\infty$.
 \edge{} guarantees that points that survive intersection will be non-empty (i.e., non-infinite) values.
 If we remove the \merge{} operator such that the expression becomes:
     \begin{align}\label{eqn:advance_step}
         T_{i, d}  = G_{s, d} \cdot^1 F_{i, s} :: \bigwedge^1 + \bigvee \text{ANY}(\cup),
     \end{align}
 then the $+$ \compute{} operator will add infinity from empty values in $F$ to the edge weights in $G$\@.
 Note that the above expression assumes the default \merge{} operator of $\mathbbm{1}$ (pass-through).
 While this will still produce a valid result in this particular scenario,  the implementation cannot take advantage of the optimization opportunity to only process effectual computations (\designgoal{goal:opt}).

 \subsubsection{Rank Variable Expressions}
 Rank variable expressions are functions that map iteration-space variables to specific points in the rank of a tensor (see \S~\ref{ssec:conditionals}, \textbf{design goal~\ref{goal:powerful}}).

 For example, ternary expressions are possible:
 \begin{align}\label{eqn:lower2}
    Z_{f \equiv (w < a ? w : a)} = G_{w} \cdot P_{a}:: \bigwedge \text{AND}(\cap) \bigvee \text{OR}(\cup)
  \end{align}
  In Equation~\eqref{eqn:lower2}, the RVE in the output is labeled ($f$), read as ``$f$ such that $f$ is equivalent ($\equiv$) to the result of the expression in parentheses'', where the Boolean expression essentially returns the minimum of $w$ and $a$ as execution walks the iteration space.
  This expression requires a \reduce{} action on the output rank, as multiple points in the iteration space may correspond to the same point in $Z$'s data space.
  For example, $(w, a) = (0, 1)$ will update $f = 0$ ($0 < 1$ so select $w=0$), and $(w, a) = (1, 0)$ will also update $f = 0$ ($1 \geq 0$, so select $w=0$).
  If the $P$ tensor is one-hot, that is, its $a$ rank has at most an occupancy of 1, then this expression returns the set of coordinates in $G$ less than the occupied $a$ coordinate in $P$, as well as the coordinate in $P$ if any coordinate in $G$ is equal to or greater than that $a$ coordinate.
This type of computation is common in algorithms like connected components,
which often includes a step where computation replaces an ancestor vertex
($a$) with a smaller-valued connected vertex ($w$)
(see Figure~\ref{fig:rve-connected-components}).

In general, we can summarize a rank variable expression on the rank of a
tensor, $Z$, as:
\begin{align}\label{eqn:rv}
  Z_{f(\text{iteration space variables})},
\end{align}
where $f$ can be any user-defined function that takes as input points in the
iteration space and outputs a single coordinate.

We could rewrite Equation~\eqref{eqn:lower2} as:
\begin{align}
Z_{f(a, w)} &= G_{w} \cdot P_{a}
:: \bigwedge \text{AND}(\cap) \bigvee \text{OR}(\cup) \\
& \text{Function definitions}:\notag \\
f(a, w) &\rightarrow
\text{if } w < a \text{ then return } w \text{ else return } a,
\end{align}
\noindent
or as:
\begin{align}
Z_{\min(a, w)} = G_{w} \cdot P_{a}
:: \bigwedge \text{AND}(\cap) \bigvee \text{OR}(\cup),
\end{align}

where $f$ is also a \emph{named RVE}.

A named RVE can be referenced in multiple tensor subscripts within the
same Einsum.
For example:
\begin{align}
Z_{f(a,w)} &= A_w \cdot B_{a,f}
\; :: \;
\bigwedge \mathrm{AND}(\cap),
\;
\bigvee \mathrm{OR}(\cup),
\\
&\qquad
f(a,w) = \max(a,w).
\end{align}
Here, the second rank of $B$ is also accessed via $f$.
The name $f$ is not itself a rank variable; it is an alias for the
corresponding rank variable expression.

Equivalently, this can be written inline as:
\begin{align}
Z_{f \equiv \max(a,w)} = A_w \cdot B_{a,f}
\; :: \;
\bigwedge \mathrm{AND}(\cap),
\;
\bigvee \mathrm{OR}(\cup).
\end{align}

Figure~\ref{fig:rve-connected-components} illustrates Equation~\eqref{eqn:lower2} with a one-hot $P_a$, non-empty at $a{=}2$ and $G_w$ non-empty at $w{=}0,1,3$.
{\tikzset{external/export=true}\begin{figure}
\centering
\begin{tikzpicture}[
  cell/.style={draw, rounded corners=2pt, minimum width=1.0cm, minimum height=0.75cm,
               font=\small, inner sep=2pt},
  gcell/.style={cell, fill=teal!15},
  pcell/.style={cell, fill=teal!15},
  wcell/.style={cell, fill=violet!15, thick},  %
  acell/.style={cell, fill=orange!20, thick},  %
  ecell/.style={cell, fill=gray!10, text=gray!50},
  lbl/.style={font=\footnotesize},
  note/.style={font=\scriptsize\itshape},
  leg/.style={draw, rounded corners=2pt, minimum width=0.35cm, minimum height=0.35cm,
              inner sep=0pt},
]

\foreach \w/\wx in {0/0, 1/1.1, 2/2.2, 3/3.3} {
  \node[lbl] at (\wx, 1.75) {$w{=}\w$};
}

\node[lbl, anchor=east] at (-0.7, 1.0) {\itshape $G_w$};
\node[gcell] at (0,   1.0) {\textbf{T}};
\node[gcell] at (1.1, 1.0) {\textbf{T}};
\node[ecell] at (2.2, 1.0) {$\emptyset$};
\node[gcell] at (3.3, 1.0) {\textbf{T}};

\draw[gray!35, thin] (-0.55, 0.6) -- (4.55, 0.6);

\foreach \a/\ay in {0/0, 1/-1.0, 2/-2.0, 3/-3.0} {
  \node[lbl, anchor=east] at (-0.7, \ay) {$a{=}\a$};
}

\foreach \wx in {0, 1.1, 2.2, 3.3} {
  \node[ecell] at (\wx,  0)   {};
  \node[ecell] at (\wx, -1.0) {};
  \node[ecell] at (\wx, -3.0) {};
}

\node[wcell] at (0,   -2.0) {\begin{tabular}{c}\scriptsize$f{=}0$\\[1pt]\textbf{T}\end{tabular}};
\node[wcell] at (1.1, -2.0) {\begin{tabular}{c}\scriptsize$f{=}1$\\[1pt]\textbf{T}\end{tabular}};
\node[ecell] at (2.2, -2.0) {\begin{tabular}{c}\scriptsize$f{=}2$\\[1pt]$\emptyset$\end{tabular}};
\node[acell] at (3.3, -2.0) {\begin{tabular}{c}\scriptsize$f{=}2$\\[1pt]\textbf{T}\end{tabular}};

\draw[gray!35, thin] (3.85, 0.6) -- (3.85, -3.4);

\node[lbl]  at (4.4, 1.65) {\itshape $P_a$};
\node[note] at (4.4, 1.3)  {(one-hot)};
\node[ecell] at (4.4,  0)   {$\emptyset$};
\node[ecell] at (4.4, -1.0) {$\emptyset$};
\node[pcell] at (4.4, -2.0) {\textbf{T}};
\node[ecell] at (4.4, -3.0) {$\emptyset$};

\draw[gray!35, thin] (4.95, 0.6) -- (4.95, -3.4);

\node[lbl] at (5.55, 1.75) {$Z_f$};
\node[lbl, anchor=east] at (5.55,  0)   {$f{=}0$};
\node[lbl, anchor=east] at (5.55, -1.0) {$f{=}1$};
\node[lbl, anchor=east] at (5.55, -2.0) {$f{=}2$};
\node[lbl, anchor=east] at (5.55, -3.0) {$f{=}3$};

\node[wcell] at (6.0,  0)   {\textbf{T}};
\node[wcell] at (6.0, -1.0) {\textbf{T}};
\node[acell] at (6.0, -2.0) {\textbf{T}};
\node[ecell] at (6.0, -3.0) {$\emptyset$};

\node[note, anchor=west] at (6.55,  0)   {vertex~0 keeps its id (smaller than ancestor)};
\node[note, anchor=west] at (6.55, -1.0) {vertex~1 keeps its id (smaller than ancestor)};
\node[note, anchor=west] at (6.55, -2.0) {vertex~3 absorbed into ancestor~2's group};

\node[font=\scriptsize, align=center] at (3.0, -4.0)
  {$Z$ = updated representative candidates: \quad
   parents $w < a$ keep their own id \quad
   parents $w > a$ are absorbed into ancestor~$a$};

\node[lbl] at (0.0, -4.8) {Legend:};
\node[leg, fill=violet!15] at (1.3,  -4.8) {};
\node[note, anchor=west]   at (1.4, -4.8) {active, $w < a$};
\node[leg, fill=orange!20] at (3.2,  -4.8) {};
\node[note, anchor=west]   at (3.3, -4.8) {active, $w \geq a$};
\node[leg, fill=teal!15]   at (5.1,  -4.8) {};
\node[note, anchor=west]   at (5.2, -4.8) {non-empty input};
\node[leg, fill=gray!10]   at (7.3,  -4.8) {};
\node[note, anchor=west]   at (7.4, -4.8) {empty / culled};

\end{tikzpicture}
\caption{Rank variable expression $f \equiv \min(w,a)$ with ancestor $P_a$ one-hot at $a{=}2$
and current parents $G_w$ non-empty at $w{=}0,1,3$.
$Z_f$ gives the updated representative candidates after one step:
parents smaller than the ancestor (vertices~0 and~1) keep their own coordinate in $Z$;
parents larger than the ancestor (vertex~3) are absorbed into the ancestor coordinate $Z[2]$,
and vertex~3 disappears from the output ($Z[3]{=}\emptyset$).
This is the ancestor replacement step common in connected-components algorithms.}
\label{fig:rve-connected-components}
\end{figure}}

\subsubsection{User-Defined Functions in \maptxt{} and \reduce{}}
\edge{} allows \compute{} operators in both \maptxt{} and \reduce{} to be user-defined functions.
For a \maptxt{} action, the \compute{} operator takes as input the current iteration-space point and the operand values at that point, and returns a single data value.
For a \reduce{} action, the \compute{} operator takes as input the current iteration-space point, the current reduce state, and the current value being reduced, and returns an updated data value.
These functions may be defined in a host language such as C or Python.
Table~\ref{tab:operator-summary} summarizes the inputs and outputs of \compute{} and \merge{} operators in \maptxt{} and \reduce{}.

\subsubsection{User-Defined Functions in Populate}
\edge{} allows \compute{} and \coordinate{} operators to be any general, user-defined function in \populate{}.
The \compute{} operator takes as input the iteration-space point, and a single data value corresponding to the computed value on the RHS of the Einsum at that point.
The output must be a single data value, thus, \compute{} within \populate{} acts as a unary operator.

The \coordinate{} operator takes as input the current iteration-space point, the \compute{} data value at that point, and the current entire partial output fiber on the LHS (which consists of the coordinates and their values) for the mutable rank on which \populate{} is operating.
This allows a user-defined function to use both the current coordinates and values of the LHS to determine which coordinates will exist in the output fiber.
The \coordinate{} operator returns a set of (a) edited and (b) empty coordinates for the output fiber (\S~\ref{ssec:populate_semantics}).
Table~\ref{tab:operator-summary} summarizes the inputs and outputs of the \coordinate{} operator.

\subsubsection{Case Statements}\label{ssec:cases}
The case statement introduced in \S~\ref{ssec:conditionals} is syntactic sugar.
It enables the expression of different Einsums for different regions of the iteration space for the same output tensor.
One can express all case statements using \maptxt{}/\reduce{}/\populate{} actions with the update operator and rank variable expressions.
For example, we can express Equation~\eqref{eqn:cases} as:
\begin{subequations}\label{eqn:sugar}
    \begin{align}
        T1_{m, n} &=  B_{k, m} \cdot C_{k, n} :: \bigwedge_k \times(\cap) \bigvee_k +(\cup) \\
        T2_{m, n: n \neq m} &= A_{m, n} \\
        Z_{m, n}  &= T1_{m, n} << T2_{m, n}
    \end{align}
\end{subequations}
\edge{} allows any Boolean predicate as conditions for each case.

\subsubsection{Shorthand Notation}
\edge{} supports several shorthand notations for informal communication; see Appendix~\ref{appendix:syntax-shorthand} for a full description and examples.

\section{Case Studies: Graph Algorithms (and Beyond!) in \edge{}}\label{ssec:examples}

Having defined the \edge{} language in Sections~\ref{ssec:walk-bfs} through~\ref{ssec:syntax}, we now present case studies of algorithms written in \edge{}.
First, we show that different algorithms for the same problem are often algebraically related in \edge{}, and that we can derive one from another through a sequence of principled manipulations (\S~\ref{ssec:variants}).
Second, we show that \edge{} expresses complex, nested iterative algorithms entirely within the tensor machinery, without resorting to a host language (\S~\ref{ssec:complex}).
Third, we show that \edge{} naturally expresses computations that arise outside traditional graph traversal, including element-wise numerical operations and randomized algorithms (\S~\ref{ssec:beyond-graphs}).
Taken together, these examples exercise \textbf{design goals~\ref{goal:sepconcerns},~\ref{goal:powerful},~\ref{goal:manipulation}, and~\ref{goal:applications}}.

We introduce the term \emph{eingebraic manipulation} to refer to the exploration of the algebraic space that \edge{} enables.
An eingebraic manipulation is an algebraic rewrite, substitution, reassociation, commutation, or mapping choice applied to an \edge{} expression to derive another expression that either preserves the value at each iteration or preserves the final value after convergence.

\subsection{Deriving Algorithmic Variants Through Manipulation}\label{ssec:variants}

One of the central promises of \edge{} is that algorithmic variants of a problem are not just isolated choices but algebraically related expressions.
Given one \edge{} expression for a problem, we can often derive other well-known variants by applying substitution, commutation, reassociation, and various mapping choices to the original expression (\textbf{design goal~\ref{goal:manipulation}}).
This reframes what ``inventing a new algorithm'' means: rather than thinking hard about a problem from scratch, a developer can start from a known expression and explore the space of algebraically equivalent alternatives.
We demonstrate this twice in this subsection, once for breadth-first search and once for single-source shortest paths.
In both cases, the derived variants correspond to algorithms that the literature historically treated as distinct contributions.

\subsubsection{Reachability \bfs{} Variants}\label{ssec:rrbfs}

Reachability \bfs{} is a variant of graph traversal that determines which nodes are reachable from a set of query nodes.
This algorithm may be used in more complex algorithms such as connected components~\cite{Hopcroft:1973:AEA}, maximum flow~\cite{edmonds:1972:TIA}, and finding the diameter of a graph~\cite{corneil:2003:OtP}.
Cascade~\ref{cascade:rbfs} provides an \edge{} expression for this problem.
Assuming all involved tensors contain Boolean values, this expression gathers the neighbors of vertices in $F$ ($G \cdot F$), filtering out those that have already been visited (intersection with $\neg P$).
It then updates the set of visited nodes by combining the newly created frontier ($F_{i+1}$) with the current set of visited nodes $(P_i)$.

\begin{cascade}%
\begin{mdframed}
{\setlength{\abovedisplayskip}{0pt}%
 \setlength{\belowdisplayskip}{0pt}%
 \setlength{\abovedisplayshortskip}{0pt}%
 \setlength{\belowdisplayshortskip}{0pt}%
 \setlength{\jot}{1pt}%
\begin{subequations}\label{eqn:rbfs}
\begin{compactalign}
\cascadecomment{$\triangleright$ Extended Einsum}
F_{i+1, d} &= (G_{s,d} \cdot^1 F_{i,s})_{i,d} \cdot^2 \neg P_{i,d} :: \bigwedge^1_{s}\text{AND}(\cap) \bigvee^1_{s}\text{OR}(\cup) \bigwedge^2_{d}\text{AND}(\cap) \\
P_{i+1, d} &= P_{i,d} \cdot F_{i+1,d} :: \bigwedge_{d}\text{OR}(\cup)
\end{compactalign}
\end{subequations}}
\end{mdframed}
\caption{Push-BFS (Reachability BFS).}
\label{cascade:rbfs}
\end{cascade}

Green introduces a variant of \bfs{} that \emph{updates the graph} on each iteration, deleting edges as they are visited~\cite{Green:2021:IDB, Zhang:2018:ABG}.
By performing a series of inductive steps on the \edge{} expression followed by substitution, we derive the \edge{} expression in Cascade~\ref{cascade:greenbfs} for Green's \bfs{} variant.

\begin{cascade}
\begin{mdframed}
\begin{subequations}
\begin{compactalign}
\cascadecomment{$\triangleright$ Initialization}
GG_{0, s, d} &= G_{s, d} \\
\cascadecomment{$\triangleright$ Extended Einsum}
F_{i+1, d}  &= GG_{i, s, d} \cdot F_{i, s}  :: \bigwedge_{s} \text{AND}(\cap) \bigvee_{s} \text{OR}(\cup) \\
GG_{i+1, s, d} &= GG_{i, s, d} \cdot \neg F_{i+1, d} :: \bigwedge_{d} \text{AND}(\cap)
\end{compactalign}
\end{subequations}
\end{mdframed}
\caption{Green's BFS variant (inverse-delete BFS).}
\label{cascade:greenbfs}
\end{cascade}

In this expression, $GG_0$ contains the original, unmodified graph.
On each iteration, we first gather the neighbors of vertices in $F_i$ to produce a new output frontier, $F_{i+1}$.
We then update the graph ($GG$) by deleting all edges whose destinations are in the newly created frontier ($GG_i \cdot \neg F_{i+1}$).

In a similar vein, through commutation, reassociation, and applying a specific mapping choice, we derive pull-\bfs{} (Cascade~\ref{cascade:pullbfs}) from push-\bfs{} (Cascade~\ref{cascade:rbfs})~\cite{Beamer:2012:DBS}.
In this cascade, we begin with push-\bfs{} (Cascade~\ref{cascade:rbfs}) and specify a loop order mapping choice where $s$ is in the outer loop nest and $d$ is in the inner loop nest.
The derivable pull-\bfs{} (Cascade~\ref{cascade:pullbfs}) can be combined with a loop order mapping choice where $d$ is in the outer loop nest and $s$ is in the inner loop nest.
This corresponds to an implementation that iterates through each unvisited vertex ($\neg P$), determines if it has parents in the frontier, and adds itself to the output frontier ($F_{i+1}$) if so.

\begin{cascade}
\begin{mdframed}
\begin{subequations}\label{eqn:pullbfs}
\begin{compactalign}
\cascadecomment{$\triangleright$ Extended Einsum}
F_{i+1, d}  &= (G_{s, d} \cdot^1 \neg P_{i, d})  \cdot^2  F_{i, s} :: \bigwedge^1_{d} \text{AND}(\cap) \bigwedge^2_{s} \text{AND}(\cap) \bigvee^2_{s} \text{OR}(\cup)  \\
P_{i+1, d} &= P_{i, d}    \cdot F_{i+1, d} :: \bigwedge_{d} \text{OR}(\cup)
\end{compactalign}
\end{subequations}
\end{mdframed}
\caption{Pull-BFS (derived from push-BFS via commutation, associativity, and a mapping choice).}
\label{cascade:pullbfs}
\end{cascade}

\subsubsection{Single Source Shortest Paths}\label{ssec:sssp}

Single-source shortest paths (SSSP) operates on weighted graphs.
It takes as input a starting source vertex, then determines the shortest path to all vertices reachable from that source.
Each vertex has a corresponding value in a \emph{Distances} array, which stores the accumulated distance of the path to that vertex.
If a vertex is not reachable, the algorithm sets its distance to infinity.
Cascade~\ref{cascade:bf} shows the corresponding \edge{} expression for Bellman-Ford.
Table~\ref{tab:sssp-tensors} in Appendix~\ref{appendix:bf-tensors} summarizes the tensors used throughout this derivation and marks which cascade each one appears in.

\begin{cascade}
\begin{mdframed}
\begin{subequations}
\begin{compactalign}
\cascadecomment{$\triangleright$ Tensors}
G^{S \equiv |V|, D \equiv |V|} &\rightarrow \text{integer, empty}=0 \\
C^{I, S \equiv |V|} &\rightarrow \text{Boolean, empty}=\False\\
N^{I, D \equiv |V|} &\rightarrow \text{integer, empty}=\infty\\
\mathsf{NewlyRelaxed}^{I, D \equiv |V|} &\rightarrow \text{integer, empty}=\infty \\
D^{I, S \equiv |V|} &\rightarrow \text{integer, empty}=\infty \\
\cascadecomment{$\triangleright$ Initialization}
G &\rightarrow \langle\text{user-specified}\rangle\\
D_{0, s \in \text{root\_id}} &= 0 \\
\cascadecomment{$\triangleright$ Extended Einsum}
&\eqcomment{For each vertex in $D$, gather its neighbors.}
&\eqcomment{Add the edge weight ($s, d$) to update the neighbor's distance.}
&\eqcomment{For each updated neighbor ($d$), select the minimum distance \\ from all sources ($s$).}
N_{i, d} &= G_{s, d} \cdot D_{i, s} :: \bigwedge_{s} +(\cap) \bigvee_{s} \min(\cup)\\
&\eqcomment{Check which new neighbor distances are less than current distances.}
C_{i, d} &= N_{i, d} \cdot D_{i, d} :: \bigwedge_d <(\cup)\\
&\eqcomment{Record the relaxed distances.}
\mathsf{NewlyRelaxed}_{i, d} &= C_{i, d} \cdot N_{i, d} :: \bigwedge_d \rightarrow(\cap) \\
&\eqcomment{Update distances ($\mathbin{<<}$) with the newly found shorter paths.}
D_{i+1, d} &= D_{i, d} \cdot \mathsf{NewlyRelaxed}_{i, d} :: \bigwedge_{d} \mathbin{<<}(\cup) \\
&\diamond: D_{i+1} \equiv D_i
\end{compactalign}
\end{subequations}
\end{mdframed}
\caption{Bellman-Ford SSSP.}
\label{cascade:bf}
\end{cascade}

Bellman-Ford requires \texttt{|V|-1} iterations to guarantee that all reachable vertices have their correct distances, and it can detect negative cycles if any distance is updated on the \texttt{|V|}th iteration~\cite{CLRS:2022:ITA}.
The Shortest Paths Faster Algorithm~\cite{Moore:1959:SPFA} is a variant of Bellman-Ford that introduces a queue to reduce the number of total edge relaxation rounds in the average case.
\edge{} can express this variant, shown in Cascade~\ref{cascade:spfa}.
We later show how SPFA can be derived from Bellman-Ford through a sequence of eingebraic manipulations.

\begin{cascade}
\begin{mdframed}
\begin{subequations}
\begin{compactalign}
\ldots \notag \\
Q^{I, S \equiv |V|} &\rightarrow \text{Boolean, empty}=\False\\
\cascadecomment{$\triangleright$ Initialization}
Q_{0, s \in \text{root\_id}} &= \True\\
\cascadecomment{$\triangleright$ Extended Einsum}
&\eqcomment{Get the distances of vertices in the queue.}
DQ_{i, s} &= Q_{i, s} \cdot D_{i,s} :: \bigwedge_{s} \rightarrow(\cap) \\
&\eqcomment{Select ANY vertex that is in the queue and has a non-infinity distance.}
&\eqcomment{This determines which vertex is the source for this iteration.}
&\eqcomment{Thus $F$ is a one-hot vector (contains only one non-empty $s$).}
F_{i, s^*} &= DQ_{i, s} :: \lll_{s^{*}} \mathbbm{1}(\text{select-any-s}) \\
&\eqcomment{Get neighbors of selected vertex and update neighbor distances \\ by adding $(s, d)$ edge weight.}
N_{i, d} &= G_{s, d} \cdot F_{i, s} :: \bigwedge_{s} +(\cap) \bigvee_{s} \min(\cup) \\
&\eqcomment{Compare neighbor distances to current distances and mark \\ which nodes have been updated.}
C_{i, d} &= N_{i, d} \cdot D_{i, d} :: \bigwedge_d <(\cup)\\
&\eqcomment{Record updated distances for neighbors.}
\mathsf{NewlyRelaxed}_{i, d} &= C_{i, d} \cdot N_{i, d} :: \bigwedge_d \rightarrow(\cap) \\
&\eqcomment{Update distances with newly relaxed vertices. This keeps the value \\ in $D$ unless the vertex was relaxed, in which case it updates to the \\ new distance.}
D_{i+1, d} &= D_{i, d} \cdot \mathsf{NewlyRelaxed}_{i, d} :: \bigwedge_{d} \mathbin{<<}(\cup) \\
&\eqcomment{Remove selected vertex from queue, store remaining vertices \\ in temp tensor $T$.}
T_{i, d} &= Q_{i, d} \cdot \neg F_{i, d} :: \bigwedge_{d} \leftarrow(\cap) \\
&\eqcomment{Update queue: old queue minus processed vertex ($T$), \\ plus newly updated vertices.}
Q_{i+1, d} &= T_{i, d} \cdot C_{i, d} :: \bigwedge_{d} \text{OR}(\cup)\\
&\diamond: D_{i+1} \equiv D_i
\end{compactalign}
\end{subequations}
\end{mdframed}
\caption{Shortest Paths Faster Algorithm (SPFA) in \edge{}.}
\label{cascade:spfa}
\end{cascade}

We can \emph{eingebraically manipulate} the Bellman-Ford expression
(Cascade~\ref{cascade:bf}) to derive SPFA.
Each step in the derivation is either a rewrite rule that preserves the value
of $D_{i+1}$ for each $i$ or a transformation that preserves the value of $D$
after convergence.
The latter may change per-iteration values but not the final answer.
We highlight the key transformations in the main text and defer the full
eingebraic manipulation steps to Appendix~\ref{appendix:bf-derivation}.

\subsubsubsection{Step 1: Promote $s$ to an Explicit Rank}\label{sssec:bf-step1}

First, we will defer the $\min$ reduction over sources until after we compute
the relaxed distances for each source.
This exposes each source's contribution to the distances tensor as a distinct
slice of the intermediate tensors, which we can then manipulate independently.
Cascade~\ref{cascade:bf-step1} shows the resulting expression; Appendix~\ref{appendix:bf-derivation} walks through each manipulation in the cascade chain in more detail.

\begin{cascade}
\begin{mdframed}
\begin{subequations}
\begin{compactalign}
\cascadecomment{$\triangleright$ Extended Einsum}
&\eqcomment{For each source, gather its neighbors and add the edge weight. \\
  Keep $s$ as an explicit rank instead of reducing immediately.}
N_{i, s, d} &= G_{s, d} \cdot D_{i, s} :: \bigwedge +(\cap)
  \label{eqn:bf-step1-N} \\
&\eqcomment{For each (source, neighbor) pair, check whether the new neighbor \\
  distance is less than the current distance.}
C_{i, s, d} &= N_{i, s, d} \cdot D_{i, d} :: \bigwedge <(\cup)
  \label{eqn:bf-step1-C} \\
&\eqcomment{For each source, record the relaxed distances to its neighbors.}
\mathsf{PerSourceRelaxed}_{i, s, d} &= C_{i, s, d} \cdot N_{i, s, d}
  :: \bigwedge \rightarrow(\cap) \label{eqn:bf-step1-persourcerelaxed} \\
&\eqcomment{Across all sources, select the best relaxed distance \\ for each
  destination.}
\mathsf{NewlyRelaxed}_{i, d} &= \mathsf{PerSourceRelaxed}_{i, s, d}
  :: \bigvee_{s} \min(\cup) \label{eqn:bf-step1-newlyrelaxed} \\
&\eqcomment{Update distances with the best relaxed distances.}
D_{i+1, d} &= D_{i, d} \cdot \mathsf{NewlyRelaxed}_{i, d}
  :: \bigwedge \mathbin{<<}(\cup) \label{eqn:bf-step1-Dnext} \\
&\diamond: D_{i+1} \equiv D_i
\end{compactalign}
\end{subequations}
\end{mdframed}
\caption{Algebraically Manipulating Bellman-Ford, Step 1.
Promote $s$ to an explicit rank and defer the min-over-sources reduction
until after relaxation.}
\label{cascade:bf-step1}
\end{cascade}

\subsubsubsection{Step 2: Collapse the Boolean-Check Update Into a $\min$ Einsum}\label{sssec:bf-step2}

Cascade~\ref{cascade:bf-step1} defers the
$\min$ reduction that occurs over sources from the gather Einsum~\eqref{eqn:bf-step1-N} to
the reconciliation Einsum~\eqref{eqn:bf-step1-newlyrelaxed}.
This separation exposes each source's contribution to the distances tensor as
a distinct slice of $\mathsf{PerSourceRelaxed}$, allowing us to manipulate each
source's contribution independently.

Additionally, notice that $C$ marks where $N < D$ (\eqref{eqn:bf-step1-C}),
$\mathsf{PerSourceRelaxed}$ keeps $N$'s value at exactly those points
(\eqref{eqn:bf-step1-persourcerelaxed}), and $\mathbin{<<}(\cup)$ then writes
those values into $D$ (\eqref{eqn:bf-step1-Dnext}).
Then, $D_{i+1}$ takes $N$ where $N < D$ and keeps $D$ otherwise.
This is exactly a minimum between $D$ and $N$ ($\min(D, N)$).
Thus, Einsums~\eqref{eqn:bf-step1-C} through~\eqref{eqn:bf-step1-Dnext} collapse
into a single Einsum that takes the $\min$ of $D$ and $N$.
Appendix~\ref{appendix:bf-collapse-rewrite} provides a more detailed
explanation of this transformation.
Cascade~\ref{cascade:bf-step2} shows the resulting expression.

\begin{cascade}
\begin{mdframed}
\begin{subequations}
\begin{compactalign}
\cascadecomment{$\triangleright$ Extended Einsum}
&\eqcomment{For each source, gather its neighbors and add the edge weight.}
N_{i, s, d} &= G_{s, d} \cdot D_{i, s} :: \bigwedge +(\cap) \\
&\eqcomment{Take $\min$ of $D$ and each source's neighbor candidates, \\
  then reduce over $s$ to pick the best per destination.}
D_{i+1, d} &= D_{i, d} \cdot N_{i, s, d}
  :: \bigwedge \min(\cup) \bigvee_{s} \min(\cup) \\
&\diamond: D_{i+1} \equiv D_i
\end{compactalign}
\end{subequations}
\end{mdframed}
\caption{Algebraically Manipulating Bellman-Ford, Step 2.
Einsums~\eqref{eqn:bf-step1-C} through~\eqref{eqn:bf-step1-Dnext} of
Cascade~\ref{cascade:bf-step1} collapse into a single Einsum that takes the
$\min$ of $D$ and $N$, with reduction over $s$.}
\label{cascade:bf-step2}
\end{cascade}

\subsubsubsection{Step 3: Replace the $s$-Reduction With Iterative Accumulation}\label{sssec:bf-step3}

Now, rather than deferring the update of $D$ until after we compute the relaxed
distances for all sources, we can update $D$ iteratively after processing each
source.
Cascade~\ref{cascade:bf-step3} shows the resulting expression.
The gather step still gathers from the original distance array
($\tilde{D}_{i, 0, k} = D_{i, k}$) rather than the iteratively updated distance
array.
To update the running distance array $\tilde{D}$ for this $k$th source, we
take the $\min$ of the current running distance and the relaxed distances
from this source ($N_{i, k, d}$).
This is possible due to the algebraic properties of $\min$ (idempotence and
associativity, see Appendix~\ref{appendix:bf-fold}).

\begin{cascade}
\begin{mdframed}
\begin{subequations}
\begin{compactalign}
\cascadecomment{$\triangleright$ Extended Einsum}
&\eqcomment{Initialize the running distance.}
\tilde{D}_{i, 0, d} &= D_{i, d} \\
&\eqcomment{For each source $k$, gather neighbors using the \\
  \emph{original} distance $\tilde{D}_{i, 0, k}$ and add the edge weight.}
N_{i, k, d} &= G_{k, d} \cdot \tilde{D}_{i, 0, k}
  :: \bigwedge +(\cap) \\
&\eqcomment{Add source $k$'s candidates into the running distance via $\min$.}
\tilde{D}_{i, k+1, d} &= \tilde{D}_{i, k, d} \cdot N_{i, k, d}
  :: \bigwedge \min(\cup) \label{eqn:bf-step3-tildeD} \\
&\eqcomment{Stop after processing all sources.}
\diamond_k &: k \equiv |V| \\
&\eqcomment{Commit the final running distance as the next iteration's $D$.}
D_{i+1, d} &= \tilde{D}_{i, |V|, d} \\
&\diamond: D_{i+1} \equiv D_i
\end{compactalign}
\end{subequations}
\end{mdframed}
\caption{Algebraically Manipulating Bellman-Ford, Step 3.
Replace the reduction over $s$ with accumulation along a new iterative
rank $k$. $\tilde{D}_{i, k, d}$ iteratively accumulates each source's
candidate distances via $\min$. All sources still gather from the original
distance array ($\tilde{D}_{i, 0, k} = D_{i, k}$).}
\label{cascade:bf-step3}
\end{cascade}

\subsubsubsection{Step 4: Sequentialize the Gather Step}\label{sssec:bf-step4}

The transformations so far have all been rewrite rules: each cascade
computes the same $D_{i+1}$ for each $i$ as previous cascades.
We now make a transformation that changes per-iteration values but
preserves the value of $D$ after convergence.

Specifically, we change source $k$'s gather to read from the running distance
$\tilde{D}_{i, k, k}$ rather than the original $\tilde{D}_{i, 0, k} = D_{i, k}$.
Each source now sees the improvements made by earlier sources within the same
outer ($i$) iteration.
Cascade~\ref{cascade:bf-step4} shows the resulting expression.

To see why this is sound, unroll $\tilde{D}_{i, k:k=2, d}$ for source $k = 2$
in a 3-vertex graph at a given $i$.
Applying the running distance recurrence twice
(Equation~\eqref{eqn:bf-step3-tildeD}), we have:
\begin{equation}
\tilde{D}_{i, k:k=2, d} = (\tilde{D}_{i, k:k=0, d} \cdot^1 N_{i, k:k=0, d})_{i, d}
  \cdot^2 N_{i, k:k=1, d}
  :: \bigwedge^1 \min(\cup) \bigwedge^2 \min(\cup).
\end{equation}

Source 2's gather intersects $\tilde{D}$'s third rank with $G$'s first rank
$s$, pinning that rank to $2$:
\begin{equation}
N_{i, k:k=2, d} = G_{s:s=2, d} \cdot \tilde{D}_{i, k:k=2, s:s=2}
  :: \bigwedge +(\cap).
\end{equation}

Substituting in the unrolled $\tilde{D}_{i, k:k=2, s:s=2}$:
\begin{equation}
N_{i, k:k=2, d} = G_{s:s=2, d} \cdot^1
  ((\tilde{D}_{i, k:k=0, s:s=2} \cdot^2 N_{i, k:k=0, s:s=2})_{i, s:s=2}
  \cdot^3 N_{i, k:k=1, s:s=2})_{i, s:s=2}
  :: \bigwedge^1 +(\cap) \bigwedge^2 \min(\cup) \bigwedge^3 \min(\cup).
\end{equation}

Now, $+$ distributes over $\min$, so we can distribute $G_{s:s=2, d} \cdot^1$
over the $\min$ reductions in the above expression:
\begin{enumerate}
  \item $G_{s:s=2, d} \cdot \tilde{D}_{i, k:k=0, s:s=2} :: \bigwedge +(\cap)$:
    this is the same gather computed by the original Bellman-Ford cascades
    (Cascades~\ref{cascade:bf} through~\ref{cascade:bf-step3}) for source 2.
  \item $G_{s:s=2, d} \cdot N_{i, k:k=0, s:s=2} :: \bigwedge +(\cap)$:
    a path through source 0 first, then source 2 to $d$.
  \item $G_{s:s=2, d} \cdot N_{i, k:k=1, s:s=2} :: \bigwedge +(\cap)$:
    a path through source 1 first, then source 2 to $d$.
\end{enumerate}
The new $N_{i, k:k=2, d}$ is a $\min$ over the above three terms, thus,
a gather on the running distance is pointwise less than or equal to a
gather on the original distances array, $D$.

Per-iteration values can differ, but the value of $D$ after convergence does not.
Both cascades terminate at $D_{i+1} \equiv D_i$, at which point no source can
improve any distance.
Appendix~\ref{appendix:bf-sequentialize} gives the formal inductive proof.
This new cascade typically converges in fewer outer iterations ($i$).

\begin{cascade}
\begin{mdframed}
\begin{subequations}
\begin{compactalign}
\cascadecomment{$\triangleright$ Extended Einsum}
\tilde{D}_{i, 0, d} &= D_{i, d} \\
&\eqcomment{Gather neighbors using the \emph{running} distance \\
  $\tilde{D}_{i, k, k}$ rather than the original $\tilde{D}_{i, 0, k}$.}
N_{i, k, d} &= G_{k, d} \cdot \tilde{D}_{i, k, k}
  :: \bigwedge +(\cap) \\
\tilde{D}_{i, k+1, d} &= \tilde{D}_{i, k, d} \cdot N_{i, k, d}
  :: \bigwedge \min(\cup) \\
\diamond_k &: k \equiv |V| \\
D_{i+1, d} &= \tilde{D}_{i, |V|, d} \\
&\diamond: D_{i+1} \equiv D_i
\end{compactalign}
\end{subequations}
\end{mdframed}
\caption{Algebraically Manipulating Bellman-Ford, Step 4.
Each source $k$ now reads from the running distance
$\tilde{D}_{i, k, k}$ rather than the original $\tilde{D}_{i, 0, k}$.
Per-iteration values change, but the value of $D$ after convergence does not.}
\label{cascade:bf-step4}
\end{cascade}

\subsubsubsection{Step 5: Collapse the Inner Enumeration Into the Outer Iteration (SPFA)}\label{sssec:bf-step5}

We now collapse the inner $k$ enumeration into the outer iterative rank $i$.
Instead of processing all $|V|$ sources within each outer iteration, we let
each outer iteration $i$ process exactly one source.
This requires a selection step that determines which source to process on a
given iteration.

To make this efficient, we want to skip sources whose distances have not
changed since their last processing as such sources would contribute no
improvement.
This requires tracking which destinations were actually relaxed during each
iteration: we re-introduce the mask $C_{i, d}$ that was
collapsed into the $\min$ Einsum in Cascade~\ref{cascade:bf-step2}.
We then track the sources that may still produce improvements in a queue $Q$,
updated each iteration based on $C$: a destination $d$ is added to $Q$
exactly when it was relaxed.
On each outer iteration, we select any source from $Q$ that has a non-infinity
distance in $D$; the selected source becomes the sole non-empty entry of a
frontier tensor $F$ for that iteration.

Note that a single outer iteration of the new cascade processes one source
rather than all $|V|$ sources, so per-iteration values differ.
The value of $D$ after convergence is preserved by the same $(\min, +)$
argument as Cascade~\ref{cascade:bf-step4}.
Cascade~\ref{cascade:bf-step5} shows the resulting expression.
This is exactly the Shortest Paths Faster Algorithm.
The closing paragraph of Appendix~\ref{appendix:bf-sequentialize} shows that the same convergence argument used for Cascade~\ref{cascade:bf-step4} extends to this queue-based variant.

\begin{cascade}
\begin{mdframed}
\begin{subequations}
\begin{compactalign}
\ldots \notag \\
Q^{I, S \equiv |V|} &\rightarrow \text{Boolean, empty}=\False\\
\cascadecomment{$\triangleright$ Initialization}
Q_{0, s \in \text{root\_id}} &= \True\\
\cascadecomment{$\triangleright$ Extended Einsum}
&\eqcomment{\textbf{Select a vertex to process.}}
&\eqcomment{Filter distances that have entries in $Q$.}
DQ_{i, s} &= Q_{i, s} \cdot D_{i,s} :: \bigwedge_{s} \rightarrow(\cap) \\
&\eqcomment{Select ANY vertex that is in the queue and has a non-infinity \\
  distance (determines which vertex is $D0$, $D1$, \ldots). Thus $F$ is a \\
  one-hot vector (contains only one non-empty $s$).}
F_{i, s^*} &= DQ_{i, s} :: \lll_{s^{*}} \mathbbm{1}(\text{select-any-s}) \\
&\eqcomment{\textbf{Update distances of relaxed edges.}}
&\eqcomment{Gather neighbors of selected vertex.}
N_{i, d} &= G_{s, d} \cdot F_{i, s}
  :: \bigwedge_{s} +(\cap) \bigvee_{s} \min(\cup) \\
&\eqcomment{Compare neighbor distances to current distances and mark \\
  which nodes will be updated.}
C_{i, d} &= N_{i, d} \cdot D_{i, d} :: \bigwedge_d <(\cup)\\
&\eqcomment{Record updated distances for neighbors.}
\mathsf{NewlyRelaxed}_{i, d} &= C_{i, d} \cdot N_{i, d}
  :: \bigwedge_d \rightarrow(\cap) \\
&\eqcomment{Update the distances array.}
D_{i+1, d} &= D_{i, d} \cdot \mathsf{NewlyRelaxed}_{i, d}
  :: \bigwedge_{d} \mathbin{<<}(\cup) \\
&\eqcomment{\textbf{Update the Queue.}}
&\eqcomment{Remove processed vertex from queue.}
T_{i, d} &= Q_{i, d} \cdot \neg F_{i, d}
  :: \bigwedge_{d} \leftarrow(\cap) \\
&\eqcomment{Add the neighbors that were just relaxed to the queue.}
Q_{i+1, d} &= T_{i, d} \cdot C_{i, d}
  :: \bigwedge_d \text{OR}(\cup)\\
&\eqcomment{Stop when the queue is empty.}
&\diamond: ||Q_{i+1}|| \equiv 0
\end{compactalign}
\end{subequations}
\end{mdframed}
\caption{Algebraically Manipulating Bellman-Ford, Step 5. Collapse the source
enumeration into the iterative rank and introduce a queue-based selector,
yielding SPFA.}
\label{cascade:bf-step5}
\end{cascade}

\subsubsubsection{Step 6: Restrict the Selector to Get Dijkstra}\label{sssec:bf-step6}

Now, in Cascade~\ref{cascade:bf-step5} we used \texttt{select-any-s}
to select which source to explore on a given iteration.
If we replace selecting any vertex with selecting the $\min$ vertex, and
constrain the graph to have no negative-weight edges, we obtain Dijkstra's
sequential algorithm.
This is shown in Cascade~\ref{cascade:dijkstra}.
By selecting the minimum-distance vertex on each iteration, Dijkstra's
algorithm ensures that each vertex is processed at most once, avoiding the
$|V|-1$ iterations that Bellman-Ford requires for worst-case correctness.
The same value-at-convergence argument given in Appendix~\ref{appendix:bf-sequentialize} applies here, since restricting the selector and forbidding negative-weight edges preserves the $(\min, +)$ algebraic structure.

\begin{cascade}
\begin{mdframed}
\begin{subequations}
\begin{compactalign}
\cascadecomment{$\triangleright$ Extended Einsum}
&\eqcomment{\textbf{Select a vertex to process.}}
&\eqcomment{Filter distances that have entries in $Q$.}
DQ_{i, s} &= Q_{i, s} \cdot D_{i,s} :: \bigwedge_{s} \rightarrow(\cap) \\
&\eqcomment{Select the vertex in $Q$ with minimum current distance.}
F_{i, s^*} &= DQ_{i, s} :: \lll_{s^{*}} \mathbbm{1}(\text{select-min-s}) \\
N_{i, d} &= G_{s, d} \cdot F_{i, s}
  :: \bigwedge_{s} +(\cap) \bigvee_{s} \min(\cup) \\
C_{i, d} &= N_{i, d} \cdot D_{i, d} :: \bigwedge_d <(\cup)\\
\mathsf{NewlyRelaxed}_{i, d} &= C_{i, d} \cdot N_{i, d}
  :: \bigwedge_d \rightarrow(\cap) \\
D_{i+1, d} &= D_{i, d} \cdot \mathsf{NewlyRelaxed}_{i, d}
  :: \bigwedge_{d} \mathbin{<<}(\cup) \\
T_{i, d} &= Q_{i, d} \cdot \neg F_{i, d}
  :: \bigwedge_{d} \leftarrow(\cap) \\
Q_{i+1, d} &= T_{i, d} \cdot C_{i, d}
  :: \bigwedge_d \text{OR}(\cup)\\
&\eqcomment{Stop when the queue is empty.}
&\diamond: ||Q_{i+1}|| \equiv 0
\end{compactalign}
\end{subequations}
\end{mdframed}
\caption{Algebraically Manipulating Bellman-Ford, Step 6. Dijkstra's SSSP,
derived from SPFA by replacing \texttt{select-any-s} with \texttt{select-min-s}.}
\label{cascade:dijkstra}
\end{cascade}

Efficiently implementing Dijkstra requires a priority queue.
At the \emph{format} level of the separation of concerns, this means storing $DQ$
and $Q$ in formats that support efficient minimum selection, such as a binary
heap.

\subsection{Expressing Complex Algorithms}\label{ssec:complex}

We now turn to concurrent connected-components algorithms, which require nested iteration, conditional updates, and bookkeeping over multiple tensors.
In this subsection we show that \edge{} expresses these algorithms entirely within the tensor machinery, using nested cascades and rank variable expressions rather than stepping into the concerns of mapping, format, and binding (\textbf{design goals~\ref{goal:sepconcerns} and~\ref{goal:powerful}}, see Nayak et al.~\cite{Nayak:2023:TDF}).
We present two variants: Liu and Tarjan's simple concurrent ``Algorithm~S''~\cite{Liu:2022:SCC}, and the Awerbuch-Shiloach variant~\cite{Awerbuch:1987:NCM, Azad:2019:LLA}.

\subsubsection{Algorithm S\@: Simple Concurrent Connected Components}\label{ssec:cc-algs}

Cascade~\ref{cascade:cc} presents the Einsum for simple concurrent connected components, specifically ``Algorithm S'' from Liu and Tarjan~\cite{Liu:2022:SCC}.
Connected-components algorithms group all vertices in an undirected graph into \emph{components}, where a component consists of all vertices reachable from every other vertex within the component.
Algorithm S tracks the ``parents'' of each vertex, altering parents until all vertices within a component share the same parent.
It consists of three steps: initialization (each vertex is its own parent), parent-connect (for each edge, assign the minimum parent of the two endpoints to the vertex with the lower vertex ID), and shortcut (assign each vertex's parent to its grandparent).
Algorithm S repeats parent-connect followed by rounds of shortcut until no components change.

\begin{cascade}[hbtp]
\begin{mdframed}
\begin{subequations}\label{eqn:concurrentcc}
\begin{compactalign}
\cascadecomment{$\triangleright$ Tensors}
G^{I, S\equiv|V|, D\equiv|V|} &\rightarrow  \text{Boolean, empty=\False} \label{seqn:Gcc}\\
P^{I, C\equiv|V|, M\equiv|V|} &\rightarrow \text{Boolean, empty=\False} \label{seqn:Pcc} \\
\cascadecomment{$\triangleright$ Initialization}
G &\rightarrow \langle\text{user-specified}\rangle \label{seqn:GIc}\\
P_{0, v, v} &= \True\\
\cascadecomment{$\triangleright$ Extended Einsum}
\cascadecomment{Begin Outer Cascade}
i &= 0\\
\cascadecomment{Parent Connect}
\textit{VAV}_{i, v, w, av} &= G_{v, w} \cdot P_{i, v, av} \quad :: \bigwedge_{v} \text{AND}(\cap)\\
\textit{WAW}_{i, av, aw} &= \textit{VAV}_{i, v, w, av} \cdot P_{i, w, aw} \quad :: \bigwedge_{w} \text{AND}(\cap) \bigvee_{v,w} \text{OR}(\cup)  \\
P_{i+1, gs, d: (gs < ad \;?\; gs : (ad: av > aw \;?\; aw : av))} &= \textit{WAW}_{i, av, aw} \cdot P_{i, as: (av > aw \;?\; av : aw), gs} \notag\\
&\quad :: \bigwedge_{av, aw} \text{AND}(\cap) \lll_d \min_1(\cup)  \label{connect} \\
\cascadecomment{Begin Inner Cascade}
\cascadecomment{Shortcut}
j &= i+1\\
P_{j+1, v, gv} &= P_{j, v, av} \cdot P_{j, av, gv} \quad :: \bigwedge_{av} \text{AND}(\cap) \bigvee_{av} \text{ANY}(\cup) \label{shortcut}\\
j &= j+1\\
\diamond_{j}: P_{j+1} &\equiv P_{j} \label{shortcutend}\\
i &= j\\
\diamond_{i}: P_{i+1} &\equiv P_{i}
\end{compactalign}
\end{subequations}
\end{mdframed}
\caption{Simple concurrent connected components (Algorithm S, Liu and Tarjan~\cite{Liu:2022:SCC}).}
\label{cascade:cc}
\end{cascade}

On the $i$'th iteration, the $P$ tensor contains the set of vertices ($M$ rank) that belong to a component ($C$ rank).
We express this algorithm as a cascade of cascades, with two generational ranks $i$ and $j$.
Cascade~\ref{cascade:cc} also contains rank variable expressions that select which ancestor should appear in the $P$ tensor.
The ``Shortcut'' step ``replaces the parent ($av$) of every vertex $v$ by its grandparent ($gv$)''~\cite{Liu:2022:SCC}.
It repeats until no more parents are changing (Equation~\eqref{shortcutend}).
There is no need to fully understand this Einsum: the key takeaway is that with the few extensions \edge{} adds to traditional Einsum notation, \emph{a complex algorithm is now expressible in a succinct and precise manner}.

\subsubsection{Awerbuch-Shiloach and the \graphblas{} Contrast}\label{ssec:ccgraphblas}

\graphblas{} can express the Awerbuch-Shiloach variant of connected components~\cite{Awerbuch:1987:NCM, Azad:2019:LLA}.
We can also express this variant in \edge{}:

\begin{cascade}[hbtp]
\begin{mdframed}
\begin{subequations}\label{eqn:concurrentas}
\begin{compactalign}
\cascadecomment{$\triangleright$ Tensors}
G^{I, S\equiv|V|, D\equiv|V|} &\rightarrow  \text{Boolean, empty=\False} \\
F^{I, C\equiv|V|, M\equiv|V|} &\rightarrow \text{Boolean, empty=\False} \\
GF^{I, C\equiv|V|, M\equiv|V|}&\rightarrow \text{Boolean, empty=\False}  \\
S^{I, S\equiv|V|} &\rightarrow \text{Boolean, empty=\False} \\
\cascadecomment{$\triangleright$ Initialization}
G &\rightarrow \langle\text{user-specified}\rangle\\
F_{0, v, v} &= \True\\
\cascadecomment{$\triangleright$ Extended Einsum}
i &= 0\\
\cascadecomment{Conditional Star Hooking}
S_{i, v} &= \langle\text{STARCHECK}\rangle\\
&\eqcomment{Get $v$'s parent ($av$) only if $v$ is in a star; $(v,w)$ is an edge}
\textit{VAV}_{i, v, w, av} &= (G_{v, w} \cdot^1 F_{i, v, av}) \cdot^2 S_{i, v} :: \bigwedge^1_{v} \text{AND}(\cap) \bigwedge^2_{v} \text{AND}(\cap)\\
&\eqcomment{Get $w$'s parent ($aw$), $v$'s grandparent ($gv$), store new parents of $av$}
TF0_{i, av, aw} &= (\textit{VAV}_{i, v, w, av} \cdot^1 F_{i, w, aw}) \cdot^2 F_{i, av: av > aw, gv} \notag\\
&:: \bigwedge^1_{w} \text{AND}(\cap) \bigvee^1_{v,w} \text{OR}(\cup) \bigwedge^2_{av} \text{AND}(\cap) \lll_{aw} \min_1(\cup)\\
F_{i+1, av, d} &= F_{i, av, d} \mathbin{<<} TF0_{i, av, d} \\
\cascadecomment{Unconditional Star Hooking}
S_{i+1, v} &= \langle\text{STARCHECK}\rangle\\
&\eqcomment{Get $v$ if in a star}
\textit{VAV2}_{i+1, v, w, av} &= (G_{v, w} \cdot^1 F_{i+1, v, av}) \cdot^2 S_{i+1, v} :: \bigwedge^1_{v} \text{AND}(\cap) \bigwedge^2_{v} \text{AND}(\cap)\\
&\eqcomment{Update $v$'s parents iff $av \neq aw$}
TF1_{i+2, av, aw} &= (\textit{VAV2}_{i+1, v, w, av: av \neq aw} \cdot^1 F_{i+1, w, aw}) \cdot^2 F_{i+1, av, gv}\notag\\
&:: \bigwedge^1_{w} \text{AND}(\cap) \bigvee^1_{v,w} \text{OR}(\cup) \bigwedge^2_{av} \text{AND}(\cap) \lll_{aw} \min_1(\cup)\\
F_{i+2, av, d} &= F_{i+1, av, d} \mathbin{<<} TF1_{i, av, d} \\
\cascadecomment{Shortcut}
S_{i+2, v} &= \langle\text{STARCHECK}\rangle\\
&\eqcomment{Update $v$'s parent to grandparent iff $v$ is not in a star}
TF2_{i+3, v, gv} &= (F_{i+2, v, av} \cdot^1 F_{i+2, av, gv}) \cdot^2 \neg S_{i+2, v} \notag\\
&:: \bigwedge^1_{av} \text{AND}(\cap) \bigvee^1_{av} \text{ANY}(\cup) \bigwedge^2_{v} \text{AND}(\cap)\label{shortcutav2}\\
i &= i+3\\
F_{i+3, v, d} &= F_{i+2, v, d} \mathbin{<<} TF2_{i+3, v, d} \\
\diamond_{i}: F_{i+3} &\equiv F_{i}
\end{compactalign}
\end{subequations}
\end{mdframed}
\caption{Awerbuch-Shiloach connected components variant (expressible in \edge{}).}
\label{cascade:as}
\end{cascade}

The ``STARCHECK'' sub-cascade is expressible as follows:

\begin{cascade}[hbtp]
\begin{mdframed}
\begin{subequations}\label{eqn:starcheck}
\begin{compactalign}
\cascadecomment{$\triangleright$ Tensors}
F^{S\equiv|V|, D\equiv|V|} &\rightarrow \text{Boolean, empty=\False}\\
GF^{S\equiv|V|, D\equiv|V|} &\rightarrow \text{Boolean, empty=\False}\\
S^{S\equiv|V|} &\rightarrow \text{Boolean, empty=\False}\\
\cascadecomment{$\triangleright$ Extended Einsum}
T1_{v} &= \True\\
GF_{v, gv} &= F_{v, av} \cdot F_{av, gv} :: \bigwedge_{av} \text{AND}(\cap)\\
T1_{v} &= T_{v} \mathbin{<<} \neg(F_{v, av} \cdot GF_{v, gv: gv \neq av}) :: \bigwedge_{v} \text{AND}(\cap) \bigvee_{gv, av} \text{OR}(\cup)\\
T2_{gv} &= T1_{v} \mathbin{<<} \neg(F_{v, av} \cdot GF_{v, gv: gv \neq av}) :: \bigwedge_{v} \text{AND}(\cap) \bigvee_{v, av} \text{OR}(\cup)\\
S_{v} &= F_{v, av} \cdot T2_{av} :: \bigwedge_{v} \text{AND}(\cap) \bigvee_{av} \text{OR}(\cup)
\end{compactalign}
\end{subequations}
\end{mdframed}
\caption{STARCHECK sub-cascade.}
\label{cascade:starcheck}
\end{cascade}

Again, note how \edge{} stays within the tensor machinery to express iteration while the \graphblas{} formulation steps out of linear algebra primitives to encode iteration as loop nests~\cite[Algorithms 3--5]{Azad:2019:LLA}.
This variant, like Algorithm S, is expressed entirely within the tensor machinery without stepping into a host language.

\subsection{Expressing Computations Beyond Graph Traversal}\label{ssec:beyond-graphs}

Although we have motivated \edge{} through graph algorithms, the extensions we introduced in \S~\ref{ssec:walk-bfs} are general: user-defined compute operators, rank variable expressions, and the map/reduce/populate decomposition apply equally well to computations that have nothing to do with vertices and edges.
This subsection collects three short examples that illustrate this reach.
Each is deliberately small, because the point is not algorithmic depth but rather that a single, consistent notation covers element-wise numerical computation, randomized algorithms, and classical sorting, in addition to the graph algorithms of the previous subsections (\textbf{design goals~\ref{goal:powerful} and~\ref{goal:applications}}).
These examples also foreshadow the broader set of domains in which \edge{} has already been applied, including neural networks, robotics, and iterative linear algebra~\cite{zhu:2026:rsu, Nayak:2023:TDF, Odemuyiwa:2025:FTF, odemuyiwa:2026:MEB, Zhang:2025:transfusion}.

\subsubsection{Element-Wise Exponentiation}\label{ssec:exp}

Suppose we want to combine two tensors through element-wise exponentiation, common in neural networks such as the sigmoid function used in GraphSAGE~\cite{Hamilton:2017:IRL}.

\begin{cascade}[hbtp]
\begin{mdframed}
\begin{subequations}
\begin{compactalign}
\cascadecomment{$\triangleright$ Extended Einsum}
Z_{m, n}  &= A_{m, n} \cdot B_{m, n} :: \bigwedge_{m, n} \text{exp}(\mathbbm{1})\label{eqn:exp}
\end{compactalign}
\end{subequations}
\end{mdframed}
\caption{Element-wise exponentiation ($A^B$).}
\label{cascade:exp}
\end{cascade}

where $\text{exp}$ is a user-defined function that computes $A_{m, n}^{B_{m, n}}$.
In shorthand notation:
\begin{align}
Z_{m, n}  &= A_{m, n}^{B_{m, n}}\label{eqn:expshorthand}
\end{align}

To express $e^B$ as a unary operation:
\begin{align}
Z_{m, n} &= \text{exp}(B_{m, n}),
\end{align}
where $\text{exp}$ applies $e^{(\cdot)}$ to every element in $B$.
We can also write this as $Z_{m,n} = e^{(B_{m,n})}$, viewing $e$ as a scalar that the Einsum automatically broadcasts to a tensor of shape $M \times N$.

\subsubsection{Randomized Vertex Exploration}\label{ssec:rand}

Suppose we want to randomly determine whether to explore a vertex, common in randomized algorithms such as random-mate connected components~\cite{Reif:1985:OPA}.

\begin{cascade}[hbtp]
\begin{mdframed}
\begin{subequations}
\begin{compactalign}
\cascadecomment{$\triangleright$ Extended Einsum}
F_{i+1, d} &= G_{s, d} \cdot F_{i, s} :: \bigwedge_{s} \text{user-rand-func}(\cap) \bigvee_{s} \text{OR}(\cup)
\end{compactalign}
\end{subequations}
\end{mdframed}
\caption{Randomized vertex exploration.}
\label{cascade:rand}
\end{cascade}

If all tensors are Boolean, then $\text{user-rand-func}$ is a user-defined function that, given the vertices that successfully intersected ($\cap$), randomly decides whether to include the corresponding $s$ vertex ($\True$) or skip exploring that vertex's neighbors ($\False$).
To make the expression deterministic under repeated evaluation, the random function may be parameterized by a seed and the current iteration-space coordinates.

\tikzexternalenable
\section{What Next?}\label{sec:5-mappings}

We have described the \edge{} language, which is the gateway to an algebraic framework for thinking about graph algorithms (see Figure~\ref{fig:factored:edge}).
Since \edge{} describes graph algorithms in an algebraic manner, one avenue of future work is exploring various algebraic manipulations both manually (see BFS Example~\ref{bfs:any}) and within an automated system.
From our preliminary explorations, algebraic manipulations enable transformations from one variant of a graph algorithm to another, potentially enabling the discovery of new graph algorithms (see \S~\ref{ssec:rrbfs}).
Manual exploration enables users to see optimizations such as replacing compute operators with more efficient operators (e.g., using \emph{ANY} instead of \emph{OR}). We hope manual exploration will enable a more systematic approach for graph algorithm developers to invent new algorithms.
Automated exploration will enable the generation of different Einsum variations.
Note that this is also useful for tensor algebra applications (\textbf{design goal}~\ref{goal:supports}).

In building this ecosystem, our current work focuses on a programmatic implementation of \edge{} with a corresponding compiler, such that a user can write graph algorithms and tensor algebra applications as \edge{} code.
Ideally, such a programmatic implementation will interface with prior work in the tensor algebra space (see Table~\ref{tab:related}) (\textbf{design goal}~\ref{goal:supports}).
Additionally, future work in this space will focus on expanding these prior tools to support general user-defined operations and to recognize constraints on computations.
There is a rich space of work in discovering new ways to implement algorithms; designing cost models to determine which Einsums and their corresponding mappings are more efficient than others given the properties of input data and backend; and refining both the mapping space and data format space to adequately express the mapping choices present in software (e.g., CUDA dynamic parallelism~\cite{Adinets:2014:APC} or dynamic load-balancing approaches such as work-stealing~\cite{Cederman:2012:AGC}).

Overall, we envision an ecosystem that allows one to succinctly describe and compare different graph algorithms for the same problem in a similar notation; enables an exploration of the possible implementation space (in both hardware and software); and enables algorithm developers to approach inventing new algorithms systematically, rather than ``thinking hard'' about new ways to solve a problem.

Beyond graph algorithms, we have been able to use extended Einsums to express various machine learning models, variations of the Cholesky computation, and various parallel primitives such as the prefix-sum.
Our hope is that the community will take this notation and apply it to their own applications to better understand the algorithmic and implementation space in their domains.

\section{Acknowledgements}

We are indebted to Chris Fletcher, Yan Zhu, Sanjana Mali, Khai Vu, Mingun Cho, and Angshuman Parashar for acting as test users, in-depth discussions on Einsums, tensor algebra, fibertrees, and accelerator modeling. We thank Charles E. Leiserson, Conal Elliott, Ayd{\i}n Bulu\c{c}, Caleb Stanford, Jason Lowe-Power, Muhammad Osama, Matthew Drescher, Sean Treichler, Jonathan Wapman, and Aamer Jaleel for feedback and technical discussions on graph algorithms and their mapping to tensor algebra.

We thank Saman Amarasinghe, Ajay Brahmakshatriya, Willow Ahrens, Teo Collins, Fredrik Kjolstad, Olivia Hsu, Jaeyoon Lee, Pravi Samaratunga, Sabrina Neuman, Courtney Golden, and Yan Gu, for discussions and feedback on \edge{} and related systems.
Finally, this work was supported in part by a Microsoft Research Fellowship, an NVIDIA Graduate Fellowship, and NSF Grant 2403389.

\bibliographystyle{./acmart-primary/ACM-Reference-Format}
\bibliography{temp}


\begin{thebibliography}{108}


\ifx \showCODEN    \undefined \def \showCODEN     #1{\unskip}     \fi
\ifx \showDOI      \undefined \def \showDOI       #1{#1}\fi
\ifx \showISBNx    \undefined \def \showISBNx     #1{\unskip}     \fi
\ifx \showISBNxiii \undefined \def \showISBNxiii  #1{\unskip}     \fi
\ifx \showISSN     \undefined \def \showISSN      #1{\unskip}     \fi
\ifx \showLCCN     \undefined \def \showLCCN      #1{\unskip}     \fi
\ifx \shownote     \undefined \def \shownote      #1{#1}          \fi
\ifx \showarticletitle \undefined \def \showarticletitle #1{#1}   \fi
\ifx \showURL      \undefined \def \showURL       {\relax}        \fi
\providecommand\bibfield[2]{#2}
\providecommand\bibinfo[2]{#2}
\providecommand\natexlab[1]{#1}
\providecommand\showeprint[2][]{arXiv:#2}

\bibitem[Gra(2013)]%
        {Graph500}
 \bibinfo{year}{2013}\natexlab{}.
\newblock \bibinfo{title}{The {G}raph 500 List}.
\newblock \bibinfo{howpublished}{\url{http://www.graph500.org/}}.
\newblock


\bibitem[hif(2023)]%
        {hifiber}
 \bibinfo{year}{2023}\natexlab{}.
\newblock \bibinfo{title}{Fibertree Project}.
\newblock
  \bibinfo{howpublished}{\url{https://github.com/Fibertree-Project/fibertree}}.
\newblock


\bibitem[Adinets(2017)]%
        {Adinets:2014:APC}
\bibfield{author}{\bibinfo{person}{Andy Adinets}.}
  \bibinfo{year}{2017}\natexlab{}.
\newblock \bibinfo{title}{Adaptive Parallel Computation with {CUDA} Dynamic
  Parallelism}.
\newblock \bibinfo{howpublished}{NVIDIA Developer Blog}.
\newblock
\urldef\tempurl%
\url{https://developer.nvidia.com/blog/introduction-cuda-dynamic-parallelism/}
\showURL{%
\tempurl}


\bibitem[Ahrens et~al\mbox{.}(2022)]%
        {Ahrens:2021:AST}
\bibfield{author}{\bibinfo{person}{Willow Ahrens}, \bibinfo{person}{Fredrik
  Kjolstad}, {and} \bibinfo{person}{Saman Amarasinghe}.}
  \bibinfo{year}{2022}\natexlab{}.
\newblock \showarticletitle{Autoscheduling for Sparse Tensor Algebra with an
  Asymptotic Cost Model}. In \bibinfo{booktitle}{\emph{Proceedings of the 43rd
  ACM SIGPLAN International Conference on Programming Language Design and
  Implementation}} \emph{(\bibinfo{series}{PLDI ’22})}.
  \bibinfo{publisher}{ACM}, \bibinfo{pages}{269--285}.
\newblock
\urldef\tempurl%
\url{https://doi.org/10.1145/3519939.3523442}
\showDOI{\tempurl}


\bibitem[Ancourt and Irigoin(1991)]%
        {Ancourt:1991:SPD}
\bibfield{author}{\bibinfo{person}{Corinne Ancourt} {and}
  \bibinfo{person}{Fran{\c{c}}ois Irigoin}.} \bibinfo{year}{1991}\natexlab{}.
\newblock \showarticletitle{Scanning Polyhedra with {DO} Loops}. In
  \bibinfo{booktitle}{\emph{Proceedings of the Third {ACM} {SIGPLAN} Symposium
  on Principles {\&} Practice of Parallel Programming}} (Williamsburg,
  Virginia, USA) \emph{(\bibinfo{series}{PPOPP 1991})},
  \bibfield{editor}{\bibinfo{person}{David~S. Wise}} (Ed.).
  \bibinfo{publisher}{{ACM}}, \bibinfo{pages}{39--50}.
\newblock
\urldef\tempurl%
\url{https://doi.org/10.1145/109625.109631}
\showDOI{\tempurl}


\bibitem[Andrulis et~al\mbox{.}(2026)]%
        {andrulis:2026:faf}
\bibfield{author}{\bibinfo{person}{Tanner Andrulis}, \bibinfo{person}{Michael
  Gilbert}, \bibinfo{person}{Vivienne Sze}, {and} \bibinfo{person}{Joel~S.
  Emer}.} \bibinfo{year}{2026}\natexlab{}.
\newblock \showarticletitle{Fast and Fusiest: An Optimal Fusion-Aware Mapper
  for Accelerator Modeling and Evaluation}.
\newblock \bibinfo{journal}{\emph{CoRR}} (\bibinfo{date}{Feb.}
  \bibinfo{year}{2026}).
\newblock
\showeprint[arxiv]{2602.15166}~[cs.AR]


\bibitem[Angles(2018)]%
        {Angles:2018:PGD}
\bibfield{author}{\bibinfo{person}{Renzo Angles}.}
  \bibinfo{year}{2018}\natexlab{}.
\newblock \showarticletitle{The Property Graph Database Model}. In
  \bibinfo{booktitle}{\emph{Proceedings of the 12th Alberto Mendelzon
  International Workshop on Foundations of Data Management}} (Cali, Colombia)
  \emph{(\bibinfo{series}{{CEUR} Workshop Proceedings},
  Vol.~\bibinfo{volume}{2100})}, \bibfield{editor}{\bibinfo{person}{Dan
  Olteanu} {and} \bibinfo{person}{Barbara Poblete}} (Eds.).
  \bibinfo{publisher}{CEUR-WS.org}.
\newblock
\urldef\tempurl%
\url{https://ceur-ws.org/Vol-2100/paper26.pdf}
\showURL{%
\tempurl}


\bibitem[Awerbuch and Shiloach(1987)]%
        {Awerbuch:1987:NCM}
\bibfield{author}{\bibinfo{person}{Baruch Awerbuch} {and}
  \bibinfo{person}{Yossi Shiloach}.} \bibinfo{year}{1987}\natexlab{}.
\newblock \showarticletitle{New Connectivity and {MSF} Algorithms for
  Shuffle-Exchange Network and {PRAM}}.
\newblock \bibinfo{journal}{\emph{{IEEE} Trans. Computers}}
  \bibinfo{volume}{36}, \bibinfo{number}{10} (\bibinfo{year}{1987}),
  \bibinfo{pages}{1258--1263}.
\newblock
\urldef\tempurl%
\url{https://doi.org/10.1109/TC.1987.1676869}
\showDOI{\tempurl}


\bibitem[Azad and Bulu{\c{c}}(2019)]%
        {Azad:2019:LLA}
\bibfield{author}{\bibinfo{person}{Ariful Azad} {and} \bibinfo{person}{Aydin
  Bulu{\c{c}}}.} \bibinfo{year}{2019}\natexlab{}.
\newblock \showarticletitle{{LACC:} {A} Linear-Algebraic Algorithm for Finding
  Connected Components in Distributed Memory}. In
  \bibinfo{booktitle}{\emph{{IEEE} International Parallel and Distributed
  Processing Symposium}} (Rio de Janeiro, Brazil) \emph{(\bibinfo{series}{IPDPS
  2019})}. \bibinfo{publisher}{{IEEE}}, \bibinfo{pages}{2--12}.
\newblock
\urldef\tempurl%
\url{https://doi.org/10.1109/IPDPS.2019.00012}
\showDOI{\tempurl}


\bibitem[Backus et~al\mbox{.}(1957)]%
        {Backus:1957:FAC}
\bibfield{author}{\bibinfo{person}{John~W. Backus}, \bibinfo{person}{Robert~J.
  Beeber}, \bibinfo{person}{Sheldon Best}, \bibinfo{person}{Richard Goldberg},
  \bibinfo{person}{Lois~M. Haibt}, \bibinfo{person}{Harlan~L. Herrick},
  \bibinfo{person}{Robert~A. Nelson}, \bibinfo{person}{David Sayre},
  \bibinfo{person}{Peter~B. Sheridan}, \bibinfo{person}{H. Stern},
  \bibinfo{person}{Irving Ziller}, \bibinfo{person}{Robert~A. Hughes}, {and}
  \bibinfo{person}{R. Nutt}.} \bibinfo{year}{1957}\natexlab{}.
\newblock \showarticletitle{The {FORTRAN} automatic coding system}. In
  \bibinfo{booktitle}{\emph{Papers presented at the 1957 Western Joint Computer
  Conference: Techniques for Reliability}} (Los Angeles, California, USA)
  \emph{(\bibinfo{series}{{IRE-AIEE-ACM} 1957 (Western)})},
  \bibfield{editor}{\bibinfo{person}{Morton~M. Astrahan}} (Ed.).
  \bibinfo{publisher}{{ACM}}, \bibinfo{pages}{188--198}.
\newblock
\urldef\tempurl%
\url{https://doi.org/10.1145/1455567.1455599}
\showDOI{\tempurl}


\bibitem[Barab\'{a}si et~al\mbox{.}(2011)]%
        {Gulbahce:2011:NMN}
\bibfield{author}{\bibinfo{person}{Albert~L\'{a}szl\'{o} Barab\'{a}si},
  \bibinfo{person}{Natali Gulbahce}, {and} \bibinfo{person}{Joseph Loscalzo}.}
  \bibinfo{year}{2011}\natexlab{}.
\newblock \showarticletitle{Network medicine: A network-based approach to human
  disease}.
\newblock \bibinfo{journal}{\emph{Nature Reviews Genetics}}
  \bibinfo{volume}{12} (\bibinfo{date}{1} \bibinfo{year}{2011}),
  \bibinfo{pages}{56--68}.
\newblock
Issue 1.
\showISSN{14710056}
\urldef\tempurl%
\url{https://doi.org/10.1038/nrg2918}
\showDOI{\tempurl}


\bibitem[Baskaran et~al\mbox{.}(2012)]%
        {muthu:2012:esc}
\bibfield{author}{\bibinfo{person}{Muthu Baskaran}, \bibinfo{person}{Benoît
  Meister}, \bibinfo{person}{Nicolas Vasilache}, {and} \bibinfo{person}{Richard
  Lethin}.} \bibinfo{year}{2012}\natexlab{}.
\newblock \showarticletitle{Efficient and scalable computations with sparse
  tensors}. In \bibinfo{booktitle}{\emph{2012 IEEE Conference on High
  Performance Extreme Computing}} (Waltham, MA, USA)
  \emph{(\bibinfo{series}{HPEC '12})}. \bibinfo{numpages}{6}~pages.
\newblock
\urldef\tempurl%
\url{https://doi.org/10.1109/HPEC.2012.6408676}
\showDOI{\tempurl}


\bibitem[Beamer et~al\mbox{.}(2012)]%
        {Beamer:2012:DBS}
\bibfield{author}{\bibinfo{person}{Scott Beamer}, \bibinfo{person}{Krste
  Asanovi\'{c}}, {and} \bibinfo{person}{David Patterson}.}
  \bibinfo{year}{2012}\natexlab{}.
\newblock \showarticletitle{Direction-Optimizing Breadth-First Search}. In
  \bibinfo{booktitle}{\emph{Proceedings of the International Conference on High
  Performance Computing, Networking, Storage and Analysis}}
  \emph{(\bibinfo{series}{SC '12})}. Article \bibinfo{articleno}{12},
  \bibinfo{numpages}{10}~pages.
\newblock
\urldef\tempurl%
\url{https://doi.org/10.1109/SC.2012.50}
\showDOI{\tempurl}


\bibitem[Blondel et~al\mbox{.}(2008)]%
        {Blondel:2008:FUC}
\bibfield{author}{\bibinfo{person}{Vincent~D Blondel},
  \bibinfo{person}{Jean-Loup Guillaume}, \bibinfo{person}{Renaud Lambiotte},
  {and} \bibinfo{person}{Etienne Lefebvre}.} \bibinfo{year}{2008}\natexlab{}.
\newblock \showarticletitle{Fast unfolding of communities in large networks}.
\newblock \bibinfo{journal}{\emph{Journal of Statistical Mechanics: Theory and
  Experiment}} \bibinfo{volume}{2008}, \bibinfo{number}{10}
  (\bibinfo{date}{Oct.} \bibinfo{year}{2008}), \bibinfo{pages}{P10008}.
\newblock
\showISSN{1742-5468}
\urldef\tempurl%
\url{https://doi.org/10.1088/1742-5468/2008/10/p10008}
\showDOI{\tempurl}


\bibitem[Boisvert et~al\mbox{.}(1997)]%
        {Boisvert:1996:MMW}
\bibfield{author}{\bibinfo{person}{Ronald~F. Boisvert}, \bibinfo{person}{Roldan
  Pozo}, \bibinfo{person}{Karin Remington}, \bibinfo{person}{Richard~F.
  Barrett}, {and} \bibinfo{person}{Jack~J. Dongarra}.}
  \bibinfo{year}{1997}\natexlab{}.
\newblock \showarticletitle{{M}atrix {M}arket: a web resource for test matrix
  collections}. In \bibinfo{booktitle}{\emph{IFIP Advances in Information and
  Communication Technology}}. \bibinfo{publisher}{Springer US},
  \bibinfo{pages}{125--137}.
\newblock
\showISBNx{9781504129404}
\showISSN{1868-422X}
\urldef\tempurl%
\url{https://doi.org/10.1007/978-1-5041-2940-4_9}
\showDOI{\tempurl}


\bibitem[Boulet and Feautrier(1998)]%
        {Boulet:1998:SPD}
\bibfield{author}{\bibinfo{person}{Pierre Boulet} {and} \bibinfo{person}{Paul
  Feautrier}.} \bibinfo{year}{1998}\natexlab{}.
\newblock \showarticletitle{Scanning Polyhedra without Do-loops}. In
  \bibinfo{booktitle}{\emph{Proceedings of the 1998 International Conference on
  Parallel Architectures and Compilation Techniques}} (Paris, France).
  \bibinfo{publisher}{{IEEE} Computer Society}, \bibinfo{pages}{4--11}.
\newblock
\urldef\tempurl%
\url{https://doi.org/10.1109/PACT.1998.727127}
\showDOI{\tempurl}


\bibitem[Brahmakshatriya et~al\mbox{.}(2021)]%
        {Brahmakshatriya:2021:CGA}
\bibfield{author}{\bibinfo{person}{Ajay Brahmakshatriya},
  \bibinfo{person}{Yunming Zhang}, \bibinfo{person}{Changwan Hong},
  \bibinfo{person}{Shoaib Kamil}, \bibinfo{person}{Julian Shun}, {and}
  \bibinfo{person}{Saman Amarasinghe}.} \bibinfo{year}{2021}\natexlab{}.
\newblock \showarticletitle{Compiling Graph Applications for {GPU}s with
  {G}raph{I}t}. In \bibinfo{booktitle}{\emph{IEEE/ACM International Symposium
  on Code Generation and Optimization}} \emph{(\bibinfo{series}{CGO 2021})}.
  \bibinfo{pages}{248--261}.
\newblock
\urldef\tempurl%
\url{https://doi.org/10.1109/CGO51591.2021.9370321}
\showDOI{\tempurl}


\bibitem[Bulu\c{c} et~al\mbox{.}(2017)]%
        {Buluc:2017:TGC}
\bibfield{author}{\bibinfo{person}{Ayd{\i}n Bulu\c{c}},
  \bibinfo{person}{Timothy Mattson}, \bibinfo{person}{Scott McMillan},
  \bibinfo{person}{Jose Moreira}, {and} \bibinfo{person}{Carl Yang}.}
  \bibinfo{year}{2017}\natexlab{}.
\newblock \bibinfo{booktitle}{\emph{The {GraphBLAS C API} Specification}}.
\newblock
\urldef\tempurl%
\url{http://graphblas.org/index.php/C\_language\_API}
\showURL{%
\tempurl}
\newblock
\shownote{Rev. 1.1.}.


\bibitem[Cederman and Tsigas(2012)]%
        {Cederman:2012:AGC}
\bibfield{author}{\bibinfo{person}{Daniel Cederman} {and}
  \bibinfo{person}{Philippas Tsigas}.} \bibinfo{year}{2012}\natexlab{}.
\newblock \showarticletitle{Dynamic Load Balancing Using Work-Stealing}.
\newblock In \bibinfo{booktitle}{\emph{{GPU} Computing Gems Jade Edition}},
  \bibfield{editor}{\bibinfo{person}{Wen mei W.~Hwu}} (Ed.).
  \bibinfo{publisher}{Morgan Kaufmann}, \bibinfo{address}{Boston, MA, USA},
  Chapter~35, \bibinfo{pages}{485--499}.
\newblock
\showISBNx{978-0-12-385963-1}
\urldef\tempurl%
\url{https://doi.org/10.1016/B978-0-12-385963-1.00035-6}
\showDOI{\tempurl}


\bibitem[Che et~al\mbox{.}(2009)]%
        {Che:2009:RAB}
\bibfield{author}{\bibinfo{person}{Shuai Che}, \bibinfo{person}{Michael Boyer},
  \bibinfo{person}{Jiayuan Meng}, \bibinfo{person}{David Tarjan},
  \bibinfo{person}{Jeremy~W. Sheaffer}, \bibinfo{person}{Sang-Ha Lee}, {and}
  \bibinfo{person}{Kevin Skadron}.} \bibinfo{year}{2009}\natexlab{}.
\newblock \showarticletitle{Rodinia: A Benchmark Suite for Heterogeneous
  Computing}. In \bibinfo{booktitle}{\emph{IEEE International Symposium on
  Workload Characterization}} \emph{(\bibinfo{series}{IISWC 2009})}.
  \bibinfo{pages}{44--54}.
\newblock
\urldef\tempurl%
\url{https://doi.org/10.1109/IISWC.2009.5306797}
\showDOI{\tempurl}


\bibitem[Chou et~al\mbox{.}(2018)]%
        {Chou:2018:FAF}
\bibfield{author}{\bibinfo{person}{Stephen Chou}, \bibinfo{person}{Fredrik
  Kjolstad}, {and} \bibinfo{person}{Saman Amarasinghe}.}
  \bibinfo{year}{2018}\natexlab{}.
\newblock \showarticletitle{Format Abstraction for Sparse Tensor Algebra
  Compilers}.
\newblock \bibinfo{journal}{\emph{Proceedings of the ACM on Programming
  Languages}} \bibinfo{volume}{2}, \bibinfo{number}{OOPSLA}
  (\bibinfo{date}{Oct.} \bibinfo{year}{2018}), \bibinfo{pages}{1--30}.
\newblock
\showISSN{2475-1421}
\urldef\tempurl%
\url{https://doi.org/10.1145/3276493}
\showDOI{\tempurl}


\bibitem[Chung(1997)]%
        {Chung:1997:SGT}
\bibfield{author}{\bibinfo{person}{Fan R.~K. Chung}.}
  \bibinfo{year}{1997}\natexlab{}.
\newblock \bibinfo{booktitle}{\emph{Spectral Graph Theory}}.
  Vol.~\bibinfo{volume}{92}.
\newblock \bibinfo{publisher}{American Mathematical Society}.
\newblock
\showISBNx{978-1-4704-2452-7}
\urldef\tempurl%
\url{https://doi.org/10.1090/cbms/092}
\showDOI{\tempurl}


\bibitem[Coimbra et~al\mbox{.}(2021)]%
        {Coimbra:2021:AGP}
\bibfield{author}{\bibinfo{person}{Miguel~E. Coimbra},
  \bibinfo{person}{Alexandre~P. Francisco}, {and} \bibinfo{person}{Lu{\'{\i}}s
  Veiga}.} \bibinfo{year}{2021}\natexlab{}.
\newblock \showarticletitle{An analysis of the graph processing landscape}.
\newblock \bibinfo{journal}{\emph{Journal of Big Data}} \bibinfo{volume}{8},
  \bibinfo{number}{1} (\bibinfo{year}{2021}), \bibinfo{pages}{55}.
\newblock
\urldef\tempurl%
\url{https://doi.org/10.1186/s40537-021-00443-9}
\showDOI{\tempurl}


\bibitem[Cooper and Torczon(2023)]%
        {Cooper:2023:IR}
\bibfield{author}{\bibinfo{person}{Keith~D. Cooper} {and}
  \bibinfo{person}{Linda Torczon}.} \bibinfo{year}{2023}\natexlab{}.
\newblock \showarticletitle{Chapter 4 - Intermediate Representations}.
\newblock In \bibinfo{booktitle}{\emph{Engineering a Compiler (Third Edition)}
  (\bibinfo{edition}{third edition} ed.)},
  \bibfield{editor}{\bibinfo{person}{Keith~D. Cooper} {and}
  \bibinfo{person}{Linda Torczon}} (Eds.). \bibinfo{publisher}{Morgan
  Kaufmann}, \bibinfo{address}{Philadelphia}, \bibinfo{pages}{159--207}.
\newblock
\showISBNx{978-0-12-815412-0}
\urldef\tempurl%
\url{https://doi.org/10.1016/B978-0-12-815412-0.00010-3}
\showDOI{\tempurl}


\bibitem[Cormen et~al\mbox{.}(2022)]%
        {CLRS:2022:ITA}
\bibfield{author}{\bibinfo{person}{Thomas~H. Cormen},
  \bibinfo{person}{Charles~E. Leiserson}, \bibinfo{person}{Ronald~L. Rivest},
  {and} \bibinfo{person}{Clifford Stein}.} \bibinfo{year}{2022}\natexlab{}.
\newblock \bibinfo{booktitle}{\emph{Introduction to Algorithms, 4th Edition}}.
\newblock \bibinfo{publisher}{{MIT} Press}.
\newblock
\showISBNx{978-0-26-236750-9}
\urldef\tempurl%
\url{https://mitpress.mit.edu/9780262367509}
\showURL{%
\tempurl}


\bibitem[Corneil et~al\mbox{.}(2003)]%
        {corneil:2003:OtP}
\bibfield{author}{\bibinfo{person}{Derek~G. Corneil},
  \bibinfo{person}{Feodor~F. Dragan}, {and} \bibinfo{person}{Ekkehard
  K{\"{o}}hler}.} \bibinfo{year}{2003}\natexlab{}.
\newblock \showarticletitle{On the power of {BFS} to determine a graph's
  diameter}.
\newblock \bibinfo{journal}{\emph{Networks}} \bibinfo{volume}{42},
  \bibinfo{number}{4} (\bibinfo{year}{2003}), \bibinfo{pages}{209--222}.
\newblock
\urldef\tempurl%
\url{https://doi.org/10.1002/net.10098}
\showDOI{\tempurl}


\bibitem[Davis(2019)]%
        {Davis:2019:ASG}
\bibfield{author}{\bibinfo{person}{Timothy~A. Davis}.}
  \bibinfo{year}{2019}\natexlab{}.
\newblock \showarticletitle{Algorithm 1000: {SuiteSparse}: {GraphBLAS}: Graph
  Algorithms in the Language of Sparse Linear Algebra}.
\newblock \bibinfo{journal}{\emph{{ACM} Trans. Math. Softw.}}
  \bibinfo{volume}{45}, \bibinfo{number}{4} (\bibinfo{date}{Dec.}
  \bibinfo{year}{2019}), \bibinfo{pages}{44:1--44:25}.
\newblock
\urldef\tempurl%
\url{https://doi.org/10.1145/3322125}
\showDOI{\tempurl}


\bibitem[Dima(2023)]%
        {dima:2023:gstaco}
\bibfield{author}{\bibinfo{person}{Alexandra Dima}.}
  \bibinfo{year}{2023}\natexlab{}.
\newblock \emph{\bibinfo{title}{{GSTACO}: A Generalized Sparse Tensor Algebra
  Compiler}}.
\newblock M.Eng. Thesis. \bibinfo{school}{Massachusetts Institute of
  Technology}, \bibinfo{address}{Cambridge, MA}.
\newblock
\urldef\tempurl%
\url{https://commit.csail.mit.edu/papers/2023/alexandra-meng-thesis.pdf}
\showURL{%
\tempurl}


\bibitem[Dong et~al\mbox{.}(2021)]%
        {Dong:2021:ESA}
\bibfield{author}{\bibinfo{person}{Xiaojun Dong}, \bibinfo{person}{Yan Gu},
  \bibinfo{person}{Yihan Sun}, {and} \bibinfo{person}{Yunming Zhang}.}
  \bibinfo{year}{2021}\natexlab{}.
\newblock \showarticletitle{Efficient Stepping Algorithms and Implementations
  for Parallel Shortest Paths}. In \bibinfo{booktitle}{\emph{33rd {ACM}
  Symposium on Parallelism in Algorithms and Architectures}} (Virtual Event,
  USA) \emph{(\bibinfo{series}{{SPAA} '21})},
  \bibfield{editor}{\bibinfo{person}{Kunal Agrawal} {and}
  \bibinfo{person}{Yossi Azar}} (Eds.). \bibinfo{publisher}{{ACM}},
  \bibinfo{pages}{184--197}.
\newblock
\urldef\tempurl%
\url{https://doi.org/10.1145/3409964.3461782}
\showDOI{\tempurl}


\bibitem[Du et~al\mbox{.}(2023)]%
        {Du:2023:MMB}
\bibfield{author}{\bibinfo{person}{Zhihui Du}, \bibinfo{person}{Oliver~Alvarado
  Rodriguez}, \bibinfo{person}{Fuhuan Li}, \bibinfo{person}{Mohammad Dindoost},
  {and} \bibinfo{person}{David Bader}.} \bibinfo{year}{2023}\natexlab{}.
\newblock \showarticletitle{Minimum-Mapping based Connected Components
  Algorithm}. The 10th Annual Chapel Implementers and Users Workshop (CHIUW).
\newblock
\urldef\tempurl%
\url{https://chapel-lang.org/CHIUW/2023/Du.pdf}
\showURL{%
\tempurl}


\bibitem[Edmonds and Karp(1972)]%
        {edmonds:1972:TIA}
\bibfield{author}{\bibinfo{person}{Jack Edmonds} {and}
  \bibinfo{person}{Richard~M. Karp}.} \bibinfo{year}{1972}\natexlab{}.
\newblock \showarticletitle{Theoretical Improvements in Algorithmic Efficiency
  for Network Flow Problems}.
\newblock \bibinfo{journal}{\emph{Journal of the {ACM}}} \bibinfo{volume}{19},
  \bibinfo{number}{2} (\bibinfo{year}{1972}), \bibinfo{pages}{248--264}.
\newblock
\urldef\tempurl%
\url{https://doi.org/10.1145/321694.321699}
\showDOI{\tempurl}


\bibitem[Gaihre et~al\mbox{.}(2022)]%
        {Gaihre:2022:SSS}
\bibfield{author}{\bibinfo{person}{Anil Gaihre}, \bibinfo{person}{Xiaoye~Sherry
  Li}, {and} \bibinfo{person}{Hang Liu}.} \bibinfo{year}{2022}\natexlab{}.
\newblock \showarticletitle{{gSoFa}: Scalable Sparse Symbolic {LU}
  Factorization on {GPU}s}.
\newblock \bibinfo{journal}{\emph{{IEEE} Transactions on Parallel and
  Distributed Systems}} \bibinfo{volume}{33}, \bibinfo{number}{4}
  (\bibinfo{year}{2022}), \bibinfo{pages}{1015--1026}.
\newblock
\urldef\tempurl%
\url{https://doi.org/10.1109/TPDS.2021.3090316}
\showDOI{\tempurl}


\bibitem[Gilbert et~al\mbox{.}(2024)]%
        {gilbert:2024:LEF}
\bibfield{author}{\bibinfo{person}{Michael Gilbert},
  \bibinfo{person}{Yannan~Nellie Wu}, \bibinfo{person}{Joel~S. Emer}, {and}
  \bibinfo{person}{Vivienne Sze}.} \bibinfo{year}{2024}\natexlab{}.
\newblock \showarticletitle{{LoopTree}: Exploring the Fused-Layer Dataflow
  Accelerator Design Space}.
\newblock \bibinfo{journal}{\emph{IEEE Transactions on Circuits and Systems for
  Artificial Intelligence}} \bibinfo{volume}{1}, \bibinfo{number}{1}
  (\bibinfo{date}{Sept.} \bibinfo{year}{2024}), \bibinfo{pages}{97--111}.
\newblock
\showISSN{2996-6647}
\urldef\tempurl%
\url{https://doi.org/10.1109/tcasai.2024.3461716}
\showDOI{\tempurl}


\bibitem[Gilbert et~al\mbox{.}(2023)]%
        {Gilbert:2023:LEE}
\bibfield{author}{\bibinfo{person}{Michael Gilbert},
  \bibinfo{person}{Yannan~Nellie Wu}, \bibinfo{person}{Angshuman Parashar},
  \bibinfo{person}{Vivienne Sze}, {and} \bibinfo{person}{Joel~S. Emer}.}
  \bibinfo{year}{2023}\natexlab{}.
\newblock \showarticletitle{{LoopTree}: Enabling Exploration of Fused-layer
  Dataflow Accelerators}.
\newblock \bibinfo{journal}{\emph{2023 IEEE International Symposium on
  Performance Analysis of Systems and Software (ISPASS)}},
  \bibinfo{pages}{316--318}.
\newblock
\urldef\tempurl%
\url{https://doi.org/10.1109/ispass57527.2023.00038}
\showDOI{\tempurl}


\bibitem[Golden et~al\mbox{.}(2025)]%
        {Golden:2025:QRD}
\bibfield{author}{\bibinfo{person}{Courtney Golden}, \bibinfo{person}{Axel
  Feldmann}, \bibinfo{person}{Joel Emer}, {and} \bibinfo{person}{Daniel
  Sanchez}.} \bibinfo{year}{2025}\natexlab{}.
\newblock \showarticletitle{Quartz: A Reconfigurable, Distributed-Memory
  Accelerator for Sparse Applications}. In
  \bibinfo{booktitle}{\emph{Proceedings of the 58th IEEE/ACM International
  Symposium on Microarchitecture}} (Seoul, South Korea)
  \emph{(\bibinfo{series}{MICRO '25})}. \bibinfo{publisher}{Association for
  Computing Machinery}, \bibinfo{address}{New York, NY, USA},
  \bibinfo{pages}{929–943}.
\newblock
\showISBNx{9798400715730}
\urldef\tempurl%
\url{https://doi.org/10.1145/3725843.3756035}
\showDOI{\tempurl}


\bibitem[Golden(2025)]%
        {golden:2025:RDM}
\bibfield{author}{\bibinfo{person}{Courtney~K Golden}.}
  \bibinfo{year}{2025}\natexlab{}.
\newblock \emph{\bibinfo{title}{A Reconfigurable, Distributed-Memory
  Accelerator for Sparse Applications}}.
\newblock \bibinfo{thesistype}{Master's\ thesis}.
  \bibinfo{school}{Massachusetts Institute of Technology}.
\newblock


\bibitem[Green(2021)]%
        {Green:2021:IDB}
\bibfield{author}{\bibinfo{person}{Oded Green}.}
  \bibinfo{year}{2021}\natexlab{}.
\newblock \showarticletitle{Inverse-Deletion {BFS} - Revisiting Static Graph
  {BFS} Traversals with Dynamic Graph Operations}. In
  \bibinfo{booktitle}{\emph{{IEEE} High Performance Extreme Computing
  Conference}} (Waltham, MA) \emph{(\bibinfo{series}{HPEC 2021})}.
  \bibinfo{publisher}{{IEEE}}, \bibinfo{pages}{1--7}.
\newblock
\urldef\tempurl%
\url{https://doi.org/10.1109/HPEC49654.2021.9622864}
\showDOI{\tempurl}


\bibitem[Guo and Qiu(2022)]%
        {Guo:2022:LRT}
\bibfield{author}{\bibinfo{person}{Wei Guo} {and} \bibinfo{person}{Jing{-}Mei
  Qiu}.} \bibinfo{year}{2022}\natexlab{}.
\newblock \showarticletitle{A low rank tensor representation of linear
  transport and nonlinear Vlasov solutions and their associated flow maps}.
\newblock \bibinfo{journal}{\emph{J. Comput. Phys.}}  \bibinfo{volume}{458}
  (\bibinfo{year}{2022}), \bibinfo{pages}{111089}.
\newblock
\urldef\tempurl%
\url{https://doi.org/10.1016/j.jcp.2022.111089}
\showDOI{\tempurl}


\bibitem[Hamilton et~al\mbox{.}(2017)]%
        {Hamilton:2017:IRL}
\bibfield{author}{\bibinfo{person}{William~L. Hamilton},
  \bibinfo{person}{Zhitao Ying}, {and} \bibinfo{person}{Jure Leskovec}.}
  \bibinfo{year}{2017}\natexlab{}.
\newblock \showarticletitle{Inductive Representation Learning on Large Graphs}.
  In \bibinfo{booktitle}{\emph{Advances in Neural Information Processing
  Systems}} (Long Beach, CA, USA) \emph{(\bibinfo{series}{NIPS 2017},
  Vol.~\bibinfo{volume}{30})}, \bibfield{editor}{\bibinfo{person}{Isabelle
  Guyon}, \bibinfo{person}{Ulrike von Luxburg}, \bibinfo{person}{Samy Bengio},
  \bibinfo{person}{Hanna~M. Wallach}, \bibinfo{person}{Rob Fergus},
  \bibinfo{person}{S.~V.~N. Vishwanathan}, {and} \bibinfo{person}{Roman
  Garnett}} (Eds.). \bibinfo{pages}{1024--1034}.
\newblock
\urldef\tempurl%
\url{https://proceedings.neurips.cc/paper_files/paper/2017/file/5dd9db5e033da9c6fb5ba83c7a7ebea9-Paper.pdf}
\showURL{%
\tempurl}


\bibitem[Harary and Gupta(1997)]%
        {Harary:1997:DGM}
\bibfield{author}{\bibinfo{person}{F. Harary} {and} \bibinfo{person}{G.
  Gupta}.} \bibinfo{year}{1997}\natexlab{}.
\newblock \showarticletitle{Dynamic Graph Models}.
\newblock \bibinfo{journal}{\emph{Math. Comput. Model.}} \bibinfo{volume}{25},
  \bibinfo{number}{7} (\bibinfo{date}{April} \bibinfo{year}{1997}),
  \bibinfo{pages}{79--87}.
\newblock
\showISSN{0895-7177}
\urldef\tempurl%
\url{https://doi.org/10.1016/S0895-7177(97)00050-2}
\showDOI{\tempurl}


\bibitem[Harris et~al\mbox{.}(2020)]%
        {Harris:2020:APN}
\bibfield{author}{\bibinfo{person}{Charles~R. Harris},
  \bibinfo{person}{K.~Jarrod Millman}, \bibinfo{person}{St{\'{e}}fan~J. van~der
  Walt}, \bibinfo{person}{Ralf Gommers}, \bibinfo{person}{Pauli Virtanen},
  \bibinfo{person}{David Cournapeau}, \bibinfo{person}{Eric Wieser},
  \bibinfo{person}{Julian Taylor}, \bibinfo{person}{Sebastian Berg},
  \bibinfo{person}{Nathaniel~J. Smith}, \bibinfo{person}{Robert Kern},
  \bibinfo{person}{Matti Picus}, \bibinfo{person}{Stephan Hoyer},
  \bibinfo{person}{Marten~H. van Kerkwijk}, \bibinfo{person}{Matthew Brett},
  \bibinfo{person}{Allan Haldane}, \bibinfo{person}{Jaime~Fern{\'{a}}ndez del
  R{\'{i}}o}, \bibinfo{person}{Mark Wiebe}, \bibinfo{person}{Pearu Peterson},
  \bibinfo{person}{Pierre G{\'{e}}rard-Marchant}, \bibinfo{person}{Kevin
  Sheppard}, \bibinfo{person}{Tyler Reddy}, \bibinfo{person}{Warren Weckesser},
  \bibinfo{person}{Hameer Abbasi}, \bibinfo{person}{Christoph Gohlke}, {and}
  \bibinfo{person}{Travis~E. Oliphant}.} \bibinfo{year}{2020}\natexlab{}.
\newblock \showarticletitle{Array programming with {NumPy}}.
\newblock \bibinfo{journal}{\emph{Nature}} \bibinfo{volume}{585},
  \bibinfo{number}{7825} (\bibinfo{date}{Sept.} \bibinfo{year}{2020}),
  \bibinfo{pages}{357--362}.
\newblock
\urldef\tempurl%
\url{https://doi.org/10.1038/s41586-020-2649-2}
\showDOI{\tempurl}


\bibitem[Hegde et~al\mbox{.}(2019)]%
        {Hegde:2019:EAA}
\bibfield{author}{\bibinfo{person}{Kartik Hegde}, \bibinfo{person}{Hadi~Asghari
  Moghaddam}, \bibinfo{person}{Michael Pellauer}, \bibinfo{person}{Neal~Clayton
  Crago}, \bibinfo{person}{Aamer Jaleel}, \bibinfo{person}{Edgar Solomonik},
  \bibinfo{person}{Joel~S. Emer}, {and} \bibinfo{person}{Christopher~W.
  Fletcher}.} \bibinfo{year}{2019}\natexlab{}.
\newblock \showarticletitle{{ExTensor}: An Accelerator for Sparse Tensor
  Algebra}. In \bibinfo{booktitle}{\emph{Proceedings of the 52nd Annual
  {IEEE/ACM} International Symposium on Microarchitecture}}
  \emph{(\bibinfo{series}{MICRO-52})}. \bibinfo{pages}{319--333}.
\newblock
\urldef\tempurl%
\url{https://doi.org/10.1145/3352460.3358275}
\showDOI{\tempurl}


\bibitem[Hegde et~al\mbox{.}(2021)]%
        {Hegde:2021:MME}
\bibfield{author}{\bibinfo{person}{Kartik Hegde}, \bibinfo{person}{Po{-}An
  Tsai}, \bibinfo{person}{Sitao Huang}, \bibinfo{person}{Vikas Chandra},
  \bibinfo{person}{Angshuman Parashar}, {and} \bibinfo{person}{Christopher~W.
  Fletcher}.} \bibinfo{year}{2021}\natexlab{}.
\newblock \showarticletitle{Mind mappings: enabling efficient
  algorithm-accelerator mapping space search}. In
  \bibinfo{booktitle}{\emph{26th {ACM} International Conference on
  Architectural Support for Programming Languages and Operating Systems}}
  (Virtual Event, USA) \emph{(\bibinfo{series}{{ASPLOS} 2021})},
  \bibfield{editor}{\bibinfo{person}{Tim Sherwood}, \bibinfo{person}{Emery~D.
  Berger}, {and} \bibinfo{person}{Christos Kozyrakis}} (Eds.).
  \bibinfo{publisher}{{ACM}}, \bibinfo{pages}{943--958}.
\newblock
\urldef\tempurl%
\url{https://doi.org/10.1145/3445814.3446762}
\showDOI{\tempurl}


\bibitem[Henry et~al\mbox{.}(2021)]%
        {Henry:2021:CSA}
\bibfield{author}{\bibinfo{person}{Rawn Henry}, \bibinfo{person}{Olivia Hsu},
  \bibinfo{person}{Rohan Yadav}, \bibinfo{person}{Stephen Chou},
  \bibinfo{person}{Kunle Olukotun}, \bibinfo{person}{Saman~P. Amarasinghe},
  {and} \bibinfo{person}{Fredrik Kjolstad}.} \bibinfo{year}{2021}\natexlab{}.
\newblock \showarticletitle{Compilation of sparse array programming models}.
\newblock \bibinfo{journal}{\emph{Proc. {ACM} Program. Lang.}}
  \bibinfo{volume}{5}, \bibinfo{number}{{OOPSLA}} (\bibinfo{year}{2021}),
  \bibinfo{pages}{1--29}.
\newblock
\urldef\tempurl%
\url{https://doi.org/10.1145/3485505}
\showDOI{\tempurl}


\bibitem[Hong et~al\mbox{.}(2023)]%
        {Hong:2023:DDM}
\bibfield{author}{\bibinfo{person}{Charles Hong}, \bibinfo{person}{Qijing
  Huang}, \bibinfo{person}{Grace Dinh}, \bibinfo{person}{Mahesh Subedar}, {and}
  \bibinfo{person}{Yakun~Sophia Shao}.} \bibinfo{year}{2023}\natexlab{}.
\newblock \showarticletitle{{DOSA}: Differentiable Model-Based One-Loop Search
  for {DNN} Accelerators}. In \bibinfo{booktitle}{\emph{56th Annual IEEE/ACM
  International Symposium on Microarchitecture}} (Toronto, ON)
  \emph{(\bibinfo{series}{MICRO ’23})}. \bibinfo{publisher}{IEEE},
  \bibinfo{pages}{209--224}.
\newblock
\urldef\tempurl%
\url{https://doi.org/10.1145/3613424.3623797}
\showDOI{\tempurl}


\bibitem[Hopcroft and Tarjan(1973)]%
        {Hopcroft:1973:AEA}
\bibfield{author}{\bibinfo{person}{John~E. Hopcroft} {and}
  \bibinfo{person}{Robert~Endre Tarjan}.} \bibinfo{year}{1973}\natexlab{}.
\newblock \showarticletitle{Algorithm 447: efficient algorithms for graph
  manipulation}.
\newblock \bibinfo{journal}{\emph{Commun. ACM}} \bibinfo{volume}{16},
  \bibinfo{number}{6} (\bibinfo{year}{1973}), \bibinfo{pages}{372--378}.
\newblock
\urldef\tempurl%
\url{https://doi.org/10.1145/362248.362272}
\showDOI{\tempurl}


\bibitem[Hsu et~al\mbox{.}(2023)]%
        {Hsu:2023:SAM}
\bibfield{author}{\bibinfo{person}{Olivia Hsu}, \bibinfo{person}{Maxwell
  Strange}, \bibinfo{person}{Ritvik Sharma}, \bibinfo{person}{Jaeyeon Won},
  \bibinfo{person}{Kunle Olukotun}, \bibinfo{person}{Joel~S. Emer},
  \bibinfo{person}{Mark~A. Horowitz}, {and} \bibinfo{person}{Fredrik
  Kj{\o}lstad}.} \bibinfo{year}{2023}\natexlab{}.
\newblock \showarticletitle{The Sparse Abstract Machine}. In
  \bibinfo{booktitle}{\emph{Proceedings of the 28th ACM International
  Conference on Architectural Support for Programming Languages and Operating
  Systems}} (Vancouver, BC, Canada) \emph{(\bibinfo{series}{ASPLOS '23},
  Vol.~\bibinfo{volume}{3})}. \bibinfo{pages}{710--726}.
\newblock
\urldef\tempurl%
\url{https://doi.org/10.1145/3582016.3582051}
\showDOI{\tempurl}


\bibitem[Huang et~al\mbox{.}(2021)]%
        {Huang:2021:CSC}
\bibfield{author}{\bibinfo{person}{Qijing Huang}, \bibinfo{person}{Aravind
  Kalaiah}, \bibinfo{person}{Minwoo Kang}, \bibinfo{person}{James Demmel},
  \bibinfo{person}{Grace Dinh}, \bibinfo{person}{John Wawrzynek},
  \bibinfo{person}{Thomas Norell}, {and} \bibinfo{person}{Yakun~Sophia Shao}.}
  \bibinfo{year}{2021}\natexlab{}.
\newblock \showarticletitle{{CoSA}: Scheduling by Constrained Optimization for
  Spatial Accelerators}. In \bibinfo{booktitle}{\emph{48th {ACM/IEEE} Annual
  International Symposium on Computer Architecture}} (Valencia, Spain)
  \emph{(\bibinfo{series}{{ISCA} 2021})}. \bibinfo{publisher}{{IEEE}},
  \bibinfo{pages}{554--566}.
\newblock
\urldef\tempurl%
\url{https://doi.org/10.1109/ISCA52012.2021.00050}
\showDOI{\tempurl}


\bibitem[Huang et~al\mbox{.}(2024)]%
        {huang:2024:MTG}
\bibfield{author}{\bibinfo{person}{Qijing Huang}, \bibinfo{person}{Po-An Tsai},
  \bibinfo{person}{Joel~S. Emer}, {and} \bibinfo{person}{Angshuman Parashar}.}
  \bibinfo{year}{2024}\natexlab{}.
\newblock \showarticletitle{Mind the Gap: Attainable Data Movement and
  Operational Intensity Bounds for Tensor Algorithms}. In
  \bibinfo{booktitle}{\emph{2024 ACM/IEEE 51st Annual International Symposium
  on Computer Architecture (ISCA)}} (Buenos Aires, Argentina)
  \emph{(\bibinfo{series}{ISCA'24})}. \bibinfo{pages}{150--166}.
\newblock
\urldef\tempurl%
\url{https://doi.org/10.1109/ISCA59077.2024.00021}
\showDOI{\tempurl}


\bibitem[Kao and Krishna(2020)]%
        {Kao:2020:GAH}
\bibfield{author}{\bibinfo{person}{Sheng{-}Chun Kao} {and}
  \bibinfo{person}{Tushar Krishna}.} \bibinfo{year}{2020}\natexlab{}.
\newblock \showarticletitle{{GAMMA:} Automating the {HW} Mapping of {DNN}
  Models on Accelerators via Genetic Algorithm}. In
  \bibinfo{booktitle}{\emph{{IEEE/ACM} International Conference On Computer
  Aided Design}} (San Diego, CA, USA) \emph{(\bibinfo{series}{ICCAD 2020})}.
  \bibinfo{publisher}{{IEEE}}, \bibinfo{pages}{44:1--44:9}.
\newblock
\urldef\tempurl%
\url{https://doi.org/10.1145/3400302.3415639}
\showDOI{\tempurl}


\bibitem[Kempe et~al\mbox{.}(2003)]%
        {Kempe:2003:MSI}
\bibfield{author}{\bibinfo{person}{David Kempe}, \bibinfo{person}{Jon
  Kleinberg}, {and} \bibinfo{person}{\'{E}va Tardos}.}
  \bibinfo{year}{2003}\natexlab{}.
\newblock \showarticletitle{Maximizing the spread of influence through a social
  network}. In \bibinfo{booktitle}{\emph{Proceedings of the Ninth ACM SIGKDD
  International Conference on Knowledge Discovery and Data Mining}}
  (Washington, D.C.) \emph{(\bibinfo{series}{KDD '03})}.
  \bibinfo{publisher}{Association for Computing Machinery},
  \bibinfo{address}{New York, NY, USA}, \bibinfo{pages}{137--146}.
\newblock
\showISBNx{1581137370}
\urldef\tempurl%
\url{https://doi.org/10.1145/956750.956769}
\showDOI{\tempurl}


\bibitem[Kepner et~al\mbox{.}(2016)]%
        {Kepner:2016:MFO}
\bibfield{author}{\bibinfo{person}{Jeremy Kepner}, \bibinfo{person}{Peter
  Aaltonen}, \bibinfo{person}{David Bader}, \bibinfo{person}{Ayd{\i}n
  Bulu\c{c}}, \bibinfo{person}{Franz Franchetti}, \bibinfo{person}{John
  Gilbert}, \bibinfo{person}{Dylan Hutchison}, \bibinfo{person}{Manoj Kumar},
  \bibinfo{person}{Andrew Lumsdaine}, \bibinfo{person}{Henning Meyerhenke},
  \bibinfo{person}{Scott McMillan}, \bibinfo{person}{Jose Moreira},
  \bibinfo{person}{John~D. Owens}, \bibinfo{person}{Carl Yang},
  \bibinfo{person}{Marcin Zalewski}, {and} \bibinfo{person}{Timothy Mattson}.}
  \bibinfo{year}{2016}\natexlab{}.
\newblock \showarticletitle{Mathematical Foundations of the {GraphBLAS}}. In
  \bibinfo{booktitle}{\emph{Proceedings of the IEEE High Performance Extreme
  Computing Conference}}.
\newblock
\urldef\tempurl%
\url{https://doi.org/10.1109/HPEC.2016.7761646}
\showDOI{\tempurl}


\bibitem[Kepner and Gilbert(2011)]%
        {Kepner:2011:GAL}
\bibfield{editor}{\bibinfo{person}{Jeremy Kepner} {and}
  \bibinfo{person}{John~R. Gilbert}} (Eds.). \bibinfo{year}{2011}\natexlab{}.
\newblock \bibinfo{booktitle}{\emph{Graph Algorithms in the Language of Linear
  Algebra}}. \bibinfo{series}{Software, environments, tools},
  Vol.~\bibinfo{volume}{22}.
\newblock \bibinfo{publisher}{{SIAM}}.
\newblock
\showISBNx{978-0-89871-990-1}
\urldef\tempurl%
\url{https://doi.org/10.1137/1.9780898719918}
\showDOI{\tempurl}


\bibitem[Kjolstad et~al\mbox{.}(2017)]%
        {Kjolstad:2017:TTA}
\bibfield{author}{\bibinfo{person}{Fredrik Kjolstad}, \bibinfo{person}{Shoaib
  Kamil}, \bibinfo{person}{Stephen Chou}, \bibinfo{person}{David Lugato}, {and}
  \bibinfo{person}{Saman Amarasinghe}.} \bibinfo{year}{2017}\natexlab{}.
\newblock \showarticletitle{The tensor algebra compiler}.
\newblock \bibinfo{journal}{\emph{Proceedings of the ACM on Programming
  Languages}} \bibinfo{volume}{1}, \bibinfo{number}{OOPSLA}
  (\bibinfo{date}{Oct.} \bibinfo{year}{2017}), \bibinfo{pages}{1--29}.
\newblock
\showISSN{2475-1421}
\urldef\tempurl%
\url{https://doi.org/10.1145/3133901}
\showDOI{\tempurl}


\bibitem[Kjolstad(2020)]%
        {Kjolstad:2020:STA}
\bibfield{author}{\bibinfo{person}{Fredrik~Berg Kjolstad}.}
  \bibinfo{year}{2020}\natexlab{}.
\newblock \emph{\bibinfo{title}{Sparse Tensor Algebra Compilation}}.
\newblock PhD thesis. \bibinfo{school}{Massachusetts Institute of Technology},
  \bibinfo{address}{Cambridge, MA}.
\newblock
\newblock
\shownote{Available at \url{https://dspace.mit.edu/handle/1721.1/128314}}.


\bibitem[Kwak et~al\mbox{.}(2010)]%
        {Kwak:2010:WTS}
\bibfield{author}{\bibinfo{person}{Haewoon Kwak}, \bibinfo{person}{Changhyun
  Lee}, \bibinfo{person}{Hosung Park}, {and} \bibinfo{person}{Sue~B. Moon}.}
  \bibinfo{year}{2010}\natexlab{}.
\newblock \showarticletitle{What is Twitter, a social network or a news
  media?}. In \bibinfo{booktitle}{\emph{Proceedings of the 19th International
  Conference on World Wide Web, {WWW} 2010, Raleigh, North Carolina, USA, April
  26-30, 2010}}. \bibinfo{publisher}{{ACM}}, \bibinfo{pages}{591--600}.
\newblock
\urldef\tempurl%
\url{https://doi.org/10.1145/1772690.1772751}
\showDOI{\tempurl}


\bibitem[Kwon et~al\mbox{.}(2019)]%
        {Kwon:2019:URP}
\bibfield{author}{\bibinfo{person}{Hyoukjun Kwon}, \bibinfo{person}{Prasanth
  Chatarasi}, \bibinfo{person}{Michael Pellauer}, \bibinfo{person}{Angshuman
  Parashar}, \bibinfo{person}{Vivek Sarkar}, {and} \bibinfo{person}{Tushar
  Krishna}.} \bibinfo{year}{2019}\natexlab{}.
\newblock \showarticletitle{Understanding Reuse, Performance, and Hardware Cost
  of {DNN} Dataflow: {A} Data-Centric Approach}. In
  \bibinfo{booktitle}{\emph{Proceedings of the 52nd Annual {IEEE/ACM}
  International Symposium on Microarchitecture}} (Columbus, OH, USA)
  \emph{(\bibinfo{series}{MICRO 2019})}. \bibinfo{publisher}{{ACM}},
  \bibinfo{pages}{754--768}.
\newblock
\urldef\tempurl%
\url{https://doi.org/10.1145/3352460.3358252}
\showDOI{\tempurl}


\bibitem[Lamport(1974)]%
        {Lamport:1974:PED}
\bibfield{author}{\bibinfo{person}{Leslie Lamport}.}
  \bibinfo{year}{1974}\natexlab{}.
\newblock \showarticletitle{The Parallel Execution of {DO} Loops}.
\newblock \bibinfo{journal}{\emph{Commun. ACM}} \bibinfo{volume}{17},
  \bibinfo{number}{2} (\bibinfo{date}{Feb.} \bibinfo{year}{1974}),
  \bibinfo{pages}{83--93}.
\newblock
\urldef\tempurl%
\url{https://doi.org/10.1145/360827.360844}
\showDOI{\tempurl}


\bibitem[Laue et~al\mbox{.}(2020)]%
        {Laue:2020:SET}
\bibfield{author}{\bibinfo{person}{Sören Laue}, \bibinfo{person}{Matthias
  Mitterreiter}, {and} \bibinfo{person}{Joachim Giesen}.}
  \bibinfo{year}{2020}\natexlab{}.
\newblock \showarticletitle{A Simple and Efficient Tensor Calculus}.
\newblock \bibinfo{journal}{\emph{Proceedings of the AAAI Conference on
  Artificial Intelligence}} \bibinfo{volume}{34}, \bibinfo{number}{04}
  (\bibinfo{date}{April} \bibinfo{year}{2020}), \bibinfo{pages}{4527--4534}.
\newblock
\urldef\tempurl%
\url{https://doi.org/10.1609/aaai.v34i04.5881}
\showDOI{\tempurl}


\bibitem[Leskovec and Faloutsos(2006)]%
        {Leskovec:2006:SLG}
\bibfield{author}{\bibinfo{person}{Jure Leskovec} {and}
  \bibinfo{person}{Christos Faloutsos}.} \bibinfo{year}{2006}\natexlab{}.
\newblock \showarticletitle{Sampling from large graphs}. In
  \bibinfo{booktitle}{\emph{Proceedings of the Twelfth {ACM} {SIGKDD}
  International Conference on Knowledge Discovery and Data Mining}}
  (Philadelphia, PA, USA) \emph{(\bibinfo{series}{KDD'06})},
  \bibfield{editor}{\bibinfo{person}{Tina Eliassi{-}Rad},
  \bibinfo{person}{Lyle~H. Ungar}, \bibinfo{person}{Mark Craven}, {and}
  \bibinfo{person}{Dimitrios Gunopulos}} (Eds.). \bibinfo{publisher}{{ACM}},
  \bibinfo{pages}{631--636}.
\newblock
\urldef\tempurl%
\url{https://doi.org/10.1145/1150402.1150479}
\showDOI{\tempurl}


\bibitem[Liu and Tarjan(2022)]%
        {Liu:2022:SCC}
\bibfield{author}{\bibinfo{person}{Sixue~Cliff Liu} {and}
  \bibinfo{person}{Robert~Endre Tarjan}.} \bibinfo{year}{2022}\natexlab{}.
\newblock \showarticletitle{Simple Concurrent Connected Components Algorithms}.
\newblock \bibinfo{journal}{\emph{{ACM} Trans. Parallel Comput.}}
  \bibinfo{volume}{9}, \bibinfo{number}{2} (\bibinfo{year}{2022}),
  \bibinfo{pages}{9:1--9:26}.
\newblock
\urldef\tempurl%
\url{https://doi.org/10.1145/3543546}
\showDOI{\tempurl}


\bibitem[Malewicz et~al\mbox{.}(2010)]%
        {Malewicz:2010:PAS}
\bibfield{author}{\bibinfo{person}{Grzegorz Malewicz},
  \bibinfo{person}{Matthew~H. Austern}, \bibinfo{person}{Aart J.~C. Bik},
  \bibinfo{person}{James~C. Dehnert}, \bibinfo{person}{Ilan Horn},
  \bibinfo{person}{Naty Leiser}, {and} \bibinfo{person}{Grzegorz Czajkowski}.}
  \bibinfo{year}{2010}\natexlab{}.
\newblock \showarticletitle{Pregel: A System for Large-scale Graph Processing}.
  In \bibinfo{booktitle}{\emph{Proceedings of the 2010 ACM SIGMOD International
  Conference on Management of Data}} (Indianapolis, Indiana, USA)
  \emph{(\bibinfo{series}{SIGMOD '10})}. \bibinfo{pages}{135--146}.
\newblock
\urldef\tempurl%
\url{https://doi.org/10.1145/1807167.1807184}
\showDOI{\tempurl}


\bibitem[McCune et~al\mbox{.}(2015)]%
        {McCune:2015:TLA}
\bibfield{author}{\bibinfo{person}{Robert~Ryan McCune}, \bibinfo{person}{Tim
  Weninger}, {and} \bibinfo{person}{Greg Madey}.}
  \bibinfo{year}{2015}\natexlab{}.
\newblock \showarticletitle{Thinking Like a Vertex: A Survey of Vertex-Centric
  Frameworks for Large-Scale Distributed Graph Processing}.
\newblock \bibinfo{journal}{\emph{ACM Comput. Surv.}} \bibinfo{volume}{48},
  \bibinfo{number}{2}, Article \bibinfo{articleno}{25} (\bibinfo{date}{Oct.}
  \bibinfo{year}{2015}), \bibinfo{numpages}{39}~pages.
\newblock
\showISSN{0360-0300}
\urldef\tempurl%
\url{https://doi.org/10.1145/2818185}
\showDOI{\tempurl}


\bibitem[Merrill et~al\mbox{.}(2012)]%
        {Merrill:2012:SGG}
\bibfield{author}{\bibinfo{person}{Duane Merrill}, \bibinfo{person}{Michael
  Garland}, {and} \bibinfo{person}{Andrew Grimshaw}.}
  \bibinfo{year}{2012}\natexlab{}.
\newblock \showarticletitle{Scalable {GPU} Graph Traversal}. In
  \bibinfo{booktitle}{\emph{Proceedings of the 17th ACM SIGPLAN Symposium on
  Principles and Practice of Parallel Programming}} (New Orleans, Louisiana,
  USA) \emph{(\bibinfo{series}{PPoPP '12})}. \bibinfo{pages}{117--128}.
\newblock
\urldef\tempurl%
\url{https://doi.org/10.1145/2145816.2145832}
\showDOI{\tempurl}


\bibitem[Moore(1959)]%
        {Moore:1959:SPFA}
\bibfield{author}{\bibinfo{person}{Edward~F Moore}.}
  \bibinfo{year}{1959}\natexlab{}.
\newblock \showarticletitle{The shortest path through a maze}. In
  \bibinfo{booktitle}{\emph{Proc. of the International Symposium on the Theory
  of Switching}}. Harvard University Press, \bibinfo{pages}{285--292}.
\newblock


\bibitem[Mu{\~{n}}oz{-}Mart{\'{\i}}nez et~al\mbox{.}(2023)]%
        {Martinez:2023:FMD}
\bibfield{author}{\bibinfo{person}{Francisco Mu{\~{n}}oz{-}Mart{\'{\i}}nez},
  \bibinfo{person}{Raveesh Garg}, \bibinfo{person}{Michael Pellauer},
  \bibinfo{person}{Jos{\'{e}}~L. Abell{\'{a}}n}, \bibinfo{person}{Manuel~E.
  Acacio}, {and} \bibinfo{person}{Tushar Krishna}.}
  \bibinfo{year}{2023}\natexlab{}.
\newblock \showarticletitle{Flexagon: {A} Multi-dataflow Sparse-Sparse Matrix
  Multiplication Accelerator for Efficient {DNN} Processing}. In
  \bibinfo{booktitle}{\emph{Proceedings of the 28th {ACM} International
  Conference on Architectural Support for Programming Languages and Operating
  Systems}} (Vancouver, BC, Canada) \emph{(\bibinfo{series}{ASPLOS 2023})},
  \bibfield{editor}{\bibinfo{person}{Tor~M. Aamodt}, \bibinfo{person}{Natalie
  D.~Enright Jerger}, {and} \bibinfo{person}{Michael~M. Swift}} (Eds.).
  \bibinfo{publisher}{{ACM}}, \bibinfo{pages}{252--265}.
\newblock
\urldef\tempurl%
\url{https://doi.org/10.1145/3582016.3582069}
\showDOI{\tempurl}


\bibitem[Nai et~al\mbox{.}(2015)]%
        {Nai:2015:GUG}
\bibfield{author}{\bibinfo{person}{Lifeng Nai}, \bibinfo{person}{Yinglong Xia},
  \bibinfo{person}{Ilie~G. Tanase}, \bibinfo{person}{Hyesoon Kim}, {and}
  \bibinfo{person}{Ching-Yung Lin}.} \bibinfo{year}{2015}\natexlab{}.
\newblock \showarticletitle{Graph{BIG}: understanding graph computing in the
  context of industrial solutions}. In \bibinfo{booktitle}{\emph{Proceedings of
  the International Conference for High Performance Computing, Networking,
  Storage and Analysis}} \emph{(\bibinfo{series}{SC'15})}.
  \bibinfo{pages}{1--12}.
\newblock
\urldef\tempurl%
\url{https://doi.org/10.1145/2807591.2807626}
\showDOI{\tempurl}


\bibitem[Nayak et~al\mbox{.}(2023)]%
        {Nayak:2023:TDF}
\bibfield{author}{\bibinfo{person}{Nandeeka Nayak},
  \bibinfo{person}{Toluwanimi~O. Odemuyiwa}, \bibinfo{person}{Shubham Ugare},
  \bibinfo{person}{Christopher Fletcher}, \bibinfo{person}{Michael Pellauer},
  {and} \bibinfo{person}{Joel Emer}.} \bibinfo{year}{2023}\natexlab{}.
\newblock \showarticletitle{{TeAAL}: {A} Declarative Framework for Modeling
  Sparse Tensor Accelerators}. In \bibinfo{booktitle}{\emph{56th Annual
  IEEE/ACM International Symposium on Microarchitecture}}
  \emph{(\bibinfo{series}{MICRO '23})}. \bibinfo{publisher}{ACM},
  \bibinfo{pages}{1255--1270}.
\newblock
\urldef\tempurl%
\url{https://doi.org/10.1145/3613424.3623791}
\showDOI{\tempurl}


\bibitem[Nayak et~al\mbox{.}(2024a)]%
        {odemuyiwa:2024:TeAALTutorial}
\bibfield{author}{\bibinfo{person}{Nandeeka Nayak},
  \bibinfo{person}{Toluwanimi~O. Odemuyiwa}, \bibinfo{person}{Yingchen Wang},
  \bibinfo{person}{Joel~S. Emer}, \bibinfo{person}{Michael Pellauer}, {and}
  \bibinfo{person}{Christopher~W. Fletcher}.} \bibinfo{year}{2024}\natexlab{a}.
\newblock \bibinfo{title}{{TeAAL} Tutorial at MICRO 2024}.
\newblock
  \bibinfo{howpublished}{\url{https://teaal.csail.mit.edu/tutorials/2024.micro-teaal/index.html}}.
\newblock


\bibitem[Nayak et~al\mbox{.}(2025a)]%
        {Odemuyiwa:2025:FTF}
\bibfield{author}{\bibinfo{person}{Nandeeka Nayak},
  \bibinfo{person}{Toluwanimi~O. Odemuyiwa}, \bibinfo{person}{Xinrui Wu},
  \bibinfo{person}{Michael Pellauer}, \bibinfo{person}{Joel~S. Emer}, {and}
  \bibinfo{person}{Christopher~W. Fletcher}.} \bibinfo{year}{2025}\natexlab{a}.
\newblock \showarticletitle{From {TeAAL} to {FuseMax}: Separation of Concerns
  for Attention Accelerator Design}.
\newblock \bibinfo{journal}{\emph{{IEEE} Micro}} \bibinfo{volume}{45},
  \bibinfo{number}{4} (\bibinfo{year}{2025}), \bibinfo{pages}{44--53}.
\newblock
\urldef\tempurl%
\url{https://doi.org/10.1109/MM.2025.3589955}
\showDOI{\tempurl}


\bibitem[Nayak et~al\mbox{.}(2025b)]%
        {odemuyiwa:2025:TeAALTutorial}
\bibfield{author}{\bibinfo{person}{Nandeeka Nayak},
  \bibinfo{person}{Toluwanimi~O. Odemuyiwa}, \bibinfo{person}{Yan Zhu},
  \bibinfo{person}{Michael Pellauer}, \bibinfo{person}{Christopher~W.
  Fletcher}, {and} \bibinfo{person}{Joel~S. Emer}.}
  \bibinfo{year}{2025}\natexlab{b}.
\newblock \bibinfo{title}{{TeAAL} Tutorial at MICRO 2025}.
\newblock
  \bibinfo{howpublished}{\url{https://teaal.csail.mit.edu/tutorials/2025.micro-teaal/index.html}}.
\newblock


\bibitem[Nayak et~al\mbox{.}(2024b)]%
        {Nayak:2024:FML_micro}
\bibfield{author}{\bibinfo{person}{Nandeeka Nayak}, \bibinfo{person}{Xinrui
  Wu}, \bibinfo{person}{Toluwanimi~O. Odemuyiwa}, \bibinfo{person}{Michael
  Pellauer}, \bibinfo{person}{Joel Emer}, {and} \bibinfo{person}{Christopher~W.
  Fletcher}.} \bibinfo{year}{2024}\natexlab{b}.
\newblock \showarticletitle{{FuseMax}: Leveraging Extended Einsums to Optimize
  Attention Accelerator Design}.
\newblock \bibinfo{journal}{\emph{{IEEE/ACM} International Symposium on
  Microarchitecture}} (\bibinfo{date}{Nov.} \bibinfo{year}{2024}).
\newblock
\urldef\tempurl%
\url{https://doi.org/10.1145/3613424.3623791}
\showDOI{\tempurl}


\bibitem[Odemuyiwa et~al\mbox{.}(2023)]%
        {Odemuyiwa:2023:ASD}
\bibfield{author}{\bibinfo{person}{Toluwanimi~O. Odemuyiwa},
  \bibinfo{person}{Hadi Asghari-Moghaddam}, \bibinfo{person}{Michael Pellauer},
  \bibinfo{person}{Kartik Hegde}, \bibinfo{person}{Po-An Tsai},
  \bibinfo{person}{Neal Crago}, \bibinfo{person}{Aamer Jaleel},
  \bibinfo{person}{John~D. Owens}, \bibinfo{person}{Edgar Solomonik},
  \bibinfo{person}{Joel Emer}, {and} \bibinfo{person}{Christopher Fletcher}.}
  \bibinfo{year}{2023}\natexlab{}.
\newblock \showarticletitle{Accelerating Sparse Data Orchestration via Dynamic
  Reflexive Tiling}. In \bibinfo{booktitle}{\emph{Proceedings of the 28th ACM
  International Conference on Architectural Support for Programming Languages
  and Operating Systems}} (Vancouver, BC, Canada)
  \emph{(\bibinfo{series}{ASPLOS '23}, Vol.~\bibinfo{volume}{3})}.
  \bibinfo{pages}{18--32}.
\newblock
\urldef\tempurl%
\url{https://doi.org/10.1145/3582016.3582064}
\showDOI{\tempurl}


\bibitem[Odemuyiwa et~al\mbox{.}(2026)]%
        {odemuyiwa:2026:MEB}
\bibfield{author}{\bibinfo{person}{Toluwanimi~O. Odemuyiwa},
  \bibinfo{person}{John~D. Owens}, \bibinfo{person}{Joel~S. Emer}, {and}
  \bibinfo{person}{Michael Pellauer}.} \bibinfo{year}{2026}\natexlab{}.
\newblock \showarticletitle{{Mambalaya: {E}insum-Based Fusion Optimizations on
  State-Space Models}}.
\newblock \bibinfo{journal}{\emph{(Under Review)}} (\bibinfo{year}{2026}).
\newblock


\bibitem[Osama et~al\mbox{.}(2022)]%
        {Osama:2022:EOP}
\bibfield{author}{\bibinfo{person}{Muhammad Osama}, \bibinfo{person}{Serban~D.
  Porumbescu}, {and} \bibinfo{person}{John~D. Owens}.}
  \bibinfo{year}{2022}\natexlab{}.
\newblock \showarticletitle{Essentials of Parallel Graph Analytics}. In
  \bibinfo{booktitle}{\emph{Proceedings of the Workshop on Graphs,
  Architectures, Programming, and Learning}} \emph{(\bibinfo{series}{GrAPL
  2022})}. \bibinfo{pages}{314--317}.
\newblock
\urldef\tempurl%
\url{https://doi.org/10.1109/IPDPSW55747.2022.00061}
\showDOI{\tempurl}


\bibitem[Osama et~al\mbox{.}(2023)]%
        {Osama:2023:PMG}
\bibfield{author}{\bibinfo{person}{Muhammad Osama}, \bibinfo{person}{Serban~D.
  Porumbescu}, {and} \bibinfo{person}{John~D. Owens}.}
  \bibinfo{year}{2023}\natexlab{}.
\newblock \showarticletitle{A Programming Model for {GPU} Load Balancing}. In
  \bibinfo{booktitle}{\emph{Proceedings of the 28th ACM SIGPLAN Symposium on
  Principles and Practice of Parallel Programming}}
  \emph{(\bibinfo{series}{PPoPP 2023})}. \bibinfo{pages}{79--91}.
\newblock
\urldef\tempurl%
\url{https://doi.org/10.1145/3572848.3577434}
\showDOI{\tempurl}


\bibitem[Parashar et~al\mbox{.}(2019)]%
        {Parashar:2019:TSA}
\bibfield{author}{\bibinfo{person}{Angshuman Parashar},
  \bibinfo{person}{Priyanka Raina}, \bibinfo{person}{Yakun~Sophia Shao},
  \bibinfo{person}{Yu-Hsin Chen}, \bibinfo{person}{Victor~A. Ying},
  \bibinfo{person}{Anurag Mukkara}, \bibinfo{person}{Rangharajan Venkatesan},
  \bibinfo{person}{Brucek Khailany}, \bibinfo{person}{Stephen~W. Keckler},
  {and} \bibinfo{person}{Joel Emer}.} \bibinfo{year}{2019}\natexlab{}.
\newblock \showarticletitle{Timeloop: A Systematic Approach to {DNN}
  Accelerator Evaluation}. In \bibinfo{booktitle}{\emph{2019 IEEE International
  Symposium on Performance Analysis of Systems and Software (ISPASS)}}.
  \bibinfo{pages}{304--315}.
\newblock
\urldef\tempurl%
\url{https://doi.org/10.1109/ISPASS.2019.00042}
\showDOI{\tempurl}


\bibitem[Paszke et~al\mbox{.}(2019)]%
        {Paszke:2019:PAI}
\bibfield{author}{\bibinfo{person}{Adam Paszke}, \bibinfo{person}{S. Gross},
  \bibinfo{person}{Francisco Massa}, \bibinfo{person}{A. Lerer},
  \bibinfo{person}{James Bradbury}, \bibinfo{person}{Gregory Chanan},
  \bibinfo{person}{Trevor Killeen}, \bibinfo{person}{Z. Lin},
  \bibinfo{person}{N. Gimelshein}, \bibinfo{person}{L. Antiga},
  \bibinfo{person}{Alban Desmaison}, \bibinfo{person}{Andreas K{\"o}pf},
  \bibinfo{person}{Edward Yang}, \bibinfo{person}{Zach DeVito},
  \bibinfo{person}{Martin Raison}, \bibinfo{person}{Alykhan Tejani},
  \bibinfo{person}{Sasank Chilamkurthy}, \bibinfo{person}{Benoit Steiner},
  \bibinfo{person}{Lu Fang}, \bibinfo{person}{Junjie Bai}, {and}
  \bibinfo{person}{Soumith Chintala}.} \bibinfo{year}{2019}\natexlab{}.
\newblock \showarticletitle{{PyTorch}: An Imperative Style, High-Performance
  Deep Learning Library}. In \bibinfo{booktitle}{\emph{Advances in Neural
  Information Processing Systems 33}} \emph{(\bibinfo{series}{NeurIPS 2019})}.
  \bibinfo{pages}{8026--8037}.
\newblock
\urldef\tempurl%
\url{https://doi.org/10.5555/3454287.3455008}
\showDOI{\tempurl}


\bibitem[Pellauer et~al\mbox{.}(2019)]%
        {Pellauer:2019:BEC}
\bibfield{author}{\bibinfo{person}{Michael Pellauer},
  \bibinfo{person}{Yakun~Sophia Shao}, \bibinfo{person}{Jason Clemons},
  \bibinfo{person}{Neal Crago}, \bibinfo{person}{Kartik Hegde},
  \bibinfo{person}{Rangharajan Venkatesan}, \bibinfo{person}{Stephen~W.
  Keckler}, \bibinfo{person}{Christopher~W. Fletcher}, {and}
  \bibinfo{person}{Joel Emer}.} \bibinfo{year}{2019}\natexlab{}.
\newblock \showarticletitle{Buffets: An Efficient and Composable Storage Idiom
  for Explicit Decoupled Data Orchestration}. In
  \bibinfo{booktitle}{\emph{International Conference on Architectural Support
  for Programming Languages and Operating Systems}}. \bibinfo{publisher}{ACM},
  \bibinfo{pages}{137--151}.
\newblock
\showISBNx{9781450362405}
\urldef\tempurl%
\url{https://doi.org/10.1145/3297858.3304025}
\showDOI{\tempurl}


\bibitem[Pouchet et~al\mbox{.}(2011)]%
        {Pouchet:2011:LTC}
\bibfield{author}{\bibinfo{person}{Louis{-}No{\"{e}}l Pouchet},
  \bibinfo{person}{Uday Bondhugula}, \bibinfo{person}{C{\'{e}}dric Bastoul},
  \bibinfo{person}{Albert Cohen}, \bibinfo{person}{J. Ramanujam},
  \bibinfo{person}{P. Sadayappan}, {and} \bibinfo{person}{Nicolas Vasilache}.}
  \bibinfo{year}{2011}\natexlab{}.
\newblock \showarticletitle{Loop transformations: convexity, pruning and
  optimization}. In \bibinfo{booktitle}{\emph{Proceedings of the 38th {ACM}
  {SIGPLAN-SIGACT} Symposium on Principles of Programming Languages}} (Austin,
  TX, USA) \emph{(\bibinfo{series}{POPL 2011})},
  \bibfield{editor}{\bibinfo{person}{Thomas Ball} {and} \bibinfo{person}{Mooly
  Sagiv}} (Eds.). \bibinfo{publisher}{{ACM}}, \bibinfo{pages}{549--562}.
\newblock
\urldef\tempurl%
\url{https://doi.org/10.1145/1926385.1926449}
\showDOI{\tempurl}


\bibitem[Pourhabibi et~al\mbox{.}(2020)]%
        {Pourhabibi:2020:FDS}
\bibfield{author}{\bibinfo{person}{Tahereh Pourhabibi},
  \bibinfo{person}{Kok{-}Leong Ong}, \bibinfo{person}{Booi Kam}, {and}
  \bibinfo{person}{Yee~Ling Boo}.} \bibinfo{year}{2020}\natexlab{}.
\newblock \showarticletitle{Fraud detection: {A} systematic literature review
  of graph-based anomaly detection approaches}.
\newblock \bibinfo{journal}{\emph{Decision Support Systems}}
  \bibinfo{volume}{133} (\bibinfo{year}{2020}), \bibinfo{pages}{113303}.
\newblock
\urldef\tempurl%
\url{https://doi.org/10.1016/J.DSS.2020.113303}
\showDOI{\tempurl}


\bibitem[Purohit et~al\mbox{.}(2021)]%
        {Purohit:2021:SPG}
\bibfield{author}{\bibinfo{person}{Sumit Purohit}, \bibinfo{person}{Nhuy Van},
  {and} \bibinfo{person}{George Chin}.} \bibinfo{year}{2021}\natexlab{}.
\newblock \showarticletitle{Semantic Property Graph for Scalable Knowledge
  Graph Analytics}. In \bibinfo{booktitle}{\emph{2021 {IEEE} International
  Conference on Big Data}} (Orlando, FL, USA) \emph{(\bibinfo{series}{Big
  Data})}, \bibfield{editor}{\bibinfo{person}{Yixin Chen},
  \bibinfo{person}{Heiko Ludwig}, \bibinfo{person}{Yicheng Tu},
  \bibinfo{person}{Usama~M. Fayyad}, \bibinfo{person}{Xingquan Zhu},
  \bibinfo{person}{Xiaohua Hu}, \bibinfo{person}{Suren Byna},
  \bibinfo{person}{Xiong Liu}, \bibinfo{person}{Jianping Zhang},
  \bibinfo{person}{Shirui Pan}, \bibinfo{person}{Vagelis Papalexakis},
  \bibinfo{person}{Jianwu Wang}, \bibinfo{person}{Alfredo Cuzzocrea}, {and}
  \bibinfo{person}{Carlos Ordonez}} (Eds.). \bibinfo{publisher}{{IEEE}},
  \bibinfo{pages}{2672--2677}.
\newblock
\urldef\tempurl%
\url{https://doi.org/10.1109/BigData52589.2021.9671547}
\showDOI{\tempurl}


\bibitem[Ragan-Kelley et~al\mbox{.}(2013)]%
        {Ragan-Kelley:2013:HAL}
\bibfield{author}{\bibinfo{person}{Jonathan Ragan-Kelley},
  \bibinfo{person}{Connelly Barnes}, \bibinfo{person}{Andrew Adams},
  \bibinfo{person}{Sylvain Paris}, \bibinfo{person}{Fr\'{e}do Durand}, {and}
  \bibinfo{person}{Saman Amarasinghe}.} \bibinfo{year}{2013}\natexlab{}.
\newblock \showarticletitle{Halide: A Language and Compiler for Optimizing
  Parallelism, Locality, and Recomputation in Image Processing Pipelines}. In
  \bibinfo{booktitle}{\emph{Proceedings of the 34th ACM SIGPLAN Conference on
  Programming Language Design and Implementation}} (Seattle, Washington, USA)
  \emph{(\bibinfo{series}{PLDI '13})}. \bibinfo{pages}{519--530}.
\newblock
\showISBNx{9781450320146}
\urldef\tempurl%
\url{https://doi.org/10.1145/2491956.2462176}
\showDOI{\tempurl}


\bibitem[Reif(1985)]%
        {Reif:1985:OPA}
\bibfield{author}{\bibinfo{person}{John~H. Reif}.}
  \bibinfo{year}{1985}\natexlab{}.
\newblock \bibinfo{booktitle}{\emph{Optimal parallel algorithms for integer
  sorting and graph connectivity}}.
\newblock \bibinfo{type}{Technical Report TR-08-85}.
  \bibinfo{institution}{Harvard University}, \bibinfo{address}{Cambridge, MA}.
\newblock


\bibitem[Roy et~al\mbox{.}(2013)]%
        {Roy:2013:XEC}
\bibfield{author}{\bibinfo{person}{Amitabha Roy}, \bibinfo{person}{Ivo
  Mihailovic}, {and} \bibinfo{person}{Willy Zwaenepoel}.}
  \bibinfo{year}{2013}\natexlab{}.
\newblock \showarticletitle{{X}-{S}tream: {E}dge-centric Graph Processing using
  Streaming Partitions}. In \bibinfo{booktitle}{\emph{Proceedings of the
  Twenty-Fourth ACM Symposium on Operating Systems Principles}} (Farmington,
  PA, USA) \emph{(\bibinfo{series}{SOSP '13})},
  \bibfield{editor}{\bibinfo{person}{Michael Kaminsky} {and}
  \bibinfo{person}{Mike Dahlin}} (Eds.). \bibinfo{publisher}{{ACM}},
  \bibinfo{pages}{472--488}.
\newblock
\urldef\tempurl%
\url{https://doi.org/10.1145/2517349.2522740}
\showDOI{\tempurl}


\bibitem[Schmidt(1997)]%
        {schmidt:1997:DSM}
\bibfield{author}{\bibinfo{person}{David~A. Schmidt}.}
  \bibinfo{year}{1997}\natexlab{}.
\newblock \bibinfo{booktitle}{\emph{Denotational Semantics: A Methodology for
  Language Development}}.
\newblock \bibinfo{publisher}{Kansas State University}.
\newblock
\urldef\tempurl%
\url{https://people.cs.ksu.edu/~schmidt/text/DenSem-full-book.pdf}
\showURL{%
\tempurl}
\newblock
\shownote{Available as a full book PDF}.


\bibitem[Sha et~al\mbox{.}(2019)]%
        {Sha:2019:GBG}
\bibfield{author}{\bibinfo{person}{Mo Sha}, \bibinfo{person}{Yuchen Li}, {and}
  \bibinfo{person}{Kian{-}Lee Tan}.} \bibinfo{year}{2019}\natexlab{}.
\newblock \showarticletitle{{GPU}-based Graph Traversal on Compressed Graphs}.
  In \bibinfo{booktitle}{\emph{Proceedings of the 2019 International Conference
  on Management of Data}} (Amsterdam, The Netherlands)
  \emph{(\bibinfo{series}{SIGMOD Conference 2019})},
  \bibfield{editor}{\bibinfo{person}{Peter~A. Boncz}, \bibinfo{person}{Stefan
  Manegold}, \bibinfo{person}{Anastasia Ailamaki}, \bibinfo{person}{Amol
  Deshpande}, {and} \bibinfo{person}{Tim Kraska}} (Eds.).
  \bibinfo{publisher}{{ACM}}, \bibinfo{pages}{775--792}.
\newblock
\urldef\tempurl%
\url{https://doi.org/10.1145/3299869.3319871}
\showDOI{\tempurl}


\bibitem[Shun and Blelloch(2013)]%
        {Shun:2013:LAL}
\bibfield{author}{\bibinfo{person}{Julian Shun} {and} \bibinfo{person}{Guy~E.
  Blelloch}.} \bibinfo{year}{2013}\natexlab{}.
\newblock \showarticletitle{Ligra: a lightweight graph processing framework for
  shared memory}. In \bibinfo{booktitle}{\emph{Proceedings of the 18th ACM
  SIGPLAN Symposium on Principles and Practice of Parallel Programming}}
  (Shenzhen, China) \emph{(\bibinfo{series}{PPoPP '13})}.
  \bibinfo{pages}{135--146}.
\newblock
\urldef\tempurl%
\url{https://doi.org/10.1145/2442516.2442530}
\showDOI{\tempurl}


\bibitem[Shwatal(2024)]%
        {shwatal:2024:ipd}
\bibfield{author}{\bibinfo{person}{Nathan~A. Shwatal}.}
  \bibinfo{year}{2024}\natexlab{}.
\newblock \emph{\bibinfo{title}{Improving the Programmability of A Distributed
  Hardware Accelerator}}.
\newblock \bibinfo{thesistype}{Master's\ thesis}.
  \bibinfo{school}{Massachusetts Institute of Technology}.
\newblock


\bibitem[Smith and Karypis(2015)]%
        {Smith:2015:TMP}
\bibfield{author}{\bibinfo{person}{Shaden Smith} {and} \bibinfo{person}{George
  Karypis}.} \bibinfo{year}{2015}\natexlab{}.
\newblock \showarticletitle{Tensor-matrix products with a compressed sparse
  tensor}. In \bibinfo{booktitle}{\emph{Proceedings of the 5th Workshop on
  Irregular Applications - Architectures and Algorithms}} (Austin, Texas, USA)
  \emph{(\bibinfo{series}{IA3 2015})},
  \bibfield{editor}{\bibinfo{person}{Antonino Tumeo}, \bibinfo{person}{John
  Feo}, {and} \bibinfo{person}{Oreste Villa}} (Eds.).
  \bibinfo{publisher}{{ACM}}, \bibinfo{pages}{5:1--5:7}.
\newblock
\urldef\tempurl%
\url{https://doi.org/10.1145/2833179.2833183}
\showDOI{\tempurl}


\bibitem[Solomonik et~al\mbox{.}(2013)]%
        {Solomonik:2013:CTF}
\bibfield{author}{\bibinfo{person}{Edgar Solomonik}, \bibinfo{person}{Devin
  Matthews}, \bibinfo{person}{Jeff~R. Hammond}, {and} \bibinfo{person}{James
  Demmel}.} \bibinfo{year}{2013}\natexlab{}.
\newblock \showarticletitle{Cyclops Tensor Framework: Reducing Communication
  and Eliminating Load Imbalance in Massively Parallel Contractions}. In
  \bibinfo{booktitle}{\emph{27th {IEEE} International Symposium on Parallel and
  Distributed Processing}} (Cambridge, MA, USA) \emph{(\bibinfo{series}{IPDPS
  2013})}. \bibinfo{publisher}{{IEEE} Computer Society},
  \bibinfo{pages}{813--824}.
\newblock
\urldef\tempurl%
\url{https://doi.org/10.1109/IPDPS.2013.112}
\showDOI{\tempurl}


\bibitem[Srivastava et~al\mbox{.}(2020)]%
        {mat:2020:sri}
\bibfield{author}{\bibinfo{person}{Nitish Srivastava}, \bibinfo{person}{Hanchen
  Jin}, \bibinfo{person}{Jie Liu}, \bibinfo{person}{David Albonesi}, {and}
  \bibinfo{person}{Zhiru Zhang}.} \bibinfo{year}{2020}\natexlab{}.
\newblock \showarticletitle{{MatRaptor:} A Sparse-Sparse Matrix Multiplication
  Accelerator Based on Row-Wise Product}. In
  \bibinfo{booktitle}{\emph{International Symposium on Microarchitecture
  ({MICRO})}}. \bibinfo{pages}{766--780}.
\newblock
\urldef\tempurl%
\url{https://doi.org/10.1109/MICRO50266.2020.00068}
\showDOI{\tempurl}


\bibitem[Srivastava and Singh(2023)]%
        {Srivastava:2023:FDD}
\bibfield{author}{\bibinfo{person}{Sakshi Srivastava} {and}
  \bibinfo{person}{Anil~Kumar Singh}.} \bibinfo{year}{2023}\natexlab{}.
\newblock \showarticletitle{Fraud detection in the distributed graph database}.
\newblock \bibinfo{journal}{\emph{Cluster Computing}} \bibinfo{volume}{26},
  \bibinfo{number}{1} (\bibinfo{year}{2023}), \bibinfo{pages}{515--537}.
\newblock
\urldef\tempurl%
\url{https://doi.org/10.1007/S10586-022-03540-3}
\showDOI{\tempurl}


\bibitem[Sze et~al\mbox{.}(2020)]%
        {sze:2020:epo}
\bibfield{author}{\bibinfo{person}{Vivienne Sze}, \bibinfo{person}{Yu{-}Hsin
  Chen}, \bibinfo{person}{Tien{-}Ju Yang}, {and} \bibinfo{person}{Joel~S.
  Emer}.} \bibinfo{year}{2020}\natexlab{}.
\newblock \bibinfo{booktitle}{\emph{Efficient Processing of Deep Neural
  Networks}}.
\newblock \bibinfo{publisher}{Springer International Publishing}.
\newblock
\showISBNx{9783031017667}
\showISSN{1935-3243}
\urldef\tempurl%
\url{https://doi.org/10.1007/978-3-031-01766-7}
\showDOI{\tempurl}


\bibitem[Wang et~al\mbox{.}(2017)]%
        {Wang:2017:GGG}
\bibfield{author}{\bibinfo{person}{Yangzihao Wang}, \bibinfo{person}{Yuechao
  Pan}, \bibinfo{person}{Andrew Davidson}, \bibinfo{person}{Yuduo Wu},
  \bibinfo{person}{Carl Yang}, \bibinfo{person}{Leyuan Wang},
  \bibinfo{person}{Muhammad Osama}, \bibinfo{person}{Chenshan Yuan},
  \bibinfo{person}{Weitang Liu}, \bibinfo{person}{Andy~T. Riffel}, {and}
  \bibinfo{person}{John~D. Owens}.} \bibinfo{year}{2017}\natexlab{}.
\newblock \showarticletitle{{G}unrock: {GPU} Graph Analytics}.
\newblock \bibinfo{journal}{\emph{ACM Transactions on Parallel Computing}}
  \bibinfo{volume}{4}, \bibinfo{number}{1} (\bibinfo{date}{Aug.}
  \bibinfo{year}{2017}), \bibinfo{pages}{3:1--3:49}.
\newblock
\urldef\tempurl%
\url{https://doi.org/10.1145/3108140}
\showDOI{\tempurl}


\bibitem[Won et~al\mbox{.}(2026)]%
        {won:2025:insum}
\bibfield{author}{\bibinfo{person}{Jaeyeon Won}, \bibinfo{person}{Willow
  Ahrens}, \bibinfo{person}{Saman Amarasinghe}, {and} \bibinfo{person}{Joel~S.
  Emer}.} \bibinfo{year}{2026}\natexlab{}.
\newblock \showarticletitle{Insum: Sparse {GPU} Kernels Simplified and
  Optimized with Indirect Einsums}. In \bibinfo{booktitle}{\emph{Proceedings of
  the 31st ACM International Conference on Architectural Support for
  Programming Languages and Operating Systems, Volume 2}} (USA)
  \emph{(\bibinfo{series}{ASPLOS '26})}. \bibinfo{pages}{993--1006}.
\newblock
\showISBNx{9798400723599}
\urldef\tempurl%
\url{https://doi.org/10.1145/3779212.3790176}
\showDOI{\tempurl}


\bibitem[Won et~al\mbox{.}(2025)]%
        {won:2024:CTA}
\bibfield{author}{\bibinfo{person}{Jaeyeon Won}, \bibinfo{person}{Willow
  Ahrens}, \bibinfo{person}{Teodoro~Fields Collin}, \bibinfo{person}{Joel~S.
  Emer}, {and} \bibinfo{person}{Saman Amarasinghe}.}
  \bibinfo{year}{2025}\natexlab{}.
\newblock \showarticletitle{The Continuous Tensor Abstraction: Where Indices
  Are Real}.
\newblock \bibinfo{journal}{\emph{Proc. ACM Program. Lang.}}
  \bibinfo{volume}{9}, \bibinfo{number}{OOPSLA2}, Article
  \bibinfo{articleno}{368} (\bibinfo{date}{Oct.} \bibinfo{year}{2025}),
  \bibinfo{numpages}{29}~pages.
\newblock
\urldef\tempurl%
\url{https://doi.org/10.1145/3763146}
\showDOI{\tempurl}


\bibitem[Won et~al\mbox{.}(2023)]%
        {Won:2023:WLW}
\bibfield{author}{\bibinfo{person}{Jaeyeon Won}, \bibinfo{person}{Charith
  Mendis}, \bibinfo{person}{Joel~S. Emer}, {and} \bibinfo{person}{Saman~P.
  Amarasinghe}.} \bibinfo{year}{2023}\natexlab{}.
\newblock \showarticletitle{{WACO:} Learning Workload-Aware Co-optimization of
  the Format and Schedule of a Sparse Tensor Program}. In
  \bibinfo{booktitle}{\emph{Proceedings of the 28th {ACM} International
  Conference on Architectural Support for Programming Languages and Operating
  Systems}} (Vancouver, BC, Canada) \emph{(\bibinfo{series}{{ASPLOS} 2023},
  Vol.~\bibinfo{volume}{2})}, \bibfield{editor}{\bibinfo{person}{Tor~M.
  Aamodt}, \bibinfo{person}{Natalie D.~Enright Jerger}, {and}
  \bibinfo{person}{Michael~M. Swift}} (Eds.). \bibinfo{publisher}{{ACM}},
  \bibinfo{pages}{920--934}.
\newblock
\urldef\tempurl%
\url{https://doi.org/10.1145/3575693.3575742}
\showDOI{\tempurl}


\bibitem[Wu et~al\mbox{.}(2022)]%
        {Wu:2022:SAA}
\bibfield{author}{\bibinfo{person}{Yannan~Nellie Wu}, \bibinfo{person}{Po-An
  Tsai}, \bibinfo{person}{Angshuman Parashar}, \bibinfo{person}{Vivienne Sze},
  {and} \bibinfo{person}{Joel~S. Emer}.} \bibinfo{year}{2022}\natexlab{}.
\newblock \showarticletitle{Sparseloop: An Analytical Approach To Sparse Tensor
  Accelerator Modeling}. In \bibinfo{booktitle}{\emph{55th IEEE/ACM
  International Symposium on Microarchitecture (MICRO)}}.
  \bibinfo{publisher}{IEEE}, \bibinfo{pages}{1377--1395}.
\newblock
\urldef\tempurl%
\url{https://doi.org/10.1109/MICRO56248.2022.00096}
\showDOI{\tempurl}


\bibitem[Yang et~al\mbox{.}(2018)]%
        {Yang:2018:IPE}
\bibfield{author}{\bibinfo{person}{Carl Yang}, \bibinfo{person}{Ayd{\i}n
  Bulu\c{c}}, {and} \bibinfo{person}{John~D. Owens}.}
  \bibinfo{year}{2018}\natexlab{}.
\newblock \showarticletitle{Implementing Push-Pull Efficiently in {GraphBLAS}}.
  In \bibinfo{booktitle}{\emph{Proceedings of the International Conference on
  Parallel Processing}} \emph{(\bibinfo{series}{ICPP 2018})}.
  \bibinfo{pages}{89:1--89:11}.
\newblock
\urldef\tempurl%
\url{https://doi.org/10.1145/3225058.3225122}
\showDOI{\tempurl}


\bibitem[Zhang et~al\mbox{.}(2018a)]%
        {Zhang:2018:ABG}
\bibfield{author}{\bibinfo{person}{Guangyan Zhang}, \bibinfo{person}{Shuhan
  Cheng}, \bibinfo{person}{Jiwu Shu}, \bibinfo{person}{Qingda Hu}, {and}
  \bibinfo{person}{Weimin Zheng}.} \bibinfo{year}{2018}\natexlab{a}.
\newblock \showarticletitle{Accelerating breadth-first graph search on a single
  server by dynamic edge trimming}.
\newblock \bibinfo{journal}{\emph{J. Parallel and Distrib. Comput.}}
  \bibinfo{volume}{120} (\bibinfo{date}{Oct.} \bibinfo{year}{2018}),
  \bibinfo{pages}{383--394}.
\newblock
\showISSN{0743-7315}
\urldef\tempurl%
\url{https://doi.org/10.1016/j.jpdc.2017.09.007}
\showDOI{\tempurl}


\bibitem[Zhang et~al\mbox{.}(2025)]%
        {Zhang:2025:transfusion}
\bibfield{author}{\bibinfo{person}{Linxuan Zhang}, \bibinfo{person}{J.~Nelson
  Amaral}, {and} \bibinfo{person}{Di Niu}.} \bibinfo{year}{2025}\natexlab{}.
\newblock \showarticletitle{{TransFusion}: End-to-End Transformer Acceleration
  via Graph Fusion and Pipelining}. In \bibinfo{booktitle}{\emph{Proceedings of
  the 58th IEEE/ACM International Symposium on Microarchitecture}} (Seoul,
  South Korea) \emph{(\bibinfo{series}{MICRO '25})}.
  \bibinfo{publisher}{Association for Computing Machinery},
  \bibinfo{address}{New York, NY, USA}, \bibinfo{pages}{1491–1504}.
\newblock
\showISBNx{9798400715730}
\urldef\tempurl%
\url{https://doi.org/10.1145/3725843.3756105}
\showDOI{\tempurl}


\bibitem[Zhang et~al\mbox{.}(2018b)]%
        {Zhang:2018:GAH}
\bibfield{author}{\bibinfo{person}{Yunming Zhang}, \bibinfo{person}{Mengjiao
  Yang}, \bibinfo{person}{Riyadh Baghdadi}, \bibinfo{person}{Shoaib Kamil},
  \bibinfo{person}{Julian Shun}, {and} \bibinfo{person}{Saman Amarasinghe}.}
  \bibinfo{year}{2018}\natexlab{b}.
\newblock \showarticletitle{{GraphIt}: A High-Performance Graph {DSL}}.
\newblock \bibinfo{journal}{\emph{Proceedings of the ACM on Programming
  Languages}} \bibinfo{volume}{2}, \bibinfo{number}{OOPSLA}
  (\bibinfo{date}{Oct.} \bibinfo{year}{2018}), \bibinfo{pages}{1--30}.
\newblock
\showISSN{2475-1421}
\urldef\tempurl%
\url{https://doi.org/10.1145/3276491}
\showDOI{\tempurl}


\bibitem[Zhou et~al\mbox{.}(2018)]%
        {Zhou:2018:FFE}
\bibfield{author}{\bibinfo{person}{Shijie Zhou}, \bibinfo{person}{Rajgopal
  Kannan}, \bibinfo{person}{Hanqing Zeng}, {and} \bibinfo{person}{Viktor~K.
  Prasanna}.} \bibinfo{year}{2018}\natexlab{}.
\newblock \showarticletitle{An {FPGA} framework for edge{-}centric graph
  processing}.
\newblock \bibinfo{journal}{\emph{Proceedings of the 15th ACM International
  Conference on Computing Frontiers}}, \bibinfo{pages}{69--77}.
\newblock
\urldef\tempurl%
\url{https://doi.org/10.1145/3203217.3203233}
\showDOI{\tempurl}


\bibitem[Zhou et~al\mbox{.}(2022)]%
        {Zhou:2022:RRA}
\bibfield{author}{\bibinfo{person}{Tong Zhou}, \bibinfo{person}{Ruiqin Tian},
  \bibinfo{person}{Rizwan~A. Ashraf}, \bibinfo{person}{Roberto Gioiosa},
  \bibinfo{person}{Gokcen Kestor}, {and} \bibinfo{person}{Vivek Sarkar}.}
  \bibinfo{year}{2022}\natexlab{}.
\newblock \showarticletitle{ReACT: Redundancy-Aware Code Generation for Tensor
  Expressions}. In \bibinfo{booktitle}{\emph{Proceedings of the International
  Conference on Parallel Architectures and Compilation Techniques}} (Chicago,
  Illinois) \emph{(\bibinfo{series}{{PACT} 2022})},
  \bibfield{editor}{\bibinfo{person}{Andreas Kl{\"{o}}ckner} {and}
  \bibinfo{person}{Jos{\'{e}} Moreira}} (Eds.). \bibinfo{publisher}{{ACM}},
  \bibinfo{pages}{1--13}.
\newblock
\urldef\tempurl%
\url{https://doi.org/10.1145/3559009.3569685}
\showDOI{\tempurl}


\bibitem[Zhu et~al\mbox{.}(2020)]%
        {Zhu:2020:WEC}
\bibfield{author}{\bibinfo{person}{Huanzhou Zhu}, \bibinfo{person}{Ligang He},
  \bibinfo{person}{Matthew Leeke}, {and} \bibinfo{person}{Rui Mao}.}
  \bibinfo{year}{2020}\natexlab{}.
\newblock \showarticletitle{{WolfGraph}: The edge-centric graph processing on
  {GPU}}.
\newblock \bibinfo{journal}{\emph{Future Generation Computer Systems}}
  \bibinfo{volume}{111} (\bibinfo{date}{Oct.} \bibinfo{year}{2020}),
  \bibinfo{pages}{552--569}.
\newblock
\showISSN{0167-739X}
\urldef\tempurl%
\url{https://doi.org/10.1016/j.future.2019.09.052}
\showDOI{\tempurl}


\bibitem[Zhu et~al\mbox{.}(2026)]%
        {zhu:2026:rsu}
\bibfield{author}{\bibinfo{person}{Yan Zhu}, \bibinfo{person}{Boru Chen},
  \bibinfo{person}{Christopher~W. Fletcher}, {and} \bibinfo{person}{Nandeeka
  Nayak}.} \bibinfo{year}{2026}\natexlab{}.
\newblock \showarticletitle{{{RTeAAL}} Sim: Using Tensor Algebra to Represent
  and Accelerate {RTL} Simulation}. In \bibinfo{booktitle}{\emph{Proceedings of
  the 31st ACM International Conference on Architectural Support for
  Programming Languages and Operating Systems}} (Pittsburgh, PA, USA)
  \emph{(\bibinfo{series}{ASPLOS '26})}. \bibinfo{publisher}{ACM},
  \bibinfo{pages}{1660--1676}.
\newblock
\urldef\tempurl%
\url{https://doi.org/10.1145/3779212.3790214}
\showDOI{\tempurl}


\bibitem[Zund(1994)]%
        {Zund:1994:FDG}
\bibfield{author}{\bibinfo{person}{Joseph Zund}.}
  \bibinfo{year}{1994}\natexlab{}.
\newblock \showarticletitle{The {R}icci Calculus}.
\newblock In \bibinfo{booktitle}{\emph{Foundations of Differential Geodesy}}.
  \bibinfo{publisher}{Springer Berlin Heidelberg}, \bibinfo{address}{Berlin,
  Heidelberg}, \bibinfo{pages}{45--72}.
\newblock
\showISBNx{978-3-642-79187-1}
\urldef\tempurl%
\url{https://doi.org/10.1007/978-3-642-79187-1_2}
\showDOI{\tempurl}


\end{thebibliography}
\clearpage
\appendix
\section{Appendix: Merge Operators}\label{appendix:merge}
Merge determines which points in the iteration space participate in a \maptxt{} or \reduce{} action, indicating which points are effectual.
It is a binary operator on two Boolean inputs.
For \maptxt{}, these two inputs indicate whether the relevant coordinates exist in the left operand ($A$) and the right operand ($B$) (\S~\ref{action:map}).
For \reduce{}, these two inputs indicate whether the current output point already contains a non-empty reduce state, and whether the current value being reduced is non-empty (\S~\ref{action:reduce}).
Thus, the same 16 Boolean merge operators---one for each possible truth table on two Boolean inputs---apply to both \maptxt{} and \reduce{}; only the interpretation of the two inputs changes.
We can describe the possible combinations of these Boolean inputs set-theoretically (\S~\ref{sec:sets})~\cite{Henry:2021:CSA}.

These merge operators describe effectual points, not implementation strategy.
For example, for some merge operators in specific Einsums, leader-follower accesses work best (e.g., in sparse-dense matrix-matrix multiplication), while for others, two-finger merge works best (e.g., in sparse-sparse matrix multiplication, see TeAAL~\cite{Nayak:2023:TDF}).

Once a point survives merge, the compute operator determines how the corresponding values contribute to the action temporary (\maptmp{} or \redtmp{}).

Given two Boolean inputs and a \merge operation on a single rank $K$, there are 16 possible \merge operators.
For merge operators that admit a point even when both Boolean inputs are false, the compute operator is still invoked at that point, with the empty values of $A$ and $B$ substituted as its operands (\S~\ref{action:map}).

In the example equations below, the compute operator is shown uniformly to emphasize the merge operator. The truth table is determined by \merge.

\begin{enumerate}

\item\label{item:pass} The pass-through operator ($\mathbbm{1}$, or $\mathbb{T}$).
\begin{equation}
Z_{k} = A_{k} \cdot B_{k} :: \bigwedge +(\mathbbm{1})
\end{equation}
\begin{center}
\begin{tabular}{||c c c||}
 \hline
 Point exists in A & Point exists in B & Point exists in Z\\ [0.5ex]
 \hline\hline
 No & No & Yes \\
 \hline
 No & Yes & Yes \\
 \hline
 Yes & No & Yes \\
 \hline
 Yes & Yes & Yes \\
 \hline
 \hline
\end{tabular}
\end{center}

\item The no-pass operator ($\mathbb{F}$).
\begin{equation}
Z_{k} = A_{k} \cdot B_{k} :: \bigwedge +(\mathbb{F})
\end{equation}
\begin{center}
\begin{tabular}{||c c c||}
 \hline
 Point exists in A & Point exists in B & Point exists in Z\\ [0.5ex]
 \hline\hline
 No & No & No \\
 \hline
 No & Yes & No \\
 \hline
 Yes & No & No \\
 \hline
 Yes & Yes & No \\
 \hline
 \hline
\end{tabular}
\end{center}

\item Intersection ($\cap$).
\begin{equation}
Z_{k} = A_{k} \cdot B_{k} :: \bigwedge +(\cap),
\end{equation}
where compute ($+$) will only occur when both values are non-empty in $A$ and $B$ at a given $k$ point.

\begin{center}
\begin{tabular}{||c c c||}
 \hline
 Point exists in A & Point exists in B & Point exists in Z\\ [0.5ex]
 \hline\hline
 No & No & No \\
 \hline
 No & Yes & No \\
 \hline
 Yes & No & No \\
 \hline
 Yes & Yes & Yes \\
 \hline
 \hline
\end{tabular}
\end{center}

\item Take left ($\leftarrow$).

In
\begin{equation}
Z_{k} = A_{k} \cdot B_{k} :: \bigwedge +(\leftarrow),
\end{equation}
elementwise addition will only occur where $A$ is non-empty.

\begin{center}
\begin{tabular}{||c c c||}
 \hline
 Point exists in A & Point exists in B & Point exists in Z\\ [0.5ex]
 \hline\hline
 No & No & No \\
 \hline
 No & Yes & No \\
 \hline
 Yes & No & Yes \\
 \hline
 Yes & Yes & Yes\\
 \hline
 \hline
\end{tabular}
\end{center}

\item Take left only ($\ominus_l$).

In the following expression,
\begin{equation}
Z_{k} = A_{k} \cdot B_{k} :: \bigwedge +(\ominus_l),
\end{equation}
addition will only occur when $A$ is non-empty and $B$ is empty. In that case, the compute operator $+$ is invoked with $A$'s value and the empty value of $B$ as its operands (\S~\ref{action:map}); for a compute operator with additive identity $0$, this simply returns the value in $A$. All other entries of $Z$ will be empty.
This differs from take left ($\leftarrow$) above, which also admits points where both $A$ and $B$ are non-empty; $\ominus_l$ excludes that case.

Note that this can also be expressed using a unary operator ($\neg$) and intersection:
\begin{equation}
    Z_{k} = A_{k} \cdot \neg B_{k} :: \bigwedge +(\cap)
\end{equation}

More generally, all 16 merge operators can be expressed using unary operators together with simpler merge patterns such as intersection or union.
However, it is still useful to name these operators explicitly, since doing so avoids extra transformations and makes optimization opportunities easier to expose.

\begin{center}
\begin{tabular}{||c c c||}
 \hline
 Point exists in A & Point exists in B & Point exists in Z\\ [0.5ex]
 \hline\hline
 No & No & No \\
 \hline
 No & Yes & No \\
 \hline
 Yes & No & Yes \\
 \hline
 Yes & Yes & No\\
 \hline
 \hline
\end{tabular}
\end{center}

\item Take right ($\rightarrow$).
\begin{equation}
Z_{k} = A_{k} \cdot B_{k} :: \bigwedge +(\rightarrow),
\end{equation}
where addition only occurs when $B$ is non-empty.

In \reduce, we can write the following expression:
\begin{equation}
Z = A_{k}  :: \bigvee +(\rightarrow).
\end{equation}

This indicates that reduction into the current reduce state occurs only when the current value from $A$ at rank variable $k$ is non-empty.

\begin{center}
\begin{tabular}{||c c c||}
 \hline
 Point exists in A & Point exists in B & Point exists in Z\\ [0.5ex]
 \hline\hline
 No & No & No \\
 \hline
 No & Yes & Yes \\
 \hline
 Yes & No & No \\
 \hline
 Yes & Yes & Yes\\
 \hline
 \hline
\end{tabular}
\end{center}

\item Take right only ($\ominus_r$).
In
\begin{equation}
Z_{k} = A_{k} \cdot B_{k} :: \bigwedge +(\ominus_r),
\end{equation}
elementwise addition will only occur when $B$ is non-empty and $A$ is empty. In that case, the compute operator $+$ is invoked with the empty value of $A$ and $B$'s value as its operands (\S~\ref{action:map}); for a compute operator with additive identity $0$, this simply returns the value in $B$. All other entries of $Z$ will be empty.
This differs from take right ($\rightarrow$) above, which also admits points where both $A$ and $B$ are non-empty; $\ominus_r$ excludes that case.

\begin{center}
\begin{tabular}{||c c c||}
 \hline
 Point exists in A & Point exists in B & Point exists in Z\\ [0.5ex]
 \hline\hline
 No & No & No \\
 \hline
 No & Yes & Yes \\
 \hline
 Yes & No & No \\
 \hline
 Yes & Yes & No\\
 \hline
 \hline
\end{tabular}
\end{center}

\item Exclusive OR ($\oplus$).
\begin{equation}
Z_{k} = A_{k} \cdot B_{k} :: \bigwedge +(\oplus),
\end{equation}
where addition only occurs when only one operand is non-empty.

\begin{center}
\begin{tabular}{||c c c||}
 \hline
 Point exists in A & Point exists in B & Point exists in Z\\ [0.5ex]
 \hline\hline
 No & No & No \\
 \hline
 No & Yes & Yes \\
 \hline
 Yes & No & Yes \\
 \hline
 Yes & Yes & No\\
 \hline
 \hline
\end{tabular}
\end{center}

\item Union ($\cup$).
\begin{equation}
Z_{k} = A_{k} \cdot B_{k} :: \bigwedge +(\cup),
\end{equation}
where addition only occurs when either operand is non-empty.

\begin{center}
\begin{tabular}{||c c c||}
 \hline
 Point exists in A & Point exists in B & Point exists in Z\\ [0.5ex]
 \hline\hline
 No & No & No \\
 \hline
 No & Yes & Yes \\
 \hline
 Yes & No & Yes \\
 \hline
 Yes & Yes & Yes\\
 \hline
 \hline
\end{tabular}
\end{center}

\item NOR ($\bar{\cup}$).
\begin{equation}
Z_{k} = A_{k} \cdot B_{k} :: \bigwedge +(\bar{\cup})
\end{equation}

\begin{center}
\begin{tabular}{||c c c||}
 \hline
 Point exists in A & Point exists in B & Point exists in Z\\ [0.5ex]
 \hline\hline
 No & No & Yes \\
 \hline
 No & Yes & No \\
 \hline
 Yes & No & No \\
 \hline
 Yes & Yes & No\\
 \hline
 \hline
\end{tabular}
\end{center}

\item Equivalence ($\equiv$, or $\leftrightarrow$).
Here, $\equiv$ denotes Boolean equivalence.
\begin{equation}
Z_{k} = A_{k} \cdot B_{k} :: \bigwedge +(\equiv)
\end{equation}

\begin{center}
\begin{tabular}{||c c c||}
 \hline
 Point exists in A & Point exists in B & Point exists in Z\\ [0.5ex]
 \hline\hline
 No & No & Yes \\
 \hline
 No & Yes & No \\
 \hline
 Yes & No & No \\
 \hline
 Yes & Yes & Yes\\
 \hline
 \hline
\end{tabular}
\end{center}

\item Not take right ($\nrightarrow$).
This keeps points for which the right operand is absent---the complement of take right ($\rightarrow$), with each row of the truth table negated.
\begin{equation}
Z_{k} = A_{k} \cdot B_{k} :: \bigwedge +(\nrightarrow)
\end{equation}

\begin{center}
\begin{tabular}{||c c c||}
 \hline
 Point exists in A & Point exists in B & Point exists in Z\\ [0.5ex]
 \hline\hline
 No & No & Yes  \\
 \hline
 No & Yes & No  \\
 \hline
 Yes & No & Yes \\
 \hline
 Yes & Yes & No \\
 \hline
 \hline
\end{tabular}
\end{center}

\item Right implies left ($\Leftarrow$).
\begin{equation}
Z_{k} = A_{k} \cdot B_{k} :: \bigwedge +(\Leftarrow)
\end{equation}

This is equivalent to
\begin{equation}
Z_{k} = A_{k} \cdot \neg B_{k} :: \bigwedge +(\cup)
\end{equation}
since right-implies-left is the Boolean identity $B \rightarrow A \equiv A \lor \neg B$.

\begin{center}
\begin{tabular}{||c c c||}
 \hline
 Point exists in A & Point exists in B & Point exists in Z\\ [0.5ex]
 \hline\hline
 No & No & Yes  \\
 \hline
 No & Yes & No  \\
 \hline
 Yes & No & Yes \\
 \hline
 Yes & Yes & Yes \\
 \hline
 \hline
\end{tabular}
\end{center}

\item Not take left ($\nleftarrow$).
This keeps points for which the left operand is absent---the complement of take left ($\leftarrow$), with each row of the truth table negated.
\begin{equation}
Z_{k} = A_{k} \cdot B_{k} :: \bigwedge +(\nleftarrow)
\end{equation}

\begin{center}
\begin{tabular}{||c c c||}
 \hline
 Point exists in A & Point exists in B & Point exists in Z\\ [0.5ex]
 \hline\hline
 No & No & Yes  \\
 \hline
 No & Yes & Yes  \\
 \hline
 Yes & No & No \\
 \hline
 Yes & Yes & No \\
 \hline
 \hline
\end{tabular}
\end{center}

\item Left implies right ($\Rightarrow$).
\begin{equation}
Z_{k} = A_{k} \cdot B_{k} :: \bigwedge +(\Rightarrow)
\end{equation}

\begin{center}
\begin{tabular}{||c c c||}
 \hline
 Point exists in A & Point exists in B & Point exists in Z\\ [0.5ex]
 \hline\hline
 No & No & Yes  \\
 \hline
 No & Yes & Yes  \\
 \hline
 Yes & No & No \\
 \hline
 Yes & Yes & Yes \\
 \hline
 \hline
\end{tabular}
\end{center}

\item NAND ($\bar{\cap}$).
\begin{equation}
Z_{k} = A_{k} \cdot B_{k} :: \bigwedge +(\bar{\cap})
\end{equation}

\begin{center}
\begin{tabular}{||c c c||}
 \hline
 Point exists in A & Point exists in B & Point exists in Z\\ [0.5ex]
 \hline\hline
 No & No & Yes  \\
 \hline
 No & Yes & Yes  \\
 \hline
 Yes & No & Yes \\
 \hline
 Yes & Yes & No \\
 \hline
 \hline
\end{tabular}
\end{center}

\end{enumerate}
\clearpage

\section{Appendix: Additional Notes on \EDGEcaps Syntax}\label{appendix:syntax-notes}

The goal of the main syntax section (\S~\ref{ssec:syntax}) is to state the grammar, defaults, and required components of the language as directly as possible.
The material here provides additional intuition, implementation notes, and reference tables.

\subsection{Tensor Declarations Matter}\label{appendix:syntax-declarations}

The tensor declaration region does more than record metadata about each tensor. It also removes several sources of ambiguity that would otherwise remain when reading an Extended Einsum in isolation. In particular, declarations determine the shapes and rank names that bound the iteration space, identify which tensor ranks are being accessed by each rank variable expression, and specify the empty value associated with each tensor.

These empty values matter as \merge{} operators interact with them directly. For example, an intersection \merge{} operator may be used to skip points whose input values are empty, while a pass-through \merge{} operator will still expose those points to the corresponding \compute{} operator. As a result, the declaration region is part of what makes an \edge{} specification precise.

\subsection{Anonymous Tensors and Evaluation Flexibility}\label{appendix:syntax-anonymous}

An \edge{} expression normally has two input operands and one output tensor. When an expression contains more than two input operands, we write it using anonymous tensors so that the order of binary operations is explicit. Parentheses therefore serve two roles: they disambiguate precedence, and they identify intermediate partial tensors that may be consumed by an enclosing expression.

Although the syntax fixes the grouping of operations, it does not require a particular evaluation strategy for those intermediate tensors. An implementation may choose to materialize a single point in a partial tensor and immediately use that point in the enclosing expression. It may also choose to materialize a larger set of points, up to and including the entire partial tensor, before continuing with the outer computation.
We treat this as part of \emph{fusion}, which is explored in the \emph{mapping} and \emph{binding} levels of abstraction in our separation of concerns~\cite{Nayak:2023:TDF}.

This flexibility is useful because different implementations may make different trade-offs depending on the target system.
One implementation may prefer fine-grained pointwise evaluation to reduce temporary storage, while another may prefer to compute and reuse larger intermediate tiles of a given tensor.

\subsection{Why Nested Cascades Are Useful}\label{appendix:syntax-nested-cascades}

A fully specified extended Einsum consists of a cascade of Einsum expressions, and \edge{} extends this model to support nested cascades. This is useful when a computation is naturally modular, with one group of expressions feeding into another larger computation. Graph algorithms often have this structure. For example, an inner computation may gather, filter, or relabel a set of vertices, while an outer computation uses the result as one step of a broader iterative procedure.

Supporting nested cascades lets the language represent that structure directly rather than forcing every computation into a flat sequence of top-level expressions.
See \S~\ref{ssec:complex} for example algorithms that benefit from nested cascades.

\subsection{Additional Intuition for Rank Variable Expressions}\label{appendix:syntax-rves}

A rank variable expression maps iteration-space variables to a coordinate in a tensor rank. Because this mapping may be an arbitrary function of the iteration-space variables, rank variable expressions can express coordinate transformations that go beyond simple renaming or projection.

For example, a conditional expression such as
\begin{align}
  f(a, w) \rightarrow \text{if } w < a \text{ then return } w \text{ else return } a
\end{align}
can be used to map multiple iteration-space points to the same output coordinate. In such cases, the output rank requires a \reduce{} action, since distinct iteration-space points may contribute to the same point in the output tensor.

This kind of behavior appears in graph algorithms such as connected components, where a computation may replace an ancestor vertex with a smaller-valued connected vertex.
A parallel implementation may require \emph{atomic} operations to reduce correctly.
The key point is that the syntax of rank variable expressions is general enough to describe these coordinate mappings directly.

\subsection{User-Defined Functions}\label{appendix:syntax-udfs}

\edge{} allows \compute{}, \merge{}, and \coordinate{} operators to be user-defined functions, with constraints on their inputs and outputs.
In practice, such functions may be written in a host language such as C or Python and then associated with the corresponding operator in the \edge{} specification.

The \coordinate{} operator in \populate{} is more expressive than the standard \merge{} because it can inspect not only the current iteration-space point and computed value, but also the current partial output fiber on the mutable rank. This makes it possible to define output coordinates using both coordinate information and data values.

\subsection{Shorthand Notation}\label{appendix:syntax-shorthand}

In practice, the authors have found several shorthand notations convenient for presenting and discussing algorithms. These shorthands are not part of the \edge{} language itself, but they make informal presentations easier to read. Whenever a shorthand expression appears, the intended interpretation is obtained by expanding it into a valid \edge{} expression and applying the operational and semantic definitions in the main text.

\begin{enumerate}
    \item Dropping the tensor declaration and initialization sections when they are already understood from context.
    \item Using infix notation to specify simple computations. For example, we may write element-wise addition as $Z_{m,n} = A_{m,n} + B_{m,n}$ instead of spelling out the fully specified \maptxt{} action.
    \item\label{ref:update} Using ``$<<$'' (update) instead of ``$=$'' (assignment) to express the assignment of a generational tensor with both the previous iteration's values and the newly computed values. This is a common computation pattern. For example, we can rewrite the update to the set of visited nodes in Equation~\eqref{seqn:updatef} as:
    \begin{align}\label{seqn:update_shorthand}
        P_{i+1, d} << F_{i+1, d}
    \end{align}
    The fully specified \edge{} expression corresponds to a \maptxt{} action:
    \begin{align}\label{seqn:update_full}
        P_{i+1, d} = P_{i, d} \cdot F_{i+1, d} :: \bigwedge_{d} <<(\cup),
    \end{align}
    where the \compute{} operator of $<<$ (``update''), combined with the union \merge{} operator, has the following properties:
    \begin{itemize}
        \item if both operands are empty, return an empty value;
        \item if the left operand ($P$) is empty or both operands contain values, return the value on the right; and
        \item if the right operand ($F$) is empty, return the value on the left.
    \end{itemize}
\end{enumerate}

We use shorthand notation for informal communication only; we do not expect an underlying compiler to support it. Regardless, whenever reading an \edge{} expression, the guiding principle is always: apply the operational definition of an Einsum to the expression.

\subsection{High-Level Summary Tables}\label{appendix:syntax-summary-tables}

The following tables provide compact summaries of the actions, operators, and required symbols used in \edge{}.

\begin{table}
  \caption{Summary of what each action does.}
  \label{tab:actions-semantics}
  \begin{tabular}{lp{0.8\textwidth}}
    \toprule
    \textbf{Action} & \textbf{What it does at each point in the iteration space} \\
    \midrule
    \maptxt{} & Retrieve the data-space values for each operand on the right-hand side and apply the \compute{} operator to them. \\
    \reduce{} & Given a computed value on the right-hand side, combine it with the current left-hand side state at the corresponding output point. \\
    \populate{} & Given a computed value on the right-hand side, update the mutable fiber on the left-hand side. \\
    \bottomrule
  \end{tabular}
\end{table}

\begin{table}
  \caption{Syntax summary of the three actions in \edge{}.}
  \label{tab:actions-summary}
  \begin{tabular}{lll}
    \toprule
    \textbf{Action} & \textbf{Symbol} & \textbf{Operators Supported} \\
    \midrule
    \maptxt{} & $\bigwedge$ & \compute{} operator (\merge{} operator) \\
    \reduce{} & $\bigvee$ & \compute{} operator (\merge{} operator) \\
    \populate{} & $\lll_\textit{mutable\_rank\_list}$ & \compute{} operator (\coordinate{} operator) \\
    \bottomrule
  \end{tabular}
\end{table}

\begin{table}[h]
  \caption{Operators in an \edge{} expression, along with their inputs and outputs at a given point in the iteration space. In addition to the inputs shown here, each operator also takes the current iteration-space point as input. $^*$The unary operator is an instance of \maptxt{}; see \S~\ref{ssec:unary}.}
  \label{tab:operator-summary}
  \resizebox{\textwidth}{!}{%
  \begin{tabular}{|l|l|l|l|l|}
  \hline
   & \textbf{Input 1} & \textbf{Input 2} & \textbf{Output} & \textbf{Computation Type} \\
  \hline
  \textbf{\begin{tabular}[c]{@{}l@{}}\merge{} \\ Operator\end{tabular}} &
    \begin{tabular}[c]{@{}l@{}}On the RHS\@: existence of a \\ non-empty data-space point \\ in the left operand\end{tabular} &
    \begin{tabular}[c]{@{}l@{}}On the RHS\@: existence of a \\ non-empty data-space point \\ in the right operand\end{tabular} &
    \begin{tabular}[c]{@{}l@{}}Valid or empty coordinate \\ for that point in the \\ iteration space\end{tabular} &
    \begin{tabular}[c]{@{}l@{}}One of 16 \\ \merge{} operators \\ (see Appendix~\ref{appendix:merge})\end{tabular} \\
  \hline
  \textbf{\begin{tabular}[c]{@{}l@{}}\compute{} \\ Operator\end{tabular}} &
    \begin{tabular}[c]{@{}l@{}}Depends on the action. \\ For \maptxt{}: left operand value. \\ For \reduce{}: current reduce state. \\ For \populate{}: computed RHS value.\end{tabular} &
    \begin{tabular}[c]{@{}l@{}}Depends on the action. \\ For \maptxt{}: right operand value. \\ For \reduce{}: value being reduced. \\ For \populate{}: none.\end{tabular} &
    A value &
    User-defined \\
  \hline
  \textbf{\begin{tabular}[c]{@{}l@{}}\coordinate{} \\ Operator\end{tabular}} &
    \begin{tabular}[c]{@{}l@{}}On the RHS\@: computed data value \\ at that point\end{tabular} &
    \begin{tabular}[c]{@{}l@{}}On the LHS\@: current partial \\ output fiber, including \\ both coordinates and values, \\ for the mutable rank\end{tabular} &
    Set of valid or empty coordinates &
    User-defined \\
  \hline
  \textbf{\begin{tabular}[c]{@{}l@{}}Unary$^*$ \\ Operator \\ (on a tensor)\end{tabular}} &
    \begin{tabular}[c]{@{}l@{}}On the RHS\@: value at the \\ data-space point of the tensor\end{tabular} &
     &
    \begin{tabular}[c]{@{}l@{}}New value at that \\ data-space point\end{tabular} &
    \begin{tabular}[c]{@{}l@{}}User-defined \\ unary function\end{tabular} \\
  \hline
  \end{tabular}%
  }
\end{table}

\begin{table}
  \caption{Summary of required symbols in an Einsum expression. When a given action or operator appears in an expression, it must accompany the required symbols listed here and may accompany the optional symbols. When a given action or operator is omitted, the defaults in the main syntax section apply.}
  \label{tab:requireds}
  \begin{tabular}{lll}
    \toprule
    \textbf{When this action/operator is present:} & \textbf{Then \edge{} also requires:} & \textbf{Optional:} \\
    \midrule
    \rowcolor{lightgray!25}\maptxt{} Action & \begin{tabular}[c]{@{}l@{}}Operation label \\ and \\ rank variable list\end{tabular} & \\
    \reduce{} Action & Rank variable list & Operation label \\
    \rowcolor{lightgray!25}\populate{} Action & Modifiable rank & \\
    \merge{} Operator & \begin{tabular}[c]{@{}l@{}}\maptxt{} action or \reduce{} action \\ and \\ \compute{} operator\end{tabular} & \\
    \rowcolor{lightgray!25}\coordinate{} Operator & \begin{tabular}[c]{@{}l@{}}\populate{} action \\ and \\ \compute{} operator\end{tabular} & \\
    \compute{} Operator & \begin{tabular}[c]{@{}l@{}}\maptxt{} action, \reduce{} action, \\ or \populate{} action\end{tabular} & \\
    \bottomrule
  \end{tabular}
\end{table}

\subsection{Macros}\label{appendix:syntax-macros}

A \textsc{macro} is a named, parameterized Einsum template.
Invoking a \textsc{macro} expands it into one or more Einsums via lexical substitution of its tensor and rank variable expression parameters.

A \textsc{macro} definition has the form:
\begin{align*}
\triangleright \mathrm{MACRO} \\
& \mathrm{NAME}(TA_{a, \ldots},\, TB_{b, \ldots},\, \ldots)
    \mapsto T_{c, \ldots},\, T2_{d, \ldots},\, \ldots \\
& \text{(body: cascade of Einsums using the parameters)}
\end{align*}
The left-hand side of $\mapsto$ lists the input tensor parameters; the
right-hand side declares the output tensors produced by the expansion.

\paragraph{Inline invocation.}
When a \textsc{macro} has a single output tensor, it can be invoked
\emph{inline} as an anonymous tensor within a surrounding Einsum expression.
The subscript on the call site binds the output tensor's rank variables and
names the anonymous tensor.

\paragraph{Motivating example.}
Consider three back-to-back GEMMs:
\begin{align}
Z_{m, n} &= A_{m, k} \times B_{k, n} \\
Y_{n, q} &= C_{n, p} \times D_{p, q} \\
X_{m, q} &= Z_{m, n} \times Y_{n, q}
\end{align}
Using \textsc{macros}, we can write this cascade as:
\begin{align}
X_{m, q} = \mathrm{GEMM}(A_{m, k}, B_{k, n})_{m, n}
         \times \mathrm{GEMM}(C_{n, p}, D_{p, q})_{n, q}
\end{align}
where $\mathrm{GEMM}$ is defined by the following template:
\begin{alignat*}{2}
\triangleright \mathrm{MACRO}& \\
& \mathrm{GEMM}(\tensor{TA}_{a, c},\, \tensor{TB}_{c, b}) && \mapsto T_{a, b}\\
& T_{a, b} && = \tensor{TA}_{a, c} \times \tensor{TB}_{c, b}.
\end{alignat*}

\textsc{Macros} are simply lexical substitutions. For example, we can write
$\mathrm{GEMM}(A_{m+1, k}, B_{k, n})_{m, n}$, which expands as follows:
\begin{itemize}
  \item $\tensor{TA}$ becomes $A$; rank variables $a, c$ become $m+1, k$.
  \item $\tensor{TB}$ becomes $B$; rank variable $b$ becomes $n$.
  \item $T$ becomes the anonymous tensor $\mathrm{GEMM}_{m, n}$.
\end{itemize}
Since the arguments are rank variable \emph{expressions} (not just
variables), substitutions such as $a \mapsto m+1$ are permitted.

Equivalently, we can invoke $\mathrm{GEMM}$ in assignment form:
\begin{align*}
Z_{m, n} = \mathrm{GEMM}(A_{m, k},\, B_{k, n})
\end{align*}
Here, the subscripts on $Z$ bind the output rank variables ($a \mapsto m$,
$b \mapsto n$), and $T$ becomes $Z$ rather than an anonymous tensor. Both
invocation forms produce identical expansions.

\paragraph{Cascade invocation.}
At times, for readability, we may want to express a cascade of Einsums as a
single \textsc{macro}.
In this case, the \textsc{macro} may have multiple
output tensors, and we invoke it in \emph{cascade form}, binding output
names explicitly on the left-hand side:
\begin{align*}
Z_{m, n},\, Y_{n, q},\, X_{m, q} = \mathrm{3\text{-}GEMM}(A_{m,k},\, B_{k,n},\, C_{n,p},\, D_{p,q})
\end{align*}
where $\mathrm{3\text{-}GEMM}$ is defined as:
\begin{alignat*}{2}
\triangleright \mathrm{MACRO}& \\
& \mathrm{3\text{-}GEMM}(\tensor{TA}_{a,b},\, \tensor{TB}_{b,c},\, \tensor{TC}_{c,d},\, \tensor{TD}_{d,e})
    && \mapsto \tensor{TZ}_{a,c},\, \tensor{TY}_{c,e},\, \tensor{TX}_{a,e} \\
& \tensor{TZ}_{a,c} &&= \mathrm{GEMM}(\tensor{TA}_{a,b},\, \tensor{TB}_{b,c}) \\
& \tensor{TY}_{c,e} &&= \mathrm{GEMM}(\tensor{TC}_{c,d},\, \tensor{TD}_{d,e}) \\
& \tensor{TX}_{a,e} &&= \mathrm{GEMM}(\tensor{TZ}_{a,c},\, \tensor{TY}_{c,e})
\end{alignat*}

The output names on the left-hand side of the invocation ($Z$, $Y$, $X$)
bind positionally to the declared outputs ($\tensor{TZ}$, $\tensor{TY}$, $\tensor{TX}$), and the
subscripts bind the corresponding rank variables. If a rank variable appears
in more than one output, the caller is responsible for providing consistent
subscripts across those outputs, following the same convention as rank
variable sharing across Einsums in a cascade.

After expansion, all three output tensors are live as named tensors in the
enclosing cascade.
A downstream Einsum may reference any subset of them;
unreferenced tensors are simply dead temporaries, and whether to eliminate
them is a concern for the implementation, not the notation.

\subsection{Parameterized Rank Groups}\label{appendix:syntax-parameterized-ranks}

A \textsc{macro} substitutes symbols one-for-one at instantiation time.
Sometimes, however, we want to write a single Einsum expression that is
generic over an unknown number of ranks, where that count $R$ is only
determined at runtime. We call such a group of ranks a \emph{parameterized
rank group}.

\paragraph{Motivating example.}
Consider the standard GEMM\@:
\begin{align}
Z_{m, n} = A_{m, k} \times B_{k, n}.
\end{align}
Suppose we want this to be generic, where $m$ can stand for any number of
uncontracted ranks whose count $R$ is only known at runtime. We need a way
to write the expression once without committing to a specific value of $R$.

\paragraph{Notation.}
We declare a parameterized rank group $M.R$ in the
tensor declaration, where $M$ is the rank group name and $R$ is its
rank-count parameter. Like the generational rank $I$, $R$ is not assigned a
fixed shape at declaration time; its value is determined at runtime. In the
Einsum body, we write $m.r$ to refer to the $r$-th rank variable in the $M$
group, where $r$ is an index ranging over $[0, R)$.

The generalized GEMM is written:
\begin{align*}
&\triangleright \text{Tensors} \\
&Z^{M.R,\, N},\quad A^{M.R,\, K},\quad B^{K,\, N}
  \rightarrow \text{integer, empty} = 0 \\
&\triangleright \text{Extended Einsum} \\
&Z_{m.r,\, n} = A_{m.r,\, k} \times B_{k,\, n}.
\end{align*}
At runtime, if $R = 4$, the rank group $M$ expands to four individual ranks,
and the Einsum becomes:
\begin{align*}
Z_{m.3,\, m.2,\, m.1,\, m.0,\, n} = A_{m.3,\, m.2,\, m.1,\, m.0,\, k} \times B_{k,\, n}.
\end{align*}
The iteration space consists of the rank variables ranged by those individual
ranks, together with $k$ and $n$.

\paragraph{Contraction within a rank group.}
After expansion, each member of the rank group participates in the standard
Einsum contraction semantics independently. If a particular expanded rank
variable appears only on the right-hand side, that rank contracts. For example,
if $R = 4$ and $m.2$ appears only on the right-hand side, then the Einsum
reduces over $m.2$. If multiple expanded rank variables appear only on the
right-hand side, then the Einsum reduces over each of those ranks. Thus, a full
rank group contracts only when every member of that group appears only on the
right-hand side.

For example:
\begin{align*}
Z_{m.3,\, m.1,\, m.0,\, n} = A_{m.3,\, m.2,\, m.1,\, m.0,\, k} \times B_{k,\, n}
\end{align*}
contracts over $m.2$ and $k$, but preserves $m.3$, $m.1$, $m.0$, and $n$ in
the output.

Likewise, if the entire expanded group appears only on the right-hand side:
\begin{align*}
Z_{n} = A_{m.r,\, k} \times B_{k,\, n},
\end{align*}
then all members of the $M$ rank group contract, along with $k$. At runtime
with $R = 2$, this is equivalent to:
\begin{align*}
Z_{n} = A_{m.1,\, m.0,\, k} \times B_{k,\, n}.
\end{align*}

\paragraph{Multiple parameterized rank groups.}
More than one rank group can
appear in the same expression. For example, generalizing GEMM so that both
sets of uncontracted ranks are parameterized:
\begin{align*}
Z_{m.r,\, n.s} = A_{m.r,\, k} \times B_{k,\, n.s},
\end{align*}
where $r \in [0, R)$ and $s \in [0, S)$ are independent rank-count
parameters, each determined at runtime.

\paragraph{Relationship to \textsc{macros}.}
A parameterized rank group is
resolved by expansion at runtime, in the same spirit as a \textsc{macro}
(\S~\ref{appendix:syntax-macros}). The distinction is one of arity: a
\textsc{macro} substitutes symbols one-for-one, while a rank group expands a
single symbol $m.r$ into an ordered list of rank variables. Both are
instances of the same underlying idea of lexical expansion, operating at
different granularities.
\clearpage
\section{Appendix: Bellman-Ford Eingebraic Manipulations}\label{appendix:bf-derivation}

This appendix provides a more detailed algebraic justification for each transformation
in \S~\ref{ssec:sssp}.

\subsection{Tensor Glossary}\label{appendix:bf-tensors}

Table~\ref{tab:sssp-tensors} summarizes the tensors used in the SSSP derivation
and marks which cascade each one appears in.

\begin{table}[hbtp]
\centering
\caption{Tensors used in the SSSP derivation. Columns mark which cascade each
tensor appears in: BF (Cascade~\ref{cascade:bf}), Steps~1--4
(Cascades~\ref{cascade:bf-step1}--\ref{cascade:bf-step4}), SPFA
(Cascade~\ref{cascade:bf-step5}), and Dijkstra (Cascade~\ref{cascade:dijkstra}).}
\label{tab:sssp-tensors}
\begin{tabular}{l p{5cm} c c c c c c c}
\toprule
Tensor & Description & BF & S1 & S2 & S3 & S4 & SPFA & Dij. \\
\midrule
$G^{S, D}$                          & Edge-weight graph; $(s, d)$ indexes the edge from source $s$ to destination $d$
                             & \checkmark & \checkmark & \checkmark & \checkmark & \checkmark
                             & \checkmark & \checkmark \\
$D^{I, S}$                          & Per-vertex distances from source; $i$ outer iteration, $s$ vertex
                             & \checkmark & \checkmark & \checkmark & \checkmark & \checkmark
                             & \checkmark & \checkmark \\
$\tilde{D}^{I, K, D}$                  & Running distance within outer iteration $i$; inner step $k$ tracks progress through sources, $d$ is the destination vertex
                             &        &        &        & \checkmark & \checkmark
                             &        &        \\
$N$                          & Candidate neighbor distances after one relaxation; indexed as $N_{i,d}$ in BF/SPFA/Dijkstra and as $N_{i,s,d}$ or $N_{i,k,d}$ in the manipulation steps
                             & \checkmark & \checkmark & \checkmark & \checkmark & \checkmark
                             & \checkmark & \checkmark \\
$C$                          & Mask where $N < D$; indexed as $C_{i,d}$ in BF/SPFA/Dijkstra and as $C_{i,s,d}$ in Step~1
                             & \checkmark & \checkmark &        &        &
                             & \checkmark & \checkmark \\
$\mathsf{PerSourceRelaxed}^{I, S, D}$  & $N$ where it improves $D$, indexed per source $s$ and destination $d$
                             &        & \checkmark &        &        &
                             &        &        \\
$\mathsf{NewlyRelaxed}^{I, D}$      & Best improvement per destination $d$, after reducing over sources
                             & \checkmark & \checkmark &        &        &
                             & \checkmark & \checkmark \\
$Q^{I, S}$                          & Marks at $(i, s)$ which sources may still produce improvements
                             &        &        &        &        &
                             & \checkmark & \checkmark \\
$DQ^{I, S}$                         & $D$ restricted to entries where $Q$ is true
                             &        &        &        &        &
                             & \checkmark & \checkmark \\
$F^{I, S}$                          & One-hot at $(i, s)$ for the single-source vertex selected to process this iteration
                             &        &        &        &        &
                             & \checkmark & \checkmark \\
$T^{I, D}$                          & Queue $Q$ with the just-processed source removed
                             &        &        &        &        &
                             & \checkmark & \checkmark \\
\bottomrule
\end{tabular}
\end{table}

\subsection{Equivalence of the Boolean-Check Update and a $\min$ Einsum}\label{appendix:bf-collapse-rewrite}

\paragraph{Claim.}
The three-Einsum chain
\begin{subequations}
\begin{compactalign}
C_{i, s, d}                  &= N_{i, s, d} \cdot D_{i, d}
  :: \bigwedge <(\cup) \\
\mathsf{PerSourceRelaxed}_{i, s, d}  &= C_{i, s, d} \cdot N_{i, s, d}
  :: \bigwedge \rightarrow(\cap) \\
D_{i+1, d}                   &= D_{i, d} \cdot \mathsf{PerSourceRelaxed}_{i, s, d}
  :: \bigwedge \mathbin{<<}(\cup) \bigvee_{s} \min(\cup)
\end{compactalign}
\end{subequations}
is equivalent to the single Einsum
\begin{equation}
D_{i+1, d} = D_{i, d} \cdot N_{i, s, d}
  :: \bigwedge \min(\cup) \bigvee_{s} \min(\cup).
\end{equation}

\paragraph{Eingebraic Manipulations.}
We proceed by substitution.

Substituting $C$'s definition into $\mathsf{PerSourceRelaxed}$:
\begin{equation}
\mathsf{PerSourceRelaxed}_{i, s, d}
  = (N_{i, s, d} \cdot^1 D_{i, d})_{i, s, d} \cdot^2 N_{i, s, d}
  :: \bigwedge^1 <(\cup)\; \bigwedge^2 \rightarrow(\cap).
\end{equation}
The inner $<(\cup)$ produces a Boolean tensor whose value at $(i, s, d)$ is
$\True$ exactly when $N_{i, s, d} < D_{i, d}$. The outer $\rightarrow(\cap)$
takes the right operand's value where both operands are non-empty, so
$\mathsf{PerSourceRelaxed}_{i, s, d}$ is non-empty exactly where $N_{i, s, d} < D_{i, d}$,
and its value at those points is $N_{i, s, d}$. Thus
$\mathsf{PerSourceRelaxed}$ is $N$ where it improves $D$.

Substituting this into the $D_{i+1}$ Einsum, the right operand of
$\mathbin{<<}(\cup)$ is non-empty exactly where $N_{i, s, d} < D_{i, d}$, with
value $N_{i, s, d}$. By the definition of $\mathbin{<<}(\cup)$, $D_{i+1, d}$ takes the right operand's value where it is
non-empty, and the left operand's value otherwise. Therefore at every $(i, s, d)$:
\begin{itemize}
  \item If $N_{i, s, d} < D_{i, d}$, then $D_{i+1, d} = N_{i, s, d} = \min(D_{i, d}, N_{i, s, d})$.
  \item Otherwise: $D_{i+1, d} = D_{i, d} = \min(D_{i, d}, N_{i, s, d})$.
\end{itemize}
Either way, the per-source operation yields $\min(D_{i, d}, N_{i, s, d})$,
which is exactly the operation specified by $\bigwedge \min(\cup)$ between
$D_{i, d}$ and $N_{i, s, d}$. The outer reduction over $s$ is unchanged.

\paragraph{Intuition.}
$\mathbin{<<}(\cup)$ is ``take right where right exists, else take left.'' When
right is gated to only exist where it is strictly less than left, the operation
reduces to ``take whichever is smaller,'' which is $\min$.

\subsection{Equivalence Between $s$-Reduction and Iterative Accumulation
Over Inner Rank $k$}\label{appendix:bf-fold}

\paragraph{Claim.}
The Einsum
\begin{equation}\label{min-min}
D_{i+1, d} = D_{i, d} \cdot N_{i, s, d}
  :: \bigwedge \min(\cup) \bigvee_{s} \min(\cup)
\end{equation}
is equivalent to the cascade
\begin{subequations}
\begin{compactalign}
\tilde{D}_{i, 0, d}   &= D_{i, d} \\
\tilde{D}_{i, k+1, d} &= \tilde{D}_{i, k, d} \cdot N_{i, k, d}
  :: \bigwedge \min(\cup) \\
\diamond_k &: k \equiv |V| \\
D_{i+1, d}            &= \tilde{D}_{i, |V|, d}
\end{compactalign}
\end{subequations}
where $N_{i, k, d}$ is $N_{i, s, d}$ with the $s$ rank renamed to $k$.

\paragraph{Eingebraic Manipulations.}
Pin the source rank $s$ at each value $0, 1, \ldots, |V|-1$. The original
Einsum's reduction over $s$ unrolls into a chain of binary $\min$ operations:
\begin{multline}
D_{i+1, d} = \bigl(\ldots\bigl(\bigl(D_{i, d} \cdot^1 N_{i, s:s=0, d}\bigr)_{i, d}
  \cdot^2 N_{i, s:s=1, d}\bigr)_{i, d} \ldots \bigr)_{i, d}
  \cdot^{|V|} N_{i, s:s=|V|-1, d} \\
  :: \bigwedge^1 \min(\cup)\; \bigwedge^2 \min(\cup) \ldots \bigwedge^{|V|} \min(\cup).
\end{multline}
$\min$ is associative and commutative. Thus, this chain equals the original Einsum
regardless of the order in which the $\bigwedge$ operations are applied.

Define $\tilde{D}_{i, 0, d} = D_{i, d}$ and
$\tilde{D}_{i, k+1, d} = \tilde{D}_{i, k, d} \cdot N_{i, k, d}
  :: \bigwedge \min(\cup)$. By induction on $k$:

\begin{itemize}
  \item Base case ($k = 0$): $\tilde{D}_{i, 0, d} = D_{i, d}$ by definition.
  \item Inductive step: assume $\tilde{D}_{i, k, d}$ equals the prefix of the
    chain up through $\bigwedge^k$. Then $\tilde{D}_{i, k+1, d}$ extends by one
    more $\min$ with $N_{i, s:s=k, d}$ (renamed to $N_{i, k, d}$), which is
    exactly $\bigwedge^{k+1}$.
\end{itemize}

After at most $|V|$ steps, $\tilde{D}_{i, |V|, d}$ equals the full chain, which equals
$D_{i+1, d}$.

\paragraph{Role of idempotence.}
Suppose we instead do the elementwise $\min$ (Equation~\eqref{min-min}) first, before reducing over the $s$ rank:
\begin{subequations}
\begin{compactalign}
\mathsf{PerSourceDist}_{i, s, d} &= D_{i, d} \cdot N_{i, s, d}
  :: \bigwedge \min(\cup) \\
D_{i+1, d} &= \mathsf{PerSourceDist}_{i, s, d}
  :: \bigvee_{s} \min(\cup).
\end{compactalign}
\end{subequations}

Substituting the $\mathsf{PerSourceDist}$ values and pinning $s$, this expands to:
\begin{multline}
D_{i+1, d} = \bigl(\ldots\bigl(\bigl(D_{i, d} \cdot^1 N_{i, s:s=0, d}\bigr)_{i, d}
  \cdot^{|V|+1} \bigl(D_{i, d} \cdot^2 N_{i, s:s=1, d}\bigr)_{i, d}\bigr)_{i, d}
  \cdot^{|V|+2} \ldots \bigr)_{i, d} \\
  \cdot^{2|V|-1} \bigl(D_{i, d} \cdot^{|V|} N_{i, s:s=|V|-1, d}\bigr)_{i, d} \\
  :: \bigwedge^1 \min(\cup)\; \bigwedge^2 \min(\cup) \ldots \bigwedge^{2|V|-1} \min(\cup).
\end{multline}
Notice that $D_{i, d}$ appears $|V|$ times as an operand in the expanded form.
Idempotence of $\min$ ($\min(D, D) = D$) collapses these duplicates, so the
result agrees with the rolled form.

\paragraph{Note on the two iterative ranks.}
The cascade introduces an inner iterative rank $k$ within each outer iteration
$i$. This is a nested cascade: $\tilde{D}$'s $k$ rank drives an inner cascade
that processes one source per step, while the outer cascade's stopping
condition $\diamond: D_{i+1} \equiv D_i$ continues to govern overall
convergence.
At a given $i$, the inner cascade always runs for exactly $|V|$
steps because all sources still read from the original $D_{i, d}$.

\subsection{Sequentializing the Gather step}\label{appendix:bf-sequentialize}

Step~3 is synchronous.
Notice that in Cascade~\ref{cascade:bf-step3}, each source $k$ within an outer
iteration $i$ gathers from the same original distance tensor:
$\tilde{D}_{i, 0, k} = D_{i, k}$. So source $k$'s gather has the form:
\begin{equation}\label{sync}
N_{i, k, d} = G_{k, d} \cdot \tilde{D}_{i, 0, k} :: \bigwedge +(\cap).
\end{equation}
No source sees the running distance updates
contributed by sources $0, 1, \ldots, k-1$. We term this a \emph{synchronous}
cascade: every source operates on a frozen snapshot of $D$ taken at the start
of outer iteration $i$.

Step~4 (Cascade~\ref{cascade:bf-step4}) breaks this synchrony by routing
source $k$'s gather to $\tilde{D}_{i, k, k}$, the running distance after
sources $0, 1, \ldots, k-1$ have added contributions:
\begin{equation}
N_{i, k, d} = G_{k, d} \cdot \tilde{D}_{i, k, k} :: \bigwedge +(\cap).
\end{equation}
We now show
constructively that this change preserves the value of $D$ at convergence.

\paragraph{At $k = 0$.}
By definition, $\tilde{D}_{i, 0, d} = D_{i, d}$. Source 0's gather is:
\begin{equation}
N_{i, 0, d} = G_{0, d} \cdot \tilde{D}_{i, 0, 0} :: \bigwedge +(\cap).
\end{equation}
Since $\tilde{D}_{i, 0, 0} = D_{i, 0}$, this matches source 0's gather in the
synchronous cascade. The running distance update is then:
\begin{equation}
\tilde{D}_{i, 1, d} = \tilde{D}_{i, 0, d} \cdot N_{i, 0, d}
  :: \bigwedge \min(\cup)
  = D_{i, d} \cdot N_{i, 0, d} :: \bigwedge \min(\cup).
\end{equation}

\paragraph{At $k = 1$.}
Source 1's gather now reads $\tilde{D}_{i, 1, 1}$ rather than $D_{i, 1}$:
\begin{equation}
N_{i, 1, d} = G_{1, d} \cdot \tilde{D}_{i, 1, 1} :: \bigwedge +(\cap).
\end{equation}
Substituting the unrolled $\tilde{D}_{i, 1, 1}$:
\begin{equation}
N_{i, 1, d} = G_{1, d} \cdot^1
  \bigl(D_{i, s:s=1} \cdot^2 N_{i, 0, s:s=1}\bigr)_{i, s:s=1}
  :: \bigwedge^1 +(\cap)\; \bigwedge^2 \min(\cup).
\end{equation}
Distributing $G_{1, d} \cdot^1$ over the inner $\bigwedge^2 \min(\cup)$ via
the $(\min, +)$ semiring:
\begin{equation}
N_{i, 1, d} = \bigl(G_{1, d} \cdot^1 D_{i, s:s=1}\bigr)_{i, d}
  \cdot^3 \bigl(G_{1, d} \cdot^2 N_{i, 0, s:s=1}\bigr)_{i, d}
  :: \bigwedge^1 +(\cap)\; \bigwedge^2 +(\cap)\; \bigwedge^3 \min(\cup).
\end{equation}
The first term matches source $k$'s gather in the synchronous cascade. By the
inductive hypothesis, each remaining term $G_{k, d} \cdot N_{i, k', k}$
appends one edge (from $k$ to $d$) to a candidate ending at $k$.
The final $N$ term is the minimum across all candidate options.

The first term matches source 1's gather in the synchronous cascade (reading
the original $D_{i, 1}$). The second term is a new candidate: a two-edge path
from vertex 0 through vertex 1 to $d$, expressed as the gather
$G_{1, d} \cdot N_{i, 0, 1}$.

The new $N_{i, 1, d}$ is therefore the $\min$ of the synchronous
$N_{i, 1, d}$ (Equation~\eqref{sync}) and one additional candidate.
Since $\min(a, b) \leq a$ pointwise, the new $N$ is pointwise less than or
equal to the synchronous $N$.

\paragraph{Generalizing at $k = 2$.}
Source 2's gather:
\begin{equation}
N_{i, 2, d} = G_{2, d} \cdot \tilde{D}_{i, 2, 2} :: \bigwedge +(\cap).
\end{equation}
Unrolling $\tilde{D}_{i, 2, 2}$ via the running-distance recurrence applied
twice:
\begin{equation}
\tilde{D}_{i, 2, s:s=2} =
  \bigl(D_{i, s:s=2} \cdot^1 N_{i, 0, s:s=2}\bigr)_{i, s:s=2}
  \cdot^2 N_{i, 1, s:s=2}
  :: \bigwedge^1 \min(\cup)\; \bigwedge^2 \min(\cup).
\end{equation}
Substituting and distributing $G_{2, d} \cdot$ over the $\min$ Einsums:
\begin{multline}
N_{i, 2, d} =
  \bigl(G_{2, d} \cdot^1 D_{i, s:s=2}\bigr)_{i, d} \\
  \cdot^4 \bigl(G_{2, d} \cdot^2 N_{i, 0, s:s=2}\bigr)_{i, d}
  \cdot^5 \bigl(G_{2, d} \cdot^3 N_{i, 1, s:s=2}\bigr)_{i, d} \\
  :: \bigwedge^1 +(\cap)\; \bigwedge^2 +(\cap)\; \bigwedge^3 +(\cap)\;
     \bigwedge^4 \min(\cup)\; \bigwedge^5 \min(\cup).
\end{multline}
Each of these three terms is itself an EDGE expression in $\bigwedge +(\cap)$
and $\bigwedge \min(\cup)$ over $D_{i, \cdot}$ and $G_{\cdot, \cdot}$. The
first matches source 2's gather in the synchronous cascade; the second and
third are new candidates that chain through sources 0 and 1 within this $i$
iteration.

\paragraph{Inductive generalization.}
For arbitrary $k$, $\tilde{D}_{i, k, k}$ unrolls into a chain of
$\bigwedge \min(\cup)$ Einsums combining $D_{i, k}$ with $N_{i, k', k}$ for
each $k' < k$. Source $k$'s gather is:
\begin{equation}
N_{i, k, d} = G_{k, d} \cdot \tilde{D}_{i, k, k} :: \bigwedge +(\cap).
\end{equation}
Distributing $G_{k, d} \cdot$ over the $\bigwedge \min(\cup)$ chain via the
$(\min, +)$ semiring at the EDGE level:
\begin{multline}
N_{i, k, d} =
  \bigl(G_{k, d} \cdot^1 D_{i, s:s=k}\bigr)_{i, d} \\
  \cdot^{k+2} \bigl(G_{k, d} \cdot^2 N_{i, 0, s:s=k}\bigr)_{i, d}
  \cdot^{k+3} \ldots
  \cdot^{2k+1} \bigl(G_{k, d} \cdot^{k+1} N_{i, k-1, s:s=k}\bigr)_{i, d} \\
  :: \bigwedge^1 +(\cap)\; \bigwedge^2 +(\cap) \ldots \bigwedge^{k+1} +(\cap)\;
     \bigwedge^{k+2} \min(\cup) \ldots \bigwedge^{2k+1} \min(\cup).
\end{multline}
The first term matches source $k$'s gather in the synchronous cascade. By the
inductive hypothesis, each remaining term $G_{k, d} \cdot N_{i, k', k}$
appends one edge (from $k$ to $d$) to a candidate ending at $k$.
The final $N$ term is the minimum across all candidate options.

\paragraph{Same result at convergence.}
Every value Step~4's cascade computes is expressible as an EDGE expression in
$\bigwedge +(\cap)$ and $\bigwedge \min(\cup)$ over $D_{i}$ and
$G$. So is every value the synchronous cascade computes.
Step~4 simply discovers more such expressions per $i$ iteration.
The stopping condition $\diamond: D_{i+1} \equiv D_i$ holds when no candidate
improves any distance. For the synchronous and sequentialized cascades in
Steps~3 and~4, this stopping condition is identical. On graphs without
negative cycles, the smallest reachable distance from any combination of
$\bigwedge +(\cap)$ and $\bigwedge \min(\cup)$ Einsums over $D$ and $G$ is
unique, so both cascades reach the same value of $D$ at convergence. The
sequentialized cascade typically converges in fewer $i$ iterations because
each $i$ explores additional candidates of the form $G_{k, d} \cdot N_{i, k', k}$
for $k' < k$ that the synchronous cascade defers to subsequent outer $i$
iterations.

\paragraph{Negative cycles.}
On graphs with negative cycles, neither cascade converges: distances along the
cycle keep decreasing each iteration, so $D_{i+1} \equiv D_i$ never holds and
the stopping condition is never met. Both the synchronous and sequentialized
cascades inherit this behavior from the underlying Bellman-Ford relaxation.

\paragraph{From the sequentialized cascade to SPFA and Dijkstra.}
The SPFA algorithm collapses $k$ into $i$ such that on each $i$, one source from
the queue $Q$ is selected and processed. The queue tracks exactly those sources
whose outgoing edges may still produce improvements. Thus, SPFA terminates when
$Q$ is empty rather than when a single iteration leaves $D$ unchanged.
Dijkstra further restricts this selector: it selects the vertex in $Q$ with
minimum current distance, and it is correct under the constraint that all edge
weights are non-negative.
\clearpage
\end{document}